\documentclass{aa}

\usepackage[varg]{txfonts}
\usepackage{graphicx}
\usepackage{epsfig}
\usepackage{color}
\usepackage{txfonts}
\usepackage{lscape}
\usepackage{natbib}
\usepackage{longtable}
\bibpunct{(}{)}{;}{a}{}{,} % to follow the AandA style
\newcommand{\am}[2]{$#1'\,\hspace{-1.7mm}.\hspace{.0mm}#2$}

\newcommand{\ad}[2]{$#1{^{\circ}}\,\hspace{-1.7mm}.\hspace{.0mm}#2$}
\newcommand{\HI}{\mbox{H\,{\sc i}}}
\newcommand{\HIbf}{\mbox{H\hspace{0.155 em}{\footnotesize \bf I}}}

\newcommand{\HIsl}{\mbox{H\hspace{0.155 em}{\footnotesize \sl I}}}

\newcommand{\MHI}{$M_{\rm HI}$}
\newcommand{\Msun}{$M_\odot$}
\newcommand{\kms}{\mbox{km\,s$^{-1}$}}
\newcommand{\nan}{Nan\c{c}ay}
\newcommand{\FHI}{\mbox{$F_{\rm HI}$}}
\newcommand{\Jykms}{\mbox{Jy~km~s$^{-1}$}}

\newcommand{\Msunpyear}{\mbox{$M_{\odot}$ yr$^{-1}$}}
\newcommand{\Speak}{\mbox{$S_{\rm peak}$}}

\newcommand{\VHI}{\mbox{$V_{\rm HI}$}}
\newcommand{\VLSR}{\mbox{$V_{\rm LSR}$}}
\newcommand{\Vopt}{\mbox{$V_{\rm opt}$}}
\newcommand{\Vexp}{\mbox{$V_{\rm exp}$}}

\newcommand{\Teff}{\mbox{$T_{\rm eff}$}}

\begin{document}

\titlerunning{NRT \HI\ line observations of 290 evolved stars I. Data}
\authorrunning{E. G\'erard, et al. }

\title{\HI\ line observations of 290 evolved stars made with the \\ \nan\ Radio Telescope} 
\subtitle{I. Data}

\author{E. G\'erard\inst{1,2}, W. van Driel\inst{1,2}, L. D. Matthews\inst{3}, T. Le Bertre\inst{4}, J.-M. Martin\inst{1,2},
N. Q. Ri{\^e}u\inst{4}\thanks{Deceased}}  
\institute{GEPI, Observatoire de Paris - PSL, 5 place Jules Janssen, 92190 Meudon, France
  \email{eric.gerard@obspm.fr}
	\and 
Observatoire Radioastronomique de \nan, Observatoire de Paris - PSL, Universit\'e d'Orl\'eans, 18330 \nan, France
\and 
Massachusetts Institute of Technology Haystack Observatory, 99 Millstone Road, Westford, MA 01886, USA
\and 
LERMA, Observatoire de Paris - PSL, 61 av. de l’Observatoire, 75014 Paris, France
}

\date{submitted September 3, 2024; accepted October 15, 2024}

\abstract
{We present a compendium of \HI\ 21-cm line observations of circumstellar 
envelopes (CSEs) of 290 evolved stars, mostly ($\sim$84\%) on the asymptotic giant branch 
(AGB), made with the 100m-class, single-dish \nan\ Radio Telescope.
The observational and data reduction procedures were optimised for separating 
genuine CSE \HI\ emission from surrounding Galactic line features. 
For most targets (254) the results have not been previously published.
Clear detections were made of 34 objects, for 33 of which the total \HI\ flux 
and the size of the CSE could be determined.
Possible detections were made of 21 objects, and upper limits could be 
determined for 95 undetected targets, while for 140 objects confusion from Galactic 
\HI\ emission along the line-of-sight precluded meaningful upper limits.
The collective results of this survey can provide guidance on detectability of 
circumstellar \HI\ gas for future mapping and imaging studies.
}

           
\maketitle

\section{Introduction} \label{sec:Intro}  % Sect. 1
Observations in the 21-cm \HI\ line provide a powerful tool for the study
of various properties of the circumstellar envelopes (CSEs) of 
mass-losing Asymptotic Giant Branch (AGB) stars and other types of evolved stars, 
which cannot be determined through other spectral lines. 
For example, such observations can trace gas beyond the CO dissociation radius, 
probe the interface between the star and the interstellar medium (ISM), 
and provide independent estimates of the total masses and ages of CSEs.

Once stars leave the main sequence and reach the giant branch, they start losing mass 
(in the form of gas and dust) through stellar winds
(see, e.g. \citealt{Marengo15, Hoefner18, Decin21, Matthews24}).
On the AGB, mass-loss rates range from 
several 10$^{-8}$ \Msunpyear\ up to 10$^{-4}$ \Msunpyear, with the bulk of the mass loss 
occuring in the form of hydrogen. The stars become surrounded by CSEs, whose gaseous 
constituents can be studied using atomic and molecular spectral lines. 
The winds are accelerated through the coupling between grains and gas, from the 
stellar atmosphere (at a few stellar radii) up to the outer CSE (at radii of thousands 
of AU) where the wind starts interacting with the interstellar medium (ISM), often 
forming bow shocks in the case where the star is moving supersonically relative to the 
ambient medium.

Under the influence of the interstellar ultraviolet (UV) radiation field, the 
chemical constituents of the CSE are progressively dissociated. 
One of the best studied is the CO molecule, which has strong rotational lines at 
low excitation temperature.
However, as it is photodissociated at small radii (see, e.g. \citealt{Saberi19}) 
it is
difficult to accurately derive a total mass for all constituents in the CSE based 
only on CO line measurements.

Thus, H$_2$ (through the CO lines) and \HI\ are essential constituents to track 
throughout the CSE (see, e.g. \citealt{Glassgold1983}). 
Whereas H$_2$ has no dipole moment and is difficult to observe through its 
vibrational transitions corresponding to high excitation temperatures, \HI\ may be either 
nascent in a hot atmosphere ($T$ $>$ 2500 K) or produced by the photodissociation 
of H$_2$ at large radii. 

This paper presents an overview of the results of \HI\ 21-cm line observations 
of evolved stars made with the 4$'$$\times$22$'$ beam of the 100m-class 
single-dish \nan\ Radio Telescope (NRT). 
The results represent an investment of $\sim$5000 hours of telescope time 
since 1992. 
A total of 290 stars were observed, and for most (254, or 88\%), the results 
have not been previously reported. These include upper limits for 
non-detections, as well as identification of cases for which any circumstellar 
\HI\ line emission was completely confused with Galactic \HI\ clouds along the 
line-of-sight and where no conclusion can be drawn about the presence of 
\HI\ in the CSE. This information is nonetheless useful for guiding 
selection of targets for future surveys of circumstellar \HI\ at higher
angular resolution.
Results of NRT \HI\ observations of 36 stars from our sample have been previously published in 
\citet{Lebertre2001, Gerard2003, Lebertre2004, Gerard2006, Gardan2006, Libert2007, 
Libert2008, Matthews2008, Libert2010a, Libert2010b, Gerard2011, Lebertre2012}.

Follow-up \HI\ imaging and mapping observations made with the Very Large Array (VLA)
and the Green Bank Telescope of 16 AGB stars and red supergiants initially observed 
with the NRT were reported in 
\citet{Matthews2007, Matthews2008, Lebertre2012, Matthews2013, Matthews2015, 
Hoai2014} -- see also Tables~\ref{table:cleardetsHI}, \ref{table:possdetsHI},
and \ref{table:confused}.
In VLA imaging studies with a resolution of $\sim$1$'$ \HI\ was detected in 13 cases. 
The morphologies of the \HI\ distributions are quite varied, 
showing detached shells around the star (IRC +10216, Y CVn, V1942 Sgr, Y UMa, $\alpha$ Ori), 
blobs of \HI\ offset from the AGB (EP Aqr, R Cas), 
head-tail structures ($o$ Cet, RS Cnc, X Her, TX Psc),
a shell-tail structure (RX Lep), and one object (R Peg) with a peculiar ``horseshoe''-shaped \HI\ morphology.
The three stars that were not detected in \HI\ with the VLA are RAFGL 3099, R Aqr and IK Tau.
Furthermore, no \HI\ was detected in W Hya at the VLA by \citet{Ha93}.

A challenge for observing circumstellar \HI\ is that the detection can be severely 
hindered, if not made downright impossible, owing to the presence of ubiquitous  
Galactic \HI\ emission along the line-of-sight.
This applies in particular to observations made with single-dish radio
telescopes that rely on emission-free off-source reference position spectra
(see, e.g. \citealt{Matthews2013} for further details).
Confusion by ubiquitous, and often strong, Galactic \HI\ signals has been a major 
issue with the NRT data throughout the project (see Sections \ref{sec:Observe} and 
\ref{sec:Discuss}). Although the relatively good 4$'$ spatial resolution of the NRT 
in right ascension was a favourable factor in dealing with this confusion, 
in declination the beam size is much larger, $\ge 22'$.

The present catalogue includes a new assessment of all NRT spectra in terms of confusion 
by Galactic \HI, including for previously published NRT results. This is based on new 
experience gained, including the analysis of additional NRT \HI\ observations 
that were made of many sources with a wider range of east-west off-source positions 
as well as of positions 11$'$ (0.5 N-S HPBW) north and south of the star, 
and the use of more recently published radial velocities. % MOD

Our ultimate aim is to contribute to the understanding of the fate of 
the major constituent ejected by evolved stars during their evolution, i.e., hydrogen. 

In the present paper we want to provide guidance for other 
studies of evolved stars in terms of detectability of \HI\ in CSEs based on our 
single-dish telescope observations, in particular for follow-up imaging and mapping 
studies made at (much) higher spatial resolution with interferometers or single-dish 
telescopes with smaller beam sizes.
For a number of undetected stars we obtained sensitive upper limits (in some cases based 
on up to 100 hours of integration) that provide robust constraints on the probability 
of detection with other telescopes.

Unless otherwise indicated, all radial velocities are in the local standard
of rest (LSR) reference frame, calculated based on the conversion from heliocentric to
LSR velocities given in \citet{Ke99} (see Section \ref{sec:Results} for details).

This paper is organised as follows. In Section~\ref{sec:Sample} the selection 
of the observed AGB stars is described and Section~\ref{sec:Observe} discusses the
observations and data reduction. Results are presented in Section~\ref{sec:Results}
and further discussed in Section~\ref{sec:Discuss}. Conclusions are summarized in 
Section~\ref{sec:Conclusions}. Comments on the determination of \HI\ profiles 
are given in the Appendix for selected sources.

\section{Sample selection} \label{sec:Sample}  % Sect. 2
Our targets were primarily chosen from lists of well-studied, nearby objects 
classified as evolved stars and related objects. 
Our final sample of 290 evolved stars consists mostly (for $\sim$84\%) of AGB stars,
but other types were observed as well, to broaden the scope of our studies
of CSEs in \HI. 
We did not apply strict criteria for the selection of stars to be observed; 
these changed as the project developed over the past three decades. 

The names of the 27 stars which are clearly not AGB are flagged with an "$n$" in the Tables.
They include 8 planetary nebulae, 6 high proper motion stars, 5 post-AGB stars and 4 red supergiants.
The 19 cases for which we consider their classification as AGB dubious are flagged with a 
"$d$". 

The stars' variability types we used are primarily based on the General Catalogue 
of Variable Stars, GCVS (\citealt{gcvs}); see Section \ref{sec:Results} for further details.
In terms of their variability classification, the main categories are:
Mira (119 = 41\%), SRb (70 = 24\%), Lb (28 = 10\%), SRc (12 =  4\%), and SRa (7 = 2\%).

The observed stars span a wide range of mass-loss rates, of three orders of magnitude, 
from 2 10$^{-8}$ to 2 10$^{-5}$ \Msunpyear, as derived from CO line observations.

All targets were selected without avoiding either low Galactic latitudes or 
small radial LSR velocities, where confusion by Galactic \HI\ clouds is the most likely: 
57 targets (20\%) have $|$b$|$ $<$ 10$^{\circ}$, and 73 (25\%) have $|$\VLSR$|$ $<$ 10 \kms.
Many are located at right ascensions outside the Galactic Centre range, where the overall
time pressure on the meridian-type NRT is lower and where (very) long integrations, 
of up to 100 hours per source, could therefore be made.
The radial velocities of most (76\%) of our targets were measured in radio spectral lines 
(CO, OH, H$_2$O, SiO), the others in optical lines.

\section{Observations and data reduction} \label{sec:Observe}  % Sect. 3
The 100m-class Nan\c{c}ay Radio Telescope is a Kraus/Ohio State meridian transit-type instrument.
It consists of a fixed spherical mirror (300~m long and 35~m high), a tiltable flat mirror 
(200$\times$40~m), and a focal carriage moving along a curved rail track. 
Its collecting area is about 7000~m$^{2}$.
Due to its elongated geometry some of the instrument's characteristics, 
such as gain and N-S beam size, depend on the observed declination. 
It can observe down to a declination of $-39^{\circ}$.
At 21 cm wavelength its half-power beam width (HPBW) is always 4$'$ in right
ascension, whereas in declination it increases from 22$'$ at 
$\delta$<20$^{\circ}$ to 34$'$ at $\delta$ = 75$^{\circ}$. 
Sources on the celestial equator can be tracked for about 60 minutes.

Our first observations were made in 1992 and 1993 of 103 sources, 67 of which
were reobserved a decade later, after a major renovation of the telescope, 
and the last observations were made in 2017.
More than 5000 hours of telescope time were used in total for the project. 
Total integration times per source varied considerably, from one hour to 100+ hours.
Telescope time allocation permitting, we spent more time on sources which
were faint (as was often the case with Miras) or/and whose large angular sizes 
required observations with multiple off-source pointing positions 
(see Sect. \ref{sec:sizes}). 

In 1992 and 1993 data were acquired using a 1024-channel autocorrelator, with half the
frequency channels used for observations in the left- or right-handed
circular polarization, respectively. The total bandwidth was 0.4 MHz, i.e., 84.4 \kms\
at the rest frequency of 1420.40 MHz. After boxcar smoothing in velocity and averaging 
of both polarizations, the final spectrum consisted of 128 channels with a velocity resolution 
of 0.66 \kms. The system temperature was 37~K at zero declination.
The observing procedure consisted of position switching using only two off-source positions,  
located one HPBW east and one HPBW west of the star.

In the late 1990s the NRT underwent a major renovation, 
including the installation of corrugated horns in an optimized dual-reflector
offset Gregorian mirror system, which increased its sensitivity by a factor of 2.2 
\citep{vanDriel1997,Granet1999}. Starting in 2001, observations were made with a new 
8192 channel autocorrelator, split into 4 banks of 2048 channels each for simultaneous 
observations in two orthogonal linear polarisations (PA = 0$^{\circ}$ and 90$^{\circ}$), 
and in left- and right-hand circular polarisations. 
The \HI\ emission is thermal and unpolarized.
For the observations presented here the Stokes total intensity parameter $I$ 
was formed by adding the two circular polarisations.
In order to identify possible receiver instabilities or radio frequency interference (RFI), 
we also formed the Stokes $I$ by adding the two orthogonal linear polarisations.
In practice, our observations were not affected by RFI.
The terrain surrounding the NRT site acts as a natural barrier against unwanted 
terrestrial radio emissions. 

The velocity coverage was 165 \kms\ and the maximum velocity resolution 0.08 \kms.
However, almost all spectra presented here were later smoothed to a resolution of 0.32 \kms.
The typical system temperature was 30~K near the celestial equator.
Linear baselines were fitted to the averaged spectra, and subtracted.

Our searches for circumstellar \HI\ emission were centred on the 
most accurate published systemic velocity of the target available at the time. 
In some cases the velocity we used for our current analysis is slightly different 
owing to the availability of updated information (see Section \ref{sec:Results} for details).
However, in no cases has this significantly impacted the interpretation of the data.

For observations made from 2001 onward in position switching mode 
we used more E-W off-source positions than we did in 1992-1993, 
and we also observed positions to the north and south of the star, 
to investigate, respectively, E-W and N-S source sizes.
Our standard procedure consisted of alternating between an on-source position and two
off-source positions, located symmetrically towards the east and west of the on-source position 
(see Section \ref{sec:Confusion} for further details). 

In all position-switching observations the same amount of integration time 
was used pointing towards each of the three positions observed: on-source, 
off-source east, and off-source west.  

Most sources were also observed in frequency-switching mode 
while pointing at the stellar position, 
which allows the accurate measurement of the background 
brightness temperature in the expected \HI\ velocity range of the CSE,
and to estimate the confusion level due to the telescope's sidelobes
(see Section \ref{sec:Confusion}).

The effective bandwidth was 1.53 MHz (323 \kms),  centred on the stellar velocity. 
The receiver bandwidth was set to 1.56 MHz (329 \kms) and a frequency switch of 
+1.53 MHz ($-$323 \kms) was used, as this corresponds to exactly three times the 0.51 MHz frequency 
of the ripple in the spectral baseline that is often caused by specular reflection 
between telescope structures (see, e.g. \citealt{Wilson09}). Choosing this offset 
thus cancels out the baseline ripple. Furthermore, the negative offset in velocity 
avoids contamination by possible extragalactic \HI\ emission.

NRT flux scale calibration is based on long-term, regular monitoring of radio sources 
by NRT staff. It consists of observations of 20 continuum calibration sources every 3 months, 
monthly observations of the primary calibrator 3C 123 and weekly observations of the continuum 
source 3C 161 and the line source W12. Calibrator flux densities are based on Baars 
et al. (1977). The scatter in the measurements on which the absolute flux density scale 
is based is $\sim$5\% at 1410~MHz.

All data were reduced using the standard NRT software packages NAPS and SIR, 
developed by NRT staff and telescope users.

\subsection{Dealing with Galactic \HIsl\ confusion} \label{sec:Confusion}  % 3.1
Detecting genuine CSE \HI\ emission among the ubiquitous Galactic clouds can be
a major challenge in terms of both observing procedure and data analysis.
This applies in particular for sources near the Galactic plane, and for 
single-dish radio telescopes like the NRT whose N-S beam size is 
relatively large ($\ge22'$). 
The \HI\ peak flux densities we measured within the central NRT beam for 
clear detections range between 0.9 and 0.007 Jy.
Since the brightness of the Galactic \HI\ emission typically spans a range between 
1 and 100 Jy, the maximum source/background contrast ratio is expected to vary 
from about 0.5 to 0.005 in practice -- the latter close to the Galactic plane. 

Please note that the \HI\ spectra presented here use data from only a single
on-source position for each target, centred on the coordinates of the star.
Previous studies have shown our use of off-source positions symmetrically 
to the east and to the west of the star to be an effective scheme for 
separating CSE emission from Galactic line signals in NRT spectra 
(e.g., \citealt{Gerard2006}).

This procedure works best when the intensity of the Galactic 
background has a linear variation with position in the region surrounding the target source. 
However, if its variation is quadratic, or of even higher order, spurious spectral 
features may appear. However, as these generally grow rapidly with increasing separation 
between the two off-source positions, this allows us to identify them.

\begin{figure*}[ht]  % Fig. 1
  \centering
  \includegraphics[width=17cm]{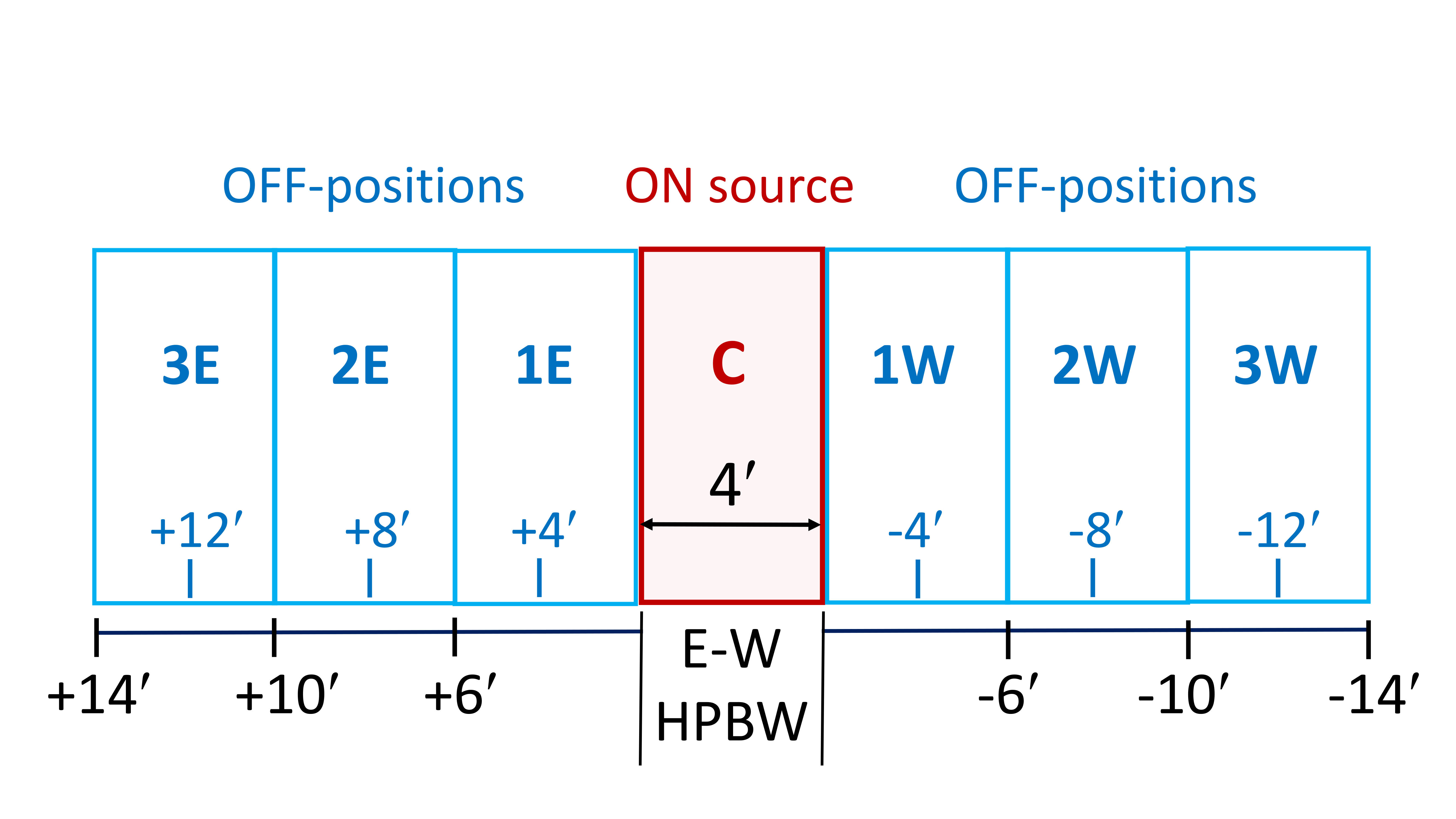}
  \caption{Diagram illustrating examples of on- and off-source NRT beam positions
which were used to estimate the E-W diameters, and peak- and total \HI\ profiles 
of the observed sources (see \ref{sec:sizes}). 
The east-west HPBW is 4$'$, but please note that in reality the NRT beam is 
relatively more elongated in the north-south direction ($\geq$22$'$) than
is shown here.}
\label{fig:throws}
\end{figure*}

The principles of our E-W on/off-source position scheme are illustrated in Fig. \ref{fig:throws}. 
We will refer to the angular separation between the on-source and an off-source position 
as a ``throw amplitude'', and use the notation $CnEW$ to indicate spectra obtained using 
a throw amplitude of $n$ HPBWs to the E and W of the target position
(denoted as $C$ in the Figure).

\begin{figure*}[ht]  % Fig. 2
  \centering
  \includegraphics[width=18.35cm]{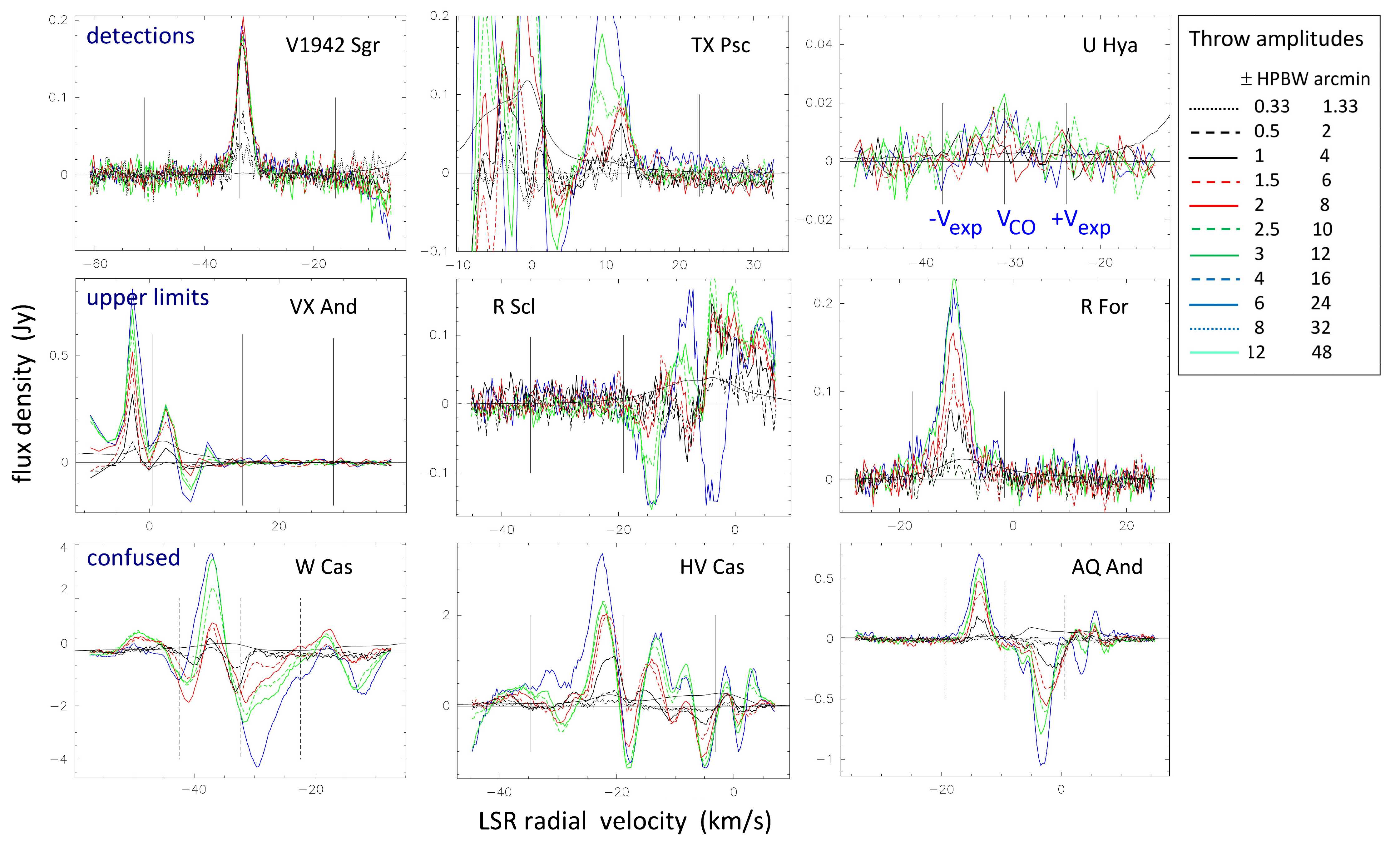}
 \caption{Examples of NRT \HI\ line proﬁles, showing flux density, $S_{\rm HI}$, in Jy as a function 
of radial velocity in the LSR frame, \VLSR, in \kms. 
The three vertical black lines indicate, respectively, the mean velocity 
and $\pm$ the expansion velocity of the CO line of each source.
The sources shown here were classified as either clear \HI\ detections (upper row), 
upper limits (middle row) or confused (lowest row).
The spectra were obtained in position-switching mode for different 
E-W offset distances between the two off-source positions (throw amplitudes) located 
symmetrically around the on-source position. 
For the match between throw amplitudes and types of plotted lines used, see the legend
on the right-hand side. Throws are given in both units of the telescope's E-W beam size
(4$'$) and in arcminutes.
The flat black horizontal lines show the 0 Jy flux density level.
Also shown (as full black lines, with a shape distinctly different from the 
position-switching profiles) are the on-source \HI\ profiles obtained in 
frequency-switching mode, with flux densities divided by a factor of 100.
The latter were used as a benchmark for the overall confusion level 
throughout the spectra induced by the telescope's near side lobes.
For detections an increase of the total flux with throw amplitude indicates that the source 
is extended and the throw amplitude where the flux converges can be used to estimate the
source diameter (see Sect. \ref{sec:Observe}).
}
\label{fig:offsets}
\end{figure*}

We have used E-W throw amplitudes ranging from \am{1}{3} to 48$'$ (i.e., from 0.3 to 12 HPBWs), 
in order to cover a large range of potential CSE sizes.
Examples of spectra obtained with a broad range of E-W throw amplitudes are shown in 
Fig. \ref{fig:offsets} for selected sources that were classified as either clear detections, 
upper limits, or confused. 

The impact on the detectability of CSEs in \HI\ caused by increasing levels of confusion 
due to Galactic lines is illustrated in the examples  shown in Fig. \ref{fig:throws}:
\vspace{-0.5mm}

\begin{enumerate}
\item When the Galactic \HI\ emission is weak, the signal from a genuine CSE detection
increases with throw amplitude and converges once both off-source positions 
lie outside the CSE: V1942 Sgr is a good example. 
\item On the other hand, when the Galactic emission dominates the \HI\ signal continues to 
increase with the throw amplitude: W Cas, HV Cas and AQ And are good examples. 
The stronger the Galactic confusion, the earlier the profiles start diverging when the
throw amplitude is increased. 
We therefore often observed with a very small throw amplitude of \am{1}{3} ($C0.33EW$),
which reduces the confusing \HI\ signal more than it does the genuine CSE line emission. 
\item In intermediate cases (e.g. TX Psc), genuine CSE \HI\ emission is seen
at the CO line velocity (12 \kms), which peaks for a throw amplitude of 
6$'$, while a Galactic emission feature grows rapidly near 10 \kms\ and 
dominates at a throw amplitude of 12$'$. %\\
\end{enumerate}
\vspace{-0.5mm}

The latter example also illustrates the advantage of the high velocity resolution 
used (typically 0.32 \kms) in distinguishing between CSE emission and a confusing,
Galactic feature along the line-of-sight that occurs close to the stellar velocity. 

In the north-south direction the NRT HPBW is $\ge22'$, considerably larger 
than the 4$'$ along the E-W axis. 
For the two reasons mentioned hereafter, we also obtained observations towards  % MOD
positions 11$'$ to the north and to the south of the star. The off-positions 
used for these N-S pointings are in general located one E-W HPBW (4$'$) to the east 
and to the west of them, but for some large sources additional off-positions with 
E-W throws of up to 24$'$ were used.
% , or even (but rarely) a full HPBW away. 

% Like for the E-W observations, for the N-S observations the same amount of integration 
% time was used pointing towards each of the three positions observed: on-position
% (to the N or the S of the star), off-source east, and off-source west.

These observations were made of all clear detections, as well as of 
many other sources, though not for the weakest as 
this would have required a prohibitive amount of additional telescope time.

The two reasons for which the N-S observations were made are to check if
the source is extended in the N-S direction, or if the detection is confused by 
ISM emission along the line-of-sight.

\begin{enumerate}
\item If a source is N-S extended, using only E-W pointings at the declination of the star 
(as we did for the fluxes given in this paper) would underestimate its measured flux. 

For an unresolved source, at both pointings 11$'$ (0.5 N-S HPBW) north and south of the star 
we would expect to measure half of the integrated line flux which was observed when pointing 
towards the star. Using this criterion, very few cases showed indications of being resolved 
in the N-S direction.
However, a difference between the north and south position fluxes could also in principle be due to 
confusing ISM emission (see below).

An indication of the expected N-S sizes could in principle be deduced from the E-W NRT 
source sizes estimated for the 33 clear detections, under the assumption that the sizes 
in both directions are similar (to be examined using VLA images).

Only four of the 33 sources (12\%) have estimated NRT E-W diameters significantly 
larger than the 22$'$ NRT N-S HPBW, whereas nine (27\%) have diameters of 20$'$, 
i.e., similar to the HPBW.

Among the 12 published VLA \HI\ maps of sources in our sample only three show 
similar E-W and N-S sizes, whereas others have quite varied morphologies which often
makes such a comparison impossible. Furthermore, none of the 12 would have been 
resolved by the NRT N-S HPBW (see Sections \ref{sec:Intro} and \ref{sec:VLANRTcomp}).

% Matthews 2013:
% RX Lep     17.5 arcmin
% Y UMa       6.8 
% R Peg     >15 
% Y CVn       9 
% V1942 Sgr   4.6 
% TX Psc      8 

\item The larger N-S beam size introduces a possible risk of confusion with ISM emission 
along the line-of-sight, but it is difficult to ascertain from NRT data where exactly these 
confusing sources are located once their presence is suspected.
Comments on confusion deduced from NRT observations can be found in Appendix
\ref{sec:Notes} for IRC +10216, S CMi, U CMI, TX Psc, and AQ Sgr.
In practice, the much smaller beam size of the VLA is required to locate confusing
sources, as was done in the case of X Her (see Sect. \ref{sec:VLANRTcomp}).

\end{enumerate}

It should be noted that the spectra obtained pointing towards the star 
and towards positions to the north and south of the star  % MOD 
have not been combined for the analysis of the NRT results (see Sect. \ref{sec:sizes}),
as they were made for different purposes, e.g. the former to estimate E-W source sizes,
and the latter to check if sources are N-S resolved.

We assume a CSE \HI\ line signal to be genuine when it is: 
(i) conﬁned within the velocity range \VLSR$\pm$\Vexp\ constrained by the expansion velocity
\Vexp\ of CO or OH line spectra, when available, or otherwise by the average expansion 
velocity of \VLSR$\pm$10 \kms, 
and (ii) symmetric with respect to the stellar radial velocity.

The latter assumption is essential because the contamination by the Galactic emission is often 
strongly asymmetric in velocity, with the stellar velocity lying on a steep slope of the background; 
in such case, only the blue or red wings of the CSE proﬁle are considered.
If no CO or OH line velocities were available we used optically-derived velocities, 
which we determined to be on average the same as the CO/OH values for 
the objects with both kinds of velocities available in our sample: 
for the clear \HI\ detections the mean CO $-$ optical velocity difference 
is -2.2$\pm$3.3 \kms.

Compared to our earlier published results, of particular significance for the interpretation of
the NRT spectra presented here were observations made (1) in position-switching mode
pointing 11$'$ to the north and south of the star's position
and (2) in frequency-switching mode 
(see Sect. \ref{sec:Observe} for further details on both). 

The latter provide a measure of the intensity of the overall \HI\ line signals observed towards the targets, 
independent of their effect on on/off-source position-switching spectra. As shown by the 
black lines in Figure \ref{fig:offsets}, we diminished the flux densities of the 
frequency-switching spectra by a factor 100 before comparing them to the on/off-source position 
switching spectra. The reason behind the application of a 0.01 scaling factor is that it 
indicates the overall confusion level throughout the spectra which was induced by the 
telescope's near side lobes, compared to the peak gain of the main lobe of the telescope. 

Therefore, we considered as unreliable any features in the position switching mode spectra that 
are fainter than the overall confusion level indicated by 0.01 times the frequency-switching spectra 
intensities at the same velocities.

\subsection{Total \HIsl\ profiles and source sizes}  \label{sec:sizes}  % 3.2
As noted above, in the case of a clear, strong detection of \HI\ in a CSE, 
the \HI\ profile converges once the throw amplitude is increased beyond the 
angular size of the CSE.

When we indicate that the spectra of a target {\it converge} for spectrum $CnEW$,
i.e., at a throw amplitude of $n$ HPBWs to the E and W of the target position $C$,
this means the following:
\vspace{-0.5mm}

\begin{enumerate}
\item the maximum line flux is reached in this spectrum; 
\item the line flux remains the same for larger throw amplitudes (i.e., in spectra $Cn$+$1EW$, etc.);
\item there is therefore no CSE \HI\ emission beyond a throw amplitude of $n-1$ HPBWs 
        from the star. Spectrum C$n-1$EW is the last to lie within the CSE, and the distances 
        between the target star and the outer edges of beam positions C$n-1$E and C$n-1$W provide 
        us with an estimate of the CSE diameter (see below). .
\end{enumerate}
\vspace{-0.5mm}

We have therefore defined two types of spectra for the measurement of \HI\ properties 
of CSEs, the ``peak profile" and the ``total profile". 
For practical reasons, the determination of the peak profile is different
for strong detections and for weaker, spatially resolved detections (see below), 
whereas the total profile can only be determined for strong detections,
whose names are flagged with a $^T$ in Tables \ref{table:cleardetsHI} and 
\ref{table:possdetsbasics}.

Please note that for the determination of both peak and the total profiles
we have used only spectra obtained using E-W off-source positions with the same declination 
as the on-source position of the central star, and not the profiles obtained pointing  
north and south of the star (see \ref{sec:Confusion}).  

\paragraph{Strong detections:}
For cases where there is a clear convergence for spectrum $CnEW$, 
obtained with a throw amplitude of $n$ HPBWs, we define this as the peak profile ($P$).
Additionally, we define a total profile ($T$) as the spectrum that contains all \HI\ emission 
within a throw amplitude of $n-1$ beams, which corresponds to the CSE size. 

The relation between the total profile and the peak profile is (cf. Fig. \ref{fig:offsets})

\begin{equation}
T = P + \sum_{j=1}^{n-1} (nE + nW) 
\end{equation}

Using the following relation between the peak profile and the profiles at throw amplitudes 
of $n$ HPBWs 

\begin{equation}
    CnEW = P - (nE+nW)/2  
\end{equation}

\noindent
Equation (1) can be rewritten as follows:

\begin{equation}
T = (2n-1) P - 2 \sum_{j=1}^{n-1} CjEW
\end{equation}
% \vspace{-0.5cm}

\noindent The diameter of the source is then (8$n$-4) arcmin, as the last profile that lies
within the CSE, $Cn$-$1EW$, samples the \HI\ out to a distance of $\pm$(4n-2) arcmin 
from the on-source position (see Fig. \ref{fig:throws}).

To give an example (see Fig. \ref{fig:throws}): if the maximum is reached in profile 
$C3EW$, with a throw amplitude of 12$'$, then $T$ = 5$P$ - 2($C1EW$+$C2EW$), 
and the source diameter is 20$'$, i.e., the maximum separation between outer edges 
of the HPBWs at off-source positions $C2E$ and $C2W$.
\vspace{-0.45cm}

\paragraph{Weaker detections:} When there was no obvious convergence towards a maximum
line flux, we use as peak profile $P$ the average of all observed $CnEW$ spectra 
(where $n$ is an integer), in order to increase the signal-to-noise ratio of the profile.
Here, we averaged the spectra, as adding them increases the noise level and makes 
weak detections invisible.

For an unresolved source this definition of the $P$ profile is the same as given above
for strong detections. However, it should be noted that for spatially resolved weak sources
it provides a lower limit to the aforementioned peak profile $P$. 
Also, for weak detections, which can only be seen in averaged spectra, 
no estimate can be made of the CSE's angular diameter.
For example, diameters could only be determined for 5 of the 21 possible detections 
(see Table \ref{table:possdetsHI}).

Details of the determination of \HI\ profiles for selected sources are given in Appendix 
\ref{sec:Notes}.

\section{Results} \label{sec:Results}  % 4
We have classified the NRT \HI\ line profiles as either clear detections, possible detections,
or upper limits. Distinguishing between clear and possible detections can be a complex process, 
and is necessarily somewhat subjective, as it is based on various factors such as obscuration by (strong) 
Galactic \HI\ lines, low signal-to-noise ratio, offset between \HI\ and CO (or optical) line velocities,
indication of confusion by a Galactic \HI\ source along the line of sight.
% No strict, purely objective criteria were applied to classify each detection as possible rather than possible. 
For each of the possible detections the reasons for which it has been classified as such  
are listed in Appendix \ref{sec:Notes}. 

Basic, non-\HI\ properties of the objects in our sample are given in 
Table \ref{table:cleardetsbasics} for those with clear detections of their circumstellar \HI\ emission, 
in Table \ref{table:possdetsbasics} for possible detections, 
in Table \ref{table:upperlimits} for objects with upper limits to their \HI\ 
non-detections, and in Table \ref{table:confused} for those whose CSE spectra are 
completely obscured by Galactic lines.

\begin{figure*}[ht]  % Fig. 4
\includegraphics[width=4.75cm]{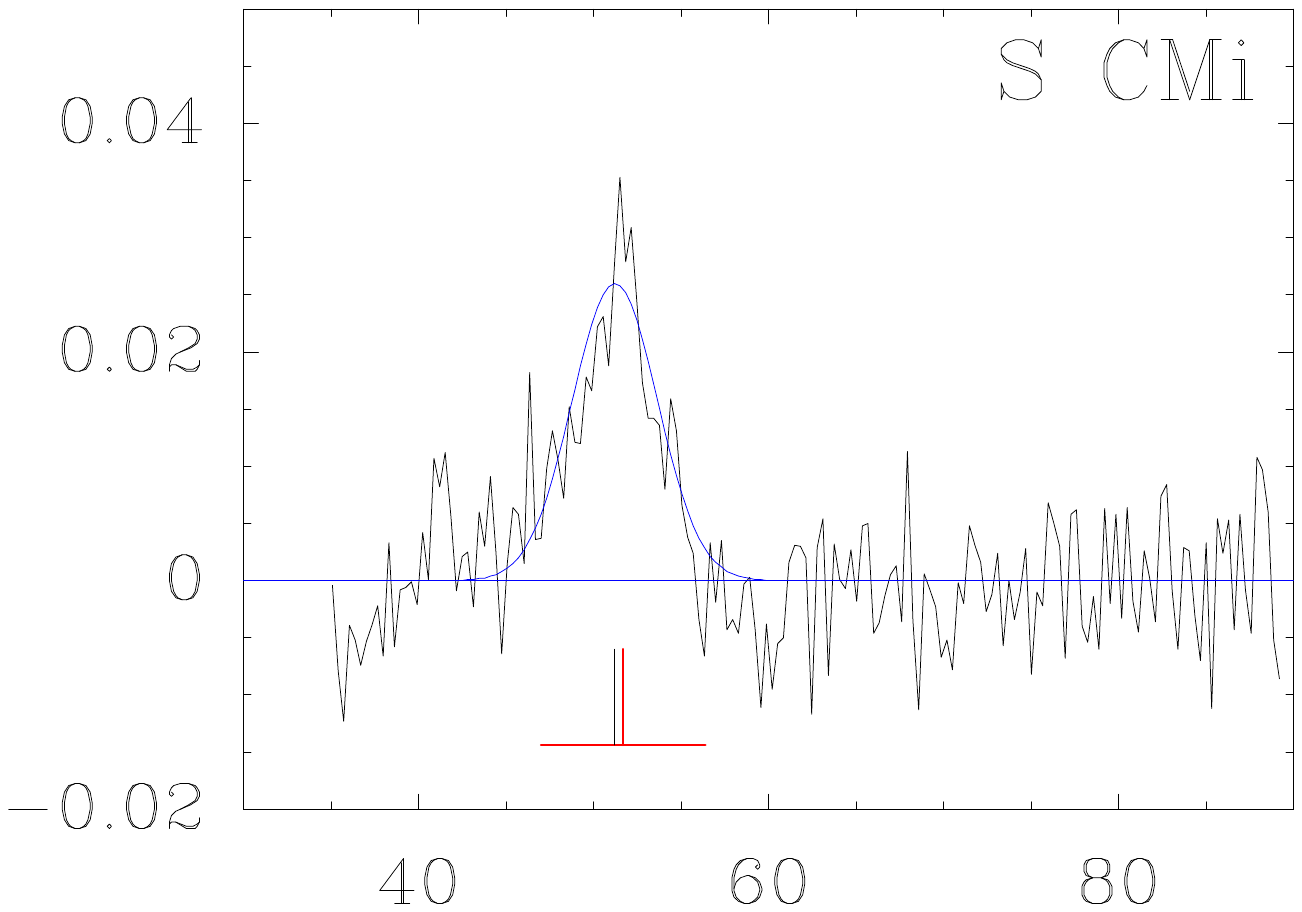}\hspace{-0.7cm}
\includegraphics[width=4.75cm]{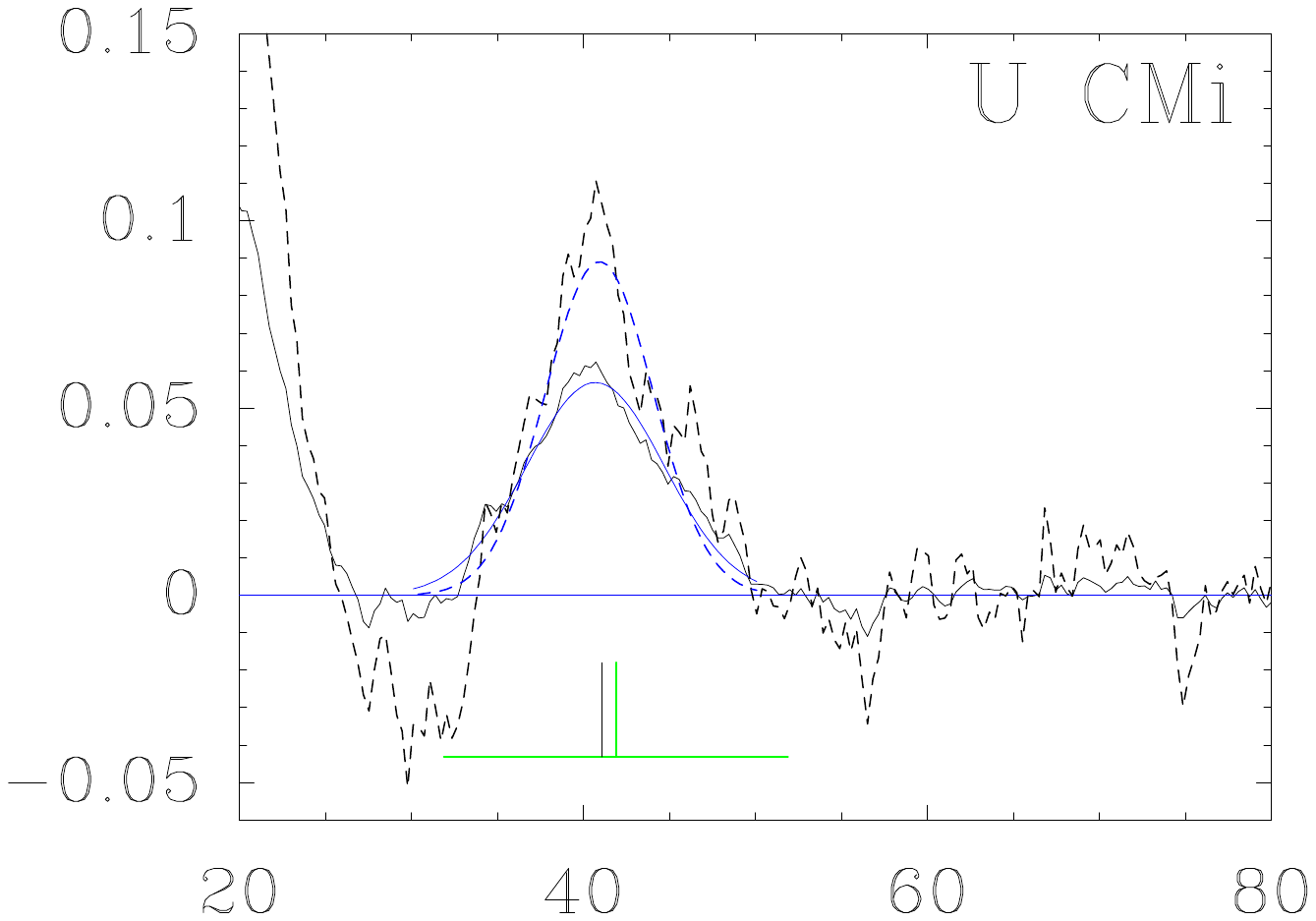}\hspace{-0.7cm}
\includegraphics[width=4.75cm]{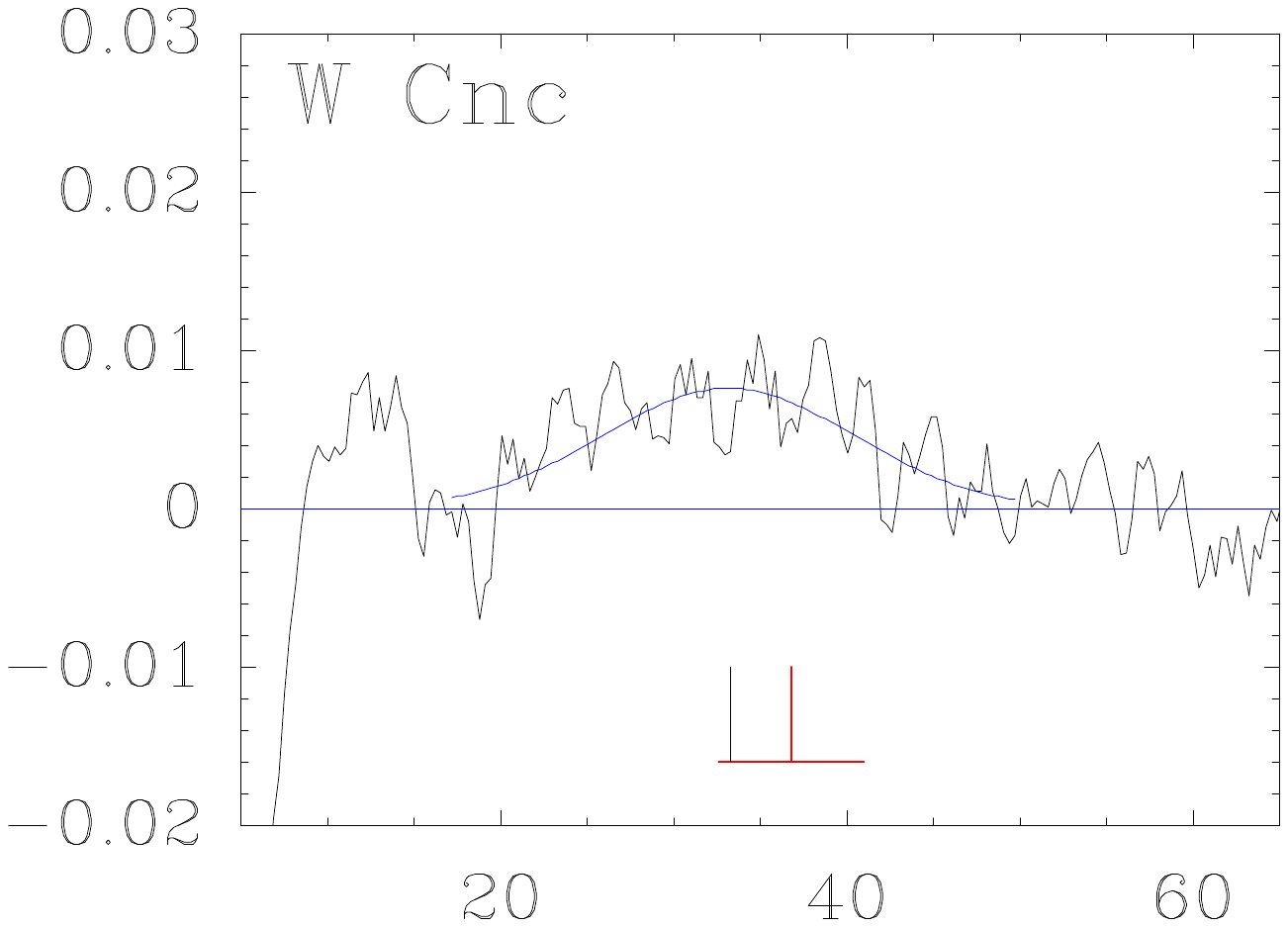}\hspace{-0.7cm}
\includegraphics[width=4.75cm]{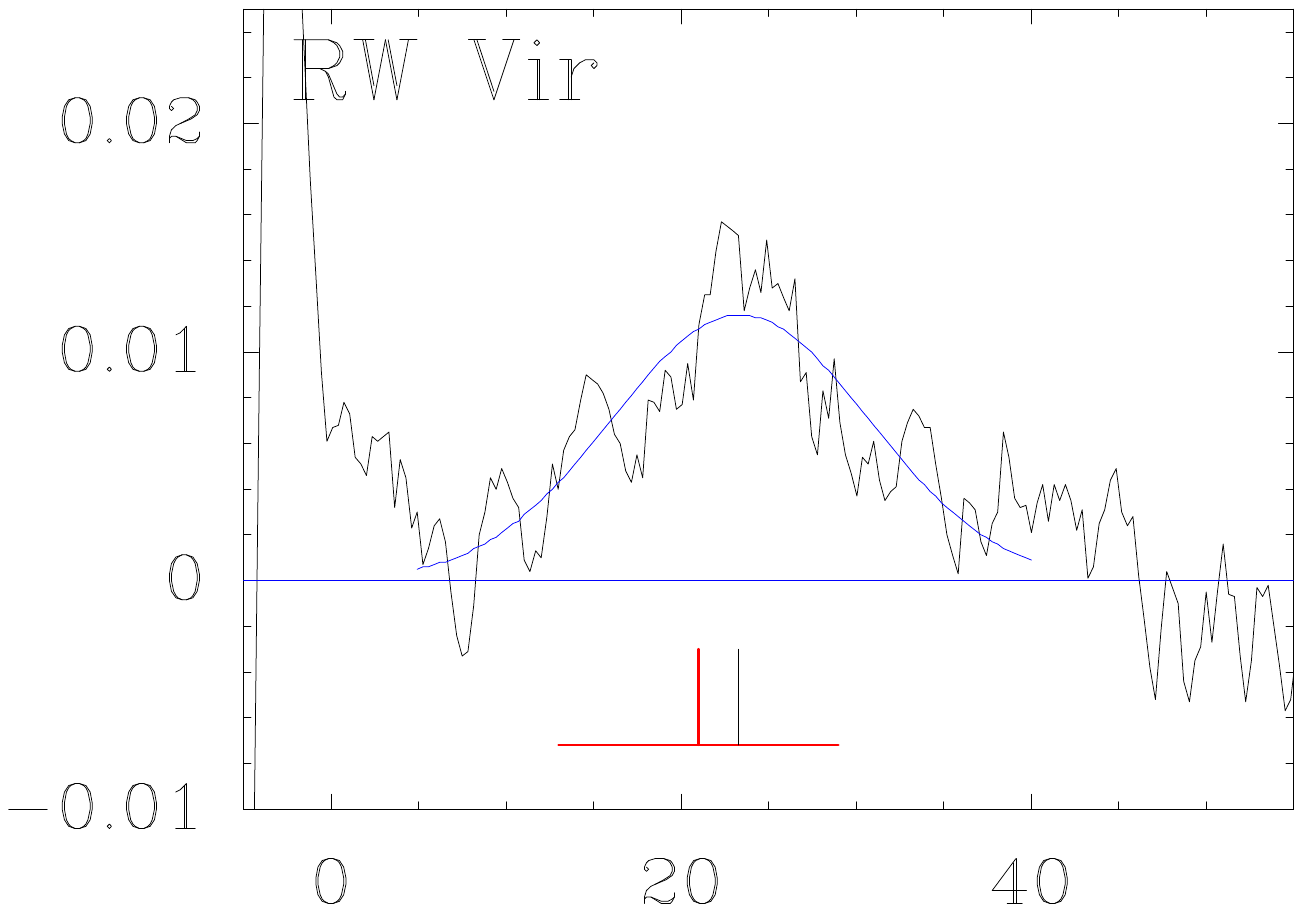}
\\ \vspace{-0.35cm}
\includegraphics[width=4.75cm]{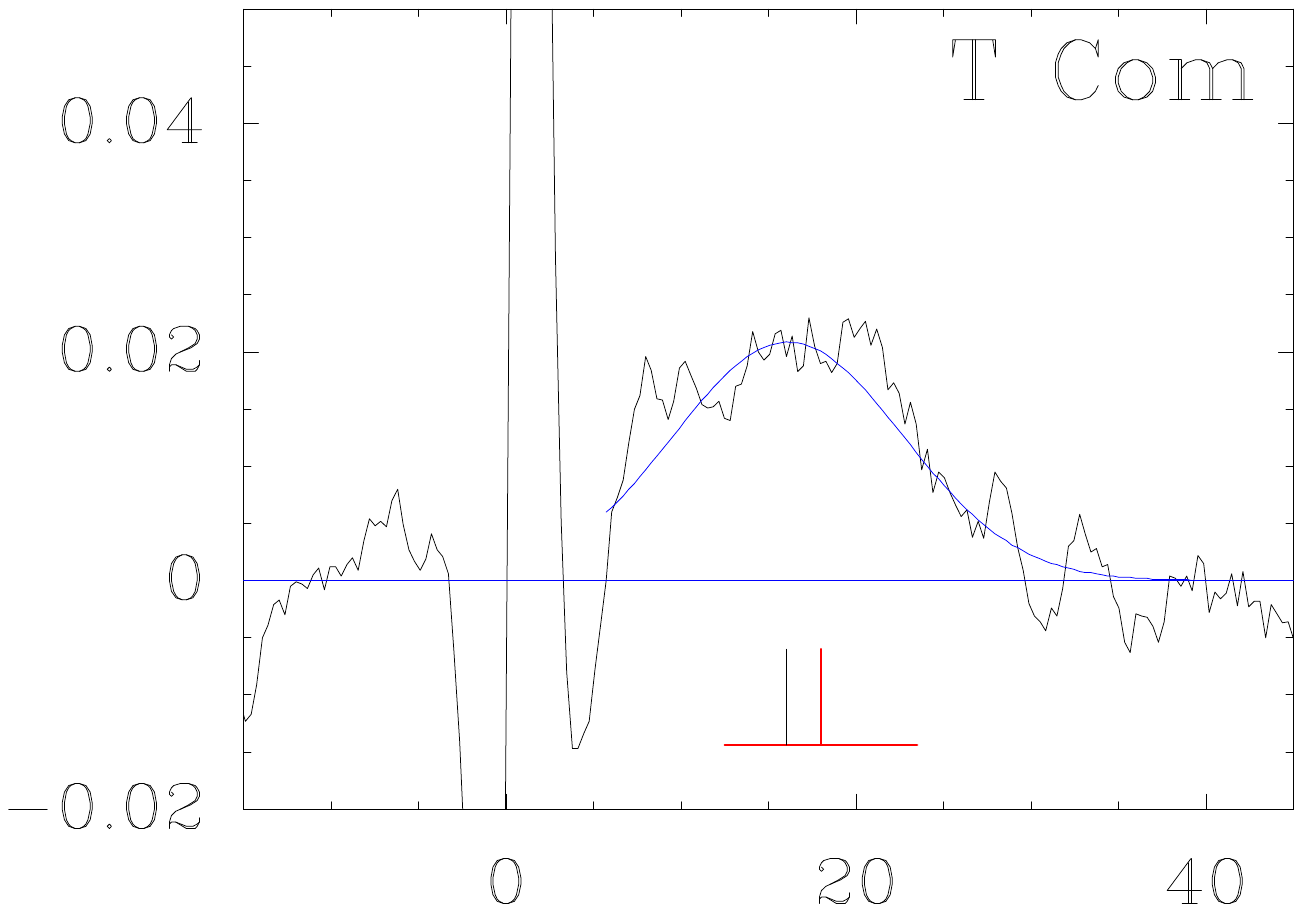}\hspace{-0.7cm}
\includegraphics[width=4.75cm]{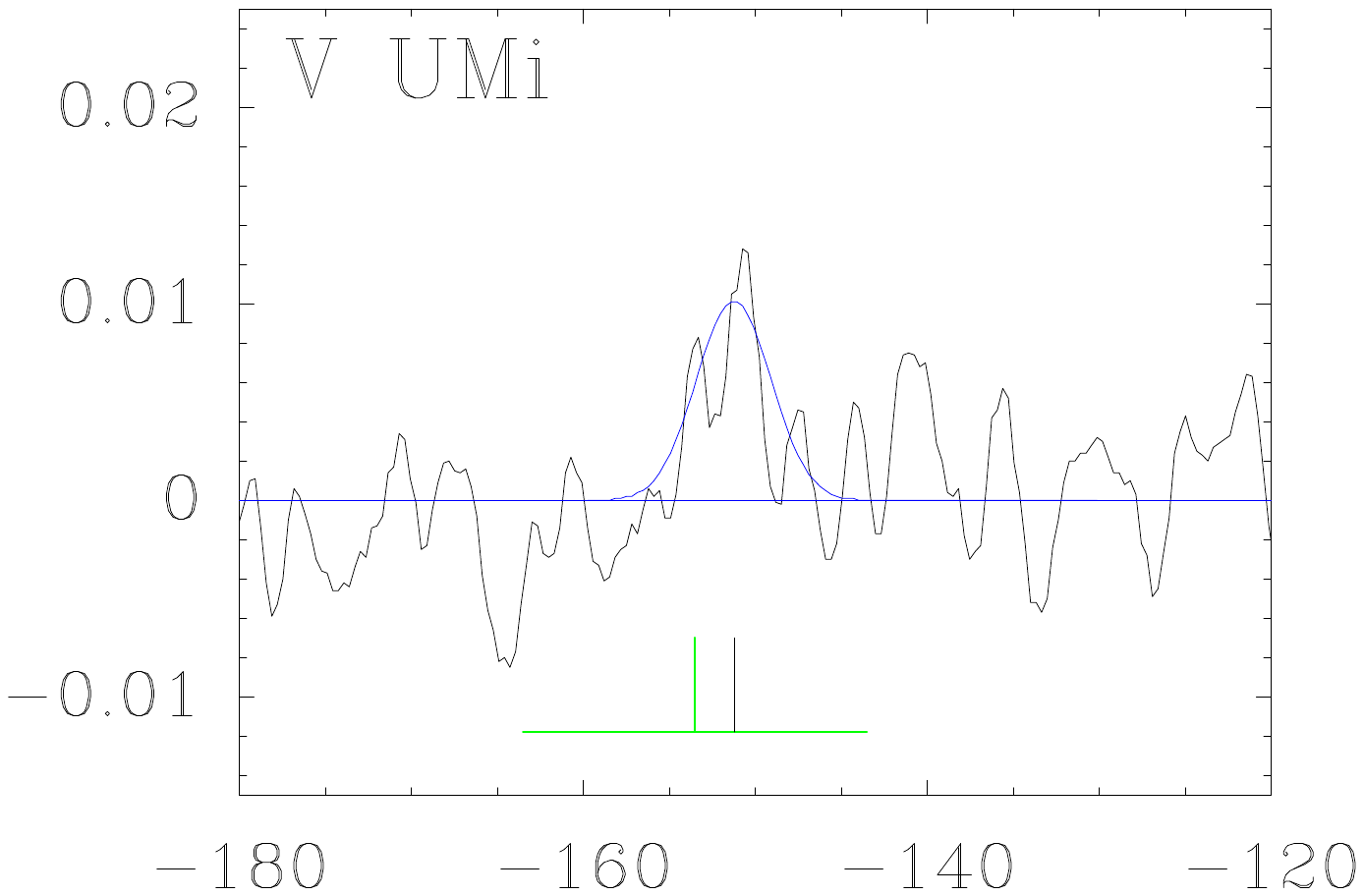}\hspace{-0.7cm}
\includegraphics[width=4.75cm]{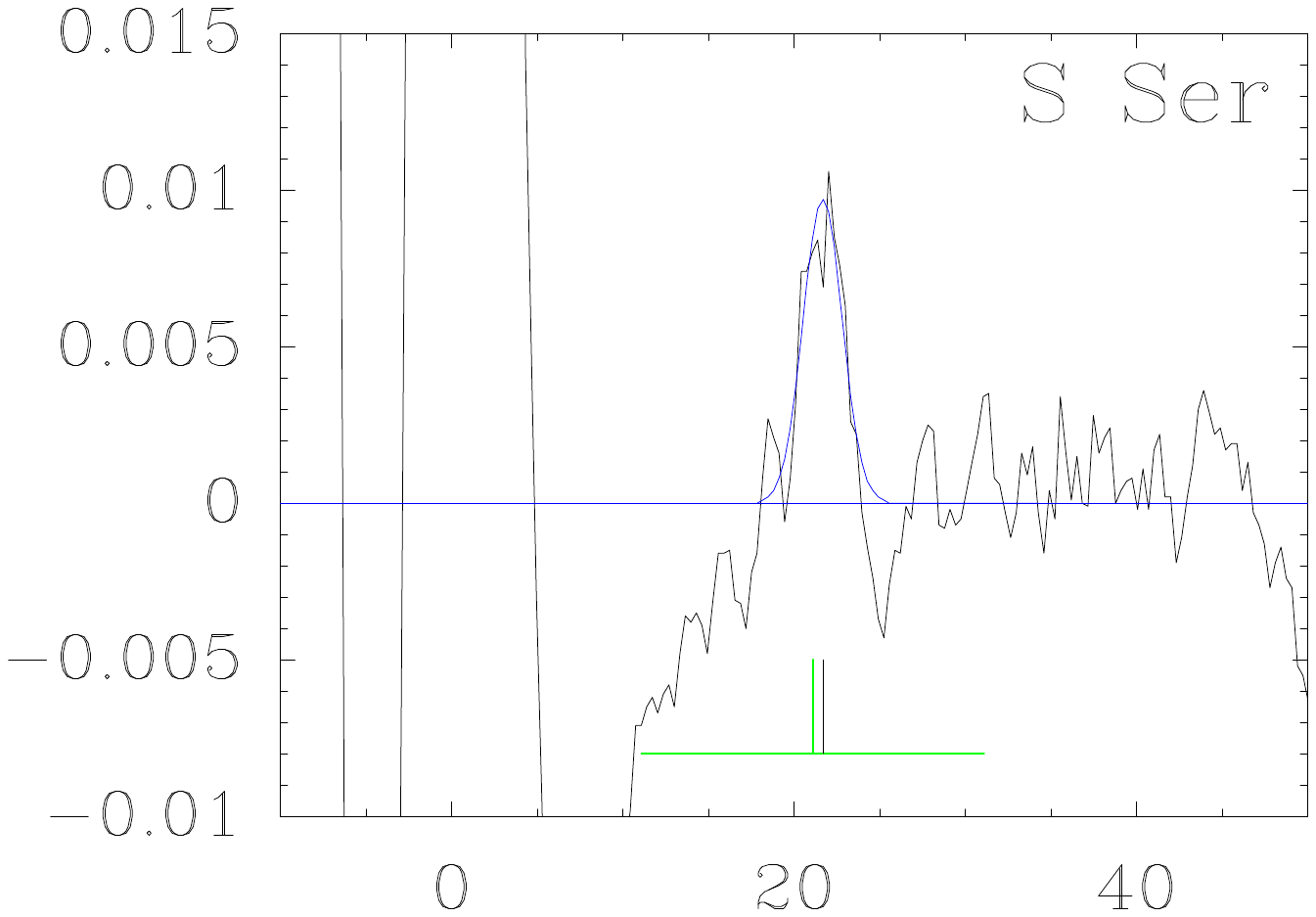}\hspace{-0.7cm}
\includegraphics[width=4.75cm]{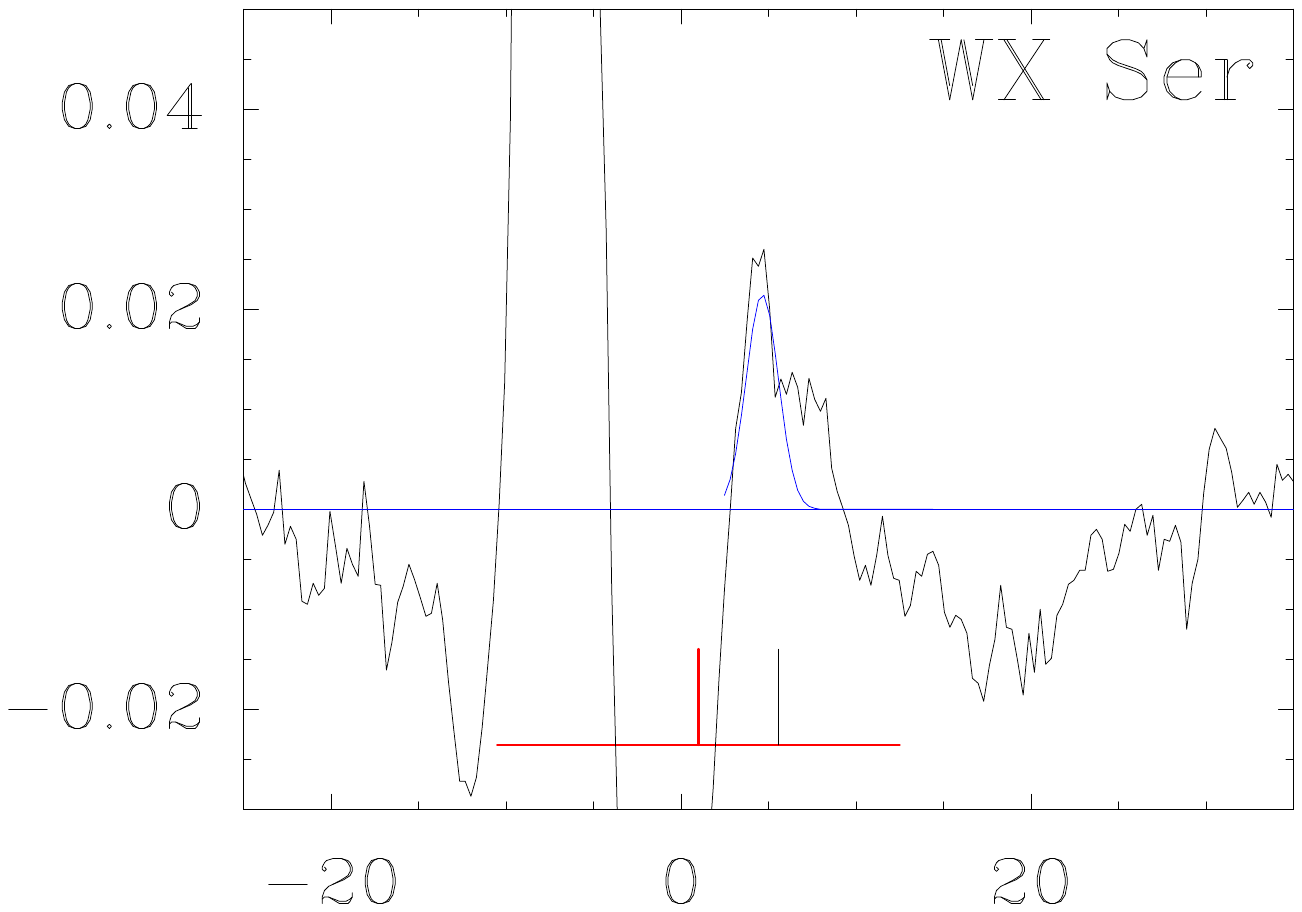}
\\ \vspace{-0.35cm}
\includegraphics[width=4.75cm]{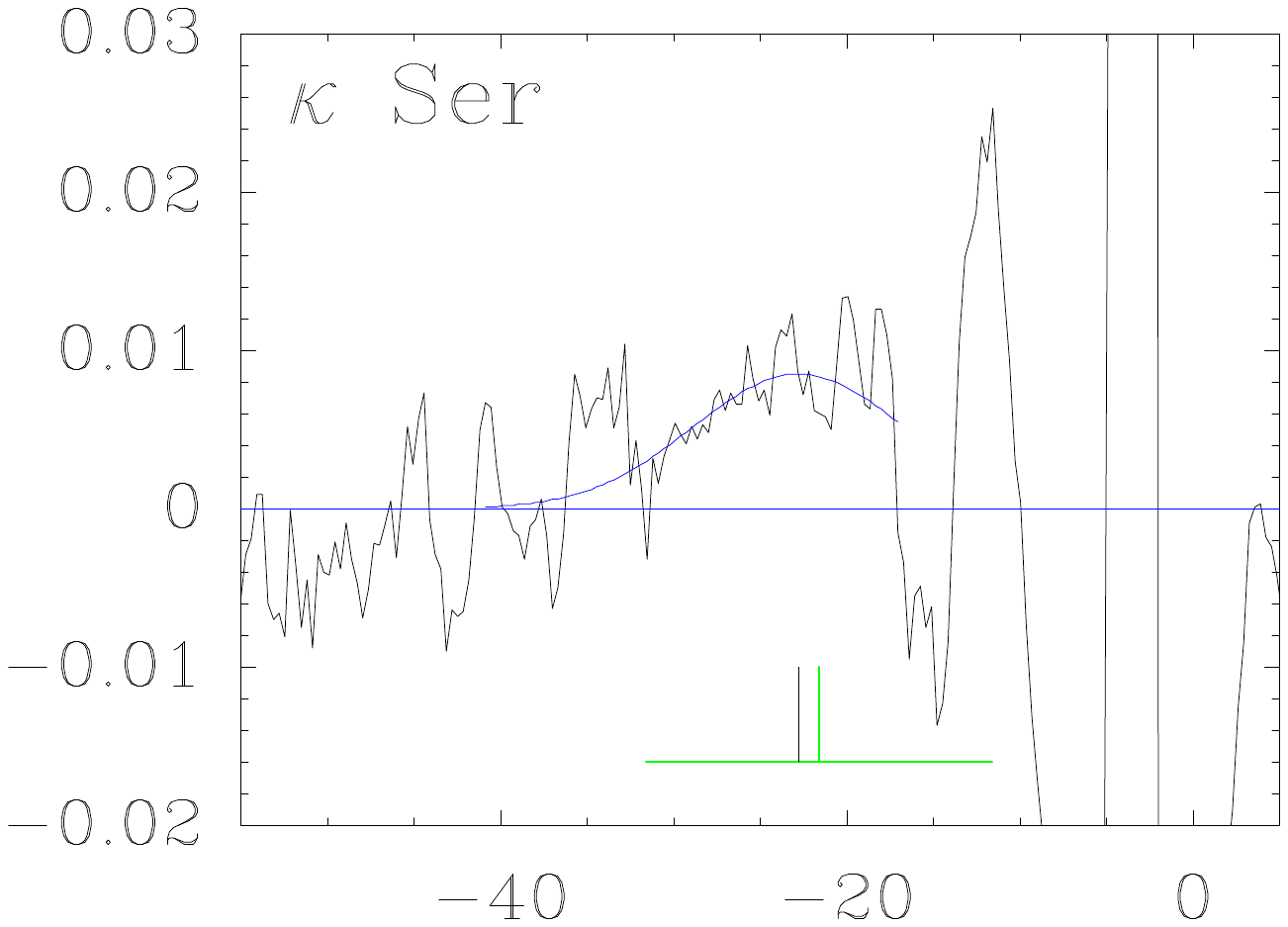}\hspace{-0.7cm}
\includegraphics[width=4.75cm]{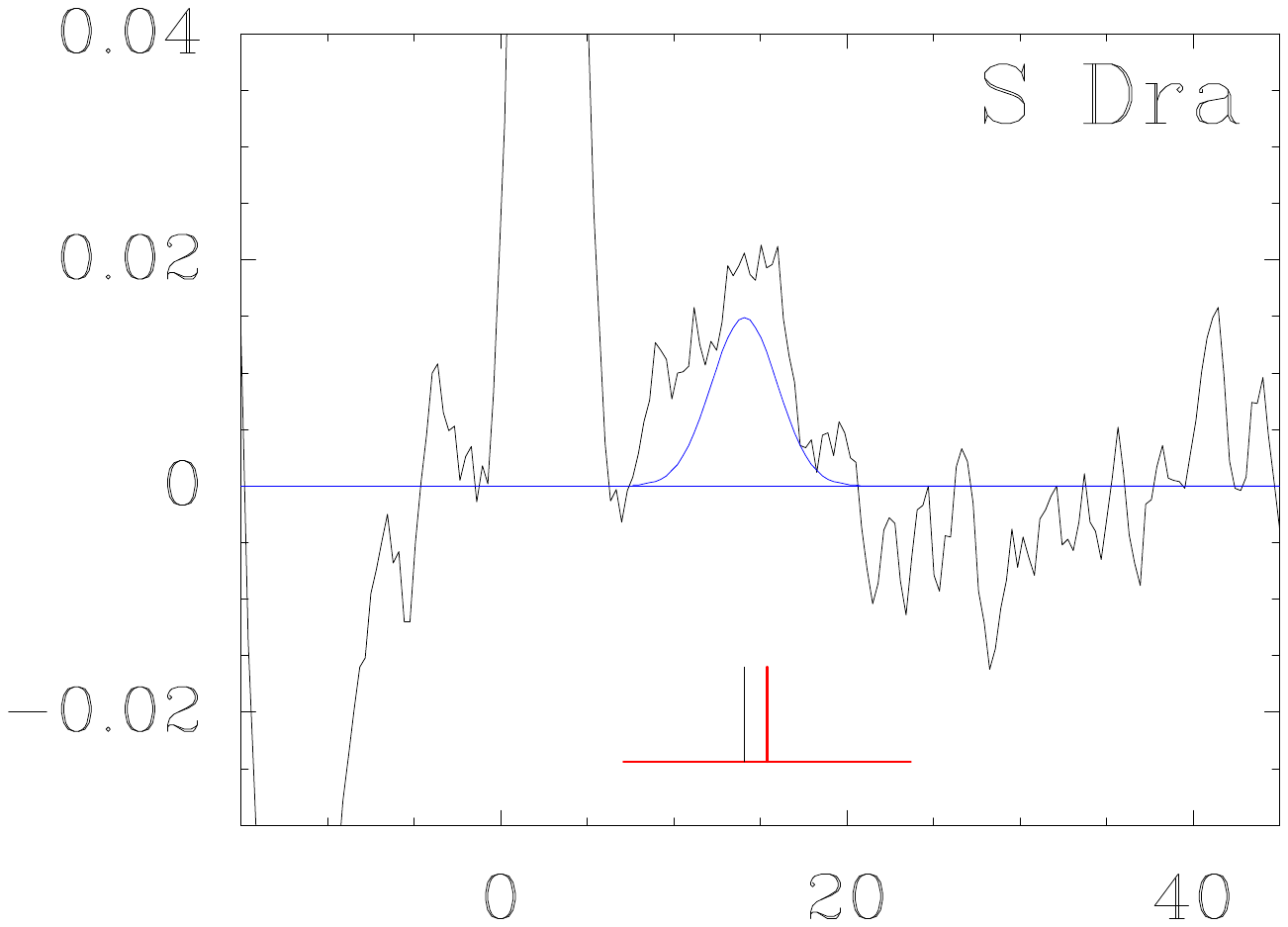}\hspace{-0.7cm}
\includegraphics[width=4.75cm]{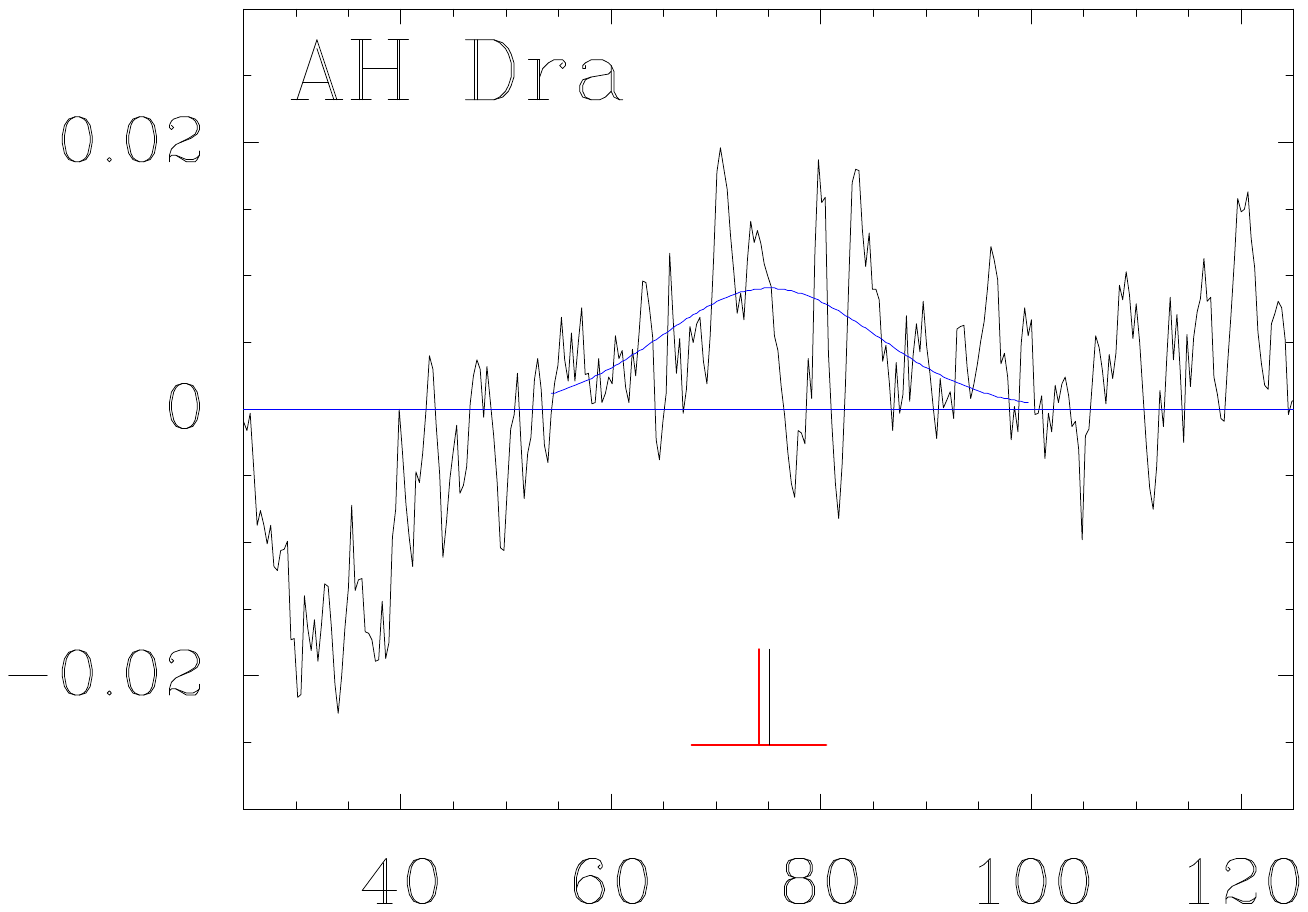}\hspace{-0.7cm}
\includegraphics[width=4.75cm]{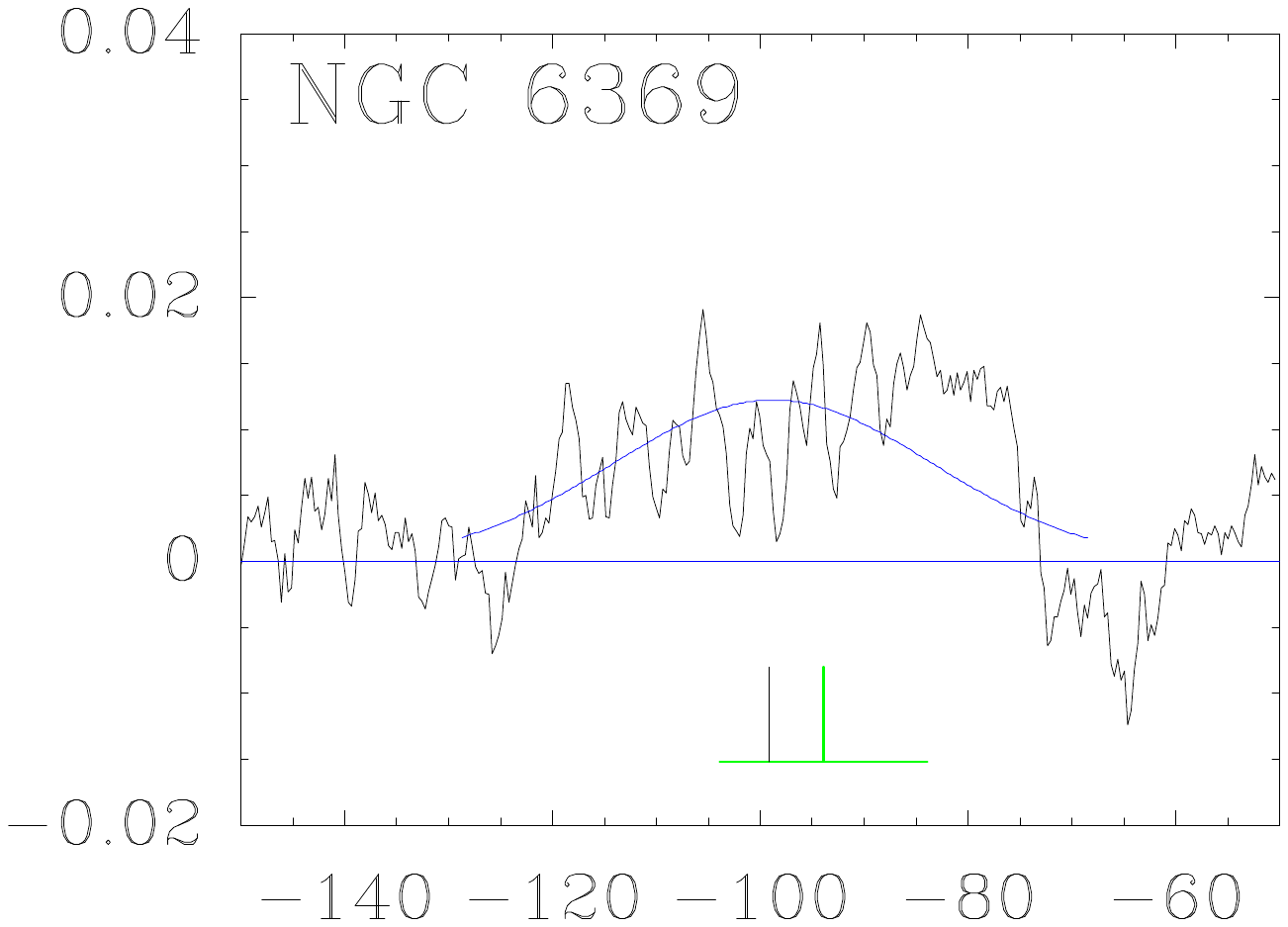}
\\ \vspace{-0.35cm}
\includegraphics[width=4.75cm]{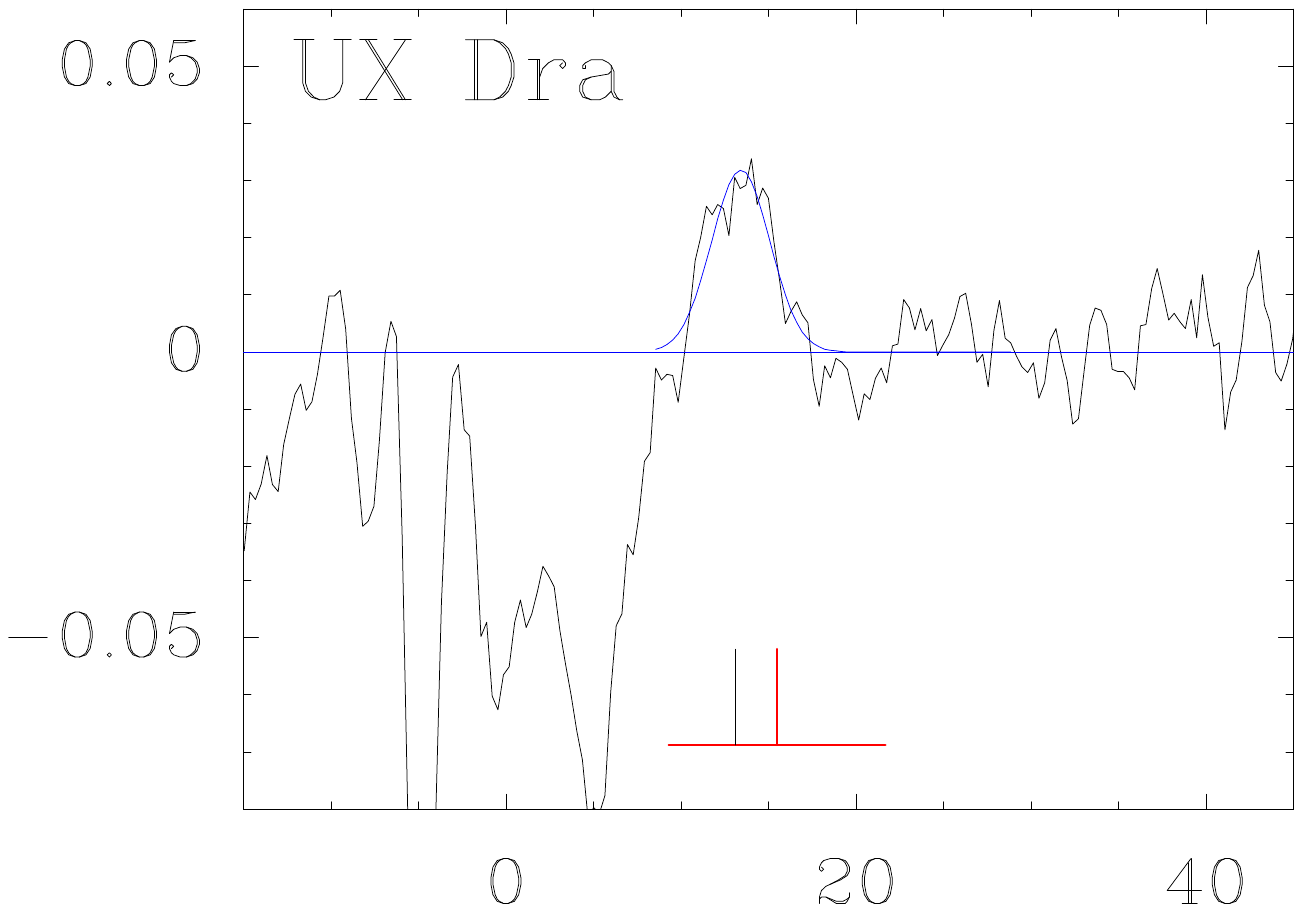}\hspace{-0.7cm}
\includegraphics[width=4.75cm]{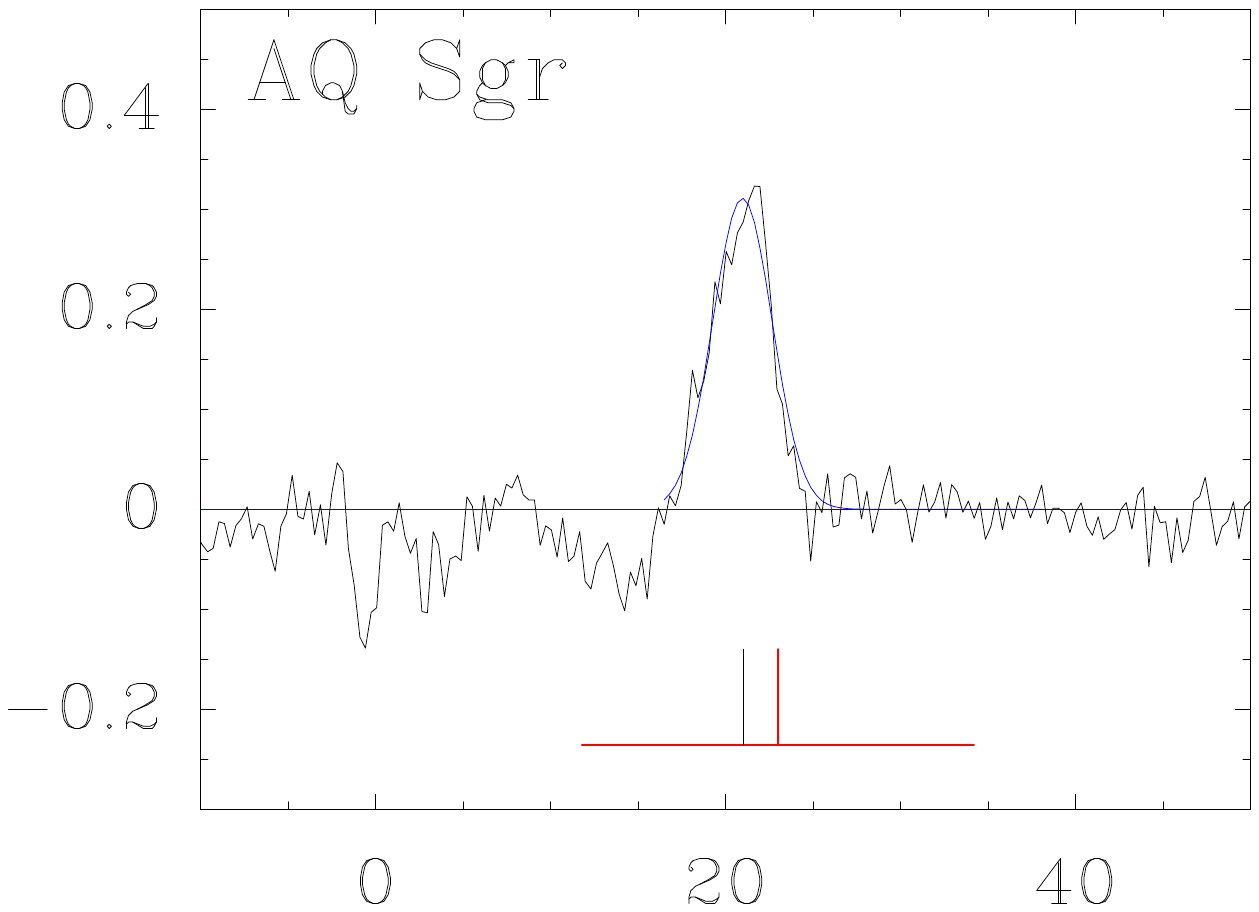}\hspace{-0.7cm}
\includegraphics[width=4.75cm]{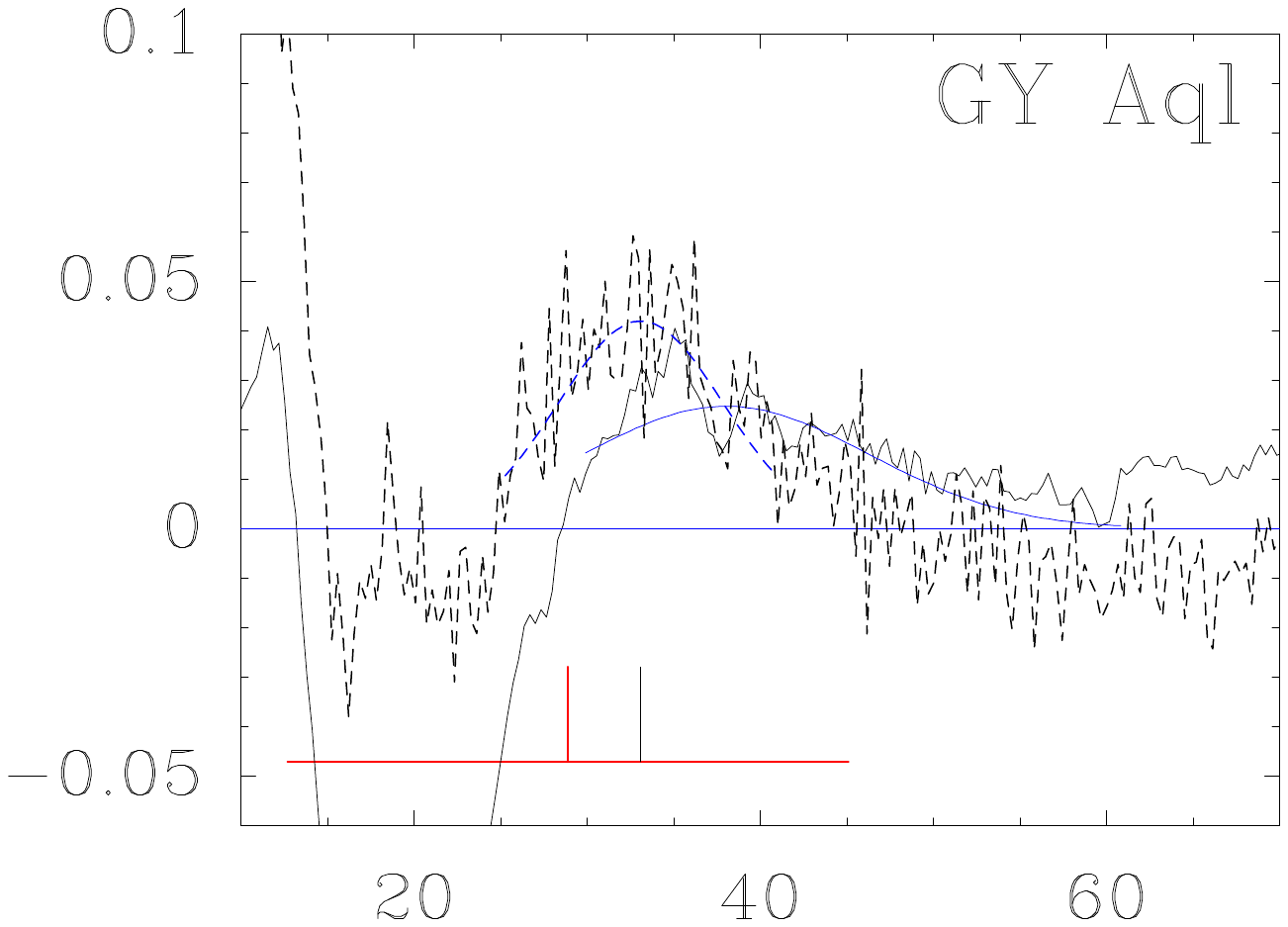}\hspace{-0.7cm}
\includegraphics[width=4.75cm]{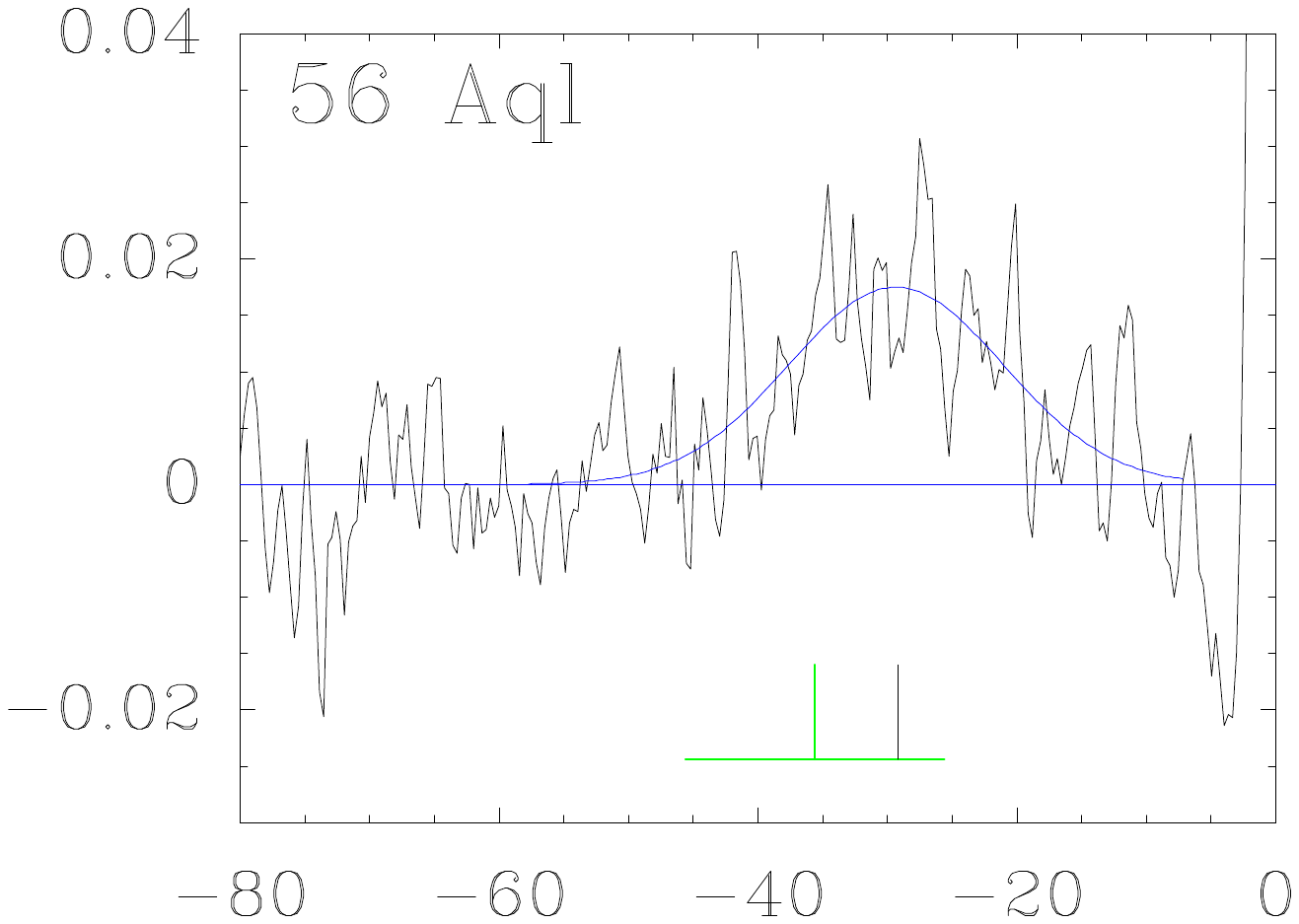}
\\ \vspace{-0.35cm}
\includegraphics[width=4.75cm]{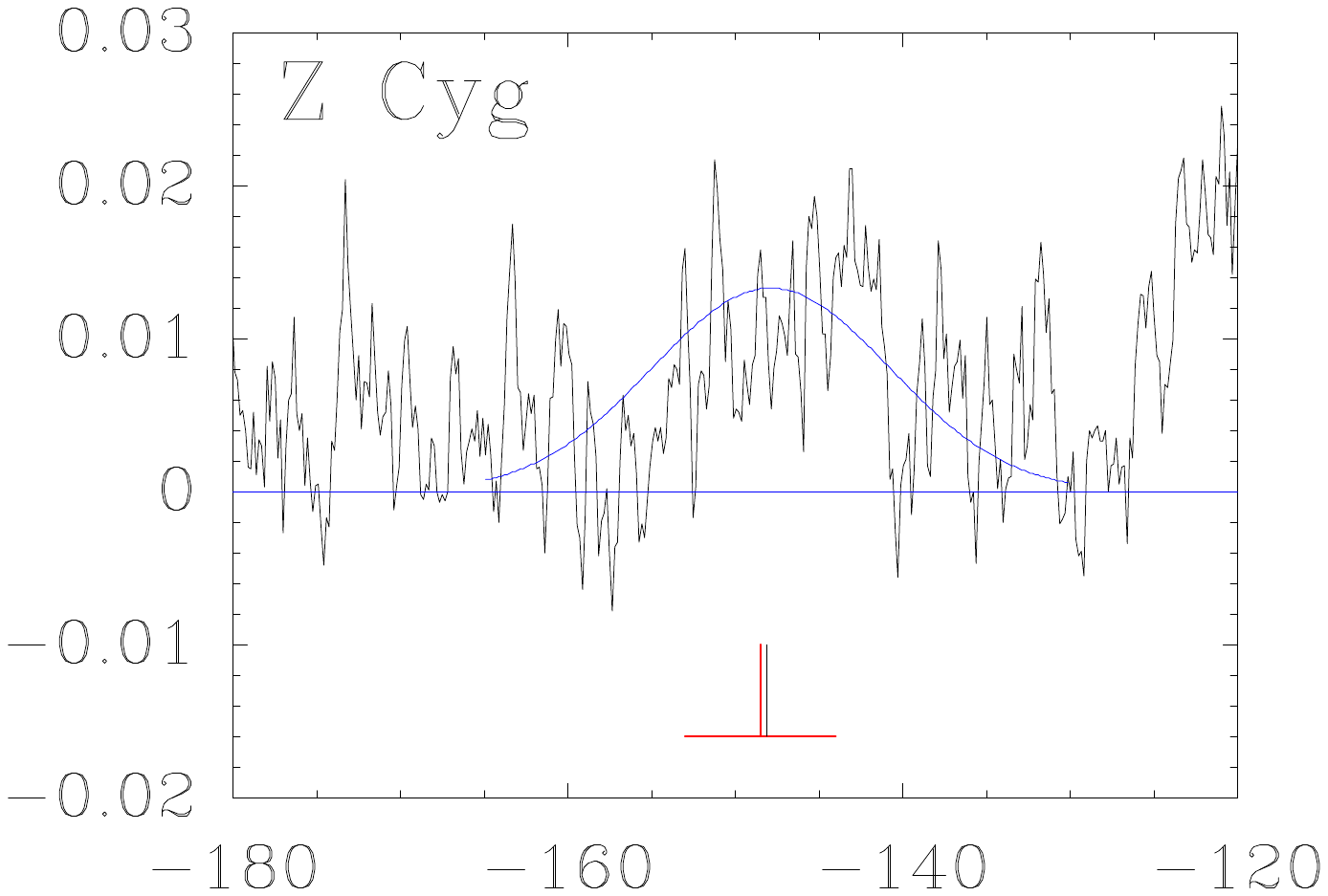}\hspace{-0.7cm}
\includegraphics[width=4.75cm]{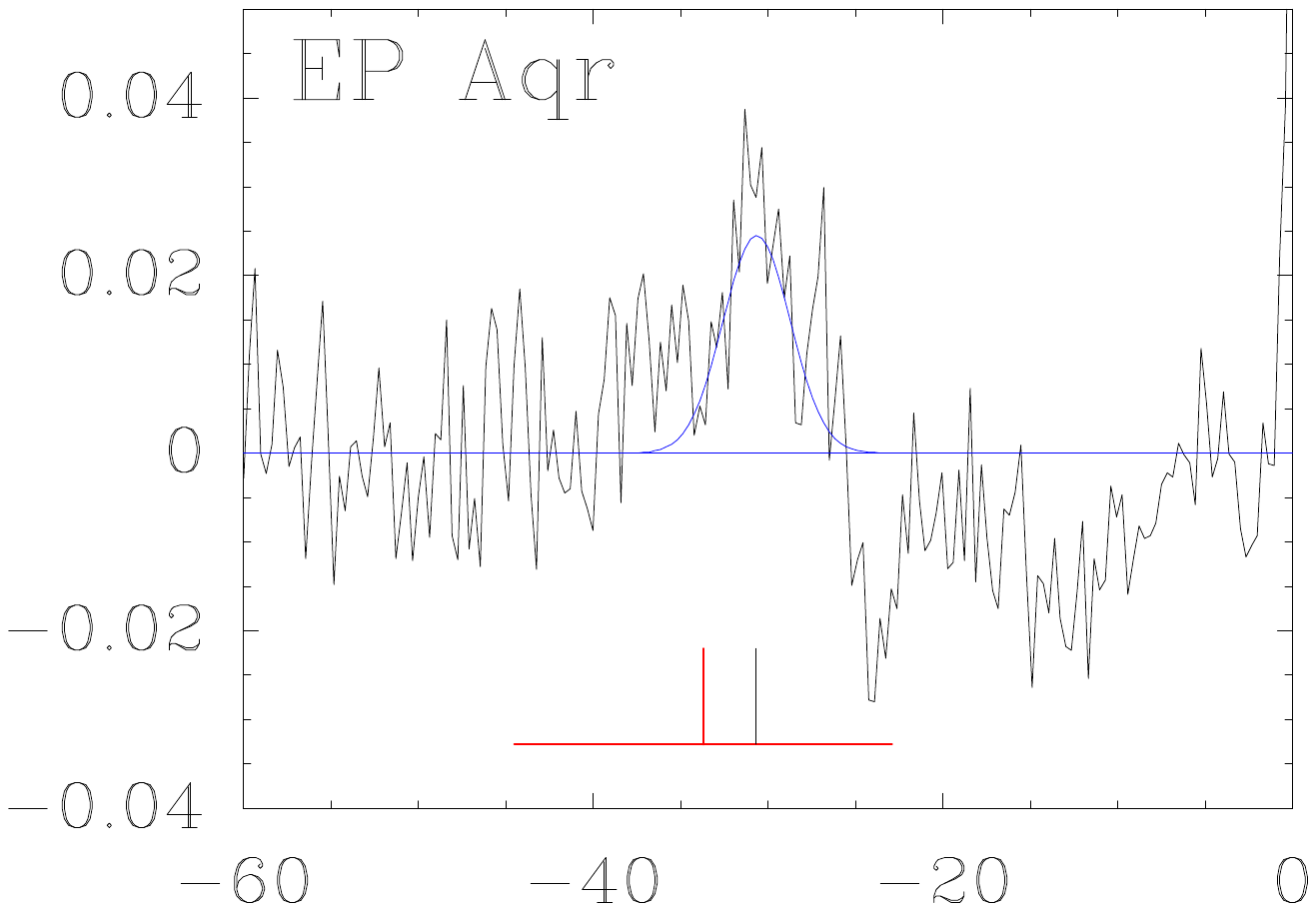}\hspace{-0.7cm}
\includegraphics[width=4.75cm]{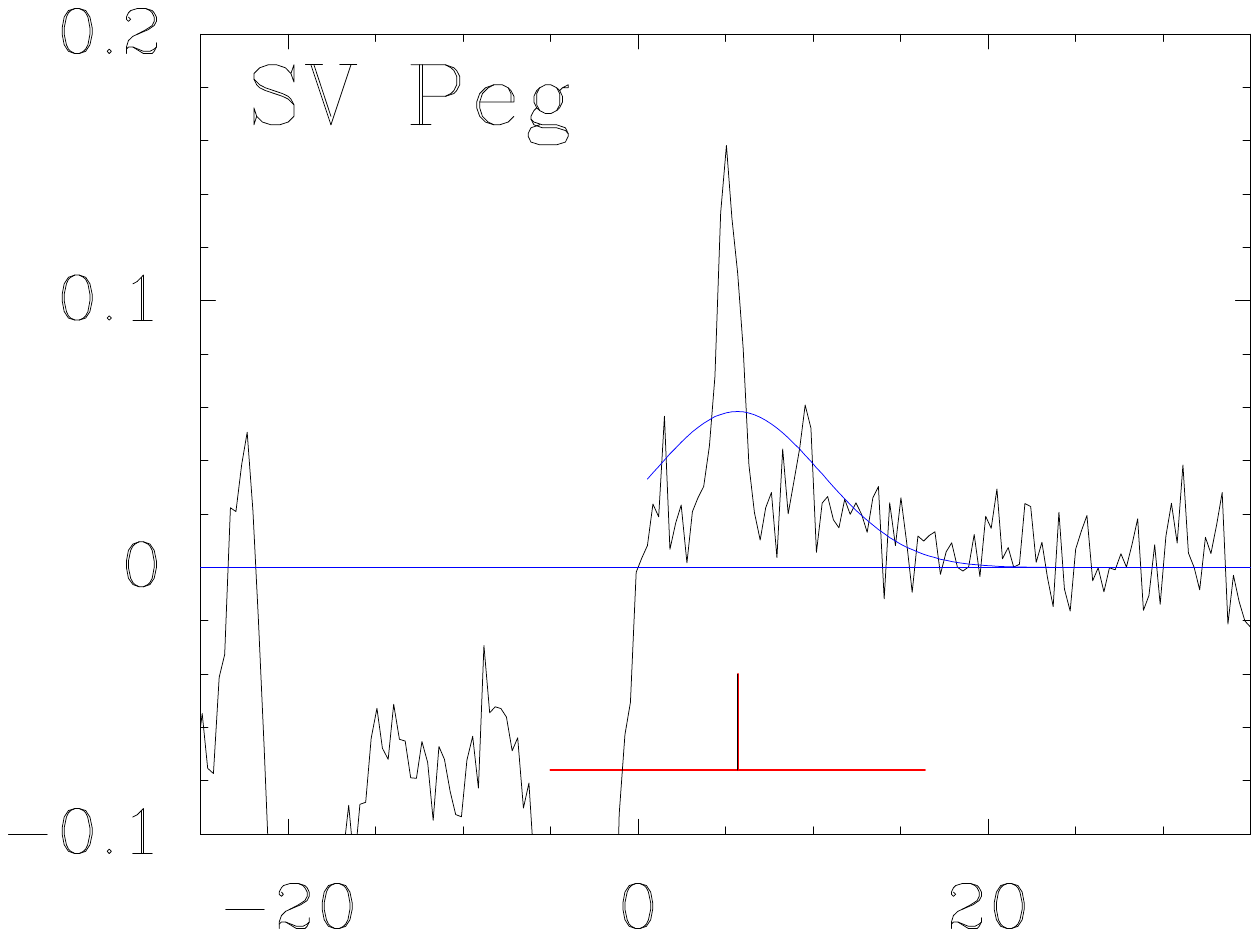}\hspace{-0.7cm}
\includegraphics[width=4.75cm]{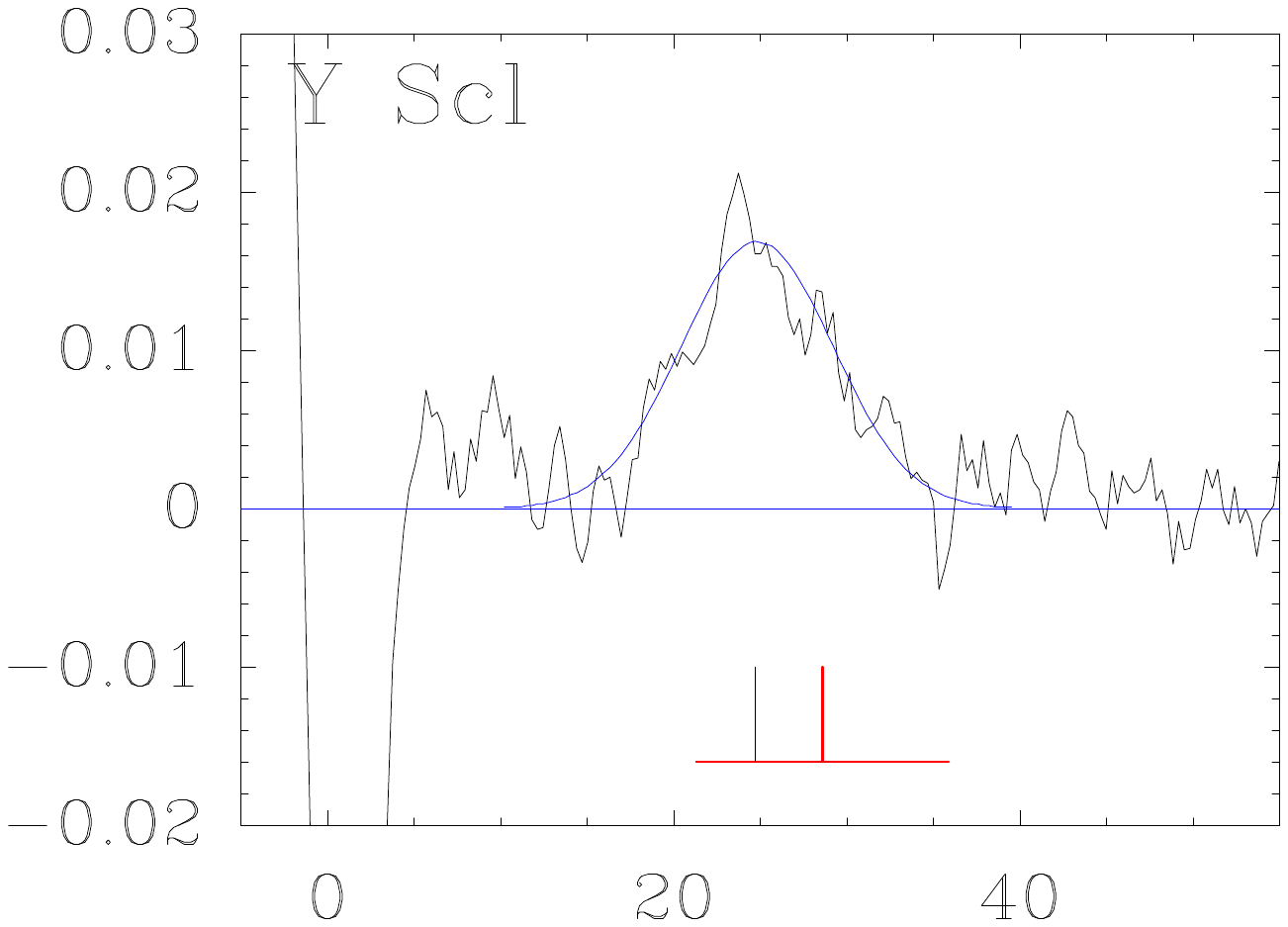}
\\ \vspace{-0.35cm}
\includegraphics[width=4.75cm]{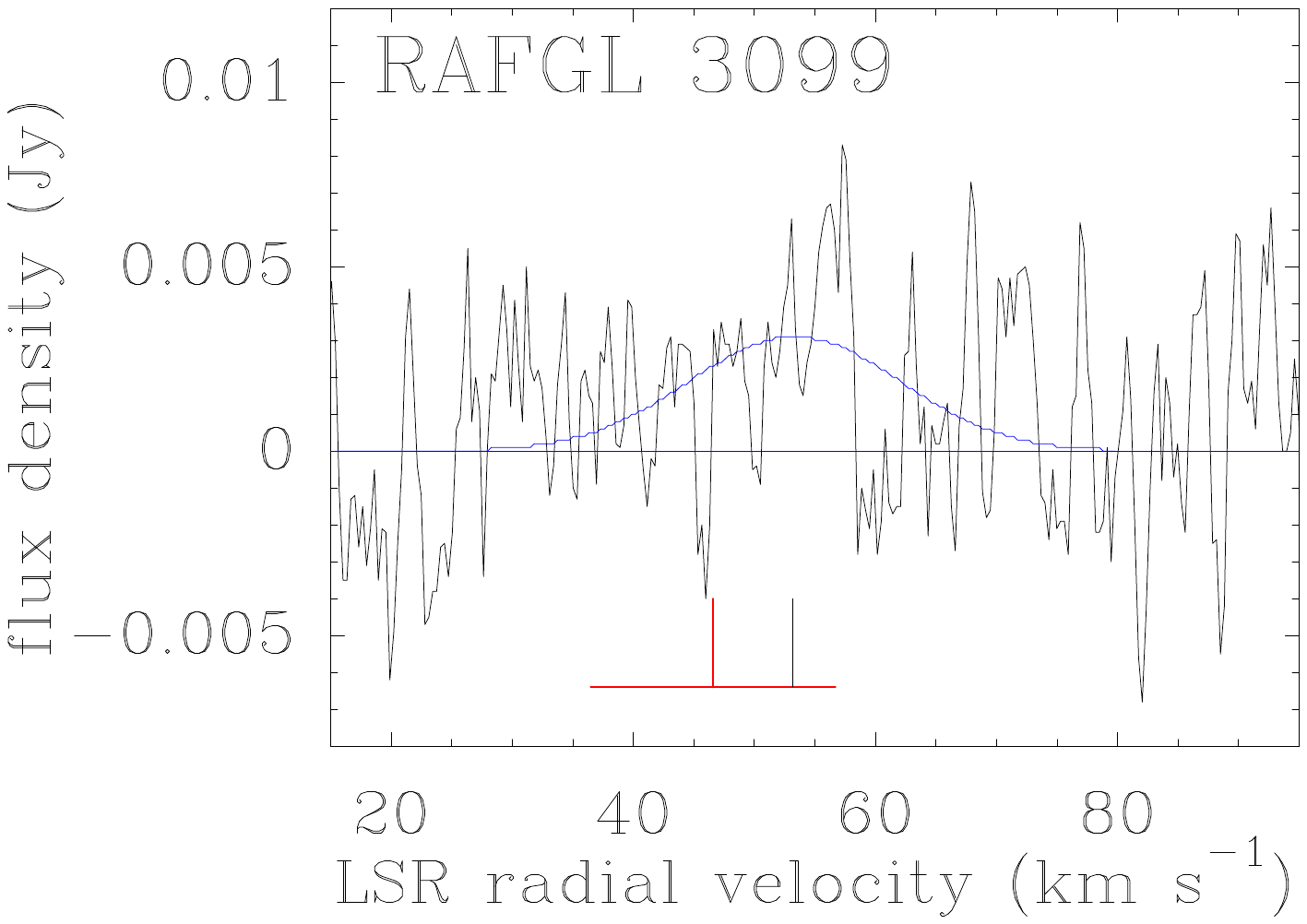}\hspace{-0.7cm}
%
% \vspace{3cm}
\\ \vspace{0.3cm}
  \caption{Possible detections. 
Shown is flux density of the 21-cm \HI\ line emission, $S_{\rm HI}$, in Jy, 
as a function of radial velocity in the LSR reference frame, \VLSR, in \kms.
% -- spectra of the 21-cm \HI\ line emission measured 
% within the beam of the NRT pointed towards the target stars.
The peak profiles (see Section \ref{sec:sizes}) are indicated by a solid black line,
while the total profiles (which could only be determined for GY Aql and U CMi) 
are indicated by a dashed black line. 
The Gaussians fitted to the profiles are shown as blue lines, 
solid for fits to the peak profiles and dashed for fits to the total profiles. 
Spectra are shown for all 21 objects with possible \HI\ detections (see Table~\ref{table:possdetsHI}).
The vertical black lines indicate the centre velocity of our Gaussian fit to the peak \HI\ 
profile (or, if available, to the total \HI\ profile), the red vertical and horizontal lines 
show respectively the central CO or OH line velocity from the literature and its corresponding expansion velocity,
whereas green lines indicate other types of literature velocities (e.g., optical or SiO lines)
and an indicative expansion velocity of 10 \kms, i.e., the average measured value. 
The flat blue horizontal lines show the 0 Jy flux density level.
The plotted total velocity range of the spectra is 60 \kms,
except for the 80 \kms\ range used for RAFGl 3099 due to its the broad profile.}
\label{fig:spectrapossibles}
\end{figure*}

For objects with clear NRT detections results of our \HI\ observations are given in 
Table \ref{table:cleardetsHI} and our spectra are shown in Figure \ref{fig:spectradetections},
for our possible detections the NRT results are listed in Tab. \ref{table:possdetsHI} 
and spectra are shown in Fig. \ref{fig:spectrapossibles}, and for objects with 
upper limits the NRT results are listed in Tab. \ref{table:upperlimits}
and spectra are shown in Fig. \ref{fig:spectralimits}, but only for objects for which 
digital spectra are available, e.g. those observed after the major NRT renovation
in 2001 (see Sect. \ref{sec:Observe}).

We fitted Gaussians to the \HI\ profiles measured at the NRT and this was found to provide 
a good representation of the line shape. The same was done with VLA global \HI\ profiles 
(see, e.g. \citealt{Matthews2013}).
A Gaussian shape is expected in case of optically thin gas in an outflow that is slowed down 
in the outer parts of a CSE by interaction with the surrounding ISM 
(see, e.g. \citealt{Lebertre2004, Libert2007, Matthews2013}). 

Our \HI\ radial velocities were measured in the same local standard of rest (LSR) 
reference frame defined based on a solar motion of 
$v_{\odot}$ = 20 \kms\ towards its apex at (B1950.0) $\alpha$ = \ad{270}{5}, 
$\delta$ = 30$^{\circ}$ \citep{Ke99}.
The same conversion was applied to convert all published heliocentric velocities 
$cz$ to the \VLSR\ values listed in the Tables.

Measured radial velocities of evolved stars can vary significantly depending on the 
spectral lines used. For each object only one LSR velocity taken from the literature 
($V$ lit.) is given in the Tables. Our order of preference for listing velocities is
(1) mm-wave CO lines, (2) maser lines (OH 1612 MHz, H$_2$O, SiO), and (3) optical lines 
-- as available. 

Mass loss rates, $\dot{\it M}$ in \Msunpyear, are based on CO line modelling 
(see the references for $\dot{\it M}$ref in the Tables), scaled to our adopted distance $d$.

The \HI\ peak flux densities, \Speak, and integrated line fluxes, \FHI, 
listed in Tables \ref{table:cleardetsHI} and \ref{table:possdetsHI}
are those of the total \HI\ profiles in the case of strong detections, 
which are flagged with a $^T$ after their names in the Tables, and of the peak \HI\ 
profiles for weaker detections, which are flagged with a $^P$ in 
the Tables (see \ref{sec:sizes} for details).
For the upper limit cases, all flux density limits listed in Table 
\ref{table:upperlimits} are for peak \HI\ profiles. 

The total \HI\ masses, in \Msun, were calculated using the standard formula 
\MHI\ = $2.356 \cdot\ 10^{-7} \, $d$^2 \, \FHI$, where the
integrated \HI\ flux \FHI\ is in \Jykms\ and distance $d$ is in pc
(see, e.g. \citealt{Kellermann88}). 

Listed in the Tables are the following elements: 

% Tables 1 and 3:
In Table \ref{table:cleardetsbasics} (clear NRT \HI\ detections: basic data) and 
Table \ref{table:possdetsbasics} (possible NRT \HI\ detections: basic data): 

\begin{itemize}
\item Name: common catalogue name of the target. 
An $^n$ after a name indicates that it is clearly not an AGB star, 
a $^d$ that we consider its classification as an AGB to be dubious,  
and a $^{\star}$ indicates that notes on the object can be found in Appendix \ref{sec:Notes}.
\item RA, Dec: literature right ascension and declination of the target from Gaia EDR 3 \citep{Gai20}, for epoch J2000.0;
\item Type: target type. 
Primarily the variability type as listed in Version 5.1 of the General Catalogue of Variable Stars, 
GCVS (\citealt{gcvs})
\footnote{A description of GCVS types is given in https://cdsarc.u-strasbg.fr/ftp/cats/B/gcvs/vartype.txt},
but if an object is not included in the GCVS, other identifiers are listed in brackets: 
HPM = high proper motion star, (OH/IR) = OH/IR maser,  
LPVc = long-period variable candidate, PN = planetary nebula, pPN = proto-planetary nebula,
and post-AGB star;
\item Spec. \& ref.: spectral type of the star, followed by its literature reference, 
as retrieved from the SIMBAD database. If none was listed there, the reference is noted as ``SIMBAD";
\item \Teff\ \& ref.: effective temperature of the star, in K, followed by its literature
reference; 
\item $d$: distance of the target, based on its parallax (mainly from the Gaia EDR3, \citealt{Gai20}), in pc.
If no Gaia parallax was available a reference to the distance we adopted is given in Appendix \ref{sec:Notes};
\item $V$ lit.: published radial velocity of the target in the LSR reference frame, in \kms;
\item \Vexp: literature expansion velocity measured from CO or OH 1612 MHz line observations, in \kms. 
If a pair of values was published for a two-velocity component CO line fit, the largest 
value is listed here;
\item ref.: literature references to the published $V$ lit. and \Vexp\ values; 
\item line: spectral line on which the published radial velocity measurement ($V$ lit.) was based;
\item $\dot{\it M}$ \& ref.: literature mass loss rates, in \Msunpyear, 
followed by its literature reference.
\end{itemize}

% Tables 2 and 4:
In Table \ref{table:cleardetsHI} (clear NRT \HI\ detections: \HI\ data) and 
Table \ref{table:possdetsHI} (possible NRT \HI\ detections: \HI\ data):

\begin{itemize}
\item Name: as in Table \ref{table:cleardetsbasics};
\item \VHI: our central radial velocity in the LSR reference frame of the Gaussian fitted 
to the \HI\ profile, in \kms.
\item $FWHM$: our full width half maximum of the Gaussian fitted to the \HI\ line profile, in \kms;
\item \Speak: our peak flux density of the \HI\ line profile, in Jy;
\item diam.: our estimated angular size of the \HI\ CSE in the east-west direction, in arcmin;
\item \FHI: our integrated line flux of the \HI\ profile, in \Jykms;
\item \MHI: our total \HI\ mass, in \Msun;
\item \HI\ ref.: references to previously published \HI\ studies.
\end{itemize}

% Table 5:
In Table \ref{table:upperlimits} (upper limits to NRT \HI\ lines):

\begin{itemize}
\item Name: as in Table \ref{table:cleardetsbasics};
\item RA, Dec: as in Table \ref{table:cleardetsbasics};
\item Type: as in Table \ref{table:cleardetsbasics};
\item Spec. \& ref.: as in Table \ref{table:cleardetsbasics};
\item \Teff\ \& ref.: as in Table \ref{table:cleardetsbasics};
\item $d$: as in Table \ref{table:cleardetsbasics};
\item $V$ lit.: as in Table \ref{table:cleardetsbasics};
\item \Vexp: as in Table \ref{table:cleardetsbasics};
\item $\dot{\it M}$: as in Table \ref{table:cleardetsbasics};
\item ref.: references for $V$ lit., \Vexp\ and $\dot{\it M}$, as applicable;
\item line: as in Table \ref{table:cleardetsbasics};
\item \Speak: peak flux density of our \HI\ line profile, in Jy;
\item \HI\ ref.: references to previously published \HI\ studies;
\item notes: blue/red side = only the blue/red side of the \HI\ profile was not confused
and could be used to measure an upper limit to the flux density level of the \HI\ line signal; 
{\it old data} = 25 sources observed only in 1992/1993, before the renovation of the NRT
(see Section \ref{sec:Observe}), for which no observations in digital form are available.
\end{itemize}

% Table 6:
In Table \ref{table:confused} (confused NRT \HI\ spectra): 

\begin{itemize}
\item Name: as in Table \ref{table:cleardetsbasics};
\item RA, Dec: as in Table \ref{table:cleardetsbasics};
\item Type: as in Table \ref{table:cleardetsbasics};
\item Spec. \& ref.: as in Table \ref{table:cleardetsbasics};
\item $V$ lit. \& ref.: as in Table \ref{table:cleardetsbasics};
\item line: as in Table \ref{table:cleardetsbasics};
\item notes: as in Table \ref{table:upperlimits};
\item \HI\ ref.: as in Table \ref{table:cleardetsHI};
\end{itemize}

Shown in Figures \ref{fig:spectradetections}, \ref{fig:spectrapossibles} and
\ref{fig:spectralimits} are all spectra obtained with the renovated NRT for, respectively, 
the 34 clear detections, the 21 possible detections, and for the 70 objects 
for which an upper limit to its \HI\ emission could be determined
from digital versions of observations, i.e., those observed from 2001 onwards 
with the renovated NRT (thus excluding the 25 objects with the 
"{\footnotesize \it old data}" note in Table~\ref{table:upperlimits}).
The peak profiles (see Section \ref{sec:sizes}) are indicated by a solid black line,
while the total profiles are indicated by a dashed black line.
Also shown are the Gaussians fitted to each \HI\ profile. 

The short vertical black line indicates the centre velocity of our Gaussian fit to 
the total profile, or to the peak profile for weaker detections, and the 
short vertical red line the \VLSR\ value taken from the literature, see 
Tables \ref{table:cleardetsHI}, \ref{table:possdetsHI}, and \ref{table:upperlimits}. 
The horizontal red line indicates the CO or OH line expansion velocity (if available  
-- if not, the average expansion velocity of $\pm$10\kms is indicated in green).

There is one object (RU Crt) without any published radial velocity.
We have listed it among the confused sources, as we could not identify any obvious
CSE \HI\ emission from it, nor put an upper flux density limit to it.

\section{Discussion} \label{sec:Discuss}  % Sect. 4

Of the 290 observed objects, within the NRT beam clear \HI\ detections 
of their CSE were made of 34 objects (11\%), 
possible detections of 21 objects (7\%) and upper limits to their \HI\ mass 
could be determined for 95 objects (36\%), whereas for 140 objects (46\%)
no conclusion could be drawn on their \HI\ mass due to confusion by Galactic \HI\ emission.
For 33 detections the total \HI\ flux and the size of the CSE could be determined.

Overall, the central velocities of our Gaussian fits to the \HI\ profiles agree well
with their CO line and optically-determined velocities. 
For the 34 clear detections the mean difference 
$\Delta$\VLSR\ = -0.2$\pm$2.0 \kms\ between CO and \HI\ line velocities,
and 0.8$\pm$5.3 \kms\ between optically-determined and \HI\ velocities; and the difference between
CO and optically-determined velocities is $-2.2\pm$3.4 \kms\ -- if we ignore the two
objects with optical values incompatible ($\Delta$\VLSR\ $\ge 20 \kms$)
with several other reported values, R Cas (\Vopt = $-$15 \kms, from \citealt{Fa05}) and 
TX Psc (\Vopt = $-$9.2 \kms, from \citealt{Wilson53}), and the one with a large reported 
uncertainty, NGC 7293 ($\pm$10 \kms, from \citealt{Wilson53}).

Since \HI\ and CO emission lines originate in quite different regions of  
CSEs their profile shapes and widths cannot be compared directly. 
However, the \HI\ lines seem to be narrower than the CO lines.

For the following 15 objects our present classification as either detection, upper limit
or confused is different from those we gave in previous publications (see Appendix \ref{sec:Notes} 
for details on individual objects). 
This difference is due to the analysis of many more, and different types of, observations made at the NRT in subsequent years.
We have sorted them into five categories, according to the differences between the new/old confusion 
and detection assessments.
In three cases the classification was upgraded to detection or possible detection, 
whereas the other 12 cases were downgraded to either possible detection, upper limit or confused: 

\begin{itemize}
\item  IRC +10216: confused, was detection \citep{Lebertre2001, Matthews2007, Matthews2015};
\item  NGC 6369: possible detection, was detection \citep{Gerard2006};
\item  RAFGL 3068: upper limit, was detection \citep{Gerard2006};
\item  RAFGL 3099: possible detection, was upper limit \citep{Gerard2006, Matthews2013};
\item  W And: confused, was detection \citep{Gerard2011};
\item  EP Aqr: possible detection, was detection \citep{Lebertre2004, Matthews2007};
\item  R Aqr: clear detection, was non-detection \citep{Matthews2007};
\item  S CMi: possible detection, was detection \citep{Gerard2006};
\item  U CMi: possible detection, was detection \citep{Gerard2006};
\item  Z Cyg: possible detection, was upper limit \citep{Gerard2006};
\item  $\alpha$1 Her: upper limit, was detection \citep{Gerard2006};
\item  OP Her: upper limit, was detection \citep{Gerard2011};
\item  $\delta$02 Lyr: confused, was detection \citep{Gerard2006};
\item  $\alpha$ Ori: confused, was detection  \citep{Bo87, Lebertre2012};
\item  $\rho$ Per: upper limit, was detection \citep{Gerard2006}.
\end{itemize}

\subsection{Comparison of VLA and \nan\ \HIsl\ detections} \label{sec:VLANRTcomp}  % 5.1

\begin{figure*}[ht]  % Fig. 6
  \centering
  \includegraphics[width=18.5cm]{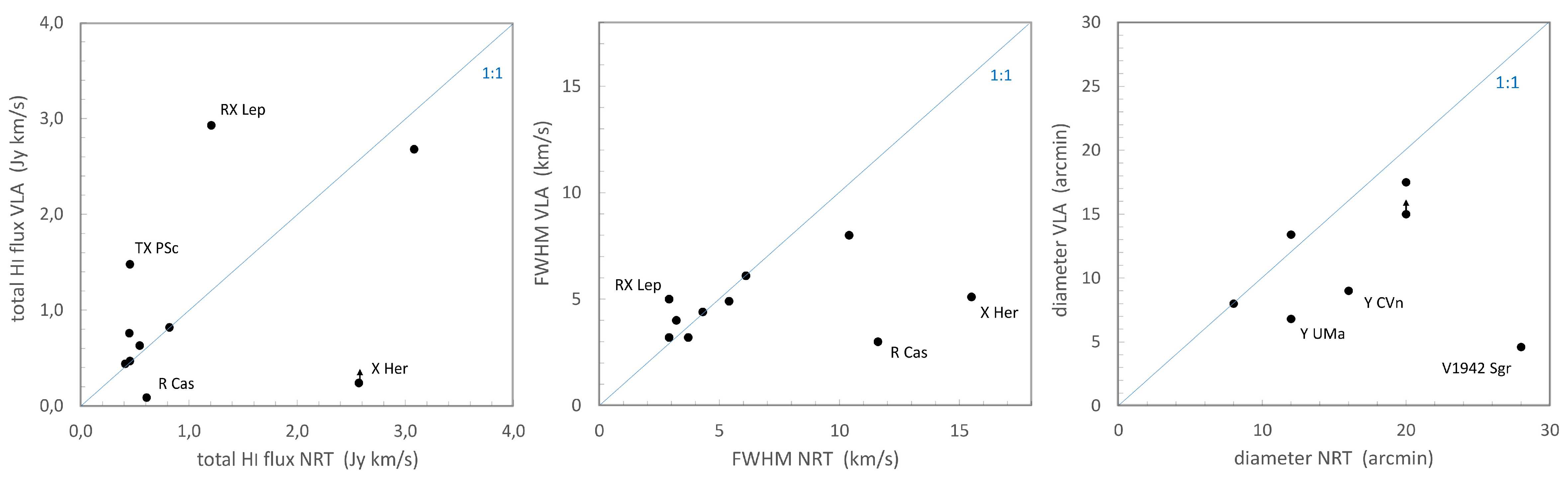}
  \caption{Comparison of results of NRT and VLA \HI\ line observations.
  Left: integrated line flux, \FHI, in \Jykms; centre: $FWHM$ width of the integrated
  line profile, in \kms; right: source diameter, in arcmin (right).
  The diagonal lines marked with ``1:1” indicate equality between the NRT and VLA
  measurements and were added to guide the eye only, but do not represent fits 
  to the data.}
  \label{fig:NRTVLAcomp} 
\end{figure*}

We compared (see Fig. \ref{fig:NRTVLAcomp}) the global \HI\ line profile parameters 
measured at the NRT and the VLA for the ten objects detected with both telescopes 
(see Sect. \ref{sec:Intro}):  
$o$ Cet, R Cas, RS Cnc, Y CVn, X Her, RX Lep, R Peg, TX Psc, V1942 Sgr, and Y UMa. 
The central line velocities are in good agreement, with an average NRT$-$VLA difference
of 0.3$\pm$0.5 \kms, cf. the typical uncertainty of $\sim$0.15 \kms\ for the VLA profiles.
When comparing the measured line fluxes and $FWHMs$, some objects show a notable difference 
between their NRT and VLA values. 

Considerably higher line fluxes measured with the NRT could, in principle, 
be ascribed to low-column density, extended \HI\ structures within a CSE, 
which remained below the VLA detectability threshold but were detectable 
with the larger NRT beam (for X Her). 

On the other hand, this effect could not explain the considerable higher line
fluxes measured with the VLA (for R Cas, RX Lep, and TX Psc). For these objects
the discrepancies may rather be due to the rather different methods used to 
separate CSE line emission from ambient Galactic lines in VLA and NRT data.

In three objects with discrepant line fluxes the line $FWHMs$ are also significantly
different (R Cas, X Her and RX Lep).

It should be noted that for the VLA data on Y UMa we used the parameters which we 
determined from the global profile published by \citet{Matthews2013}, who fitted
two components to their profile (see \ref{sec:Notes} for further details).

\begin{itemize}
\item R Cas: the VLA line flux \citep{Matthews2007} is seven times lower than 
our NRT value of 0.6 \Jykms\ and the VLA $FWHM$ is four times smaller than the NRT 
value of 12 \kms. The source diameters are similar, \am{13}{4} (VLA) and $\sim$12$'$ (NRT).
Both NRT and VLA detections are faint, at, respectively, \MHI\ = 0.00063 and 0.00047 \Msun. 
The VLA$-$NRT discrepancy may simply be due to the low signal-to-noise ratios 
in both observations.
\item  X Her: the VLA line flux \citet{Matthews2011} is 11 times lower than our NRT 
value of 1.2 \Jykms\ and the VLA $FWHM$ is three times smaller than the NRT value of 15 \kms.
The \HI\ cartography by \citet{Matthews2011} of X Her and its surroundings with 
the VLA and the GBT revealed the presence of a nearby, compact HVC cloud 
seen in superposition on the CSE emission of X Her, which they named Cloud {\sc I}.
The VLA$-$NRT discrepancy is likely due to confusion with Cloud {\sc I} -- 
see also the discussion in \citet{Matthews2011}.
\item  RX Lep: the VLA line flux \citep{Matthews2013} is 2.4 times higher than our 
NRT value of 1.2 \Jykms, which is the same as the NRT value from \citet{Libert2008}, 
and the VLA $FWHM$ is 1.7 times larger than our NRT value of 2.9 \kms. 
As noted by \citet{Matthews2013}, the reason for the discrepancy is unclear, but may be 
partly due to the different methods used to separate CSE and surrounding Galactic signals
in VLA and NRT data;
\item TX Psc: the VLA line flux \citet{Matthews2013} is 3.3 times higher than our 
NRT value of 0.45 \Jykms\ and the VLA $FWHM$ is 1.2 times larger than the NRT value of 3.2 \kms. 
% The NRT detection reported previously by \citet{Gerard2006} has a 
% 12 times higher flux of 5.2 \Jykms\ and FWHM = 4.2 \kms. 
The VLA data show the presence of large, nearby \HI\ structures which are also close in
velocity to the CSE emission from TX Psc, but which do not appear to be related to the target star.
The VLA$-$NRT discrepancy is likely due to the presence of these Galactic clouds -- see also
\citet{Matthews2013}.
  % ratios VLA/NRT: FHI 3.3 FWHM 1.2
  % NRT diameter = 8 arcmin 
  
\end{itemize}

Concerning the diameters of the NRT clear detections, 8 sources were unresolved, with diameters $<4'$,
and the diameters of the 26 resolved sources range from 4$'$ to 44$'$, with an average of
20$'$$\pm$10$'$. 
We could compare these values with source diameters measured with the VLA for seven objects.
For four (R Cas, RX Lep, R Peg, TX Psc) the diameters are comparable, the NRT diameter is 
about twice as large for two others (Y CVn and Y UMa), and for V1942 Sgr the NRT diameter is 
six times larger.

\section{Conclusions} \label{sec:Conclusions}  % Sect. 6

A search was made during the past 30 years for \HI\ line emission from circumstellar 
envelopes (CSEs) of mass-losing Asymptotic Giant Branch (AGB) stars and other types 
of evolved stars. Such observations can e.g. trace gas far outside the CO dissociation radius, 
and probe the interface between the star and the interstellar medium. 

Using the single-dish \nan\ Radio Telescope total \HI\ masses could be measured
for CSEs, and for clear detections their E-W sizes could be estimated.

Of the 290 targets, 34 were clearly detected, 21 had possible detections, 
for 95 an upper limit to their \HI\ line emission could be determined, and for 
140 objects no conclusion could be drawn about their \HI\ content due to confusion 
with surrounding Galactic \HI\ clouds. 

Of the 63 Miras only 11\% have clear \HI\ detections, whereas of the 43 SRb-types 37\% 
have clear detections.

The total \HI\ masses of the clearly detected AGB stars cover a wide range 
of about a factor 150, from $\sim$0.001 to 0.15 \Msun.
The different types of AGB stars all cover about the same range. 

The linear E-W diameters of the 28 resolved \HI\ sources as determined with the NRT 
cover a range of a factor 30, from 0.17 to 5.2 pc, with an average of 1.6$\pm$1.4 pc. 
The 10 unresolved sources have upper limits ranging from 0.5 to 2.6 pc (average 1.0$\pm$0.8 pc).
On average, the radii of the 21 resolved Miras and SR-type AGBs are comparable,
1.0$\pm$0.8 pc and 1.4$\pm$1.2 pc, respectively. 
The two outliers with large \HI\ diameter of order 5.2 pc are RY Dra (SRb:) and V1942 Sgr (Lb).

The stars in our sample cover two orders of magnitude in mass loss rate, from about
0.2 to 20 10$^{-7}$ \Msunpyear. There is no clear overall trend of \HI\ or CO expansion 
velocity with mass loss rate. 

We will present an analysis of our \HI\ results in a future study (G\'erard et al., Paper II, 
{\it in prep.}), where we will focus on the differences in atomic and molecular hydrogen masses 
of AGB CSEs.

\begin{acknowledgements} 
This paper is dedicated to the memory of 
Nguy{\^e}n Quang Ri{\^e}u, who initiated this research.
We wish to thank the staff of the \nan{} Radio Telescope for
their support with the observations over the past 30 years.
The \nan{} Radio Observatory is operated by the Paris Observatory, 
associated with the French Centre National de la Recherche Scientifique.
This research has made use of the SIMBAD database, operated at CDS, 
Strasbourg, France.
LDM was supported by award AST-2107681 from the National Science Foundation.
\end{acknowledgements}

% \begin{thebibliography}
\bibpunct{(}{)}{;}{a}{}{,} % to follow the AandA style
\bibliographystyle{aa}
\bibliography{AGBNRTrefs.bib}
% \end{thebibliography}

%  ===== FIGURES =========================================

\begin{figure*}[ht]  % Fig. 3a
  \centering
\includegraphics[width=4.75cm]{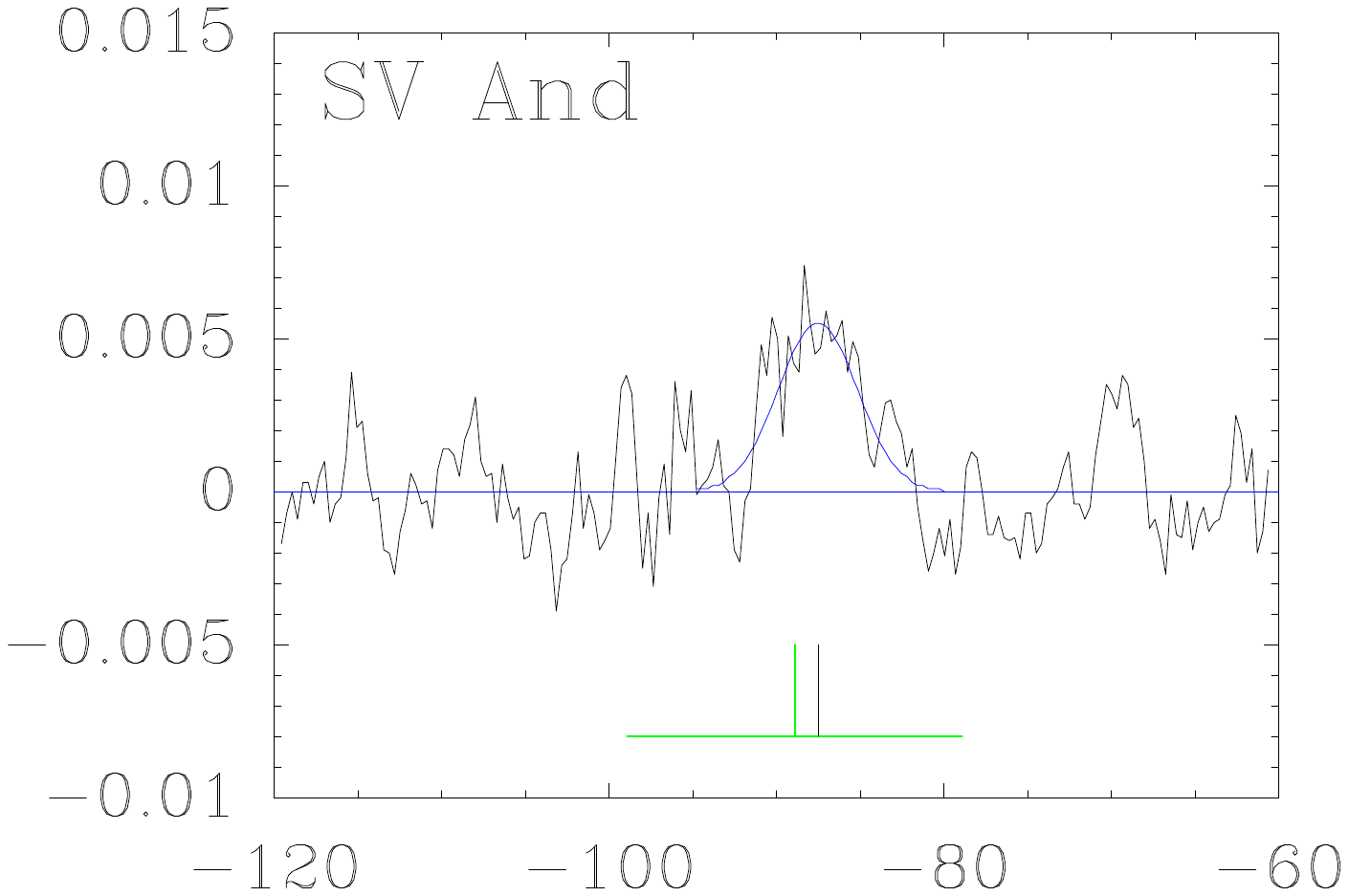}\hspace{-0.7cm}
\includegraphics[width=4.75cm]{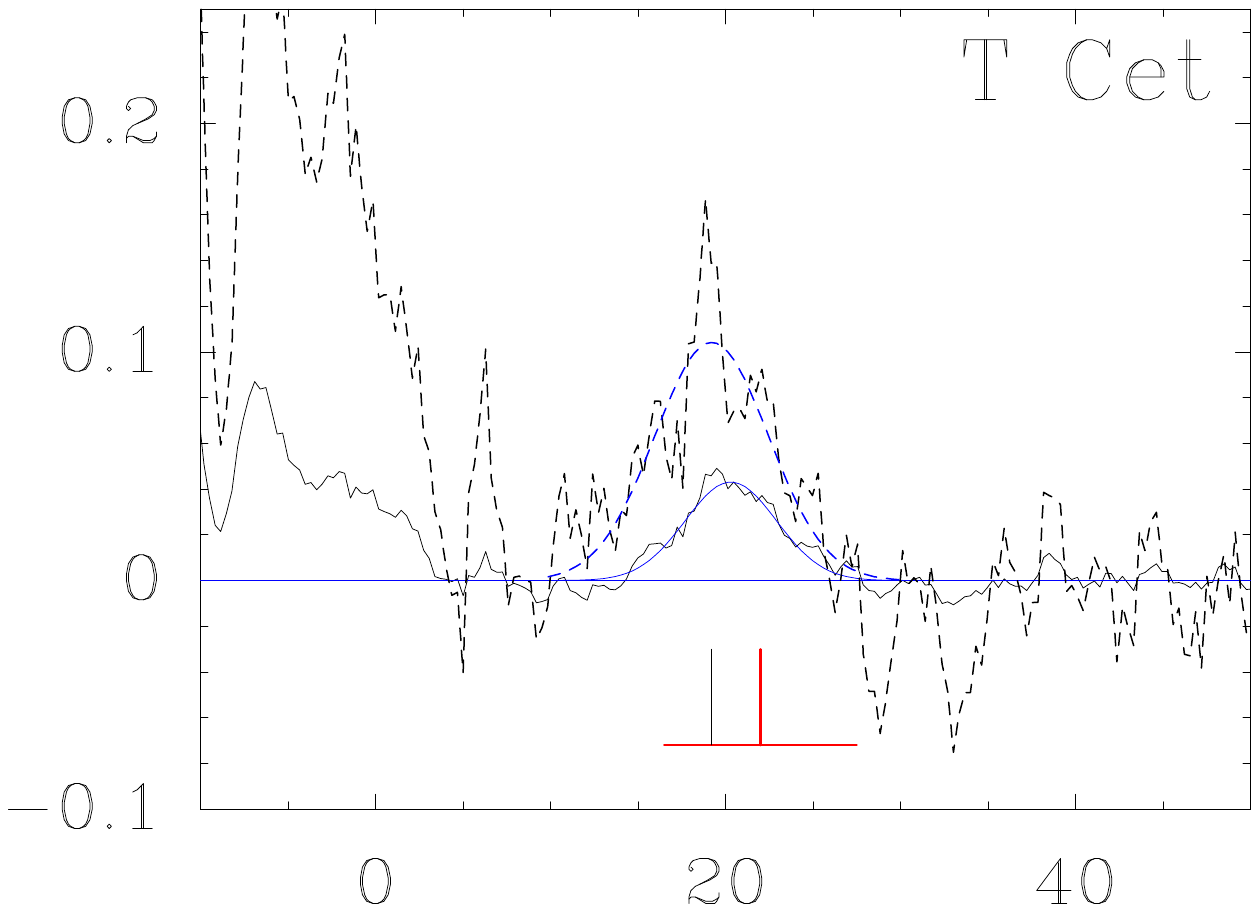}\hspace{-0.7cm}
\includegraphics[width=4.75cm]{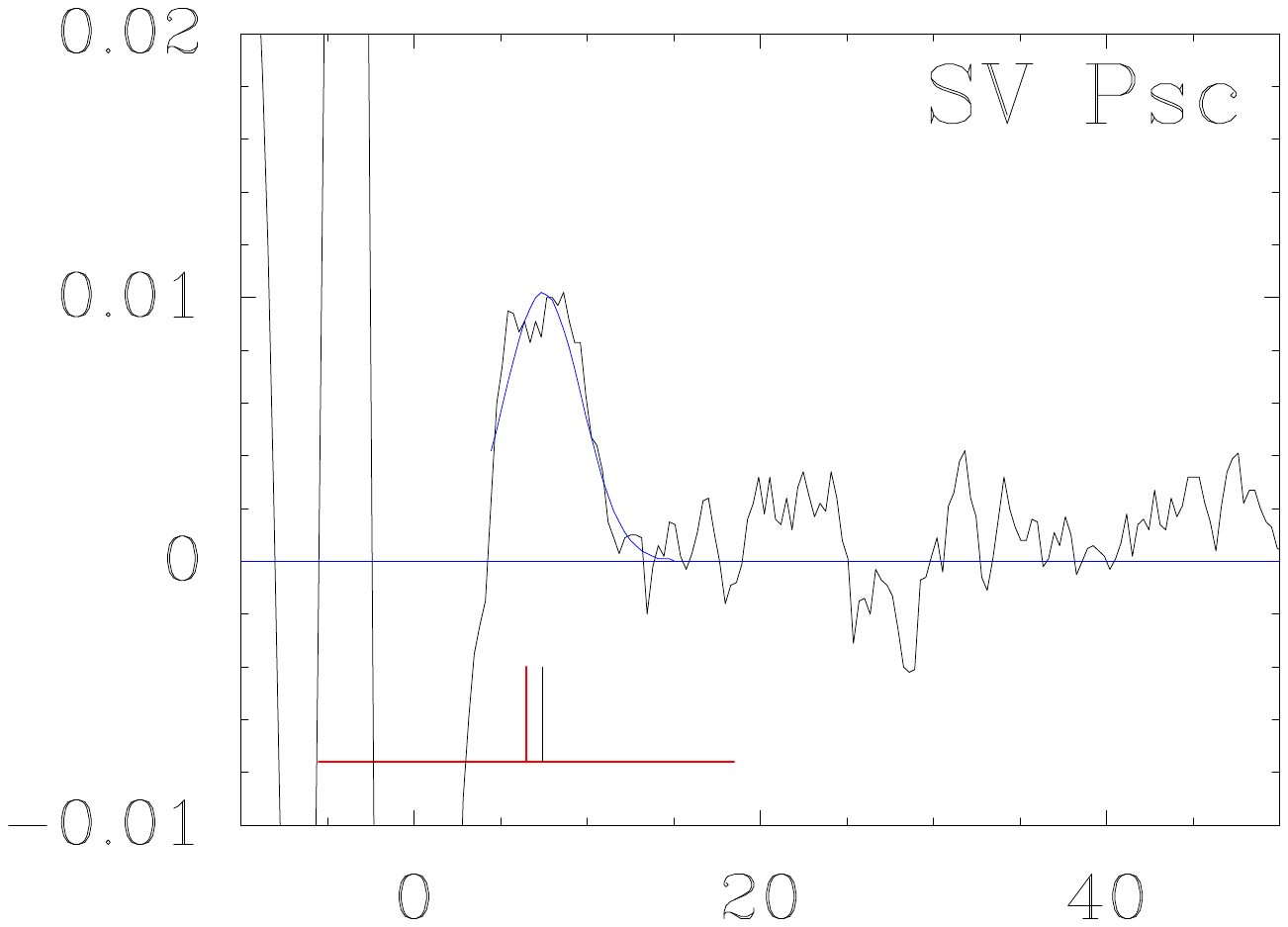}\hspace{-0.7cm}
\includegraphics[width=4.75cm]{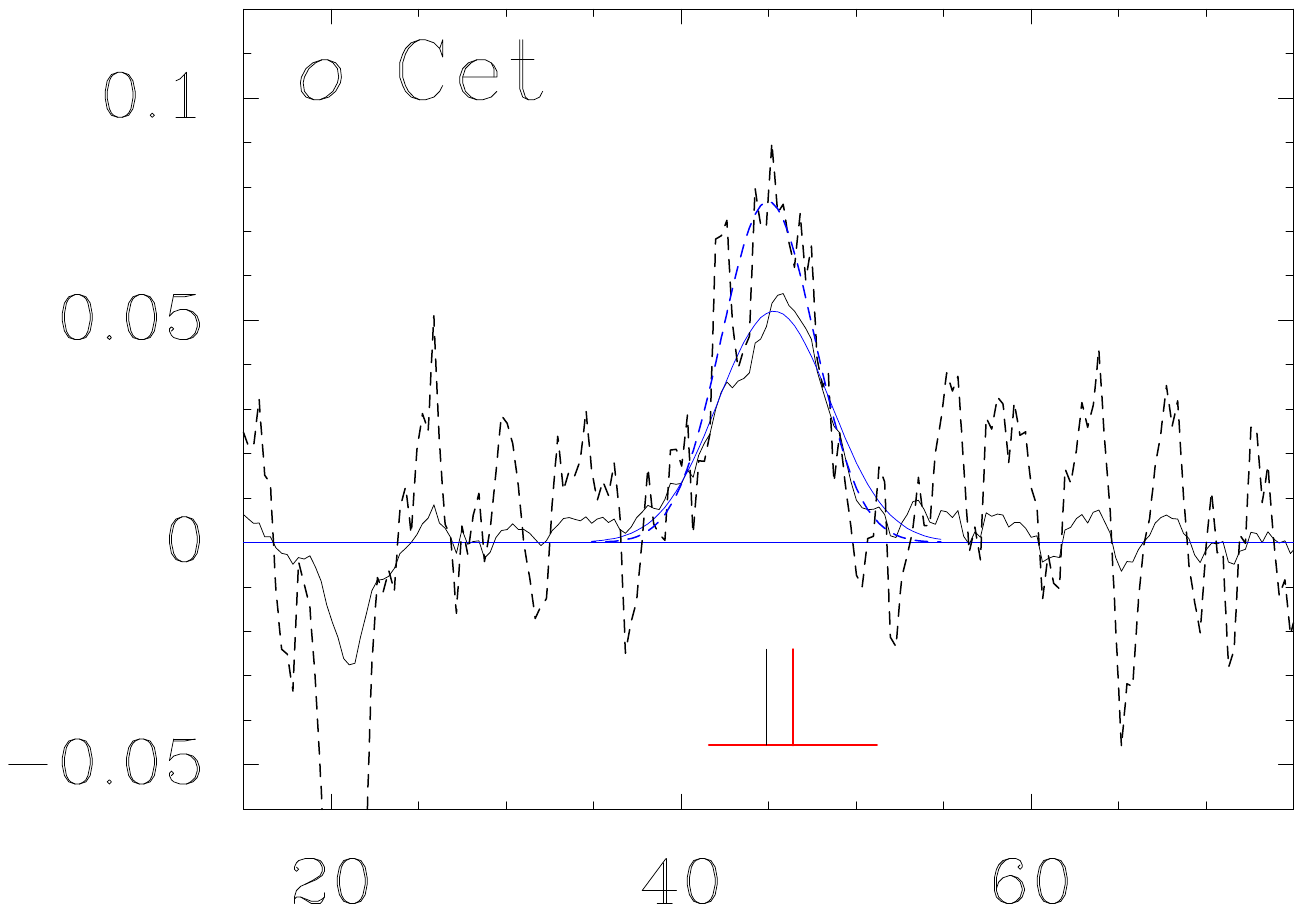}
\\ \vspace{-0.35cm}
\includegraphics[width=4.75cm]{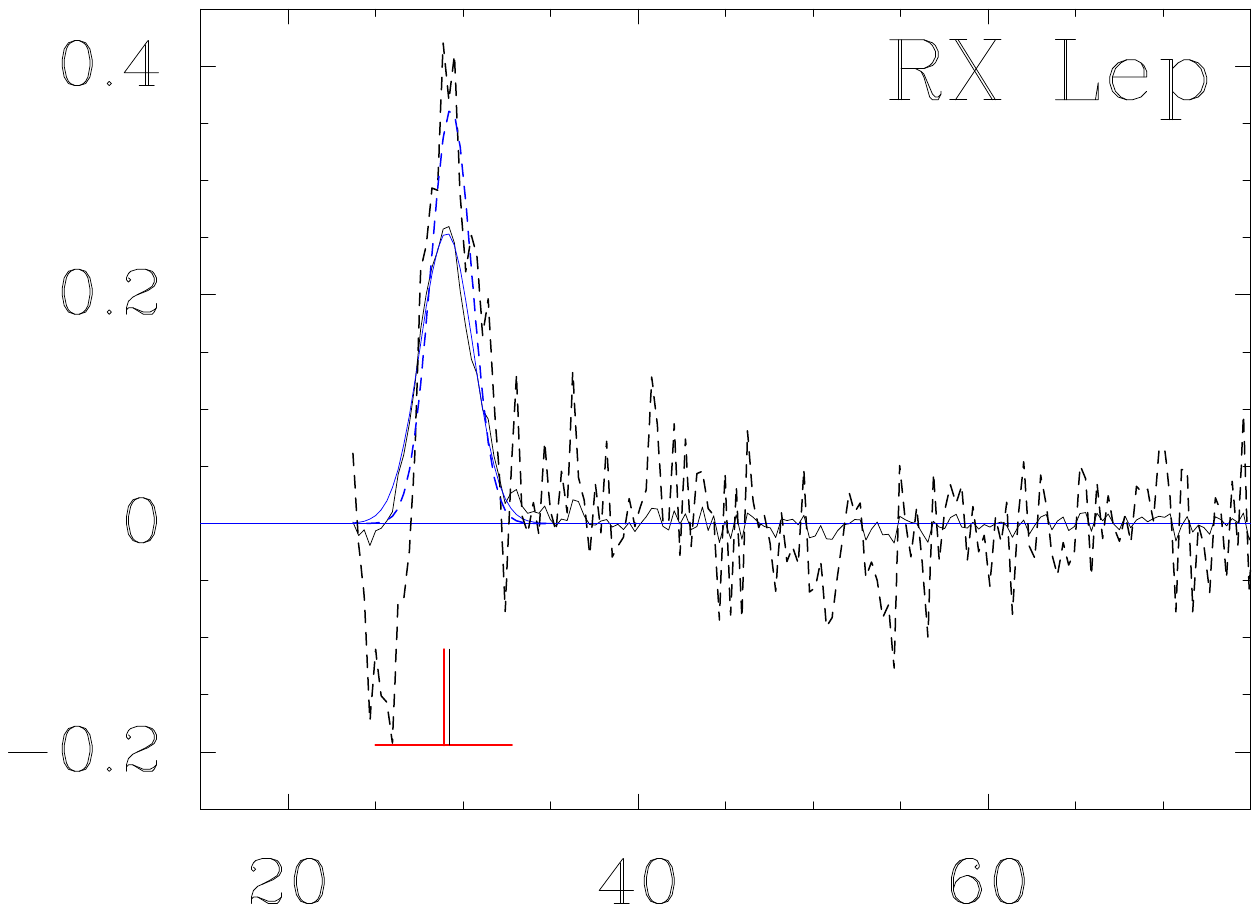}\hspace{-0.7cm}
\includegraphics[width=4.75cm]{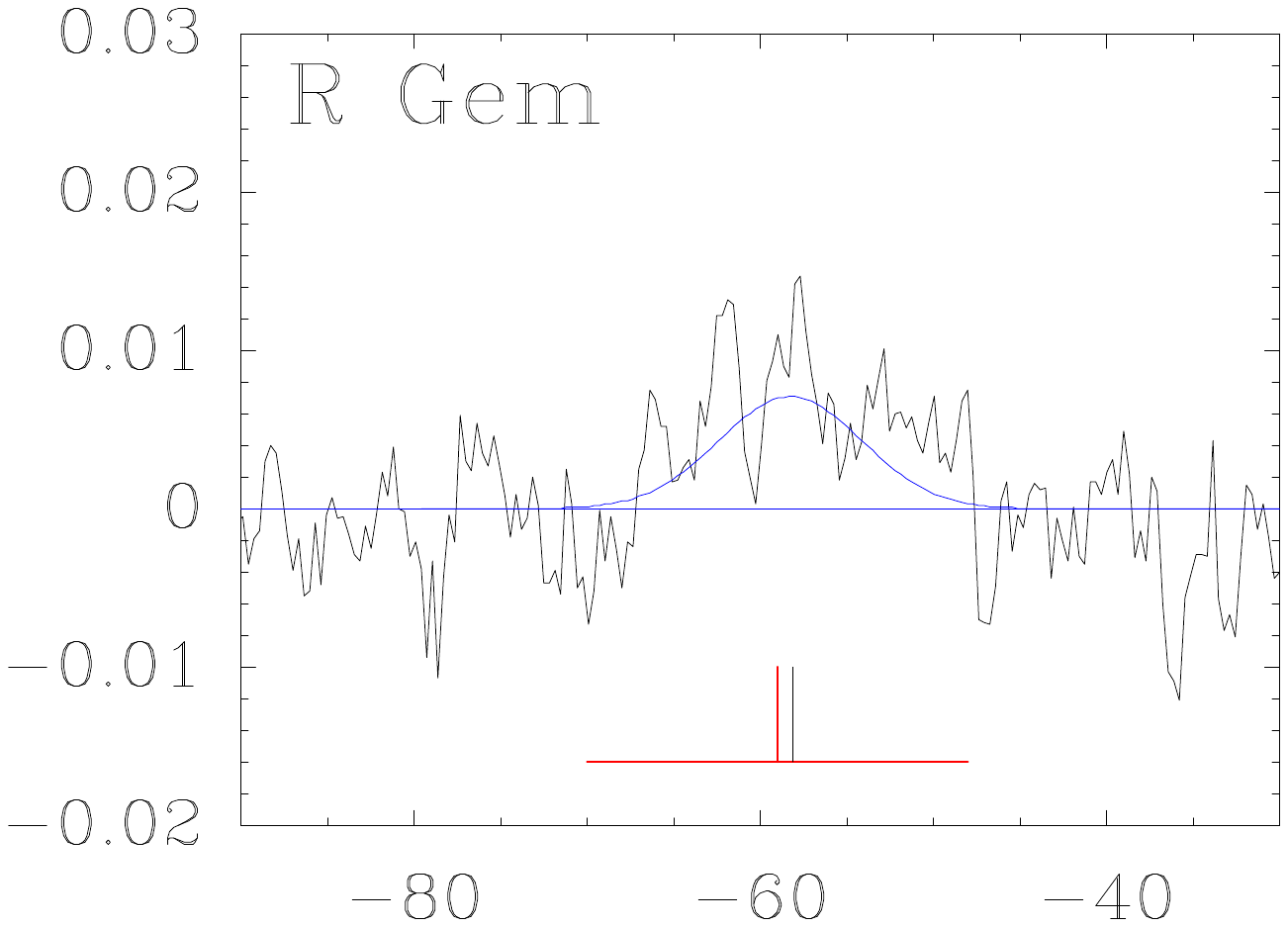}\hspace{-0.7cm}
\includegraphics[width=4.75cm]{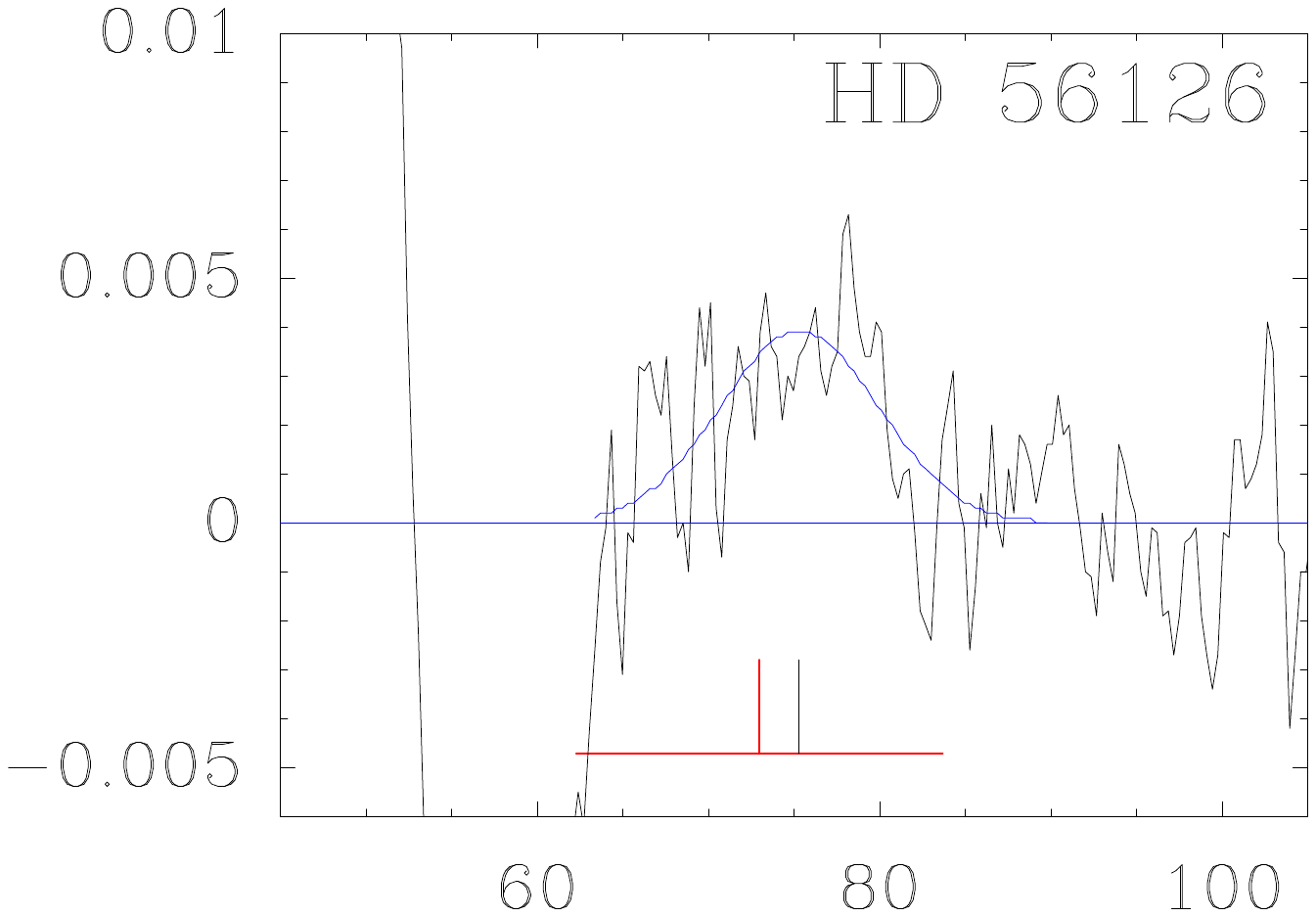}\hspace{-0.7cm}
\includegraphics[width=4.75cm]{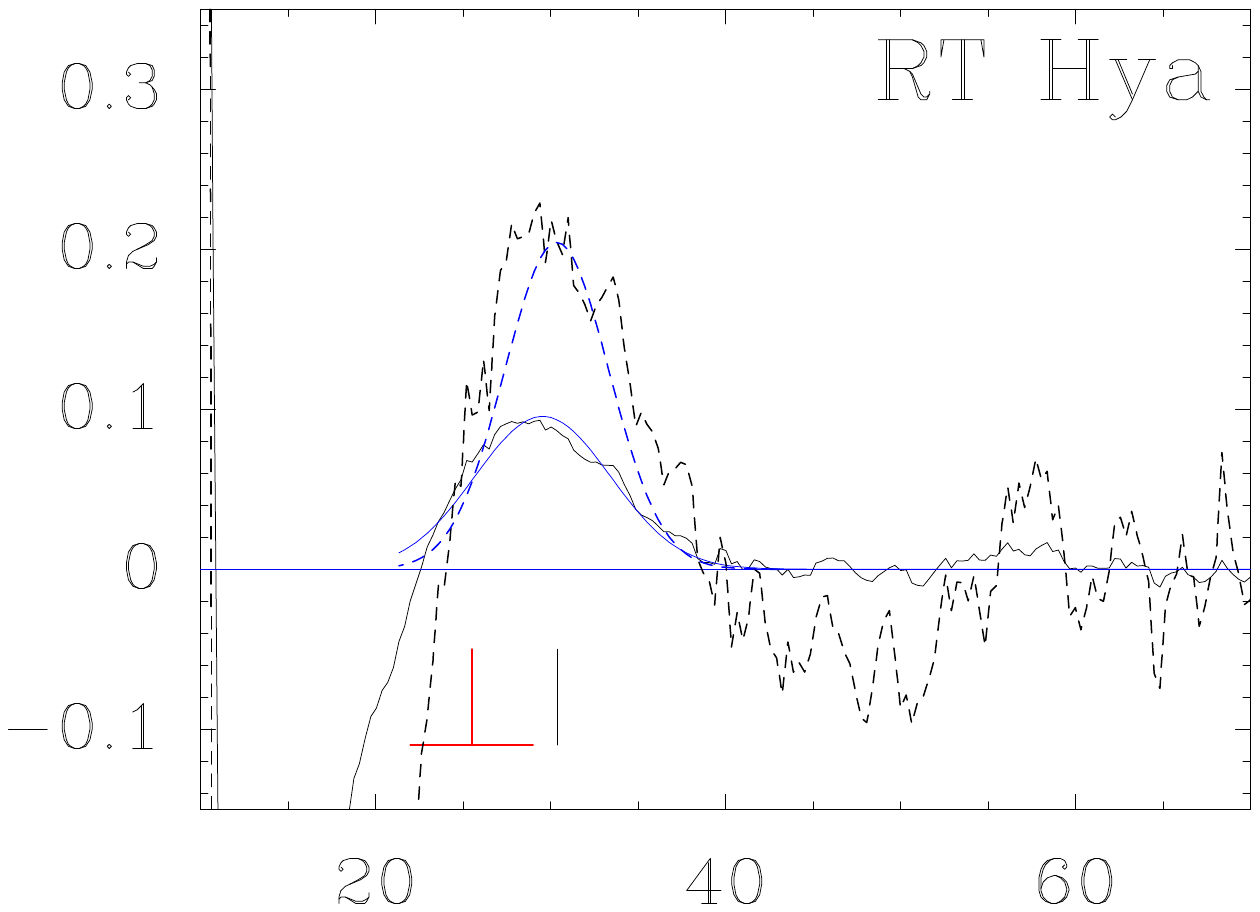}
\\ \vspace{-0.35cm}
\includegraphics[width=4.75cm]{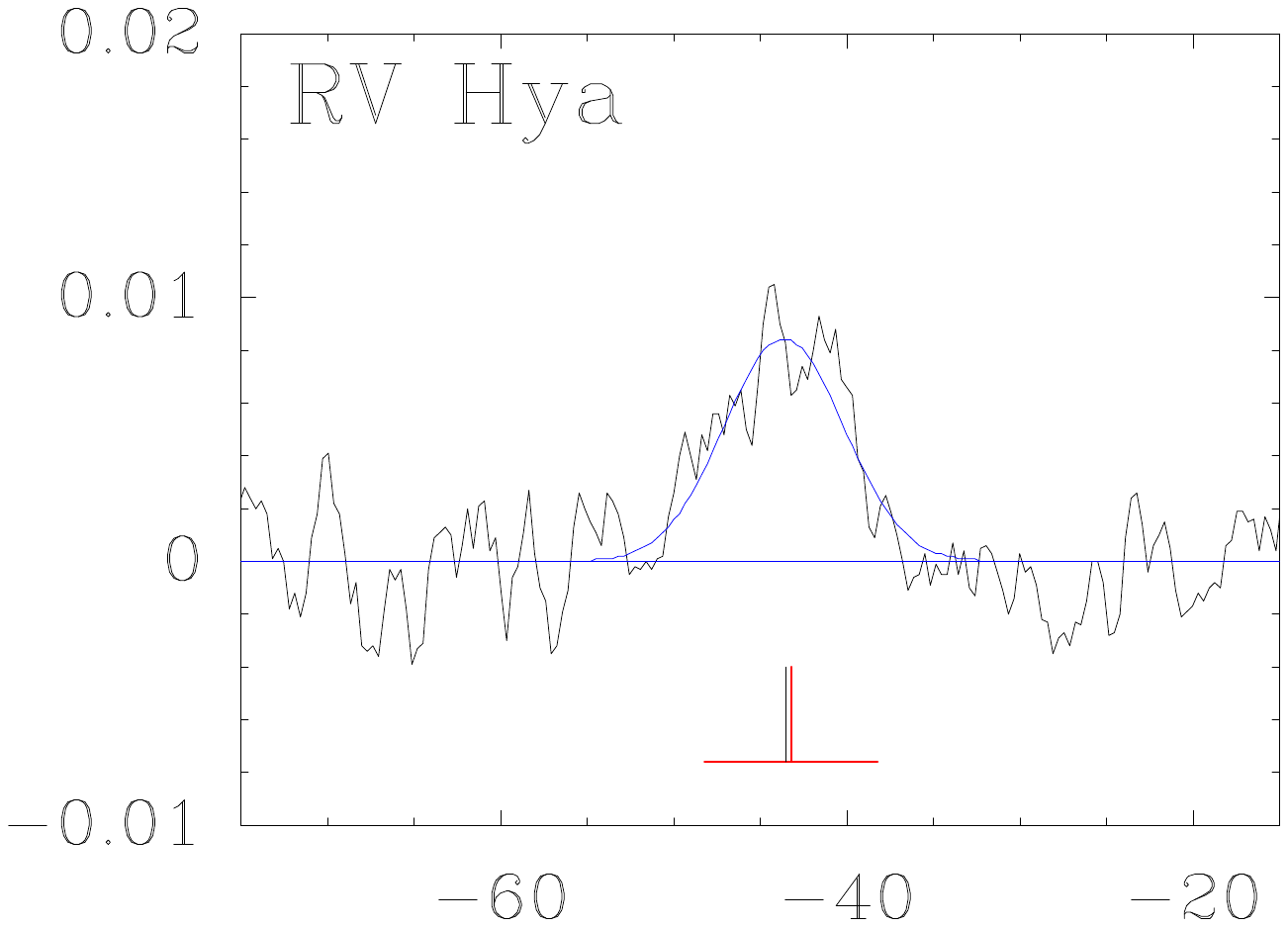}\hspace{-0.7cm}
\includegraphics[width=4.75cm]{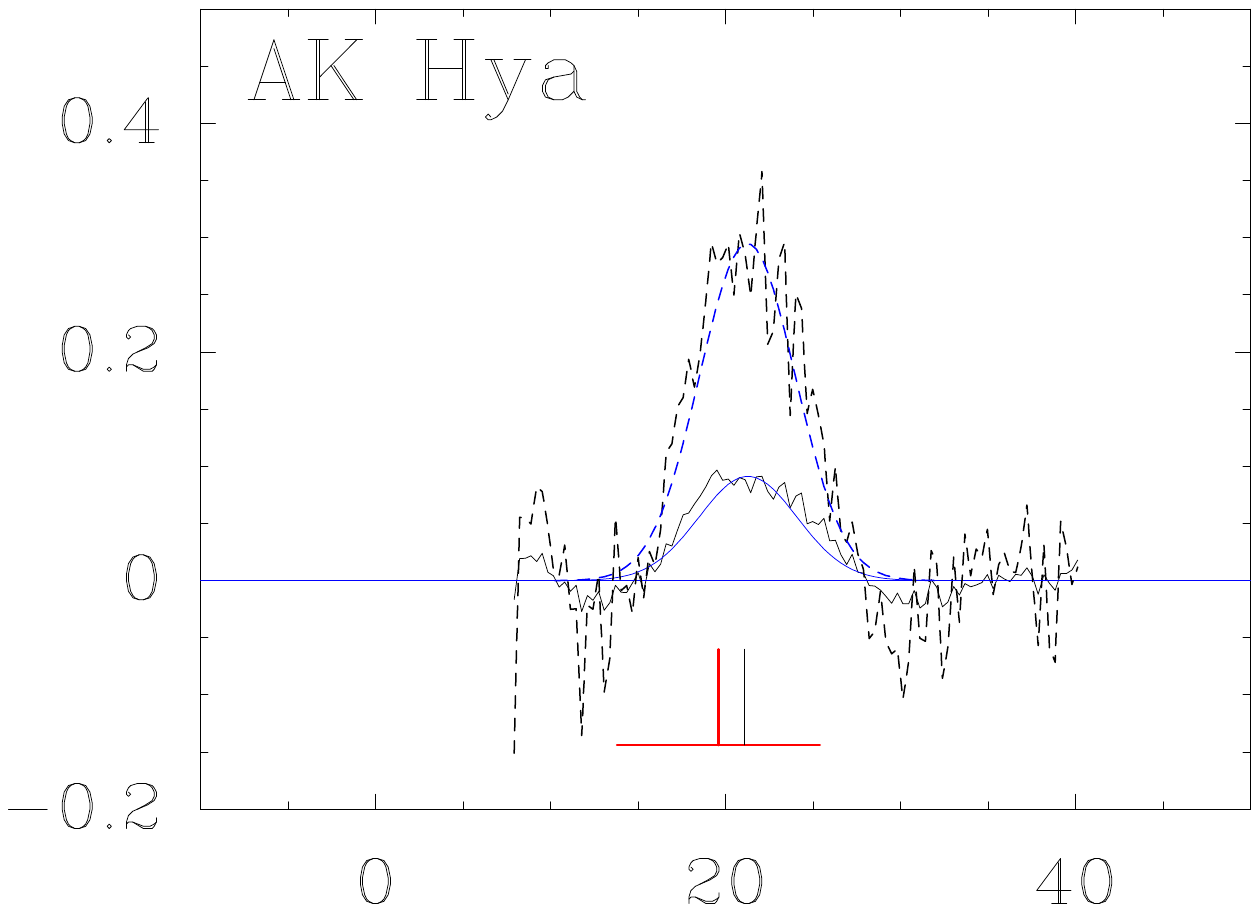}\hspace{-0.7cm}
\includegraphics[width=4.75cm]{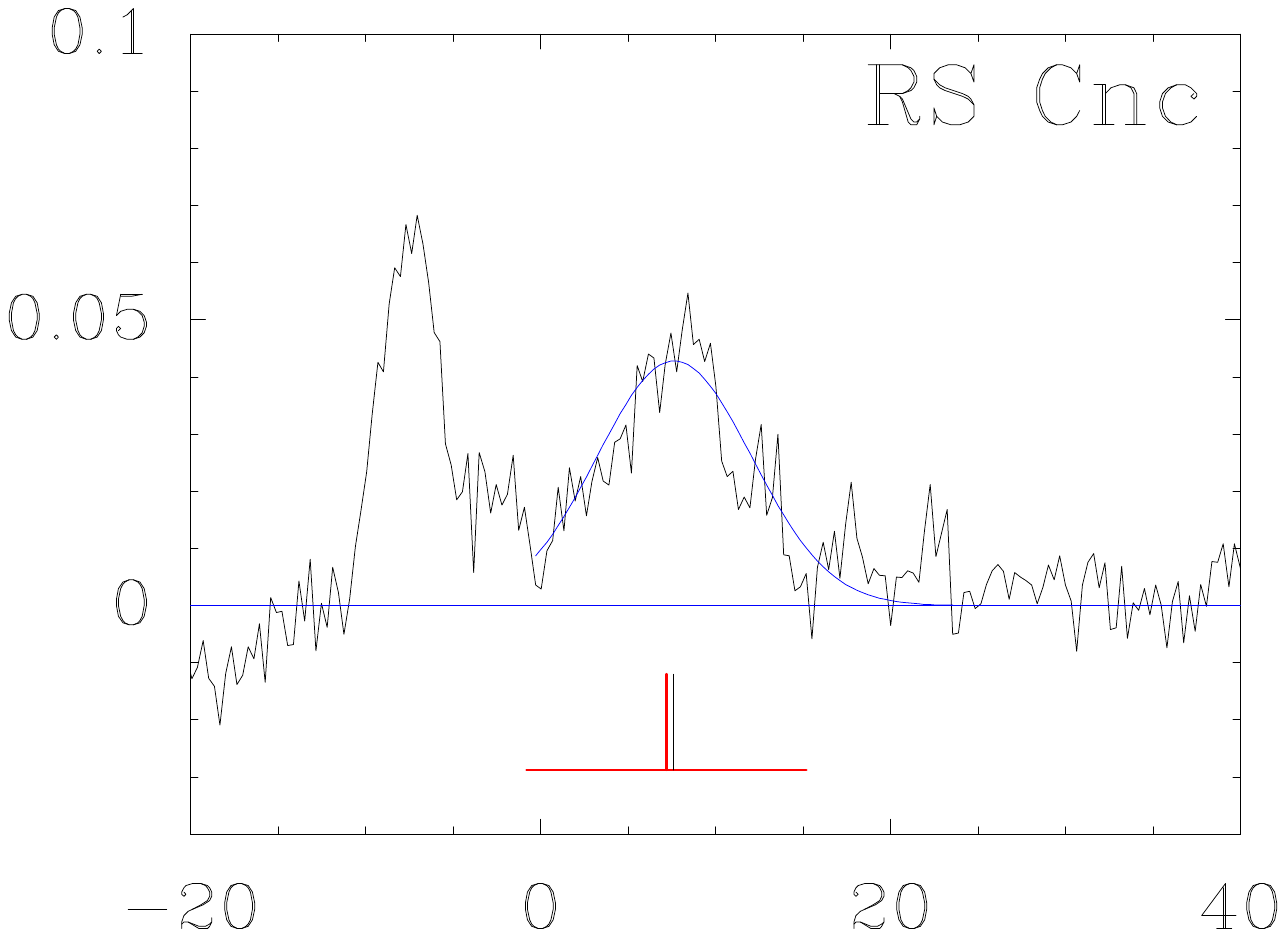}\hspace{-0.7cm}
\includegraphics[width=4.75cm]{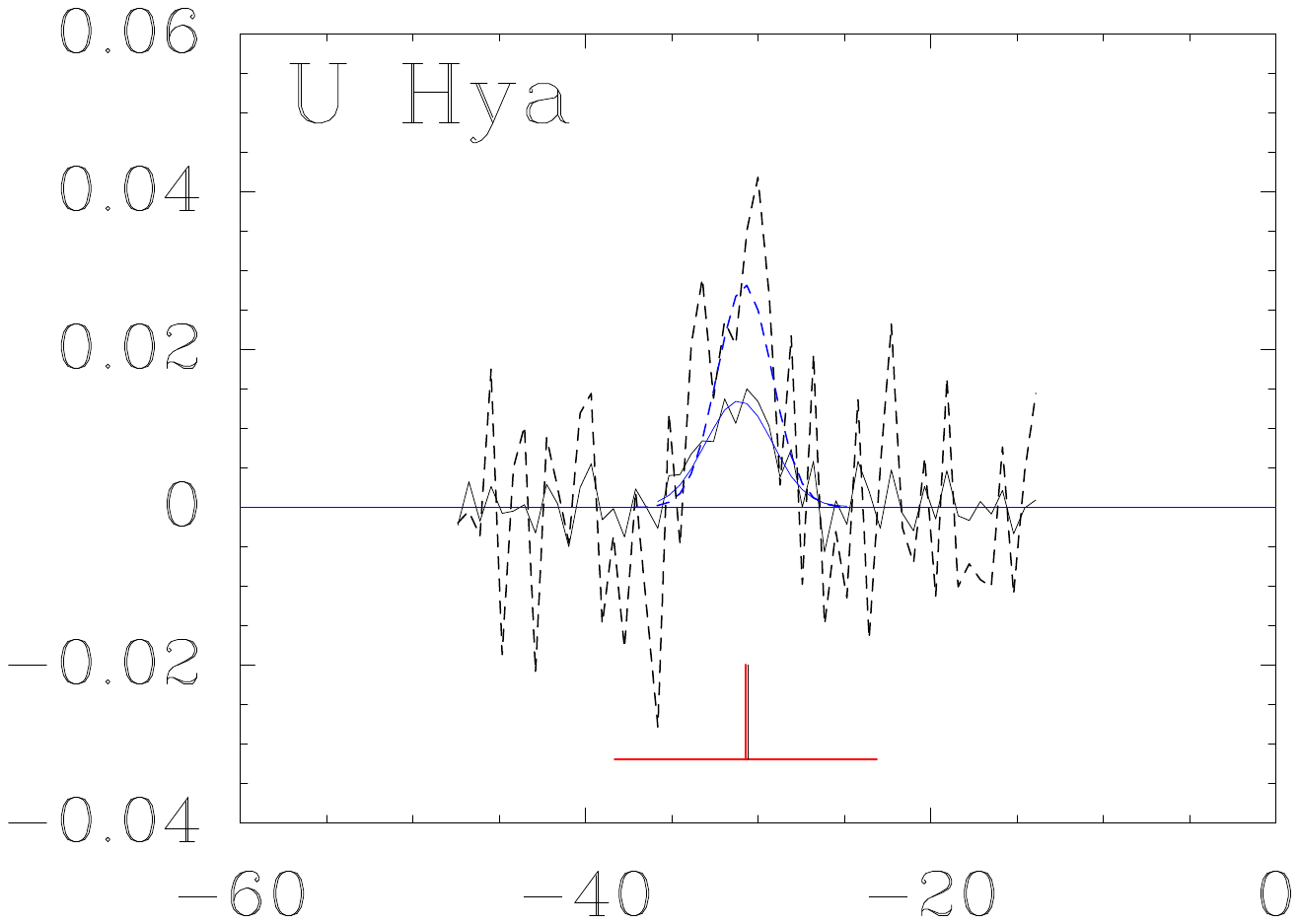}
\\ \vspace{-0.35cm}
\includegraphics[width=4.75cm]{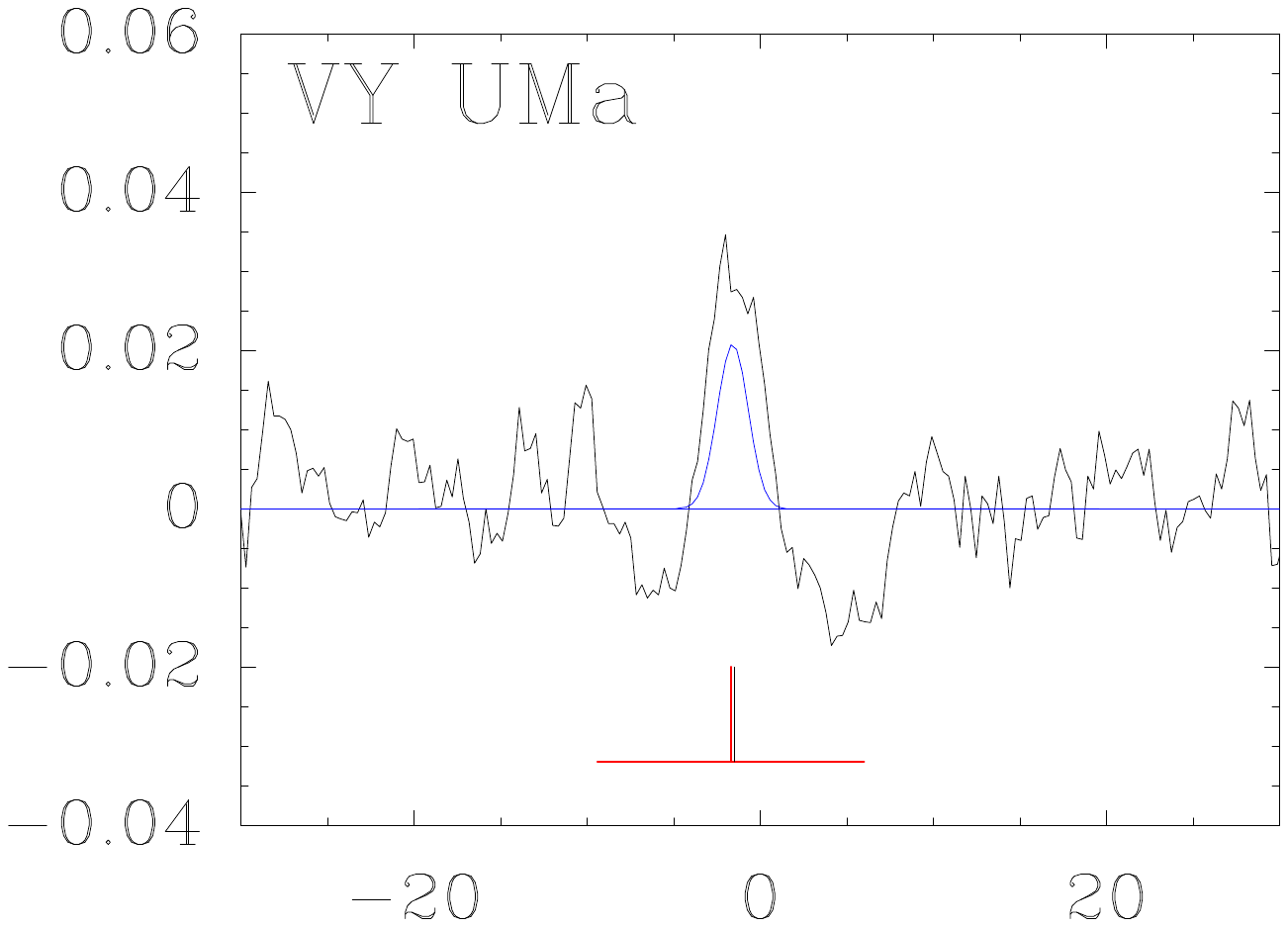}\hspace{-0.7cm}
\includegraphics[width=4.75cm]{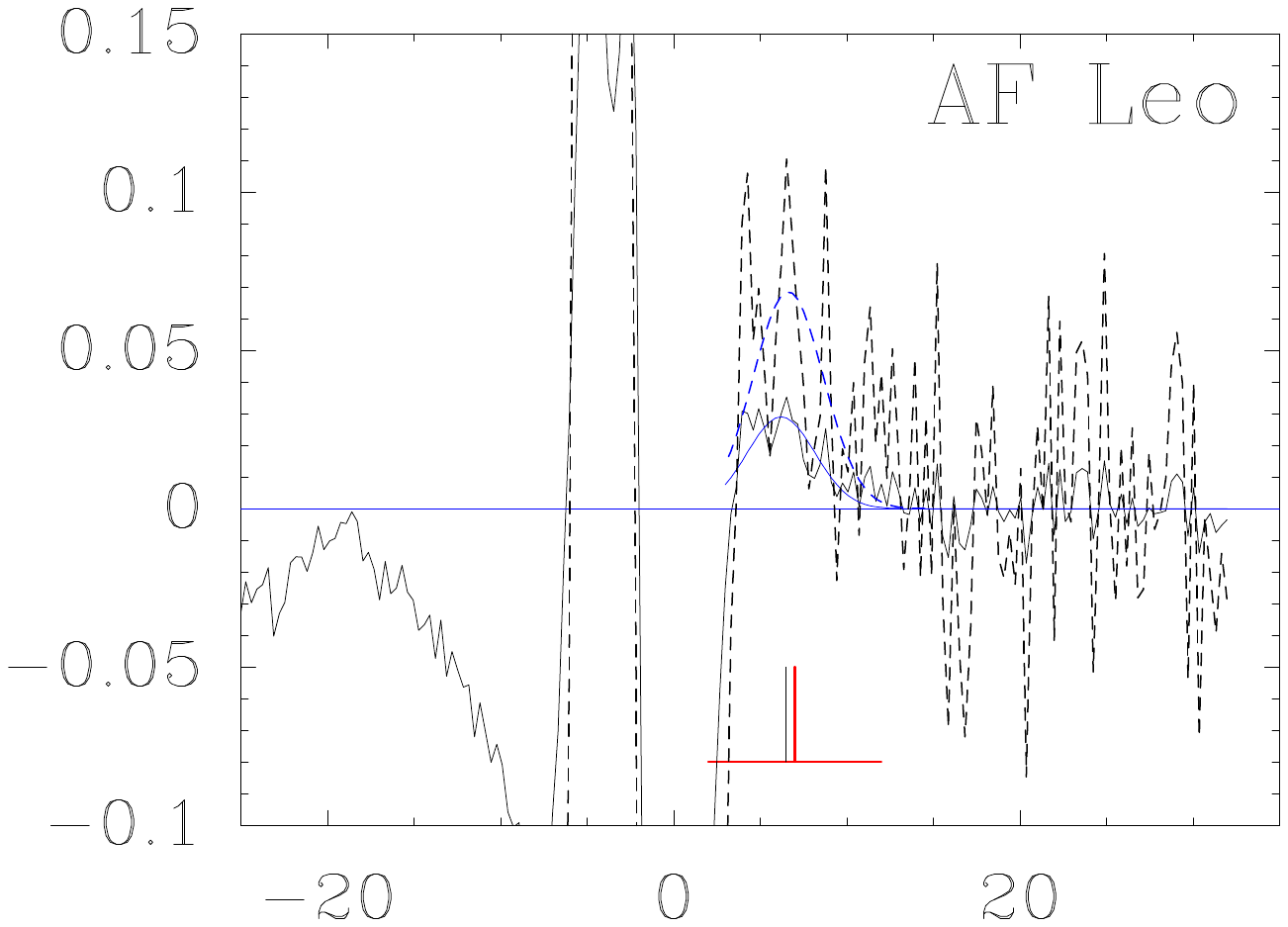}\hspace{-0.7cm}
\includegraphics[width=4.75cm]{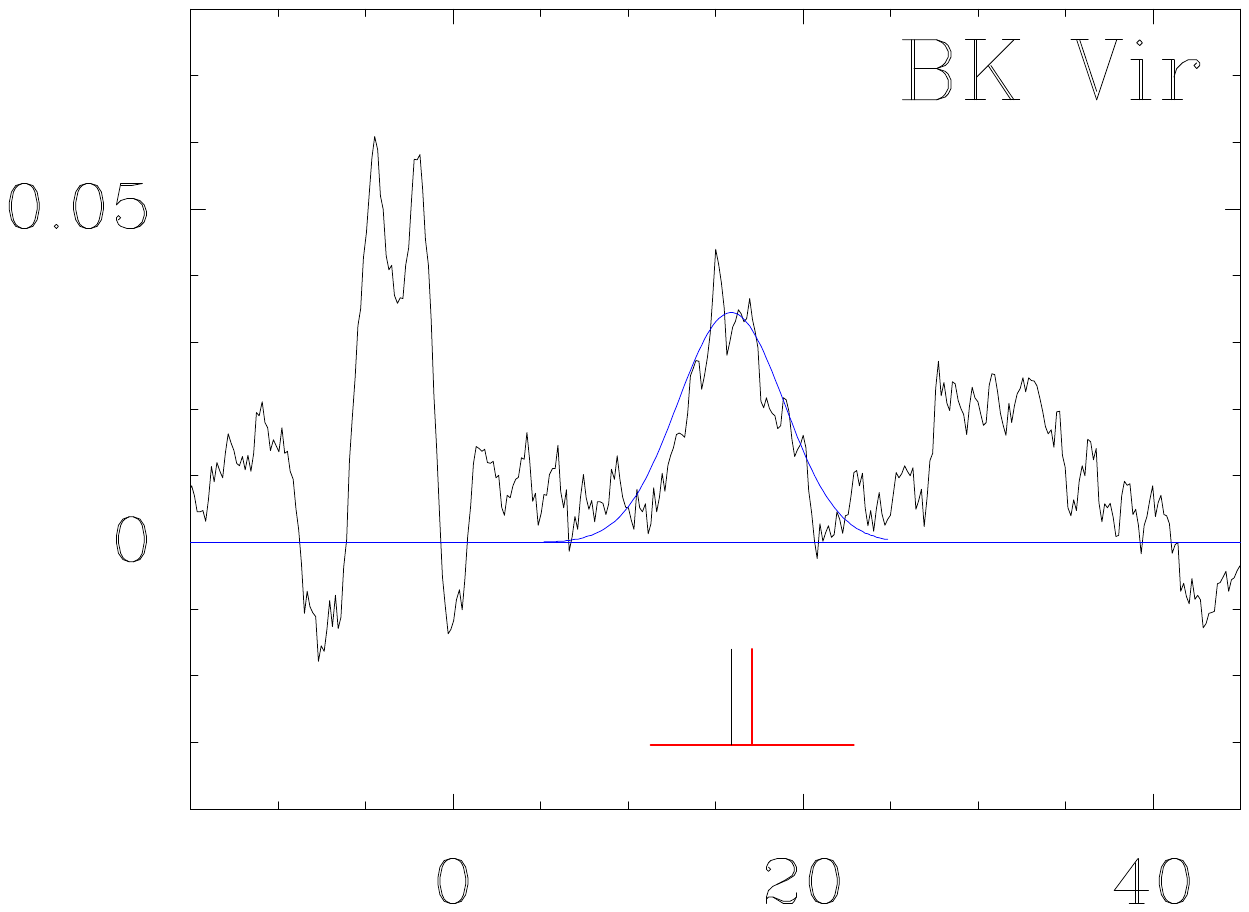}\hspace{-0.7cm}
\includegraphics[width=4.75cm]{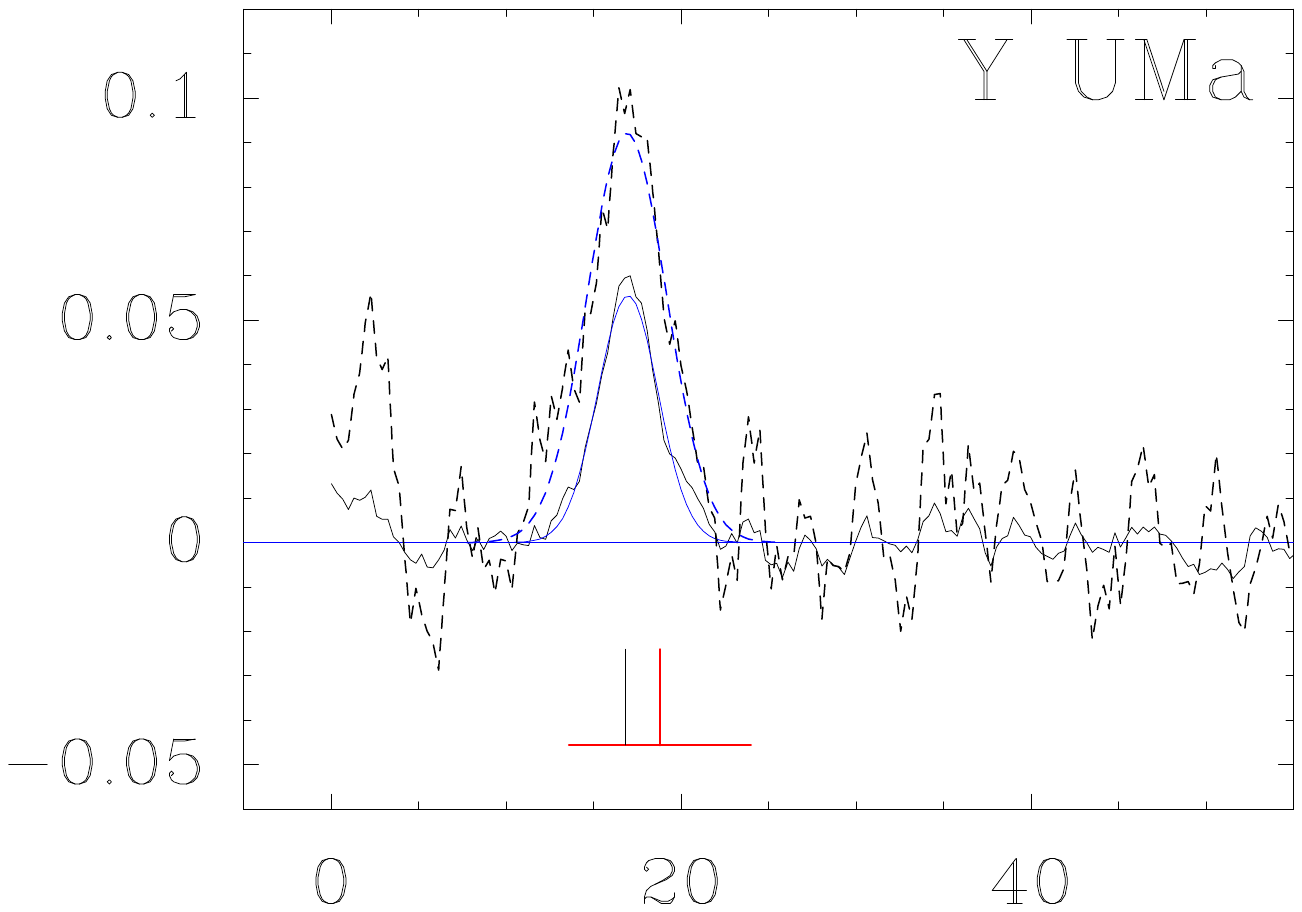}
\\ \vspace{-0.35cm}
\includegraphics[width=4.75cm]{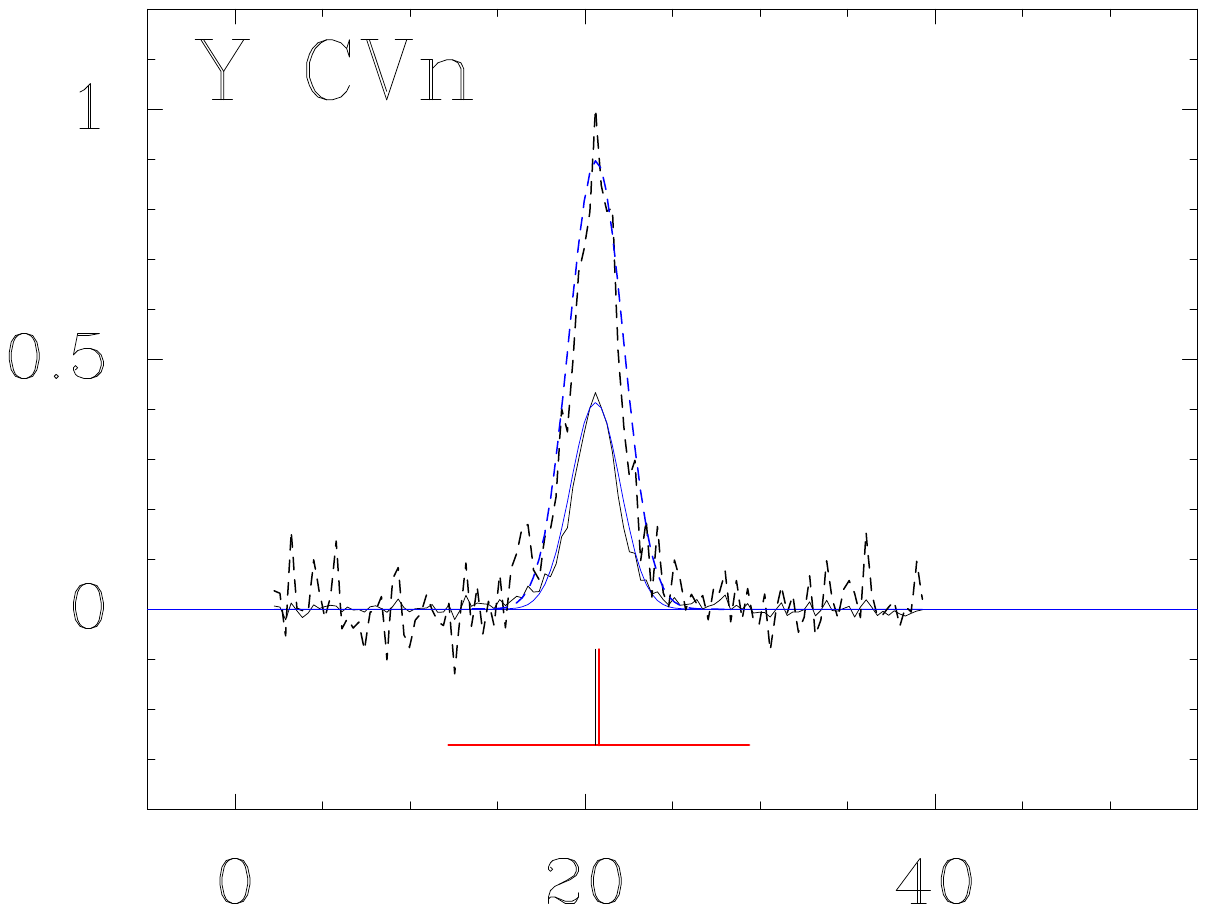}\hspace{-0.7cm}
\includegraphics[width=4.75cm]{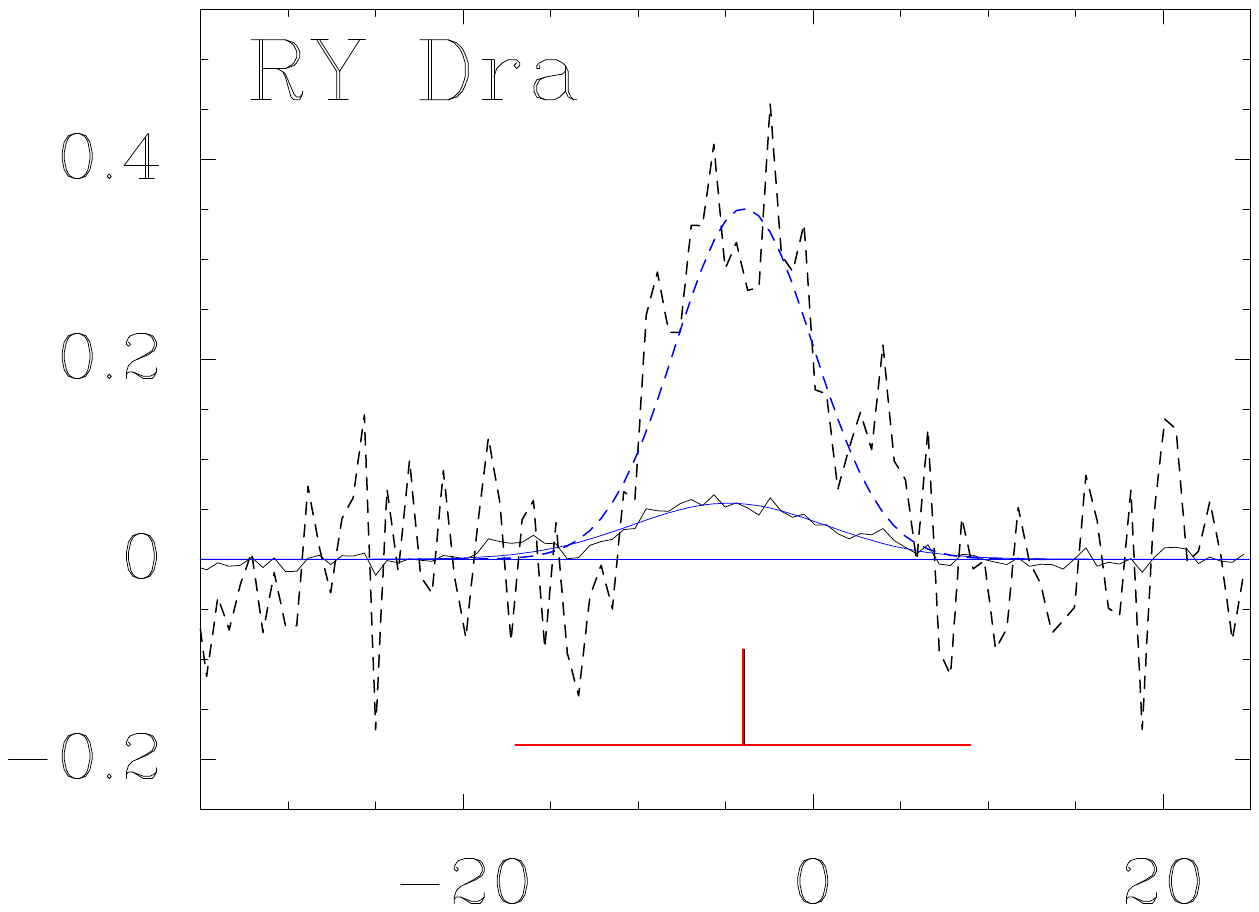}\hspace{-0.7cm}
\includegraphics[width=4.75cm]{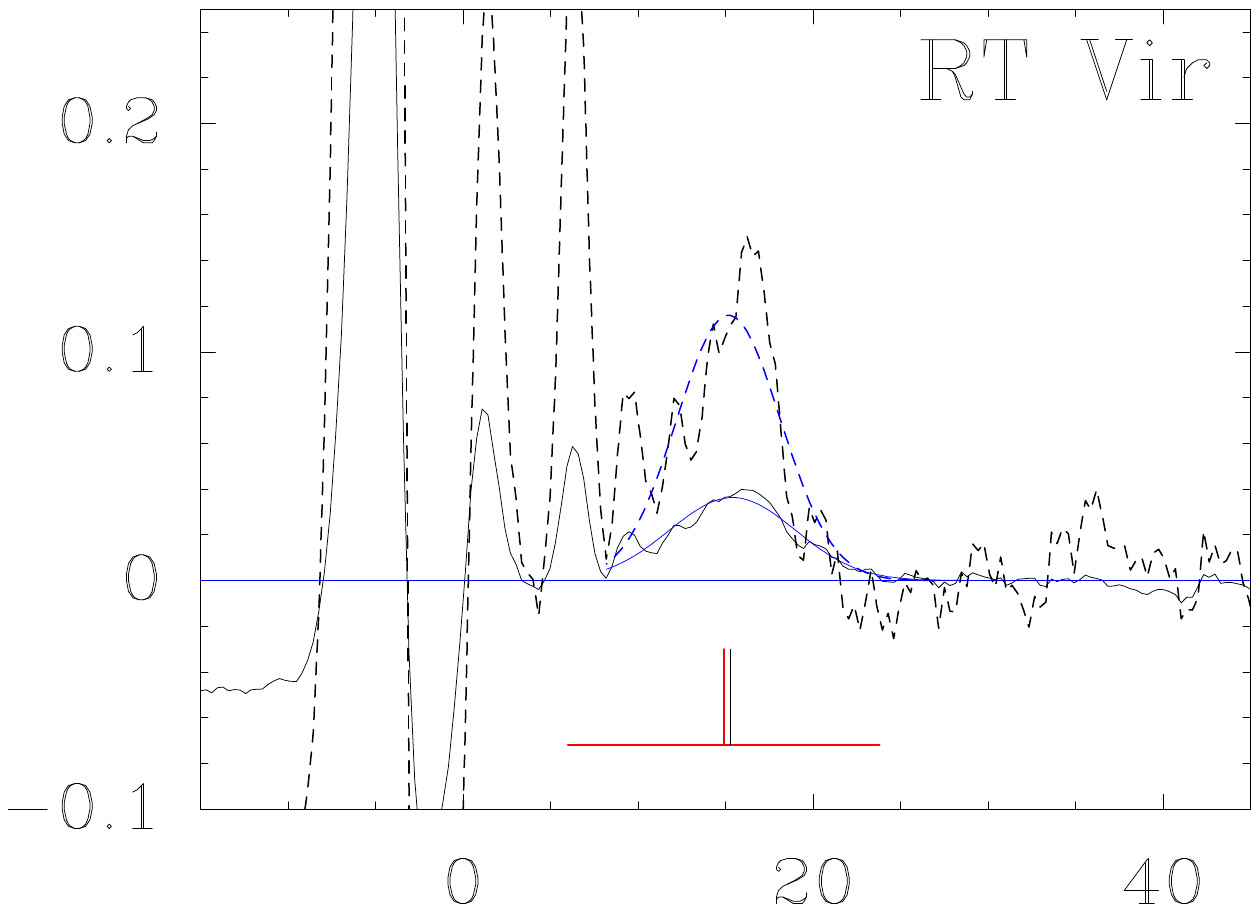}\hspace{-0.7cm}
\includegraphics[width=4.75cm]{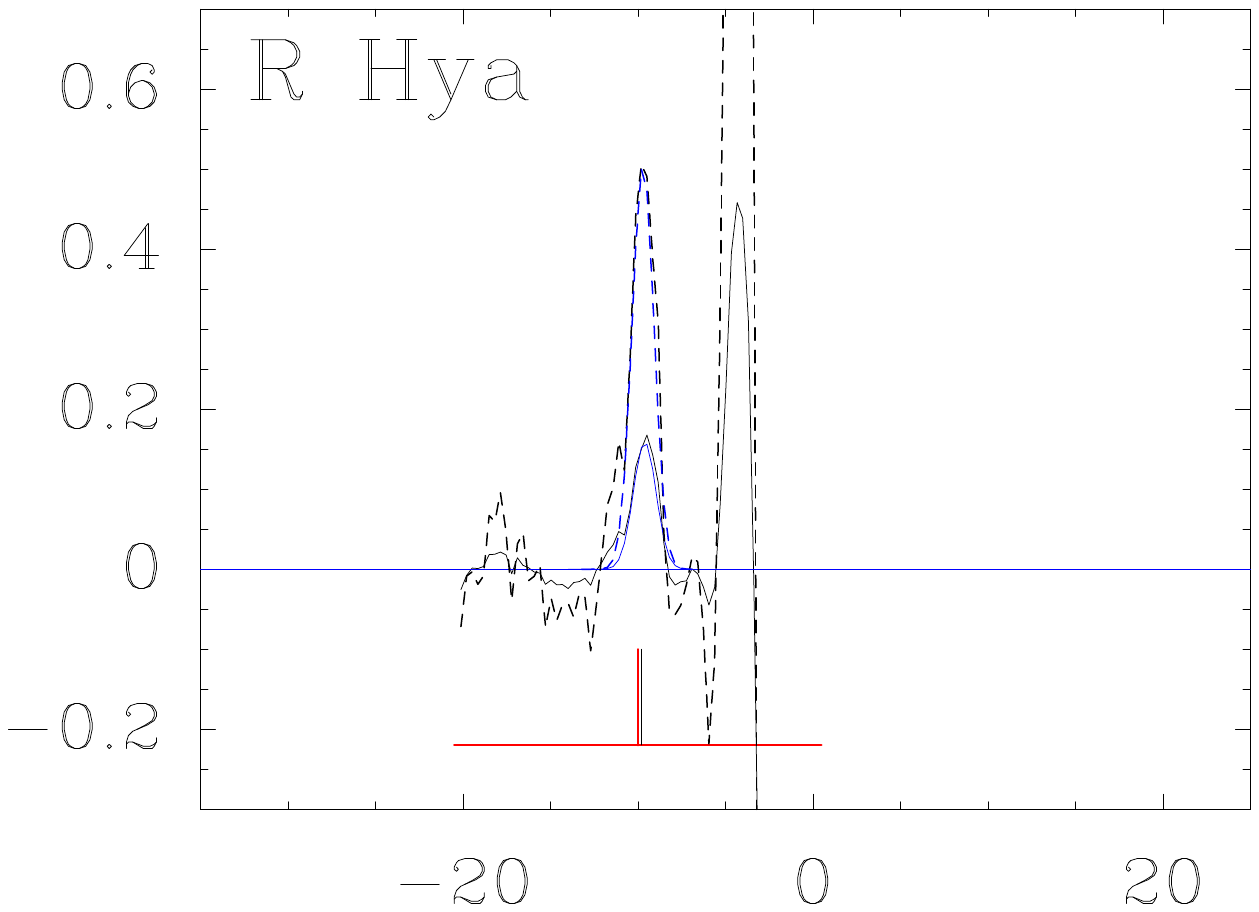}
\\ \vspace{-0.35cm}
\includegraphics[width=4.75cm]{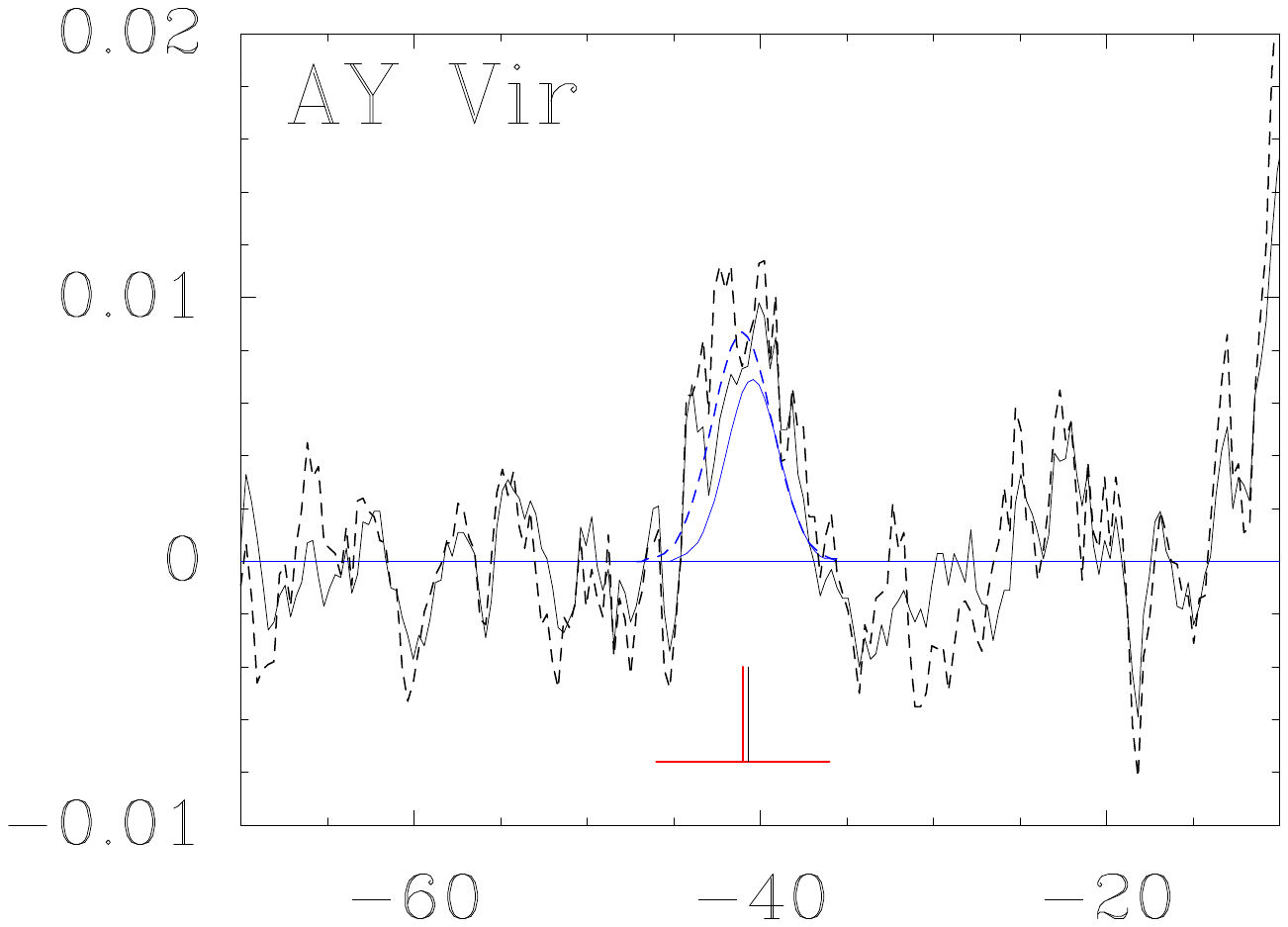}\hspace{-0.7cm}
\includegraphics[width=4.75cm]{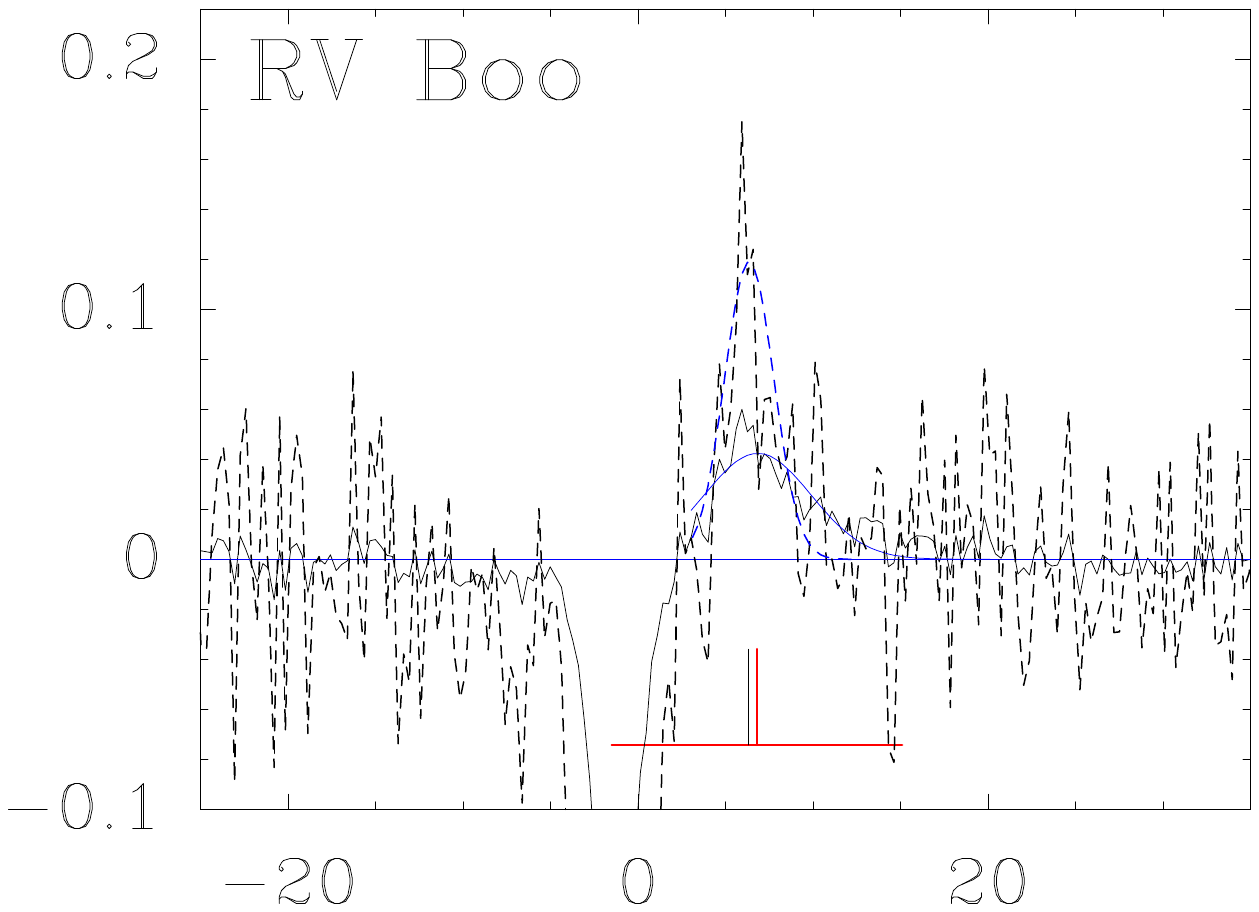}\hspace{-0.7cm}
\includegraphics[width=4.75cm]{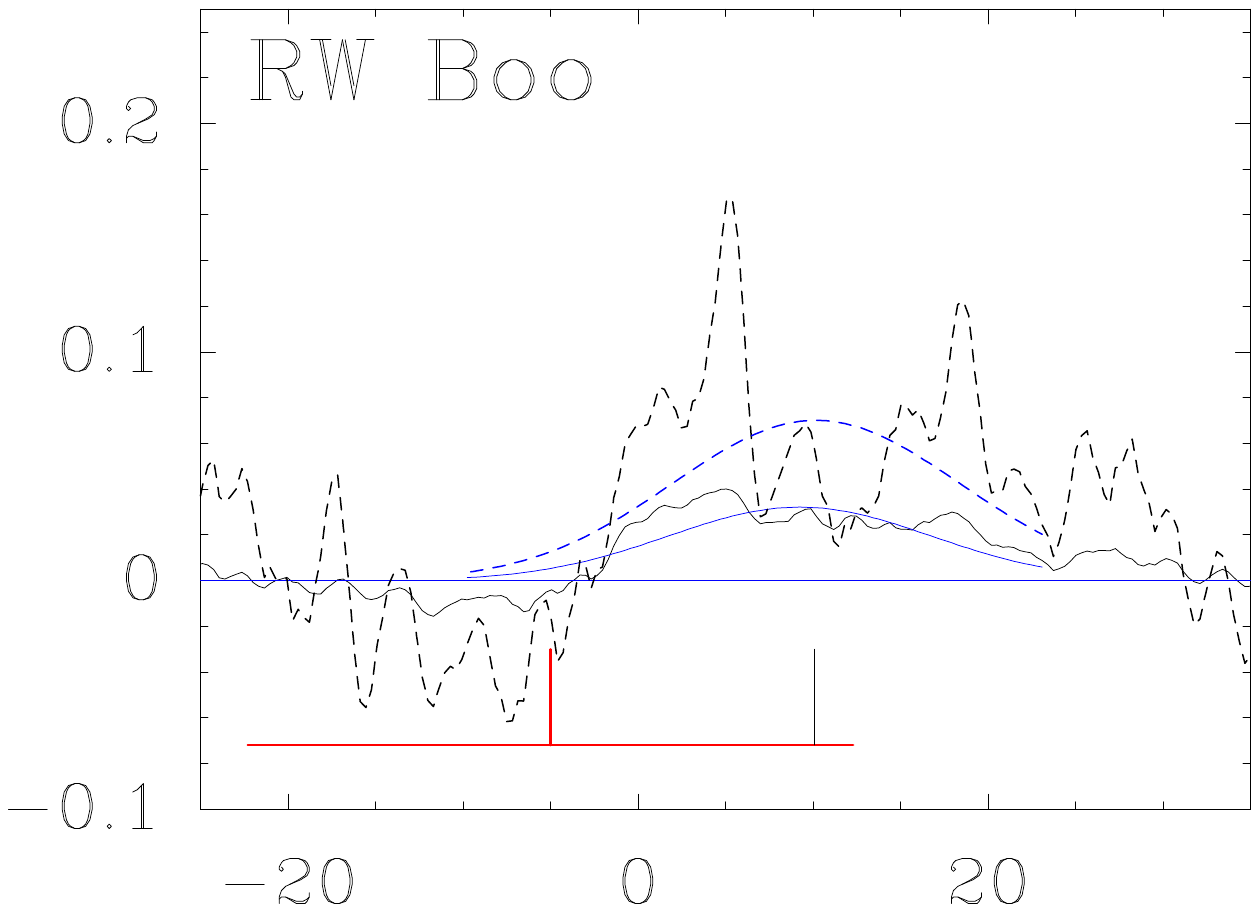}\hspace{-0.7cm}
\includegraphics[width=4.75cm]{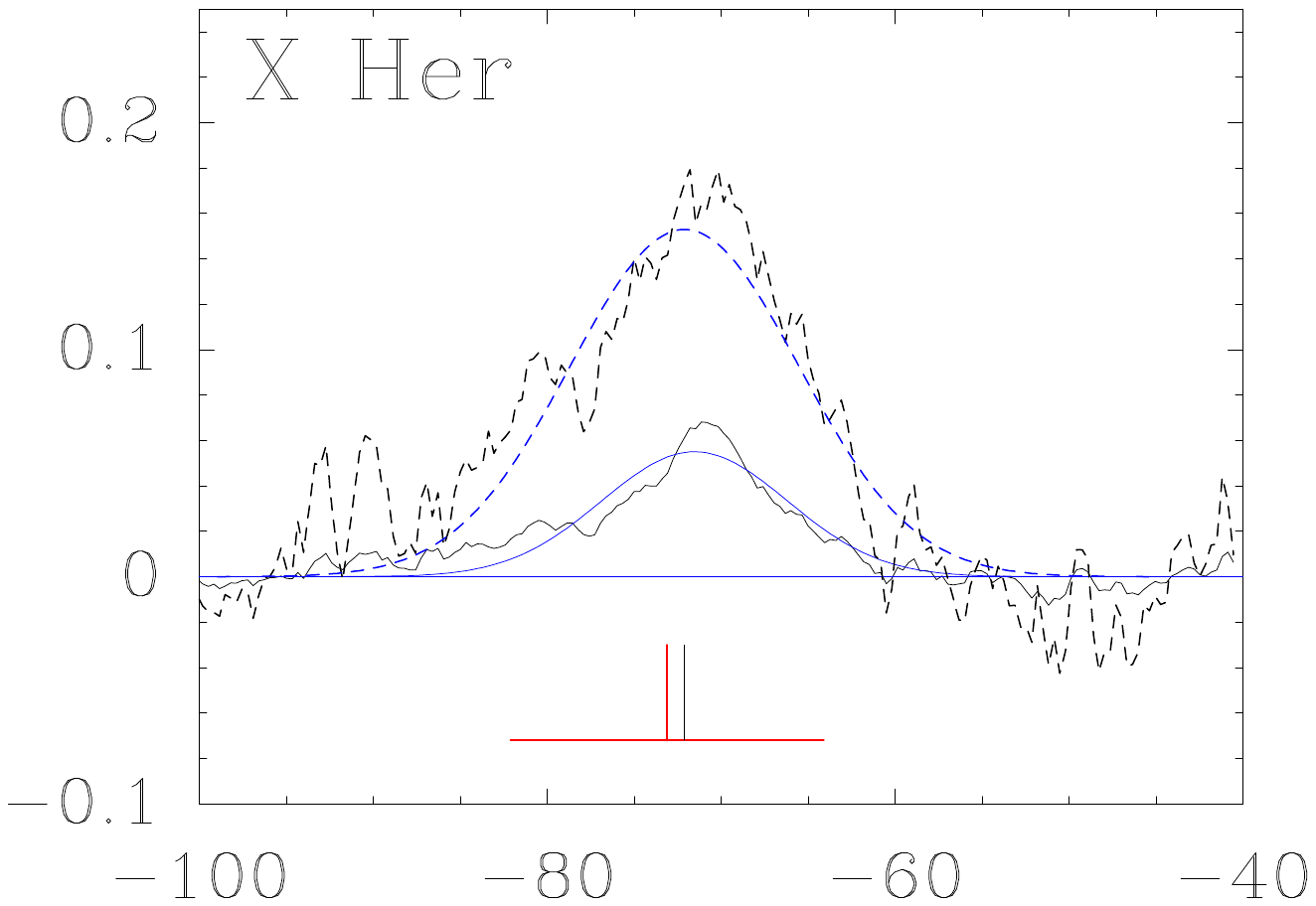}
\\ \vspace{-0.35cm}
\includegraphics[width=4.75cm]{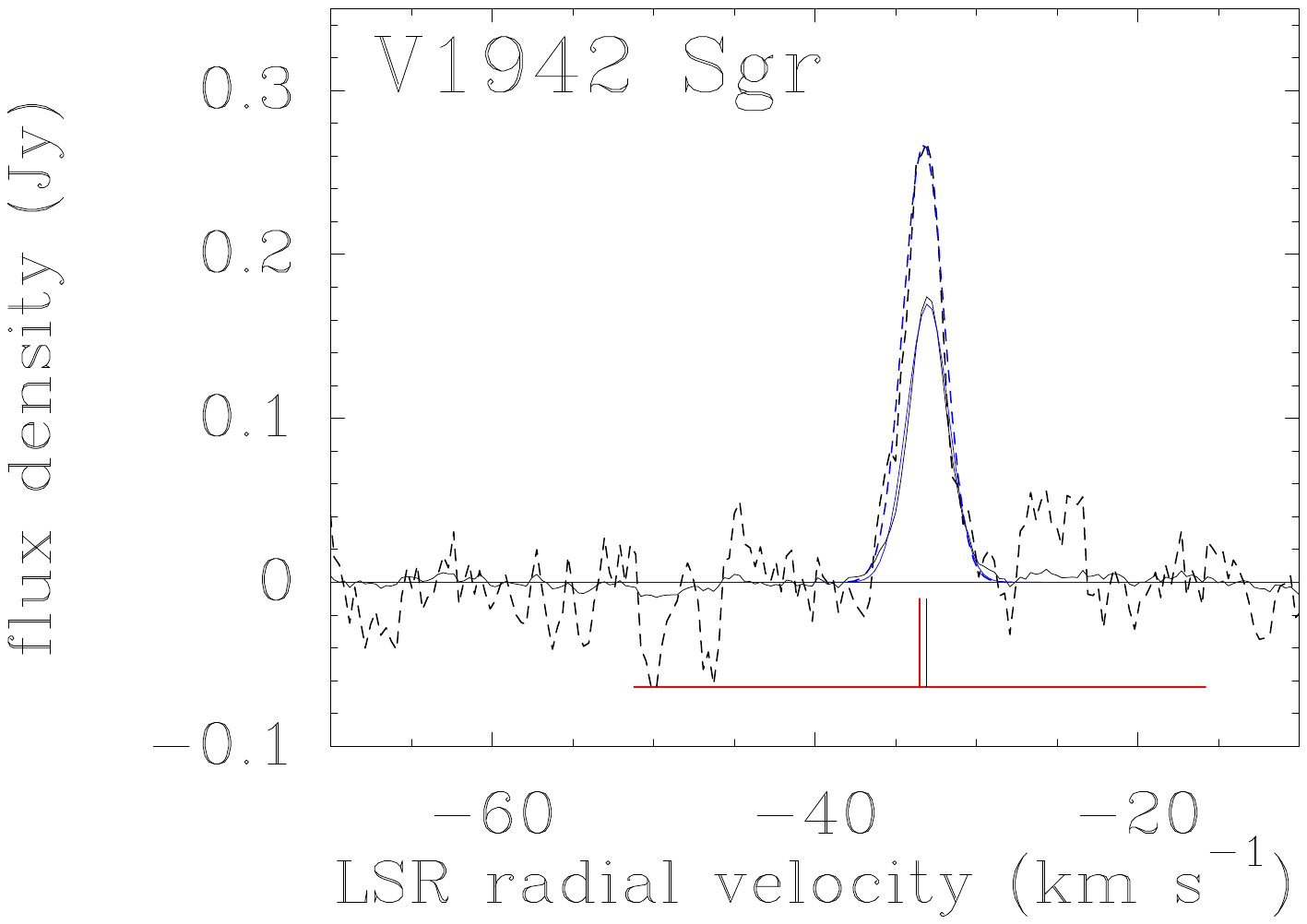}\hspace{-0.7cm}
\includegraphics[width=4.75cm]{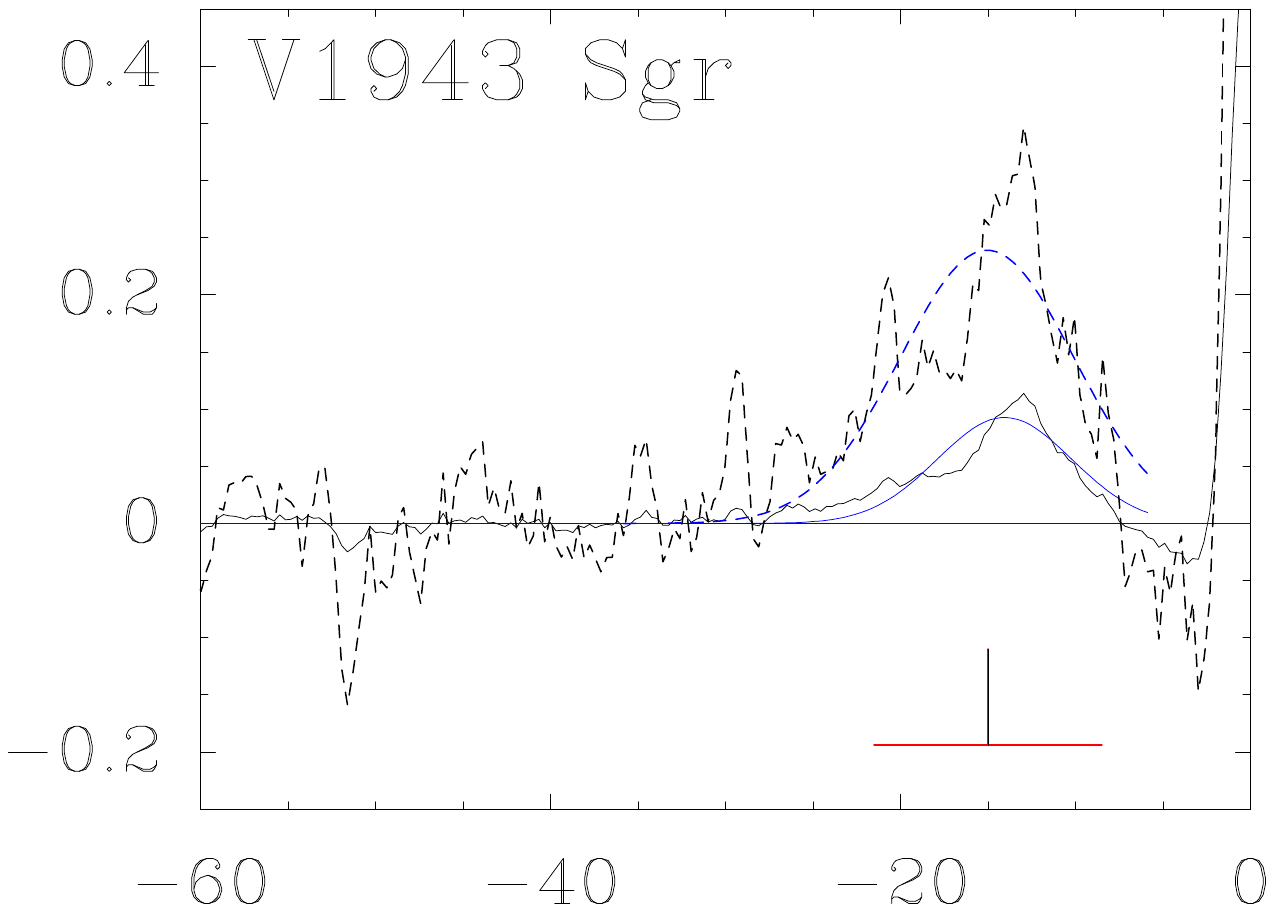}\hspace{-0.7cm}
\includegraphics[width=4.75cm]{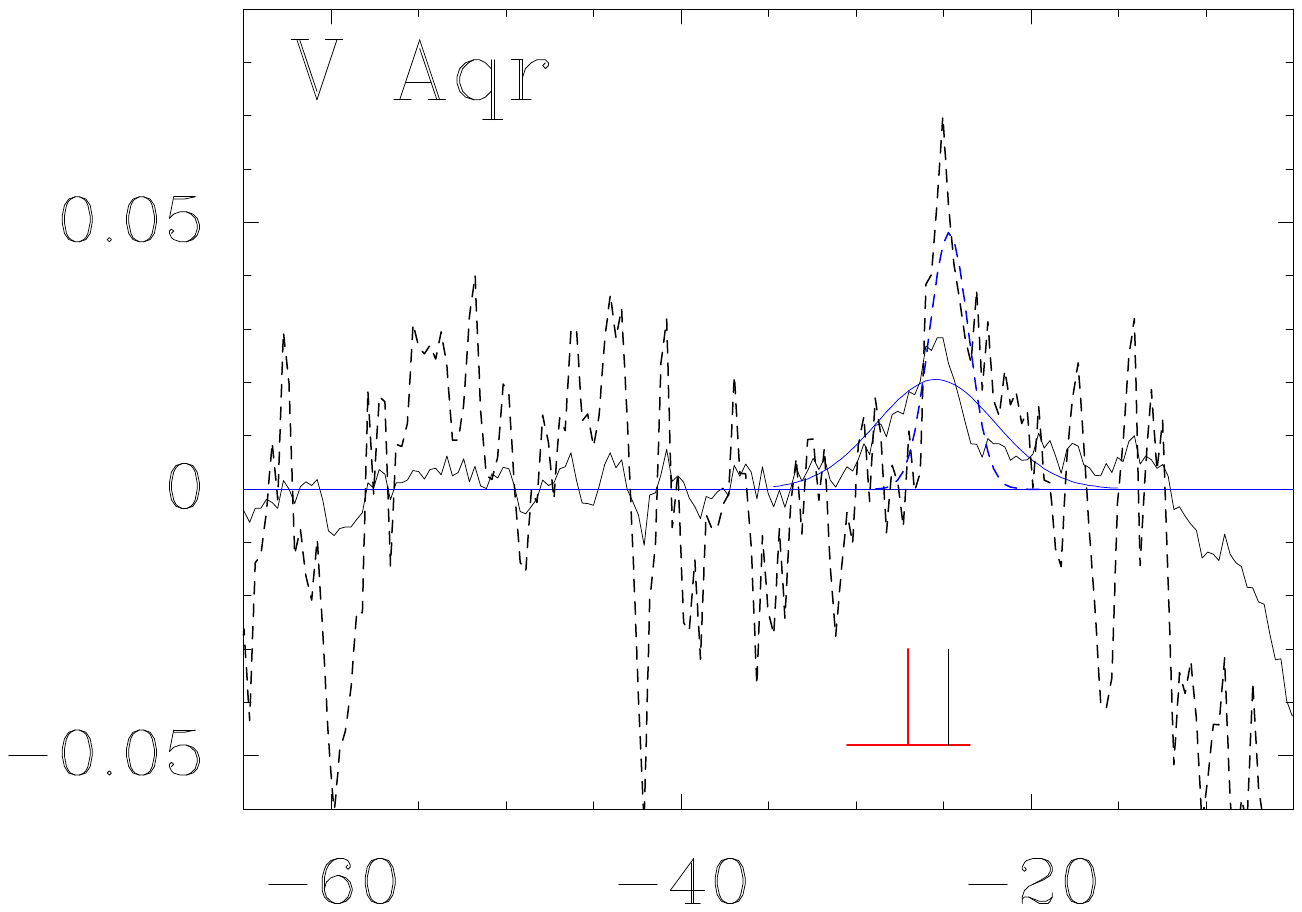}\hspace{-0.7cm}
\includegraphics[width=4.75cm]{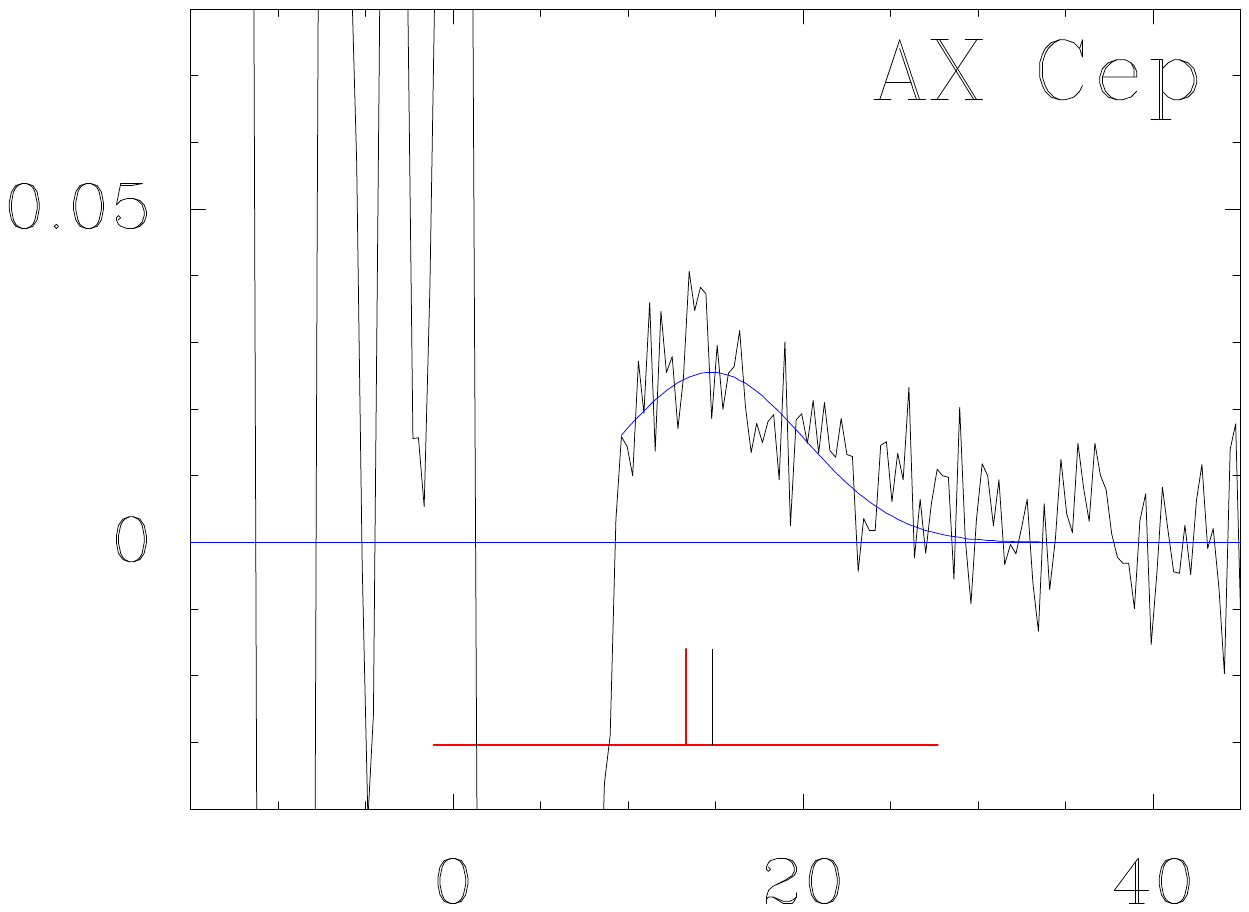}
%
% \includegraphics[width=4.75cm]{.pdf}\hspace{-0.7cm}
% \includegraphics[width=4.75cm]{.pdf}\hspace{-0.7cm}
% \includegraphics[width=4.75cm]{.pdf}\hspace{-0.7cm}
% \includegraphics[width=4.75cm]{.pdf}
% \\ \vspace{-0.35cm}
  \caption{{\bf a.} Clear detections.
% -- spectra of the 21-cm \HI\ line emission measured 
% within the beam of the NRT pointed towards the target stars.
Shown is flux density of the 21-cm \HI\ line emission, $S_{\rm HI}$, in Jy, 
as a function of radial velocity in the LSR reference frame, \VLSR, in \kms.
The peak profiles (see Section \ref{sec:sizes}) are indicated by solid black lines,
while the total profiles are indicated by dashed black lines. 
The Gaussians fitted to the profiles are shown as blue lines, 
solid for fits to the peak profiles and dashed for fits to the total profiles. 
Spectra are shown for all 34 objects with clear \HI\ detections (see Table~\ref{table:cleardetsHI}).
The vertical black lines indicate the centre velocity of our Gaussian fit to the total \HI\ 
profile (or, if not available, to the peak \HI\ profile), the red vertical and horizontal lines 
show respectively the central CO or OH line velocity from the literature and its corresponding 
expansion velocity, whereas green lines indicate other types of literature velocities 
(e.g., optical or SiO lines) and an indicative expansion velocity of 10 \kms, i.e., 
the average measured value. 
The flat blue horizontal lines show the 0 Jy flux density level.
The plotted total velocity range of all spectra is 60 \kms.}
\label{fig:spectradetections}
\end{figure*}

\begin{figure*}[ht]  % \label{fig:spectra_detections}  % Fig. 3b
\addtocounter{figure}{-1}
%  \centering
%  \includegraphics[width=18.35cm]{AGB_Legacy_cleardets_12_two_wrkn3.png}
\includegraphics[width=4.75cm]{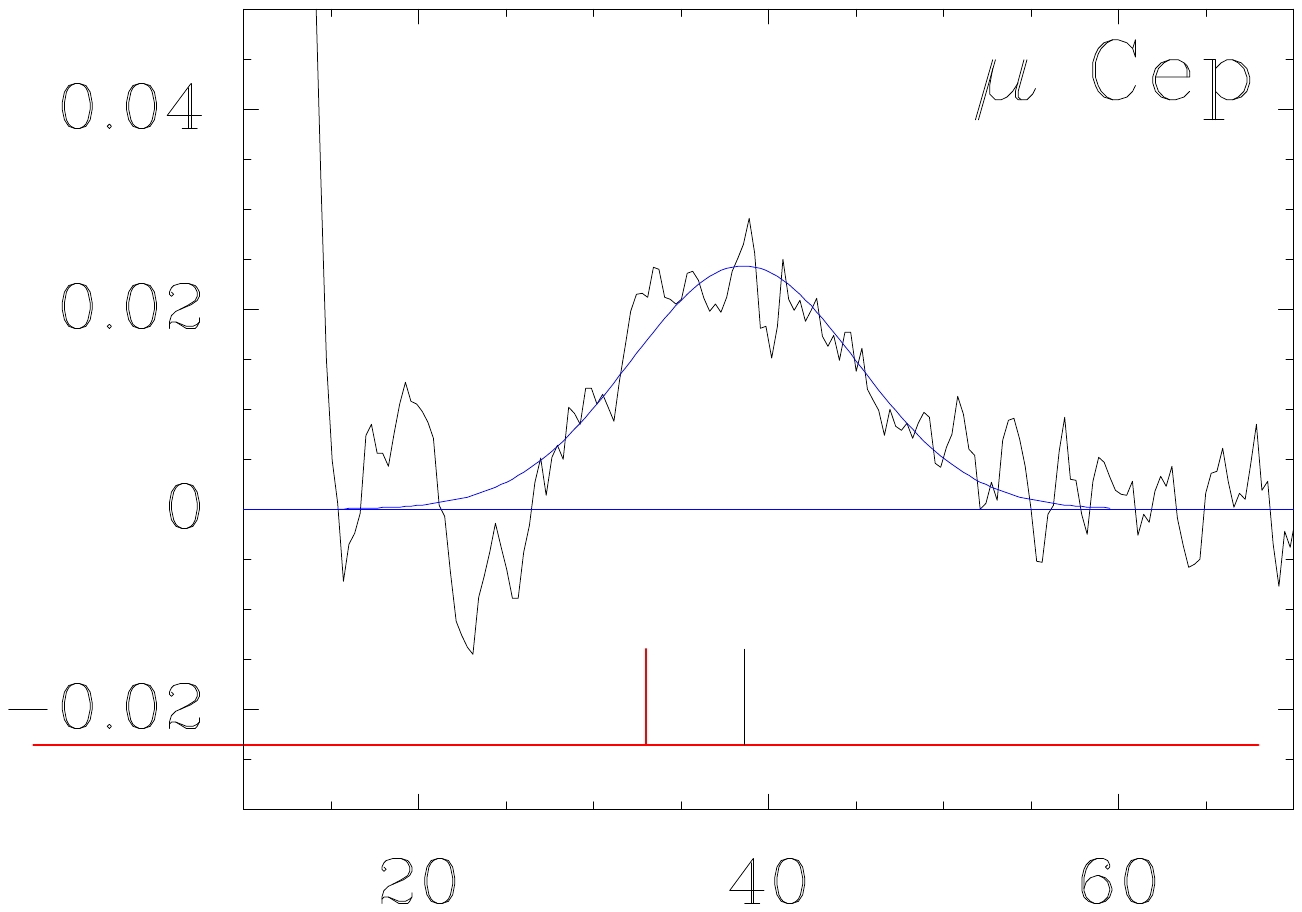}\hspace{-0.7cm}
\includegraphics[width=4.75cm]{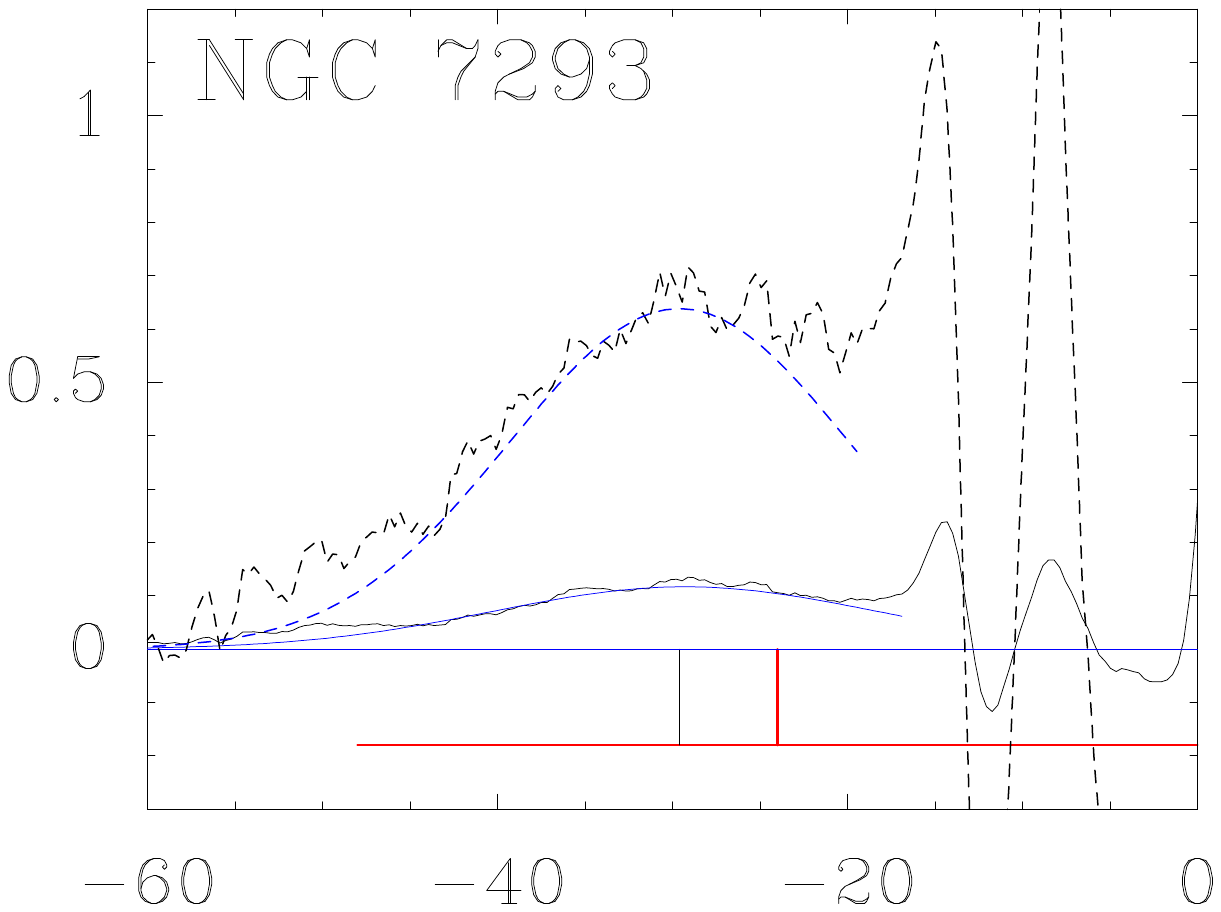}\hspace{-0.7cm}
\includegraphics[width=4.75cm]{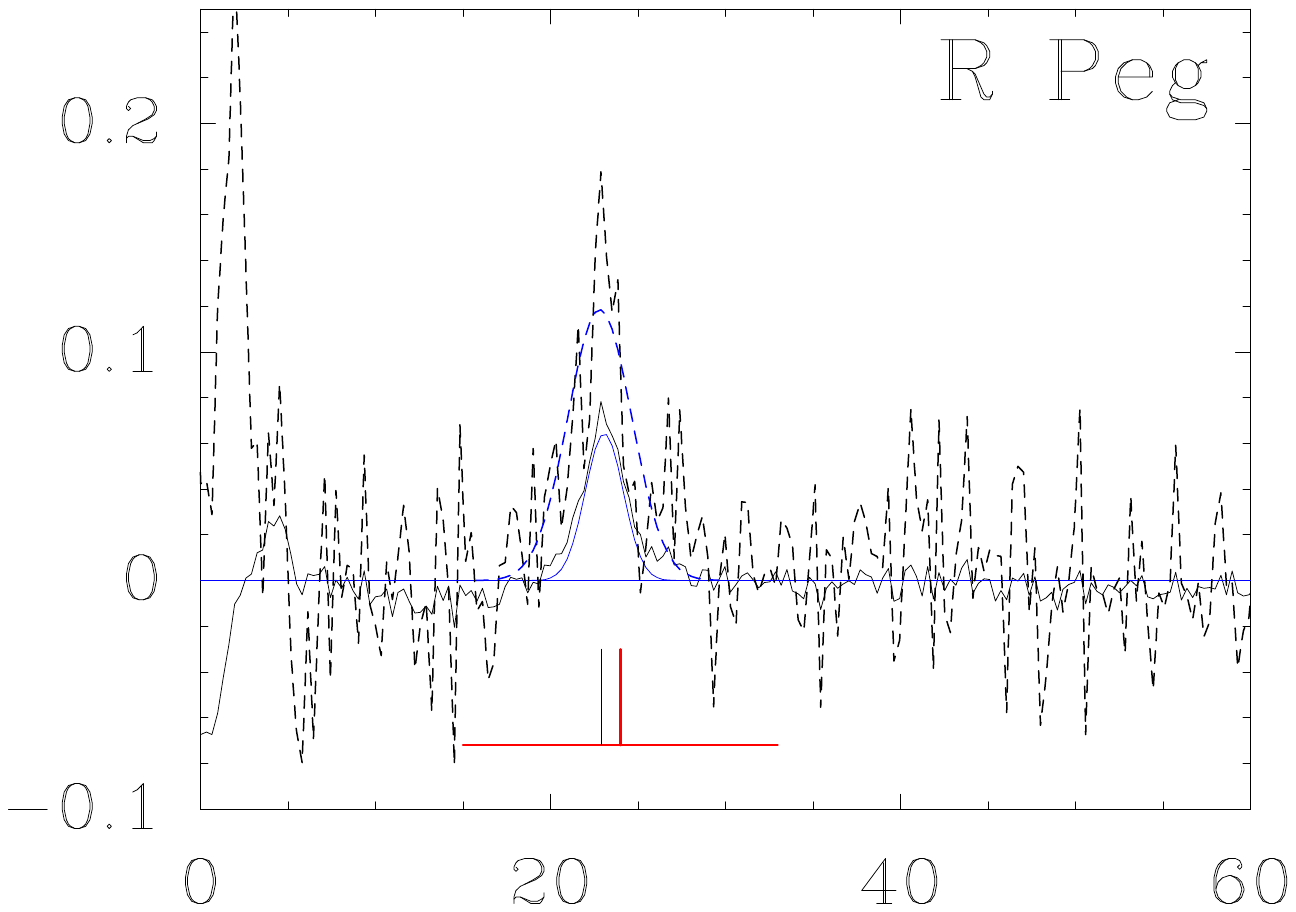}\hspace{-0.7cm}
\includegraphics[width=4.75cm]{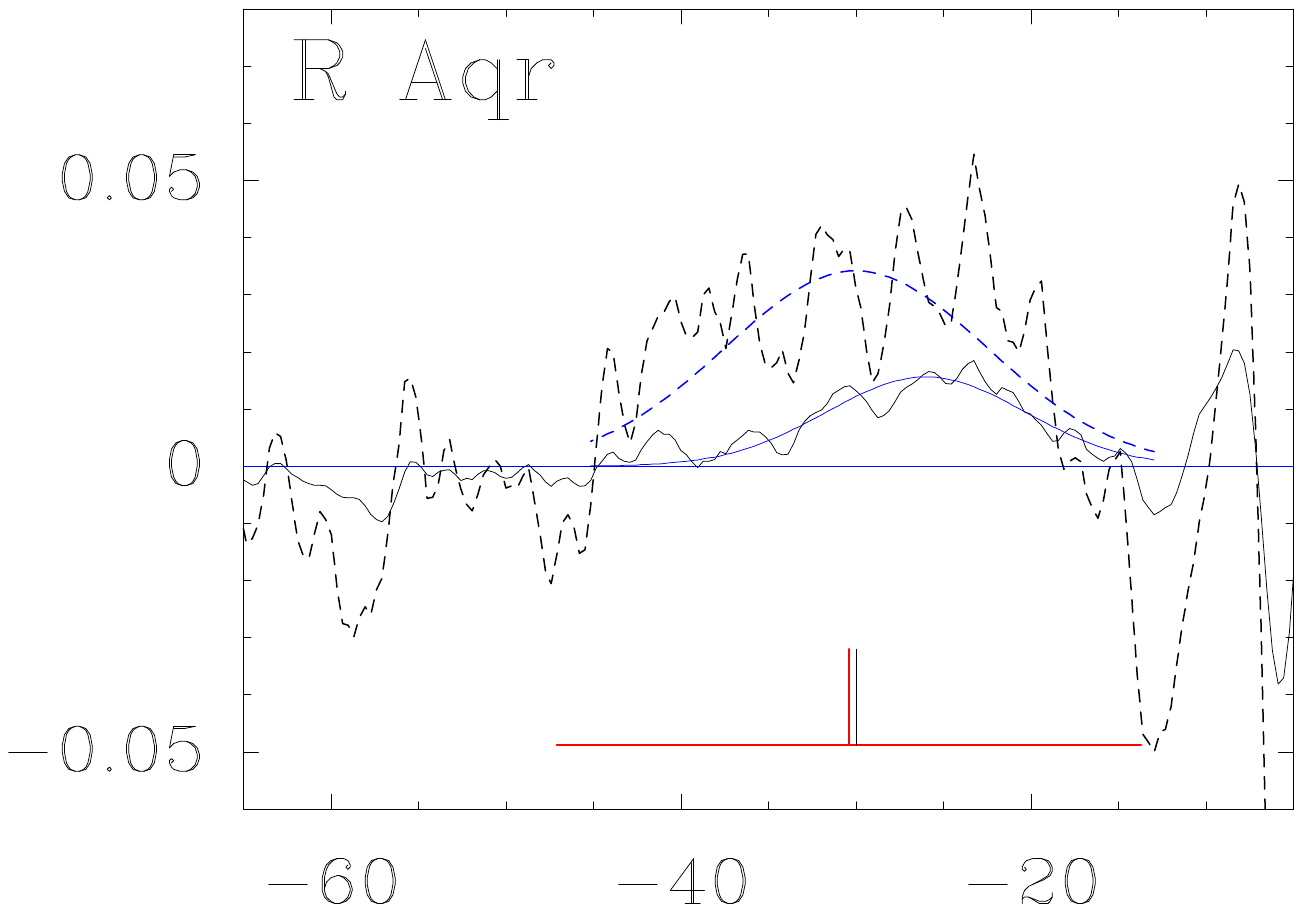}
\\ \vspace{-0.35cm}
\includegraphics[width=4.75cm]{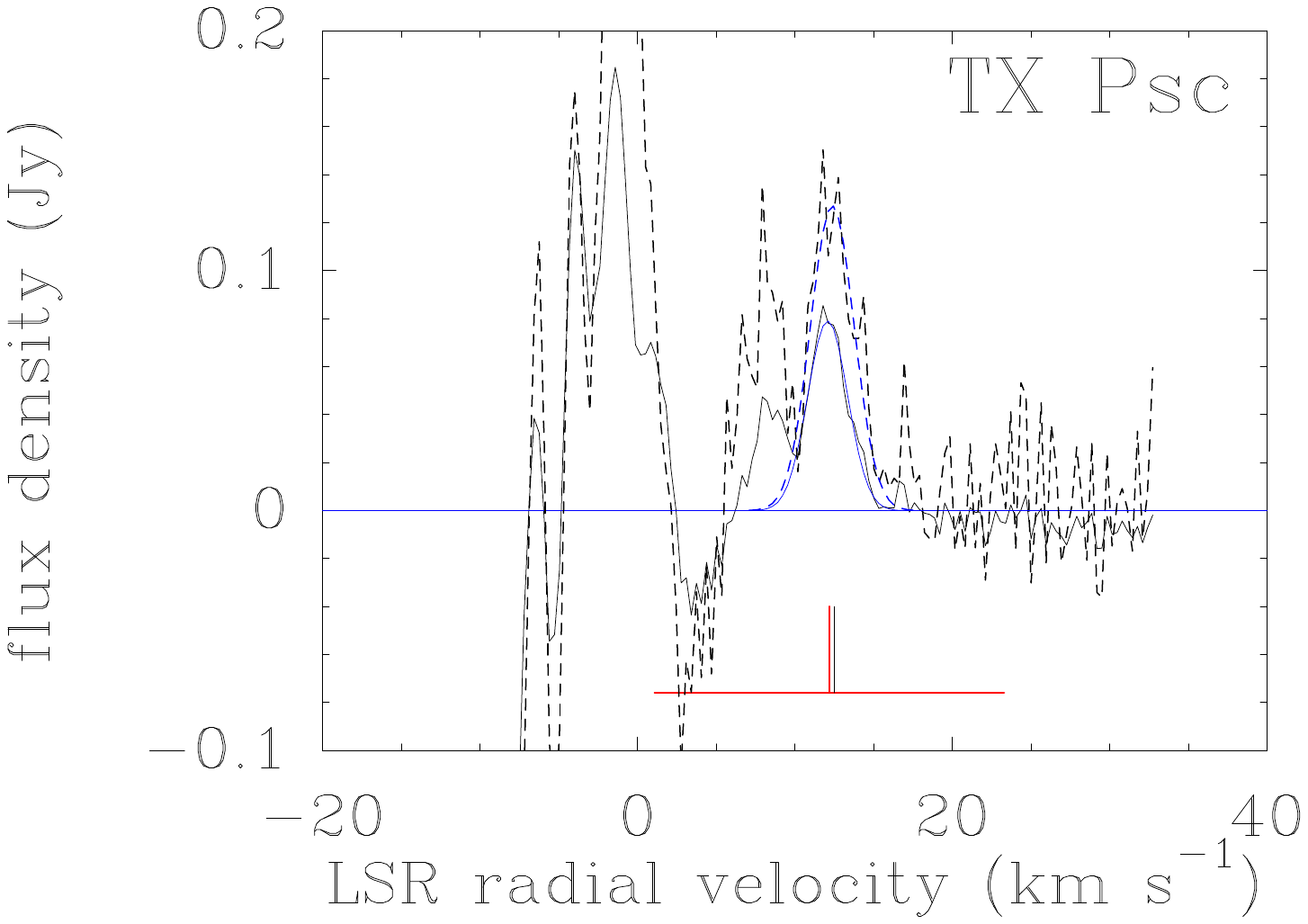}\hspace{-0.7cm}
\includegraphics[width=4.75cm]{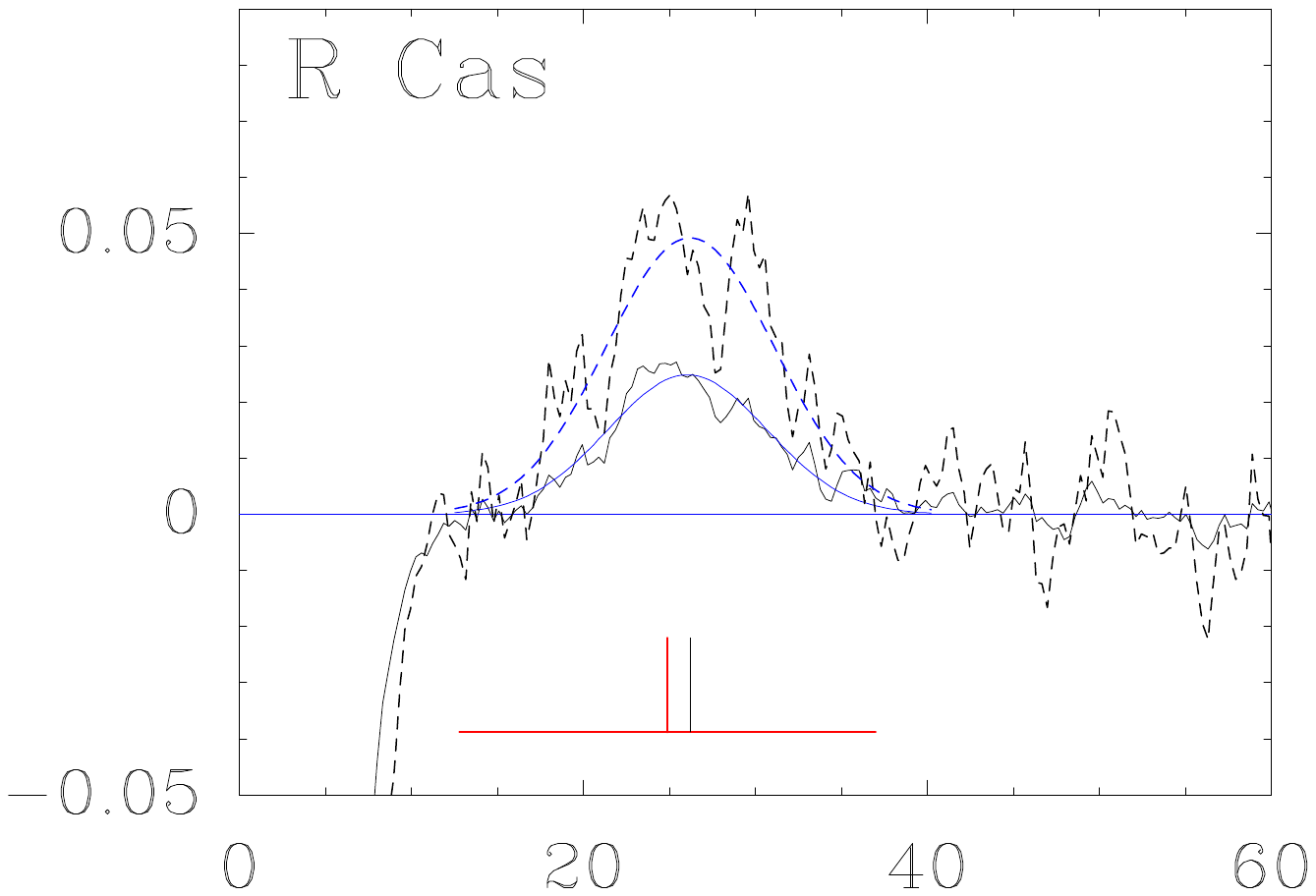}\hspace{-0.7cm}
\\ \vspace{0.3cm}
  \caption{{\bf b.} Clear detections -- continued. }
\end{figure*}

\begin{figure*}[ht]  % Fig. 5a
  \centering
\includegraphics[width=4.75cm]{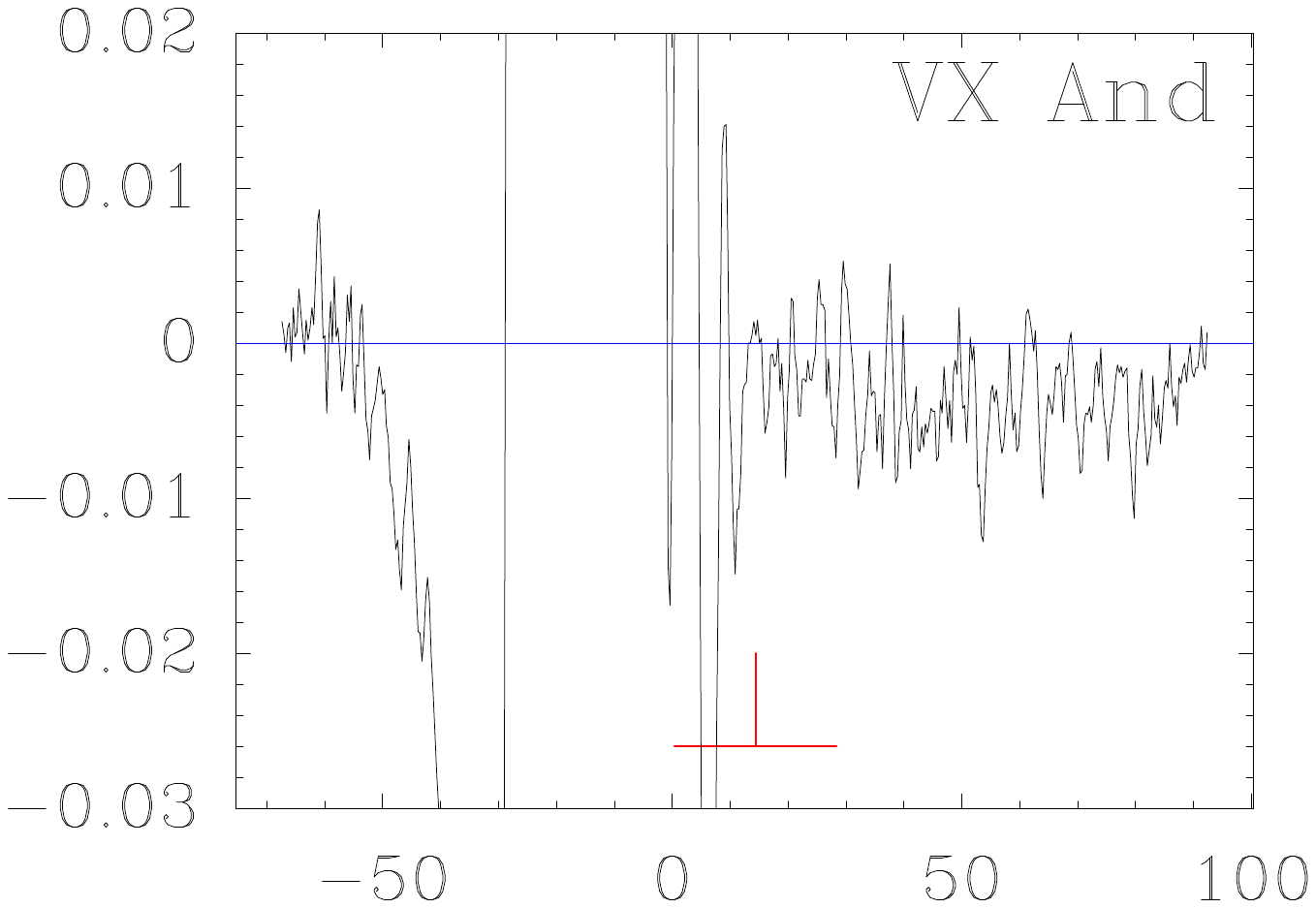}\hspace{-0.7cm}
\includegraphics[width=4.75cm]{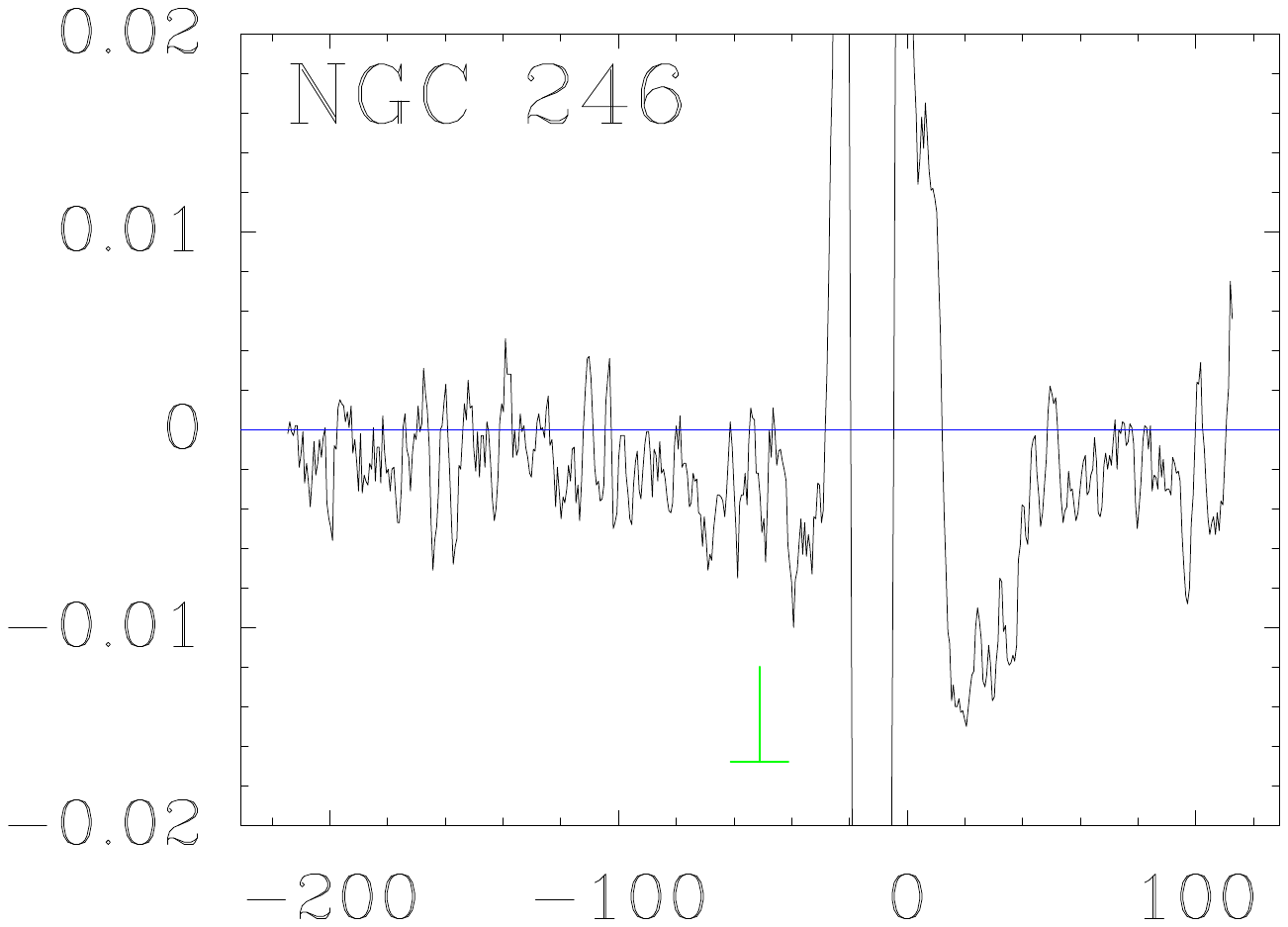}\hspace{-0.7cm}
\includegraphics[width=4.75cm]{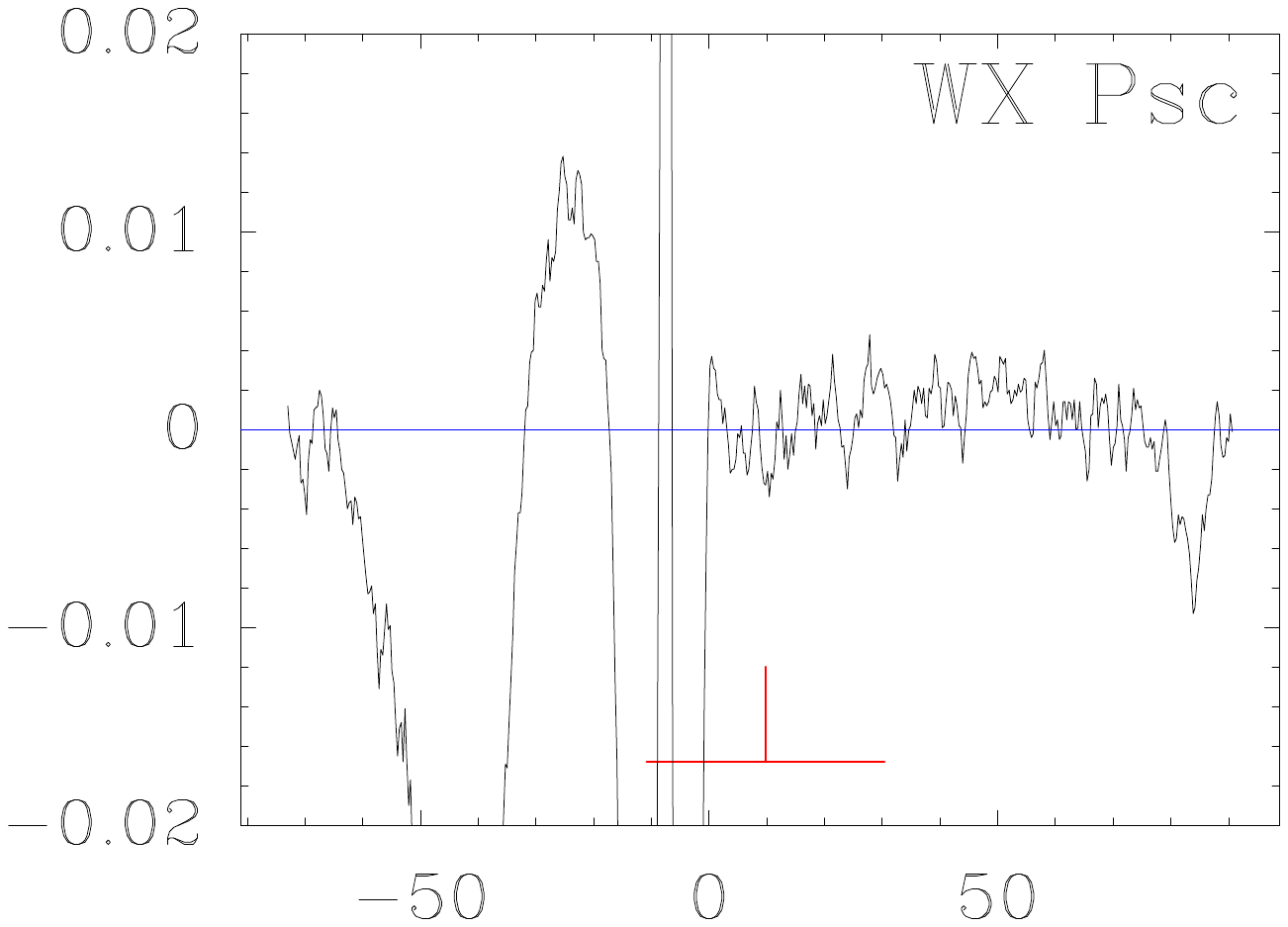}\hspace{-0.7cm}
\includegraphics[width=4.75cm]{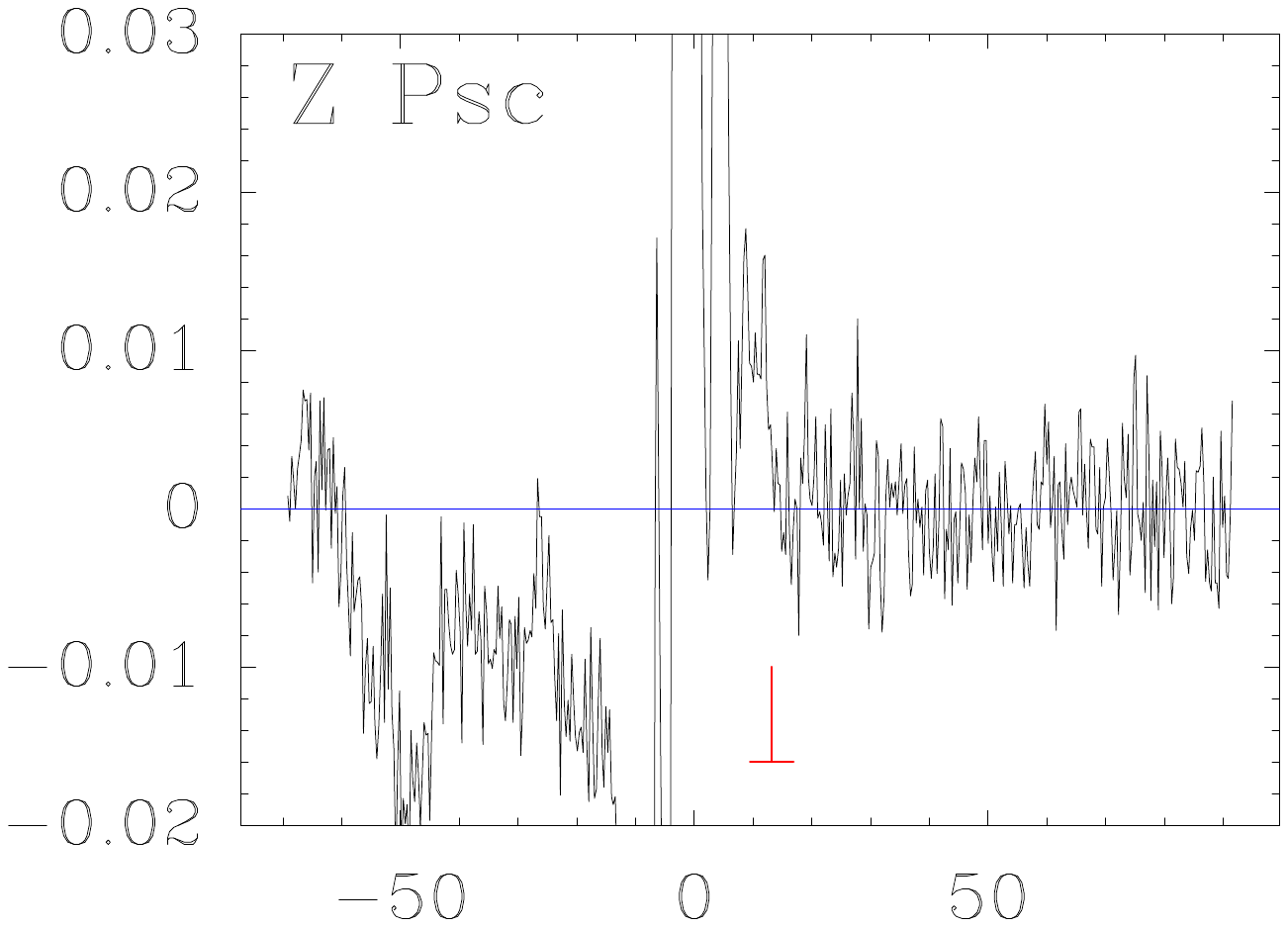}
\\ \vspace{-0.35cm}
\includegraphics[width=4.75cm]{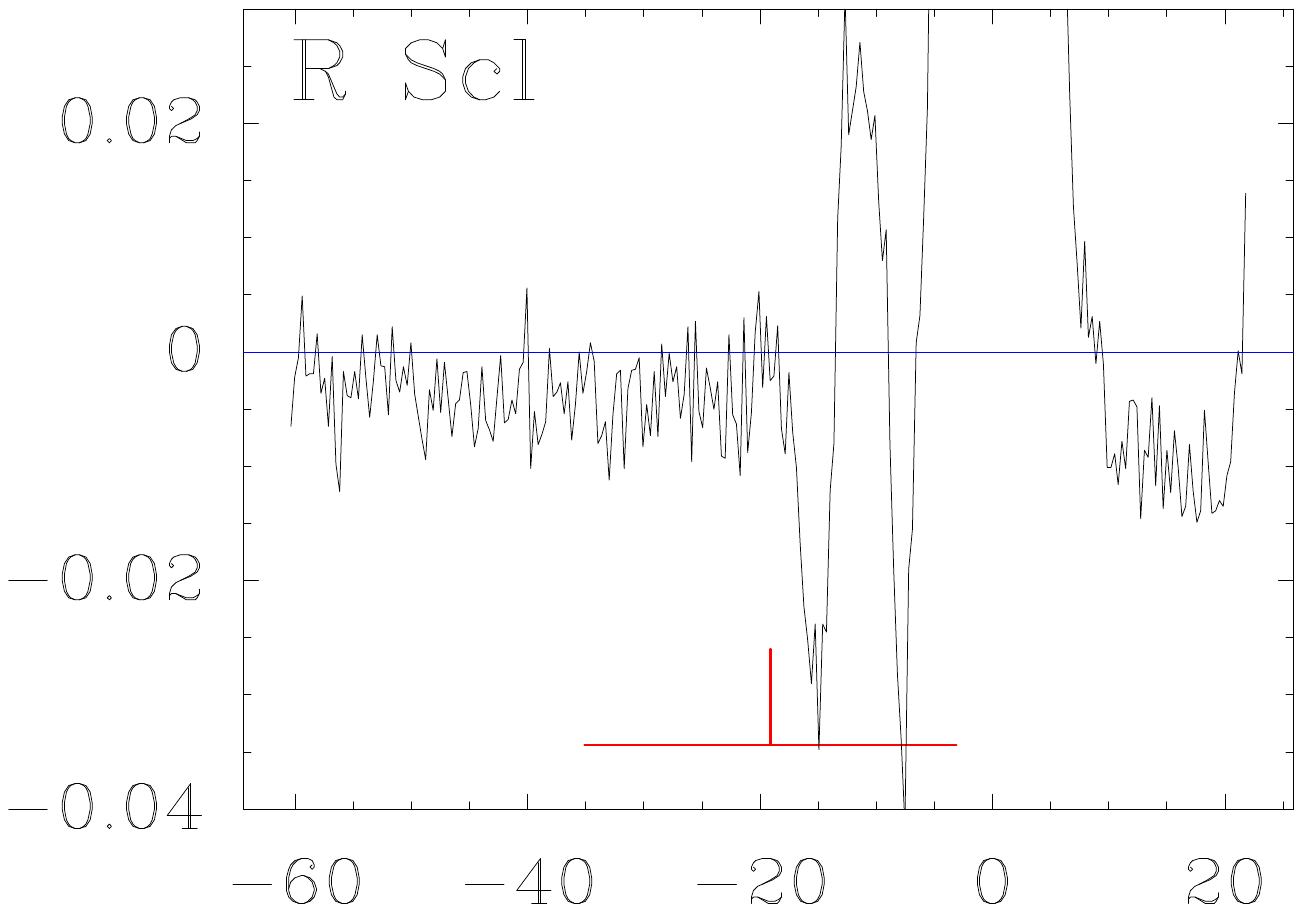}\hspace{-0.7cm}
\includegraphics[width=4.75cm]{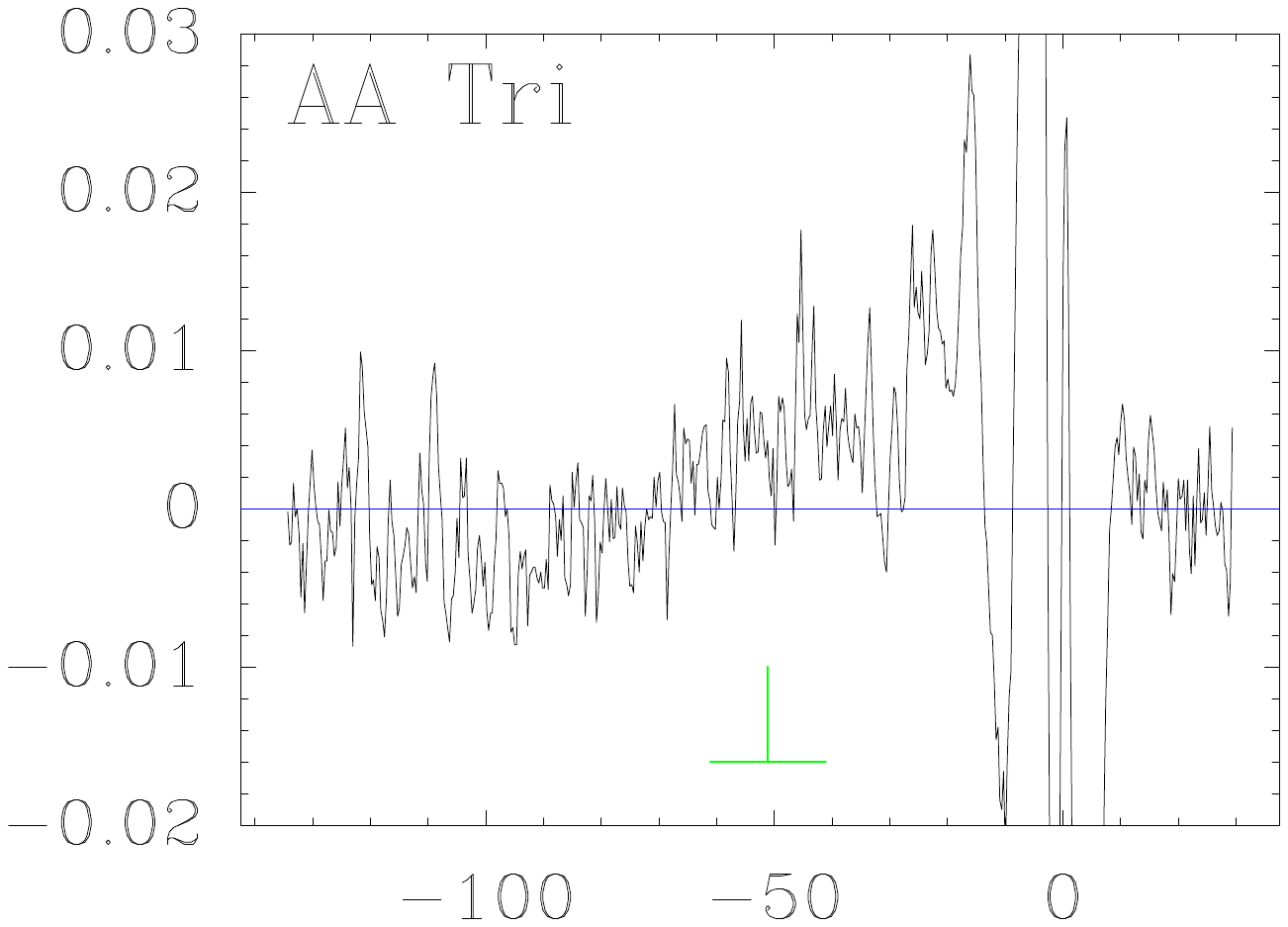}\hspace{-0.7cm}
\includegraphics[width=4.75cm]{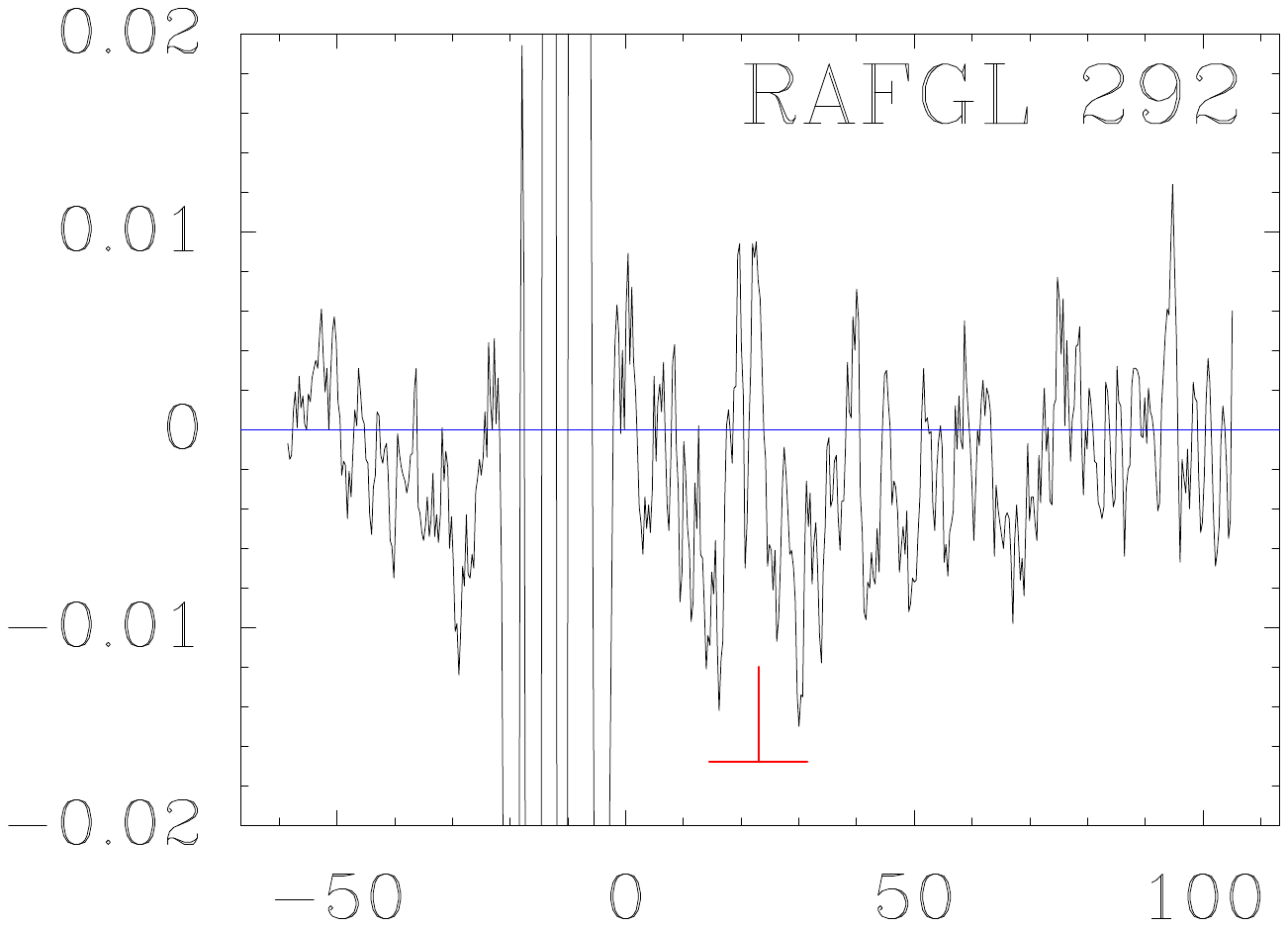}\hspace{-0.7cm}
\includegraphics[width=4.75cm]{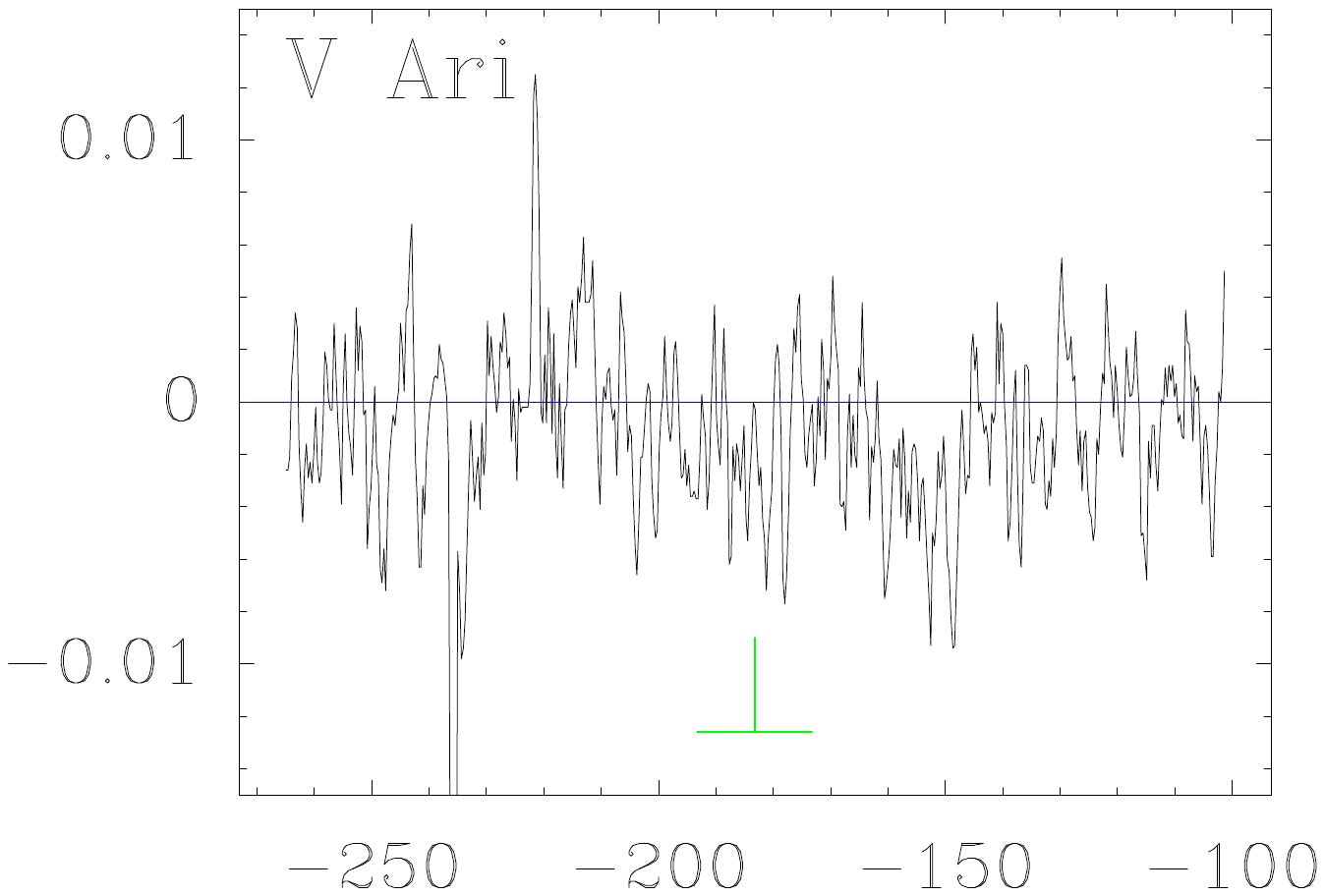}
\\ \vspace{-0.35cm}
\includegraphics[width=4.75cm]{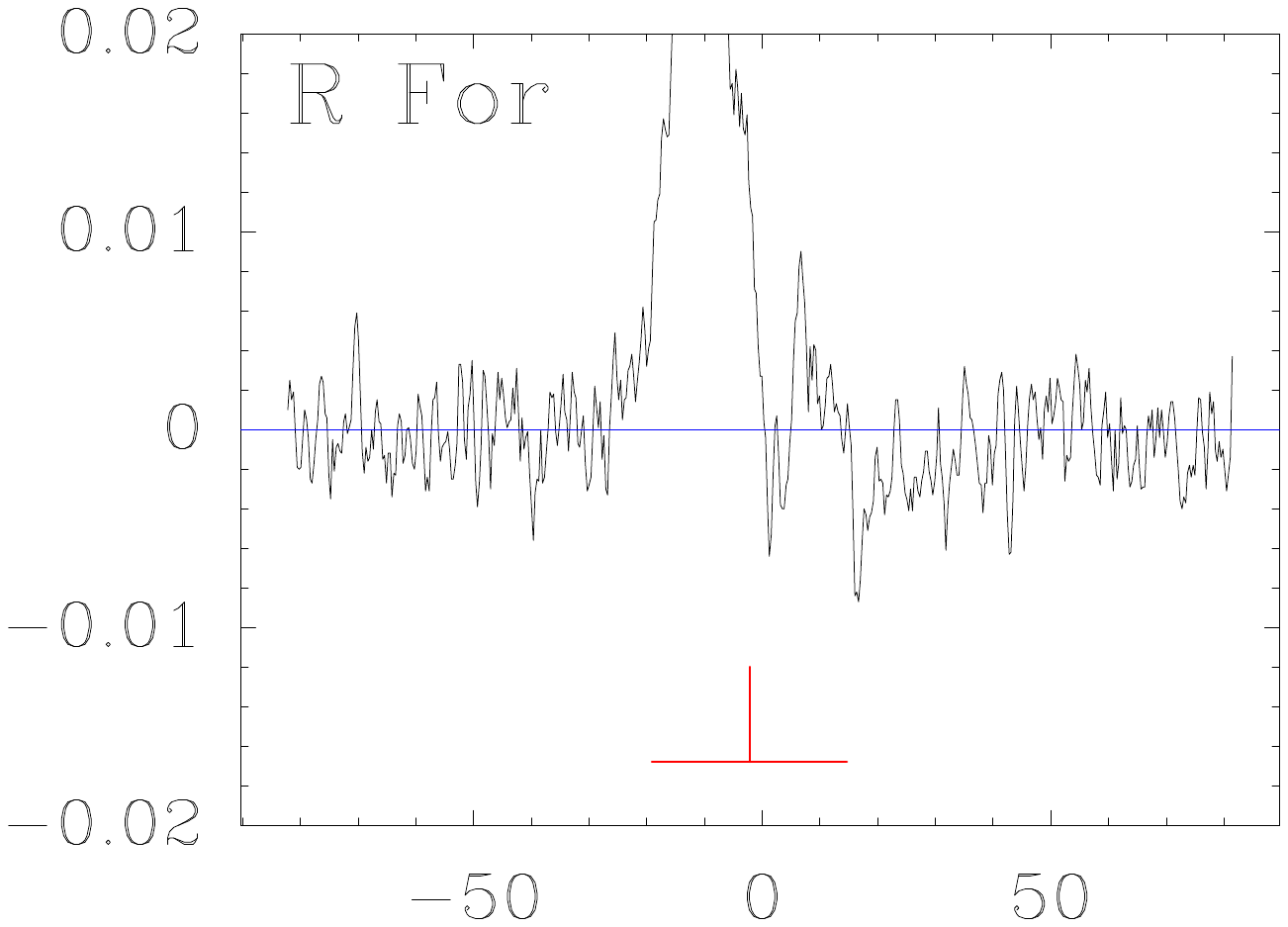}\hspace{-0.7cm}
\includegraphics[width=4.75cm]{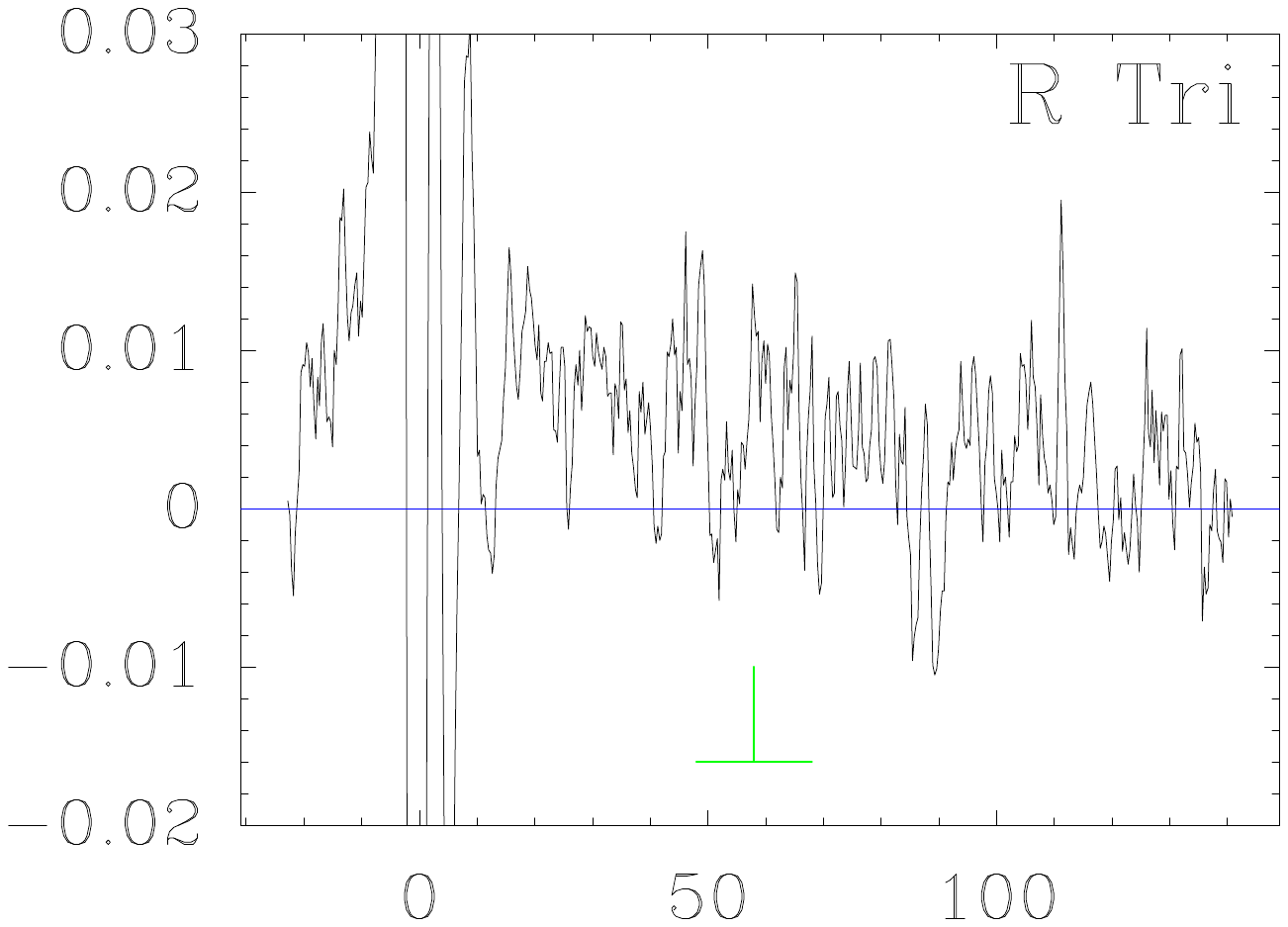}\hspace{-0.7cm}
\includegraphics[width=4.75cm]{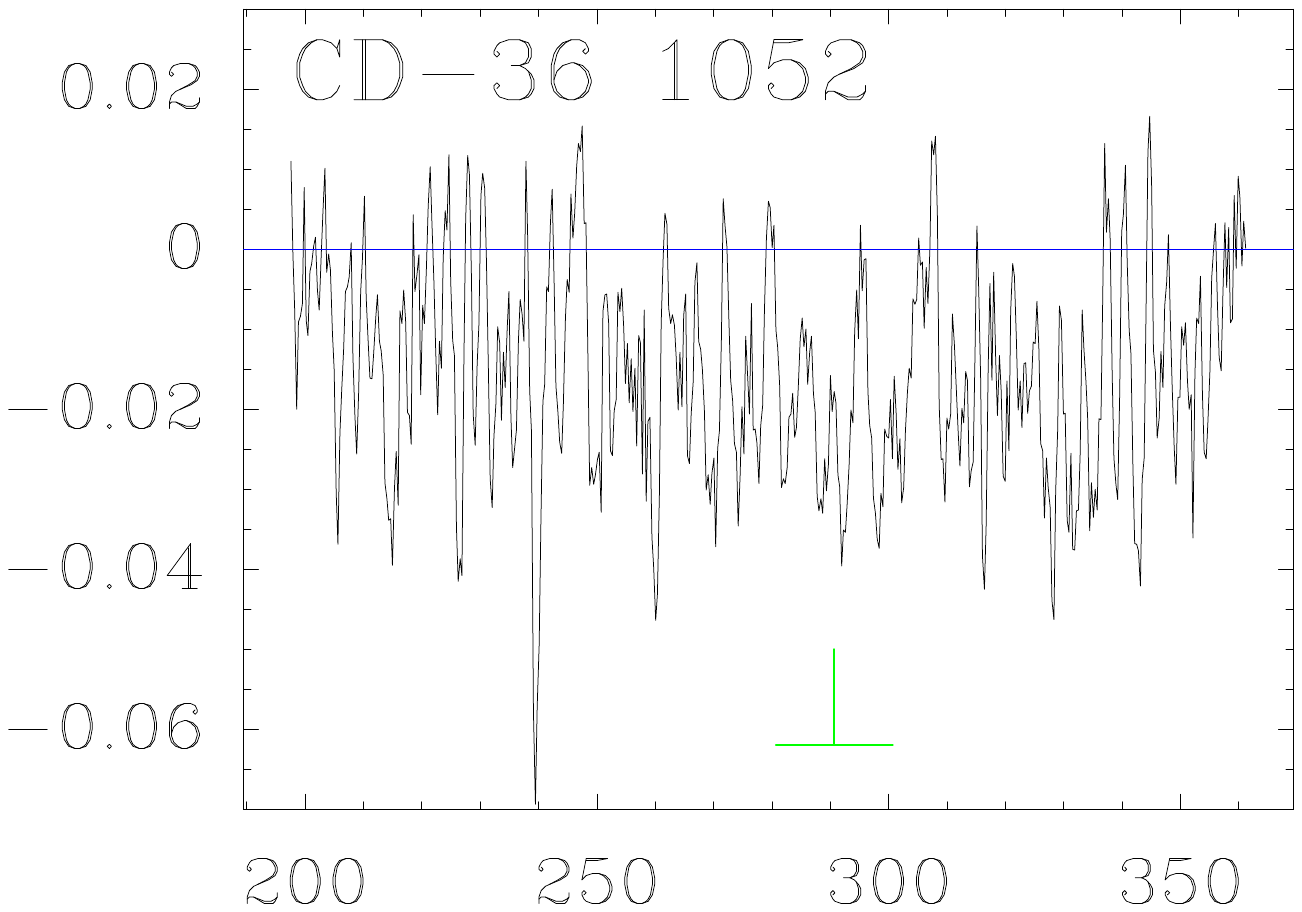}\hspace{-0.7cm}
\includegraphics[width=4.75cm]{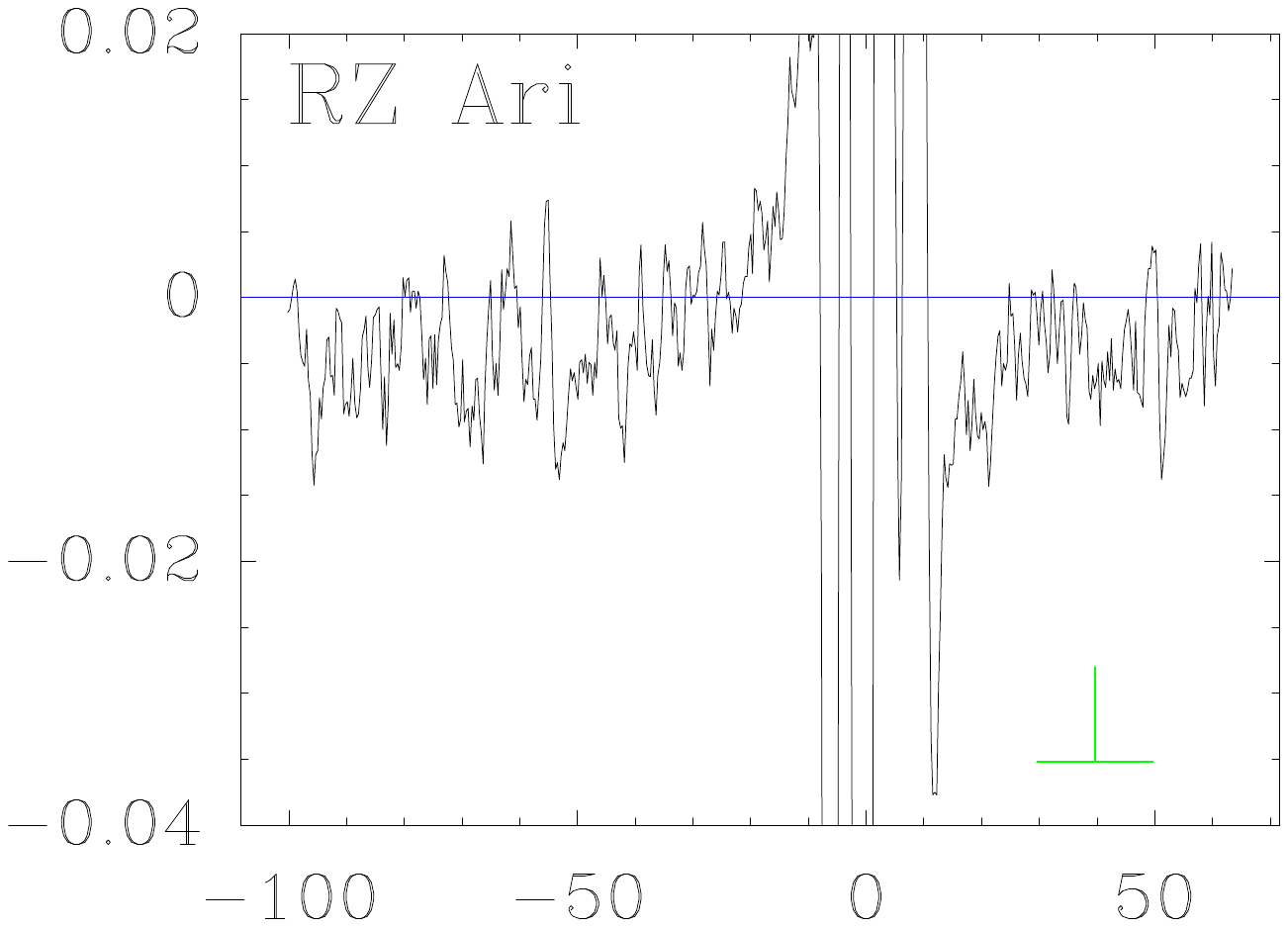}
\\ \vspace{-0.35cm}
\includegraphics[width=4.75cm]{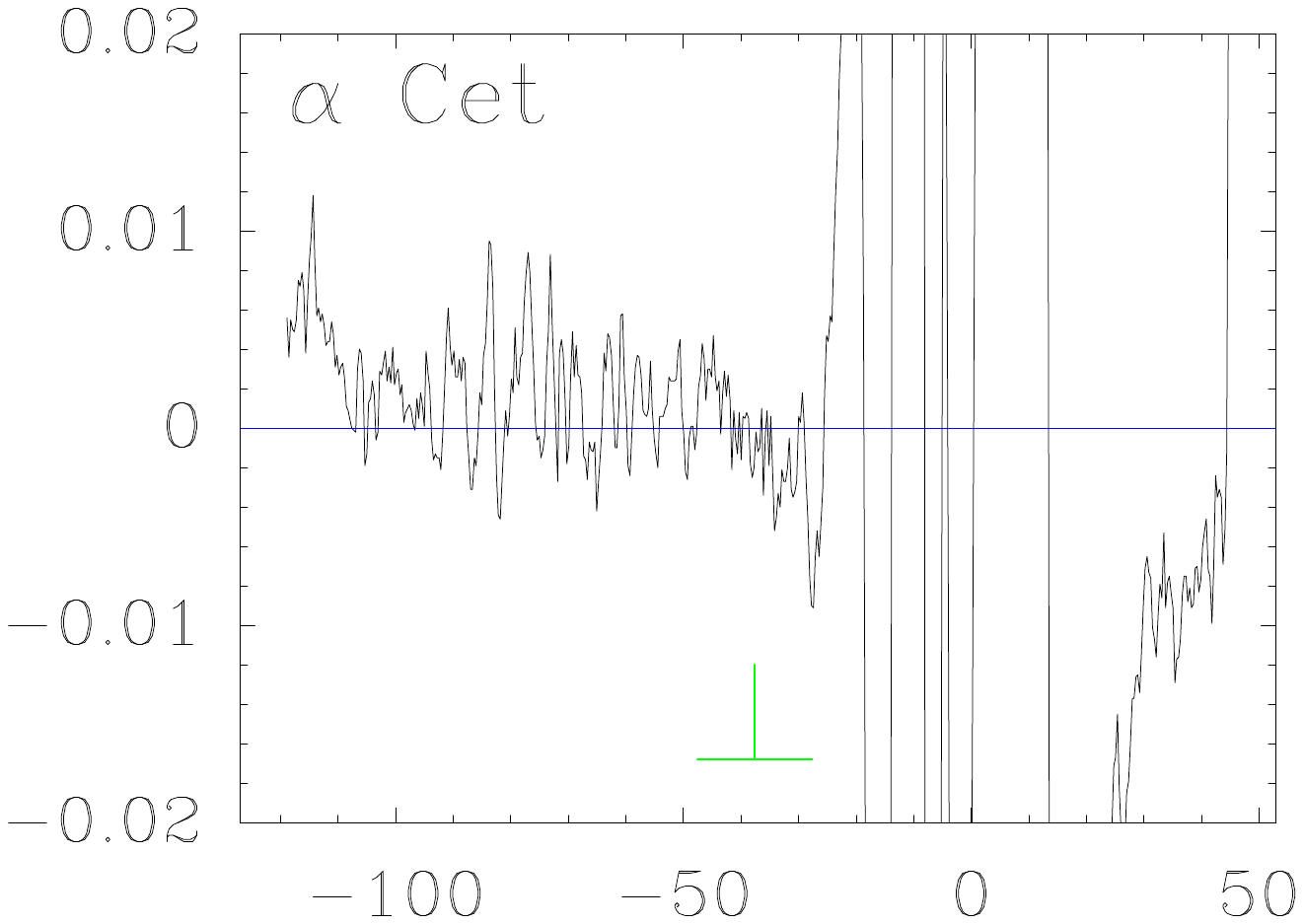}\hspace{-0.7cm}
\includegraphics[width=4.75cm]{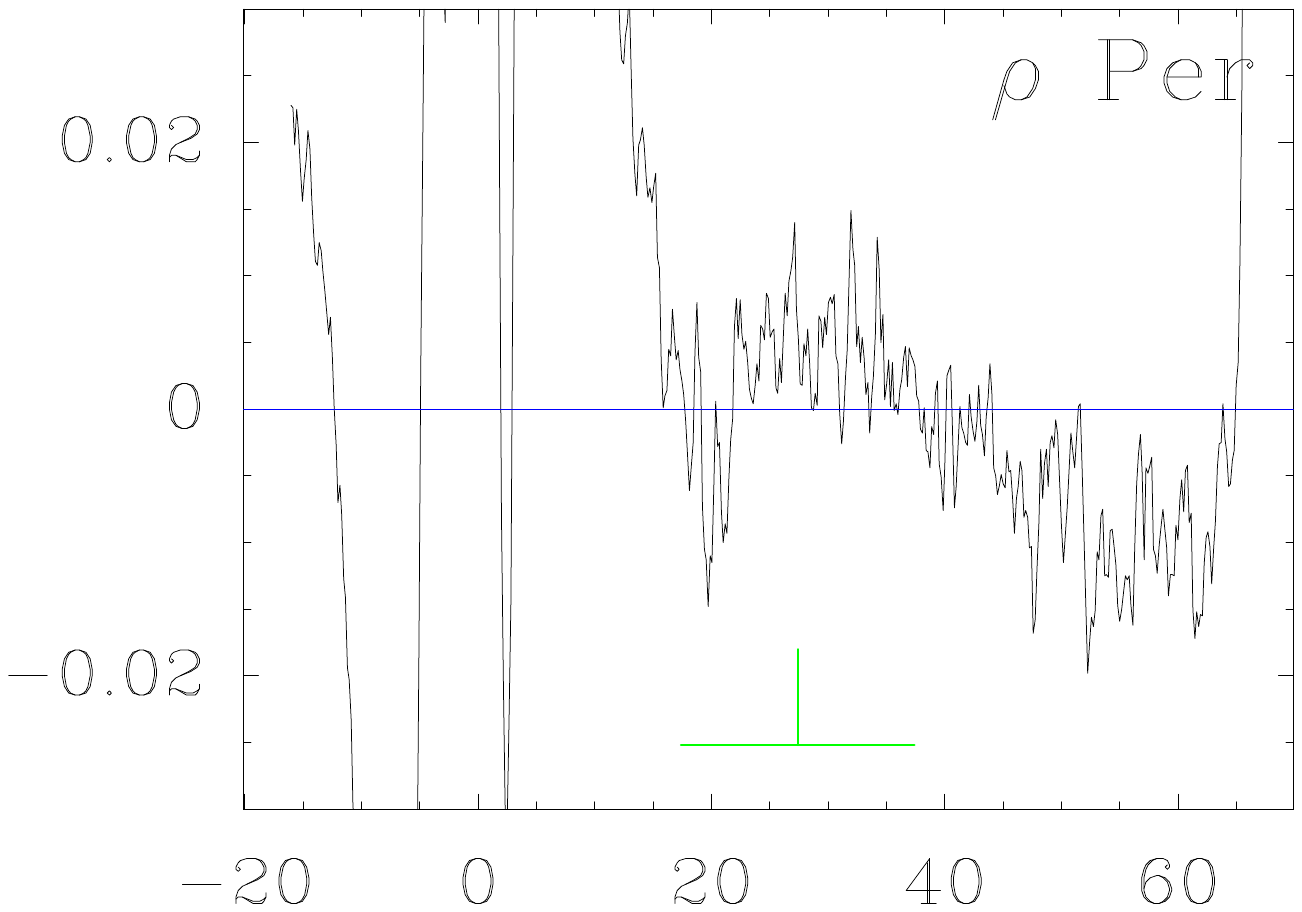}\hspace{-0.7cm}
\includegraphics[width=4.75cm]{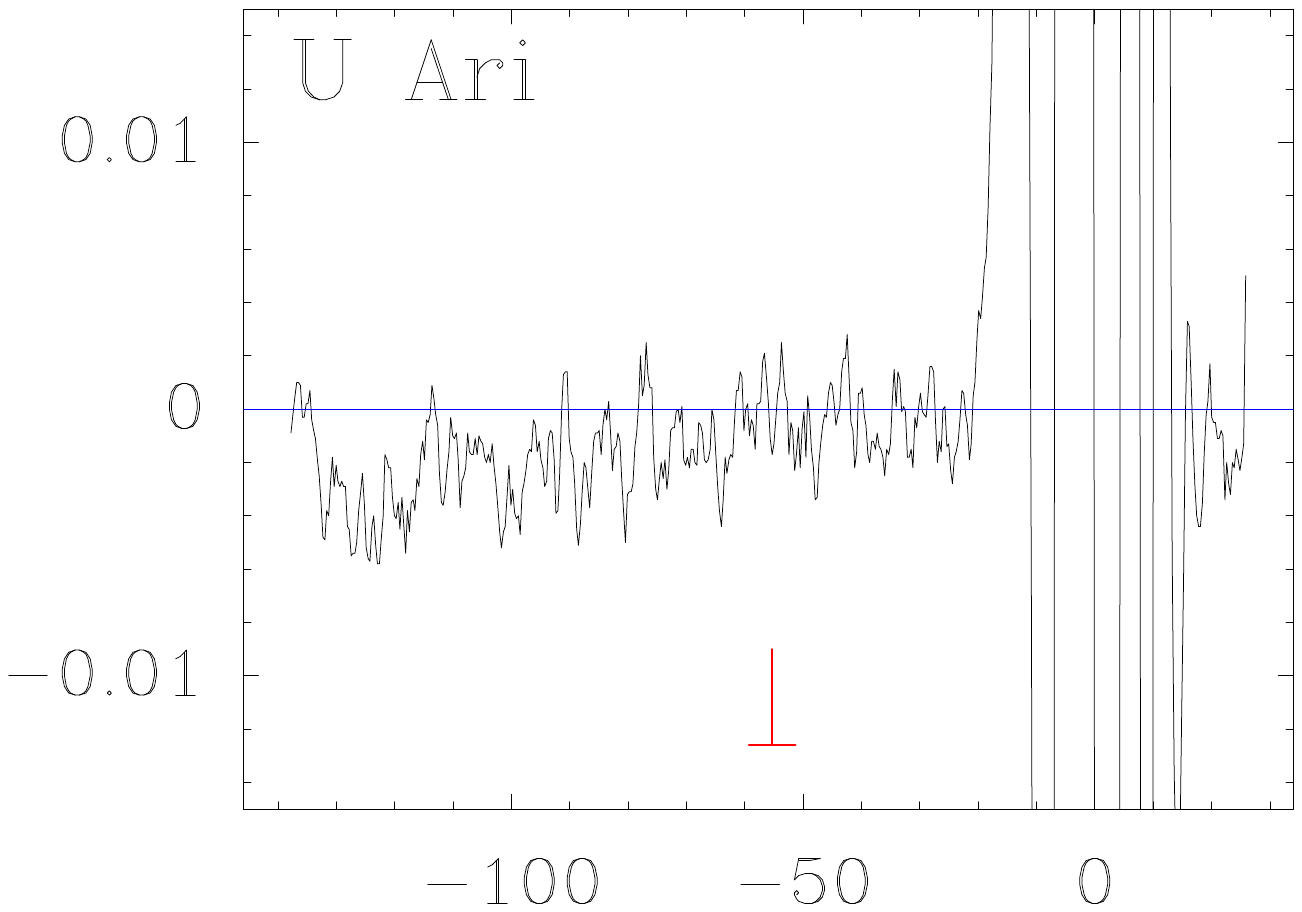}\hspace{-0.7cm}
\includegraphics[width=4.75cm]{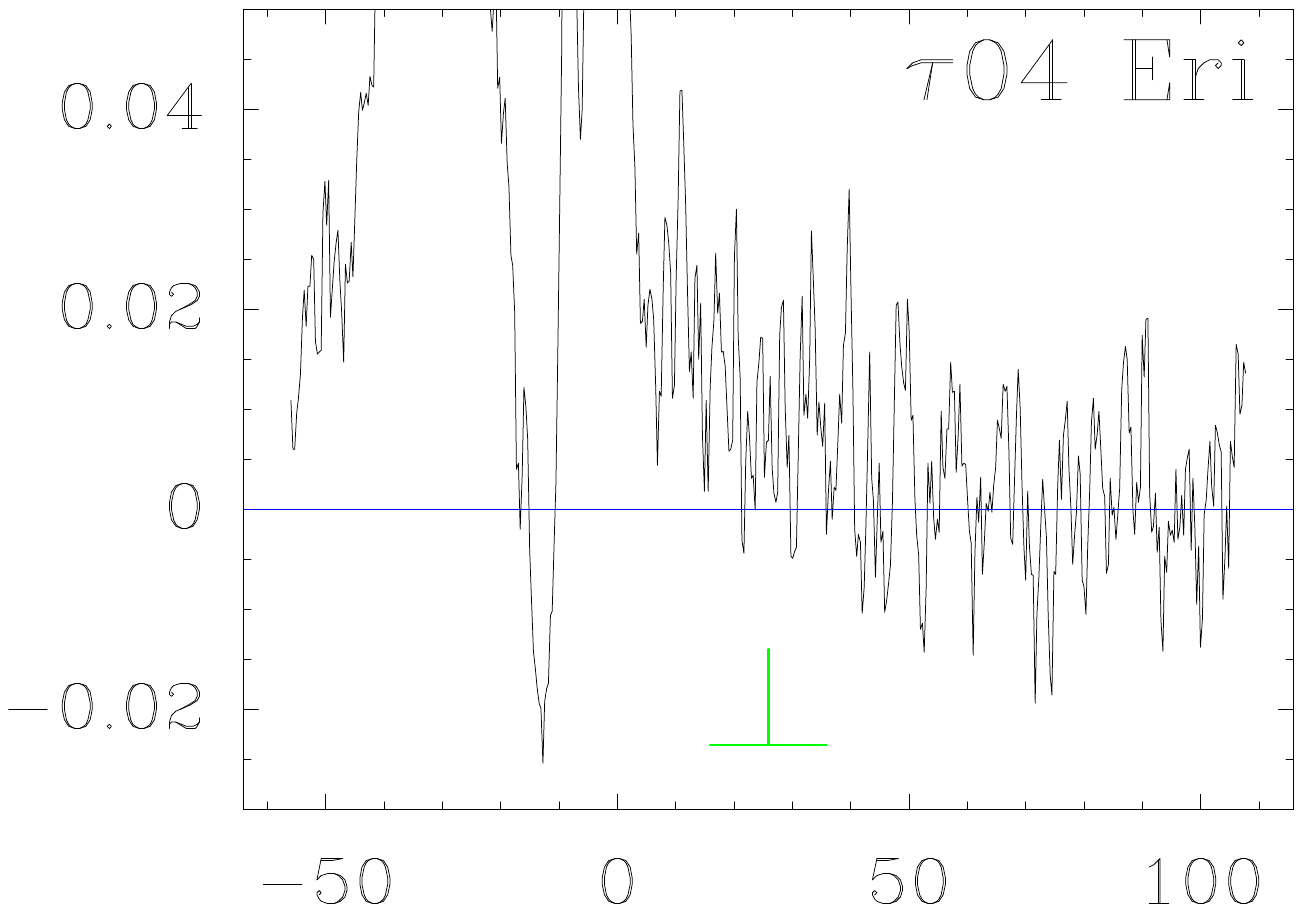}
\\ \vspace{-0.35cm}
\includegraphics[width=4.75cm]{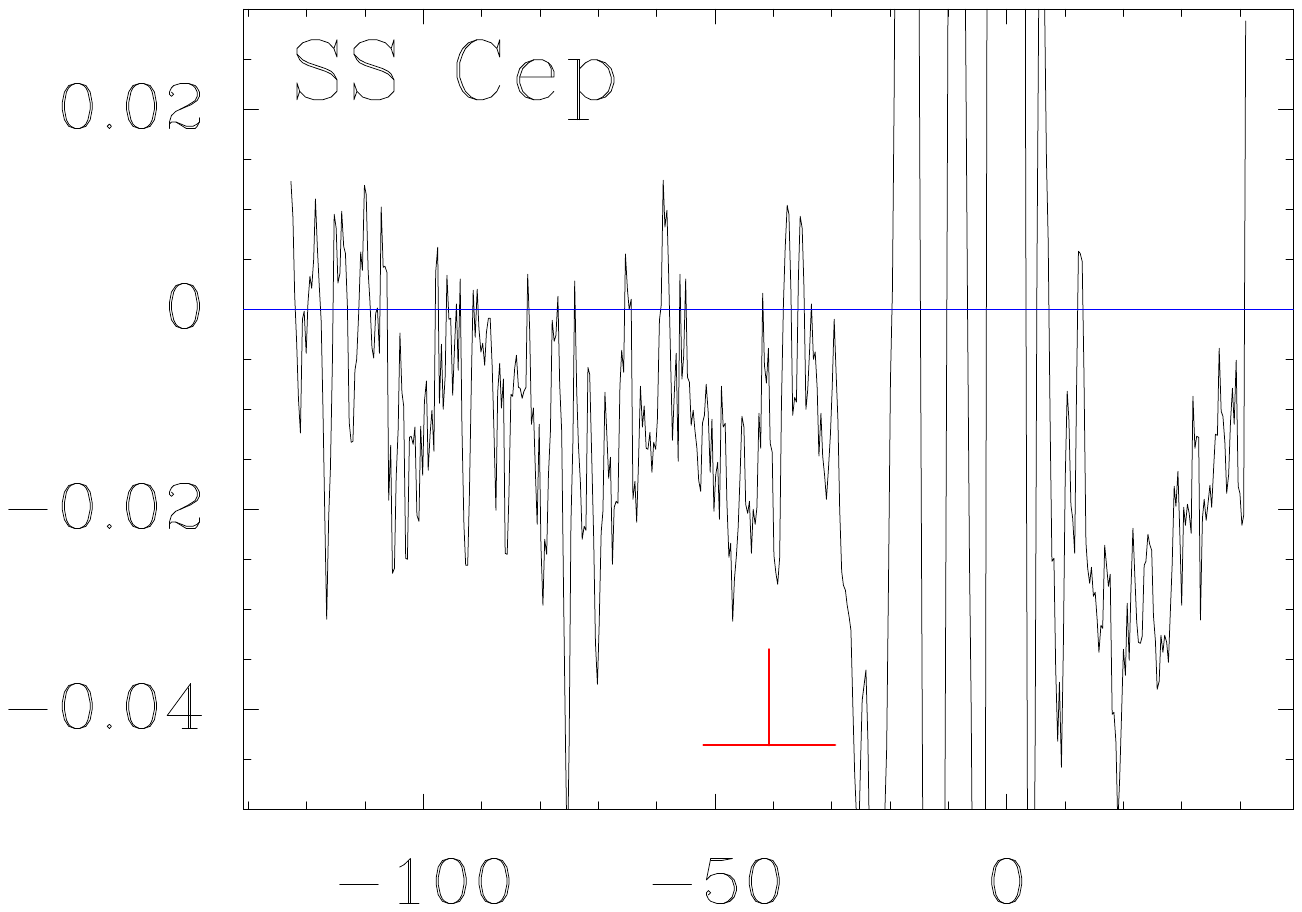}\hspace{-0.7cm}
\includegraphics[width=4.75cm]{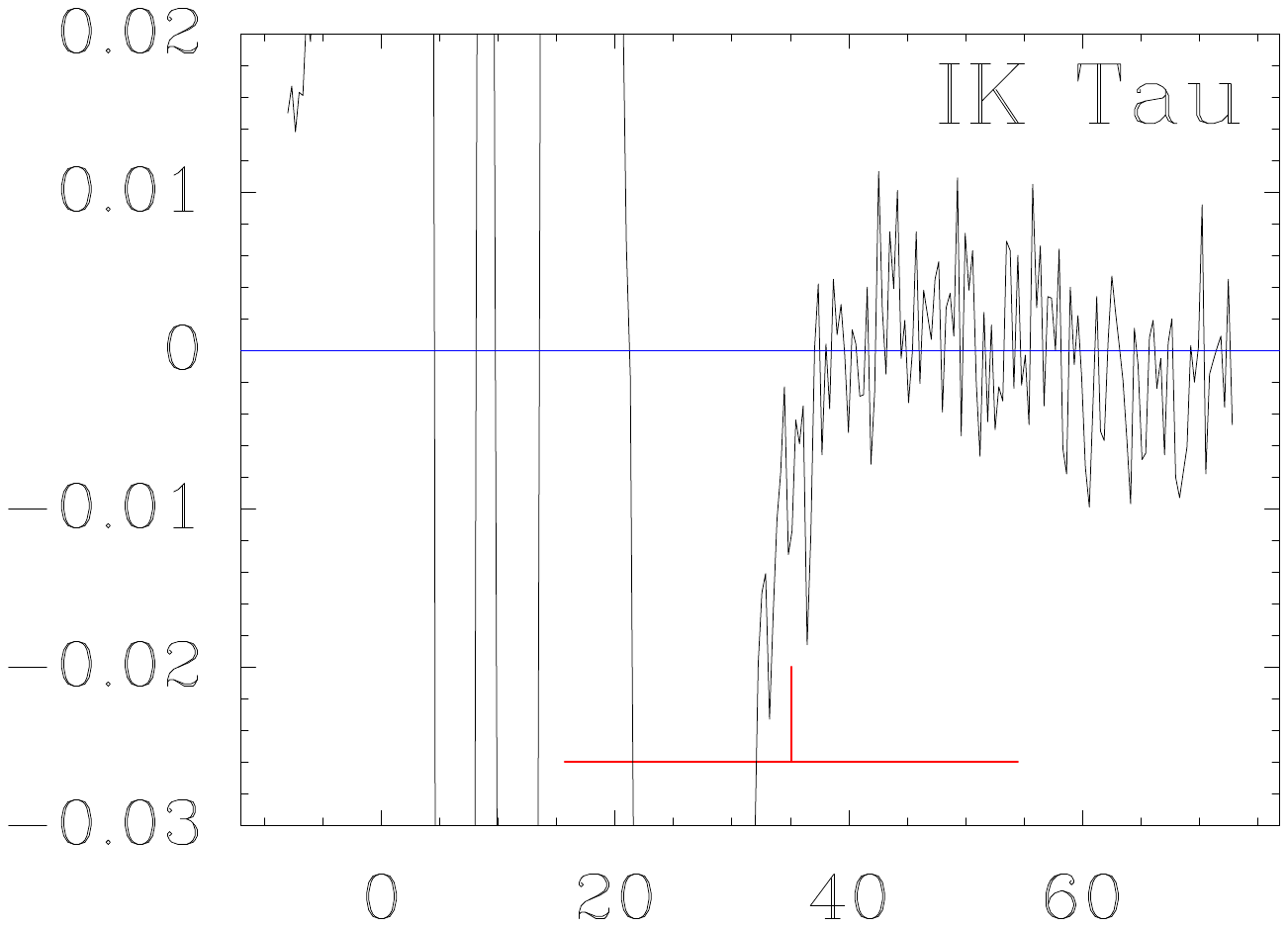}\hspace{-0.7cm}
\includegraphics[width=4.75cm]{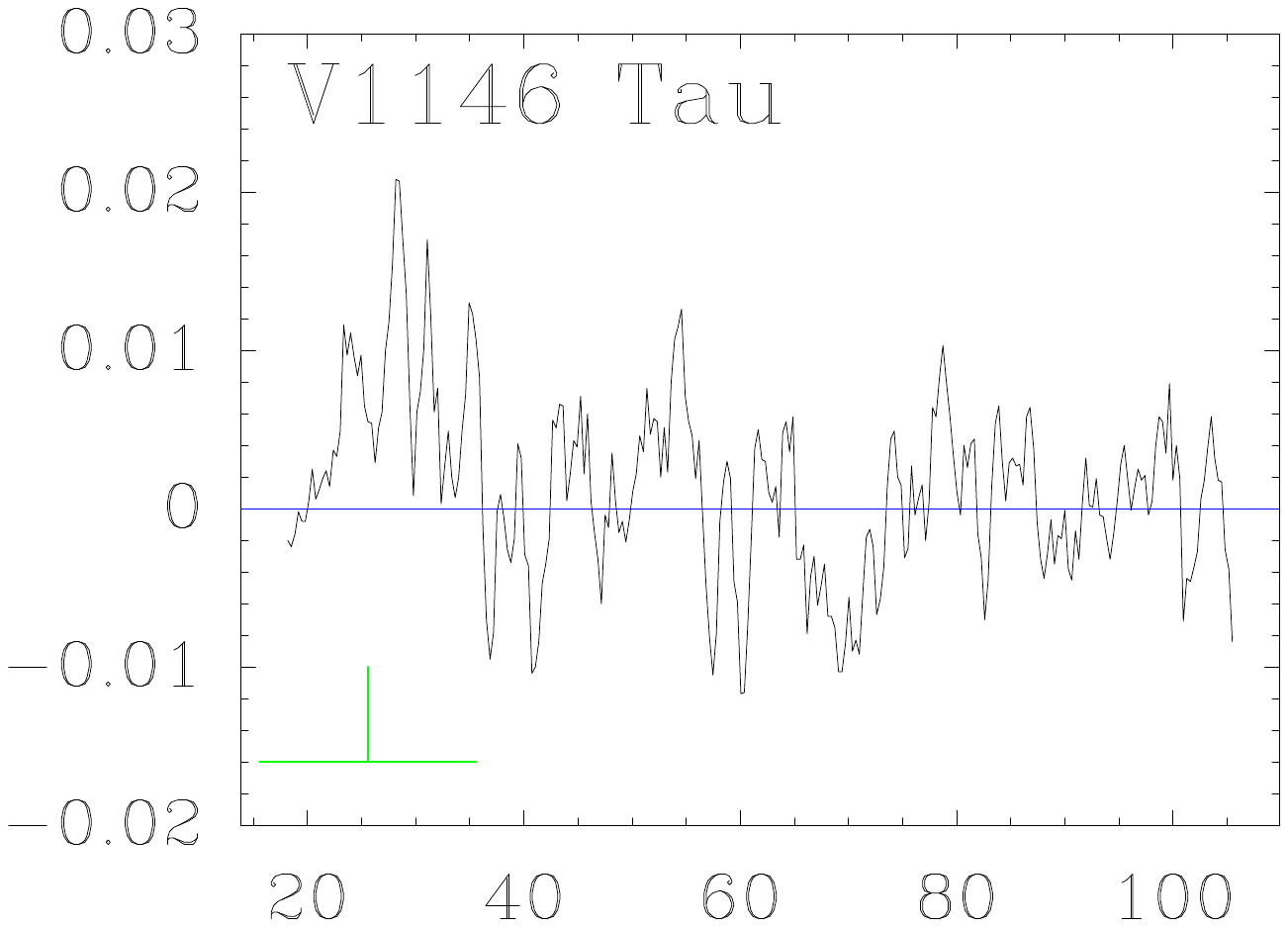}\hspace{-0.7cm}
\includegraphics[width=4.75cm]{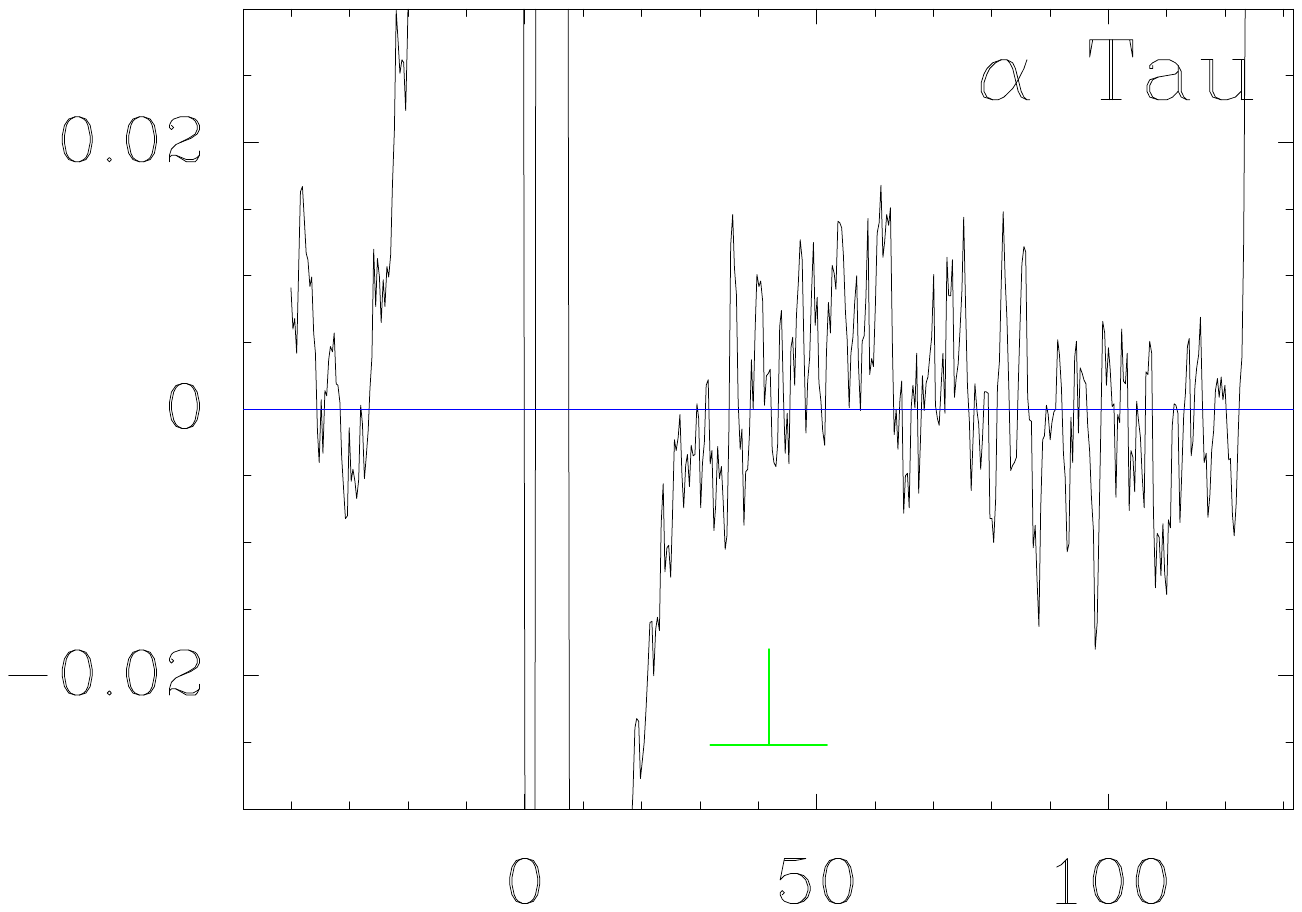}
\\ \vspace{-0.35cm}
\includegraphics[width=4.75cm]{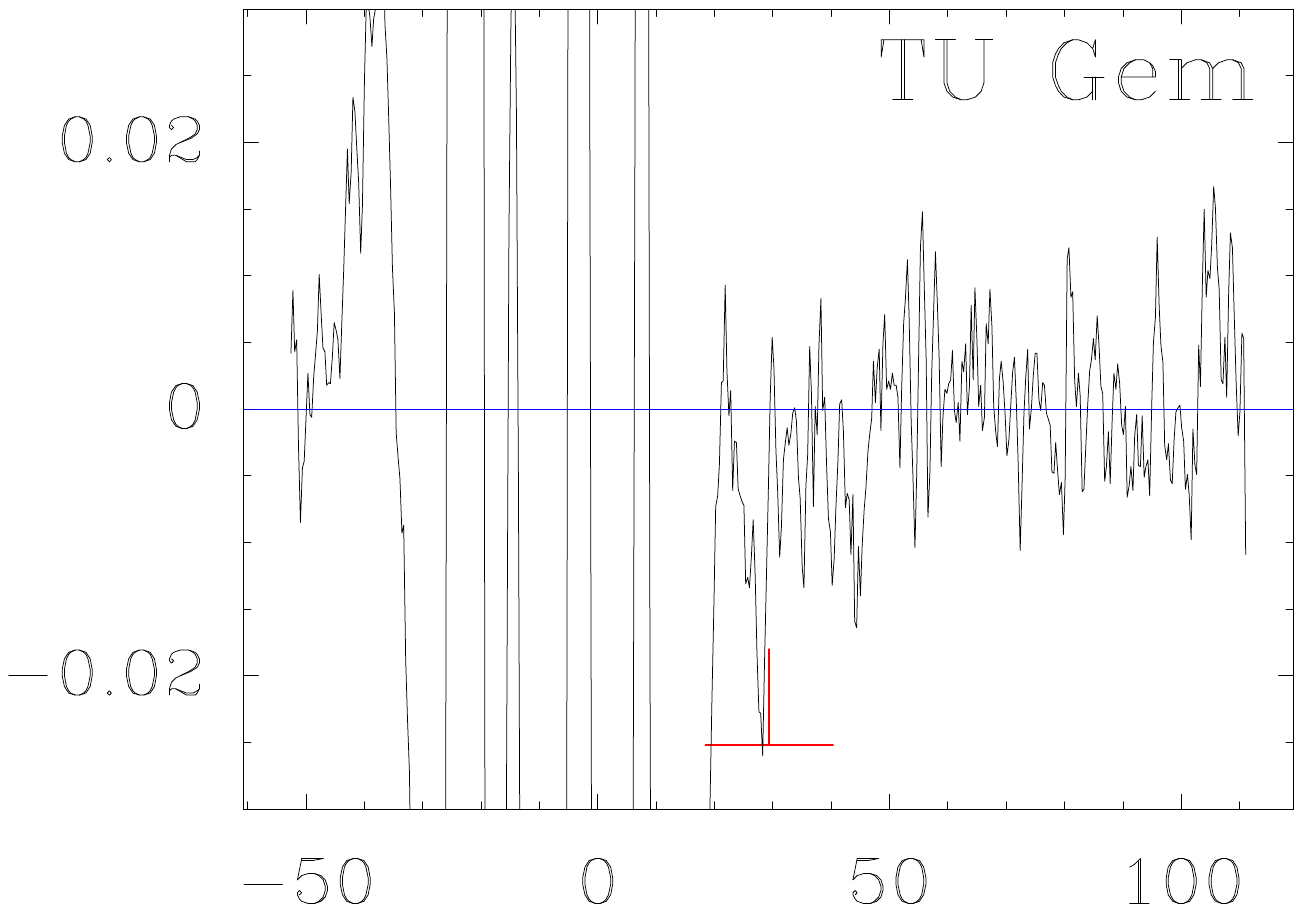}\hspace{-0.7cm}
\includegraphics[width=4.75cm]{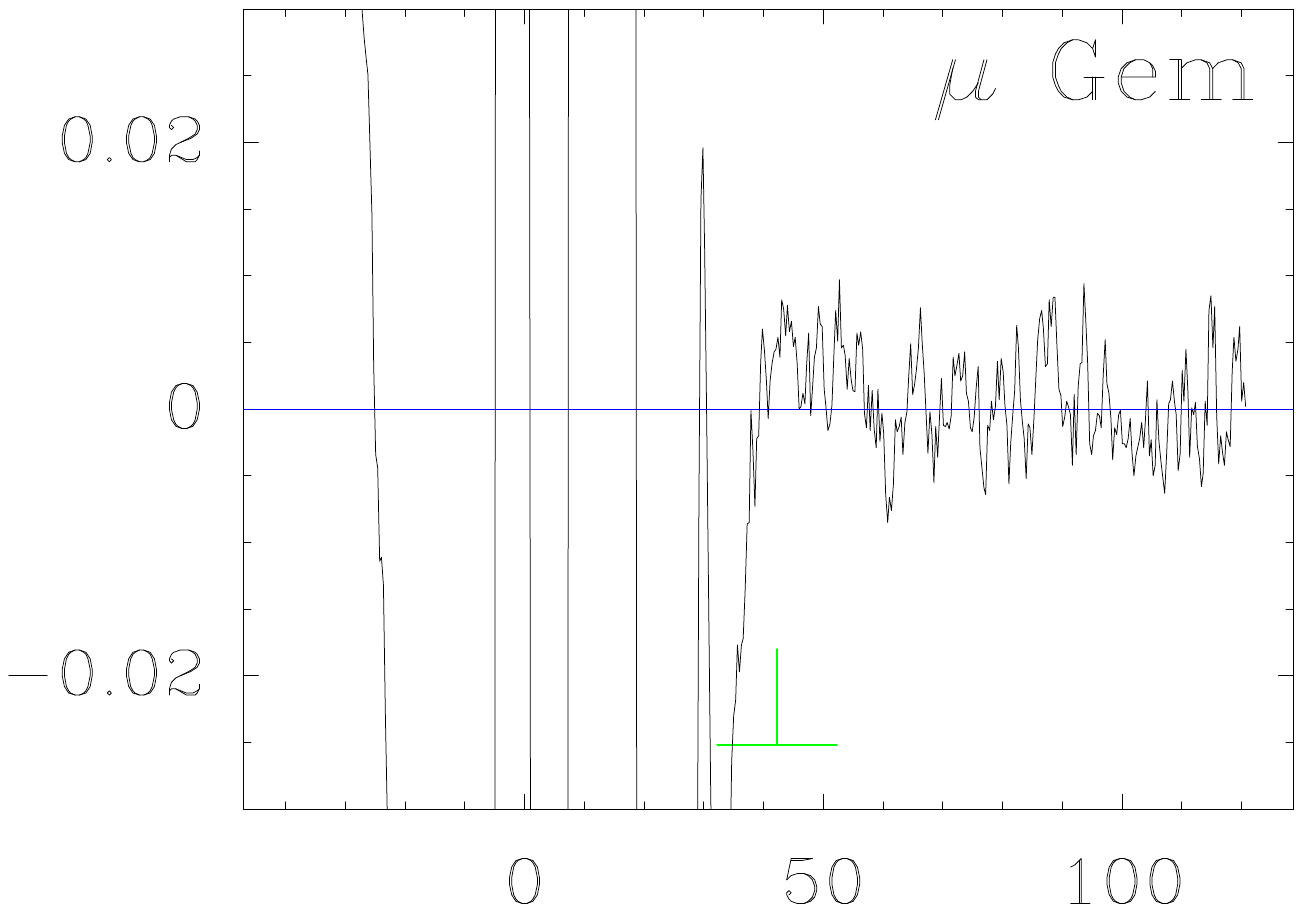}\hspace{-0.7cm}
\includegraphics[width=4.75cm]{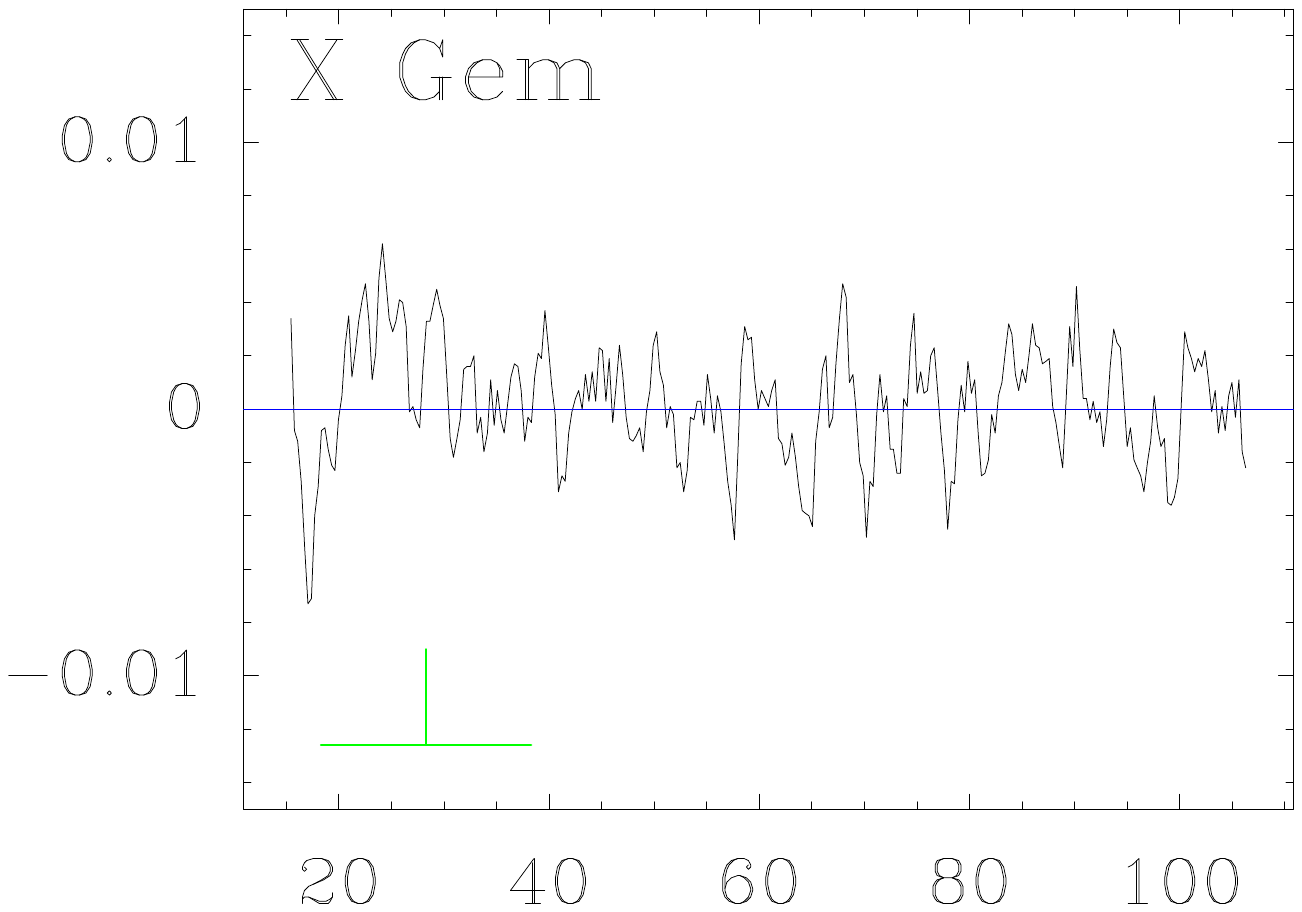}\hspace{-0.7cm}
\includegraphics[width=4.75cm]{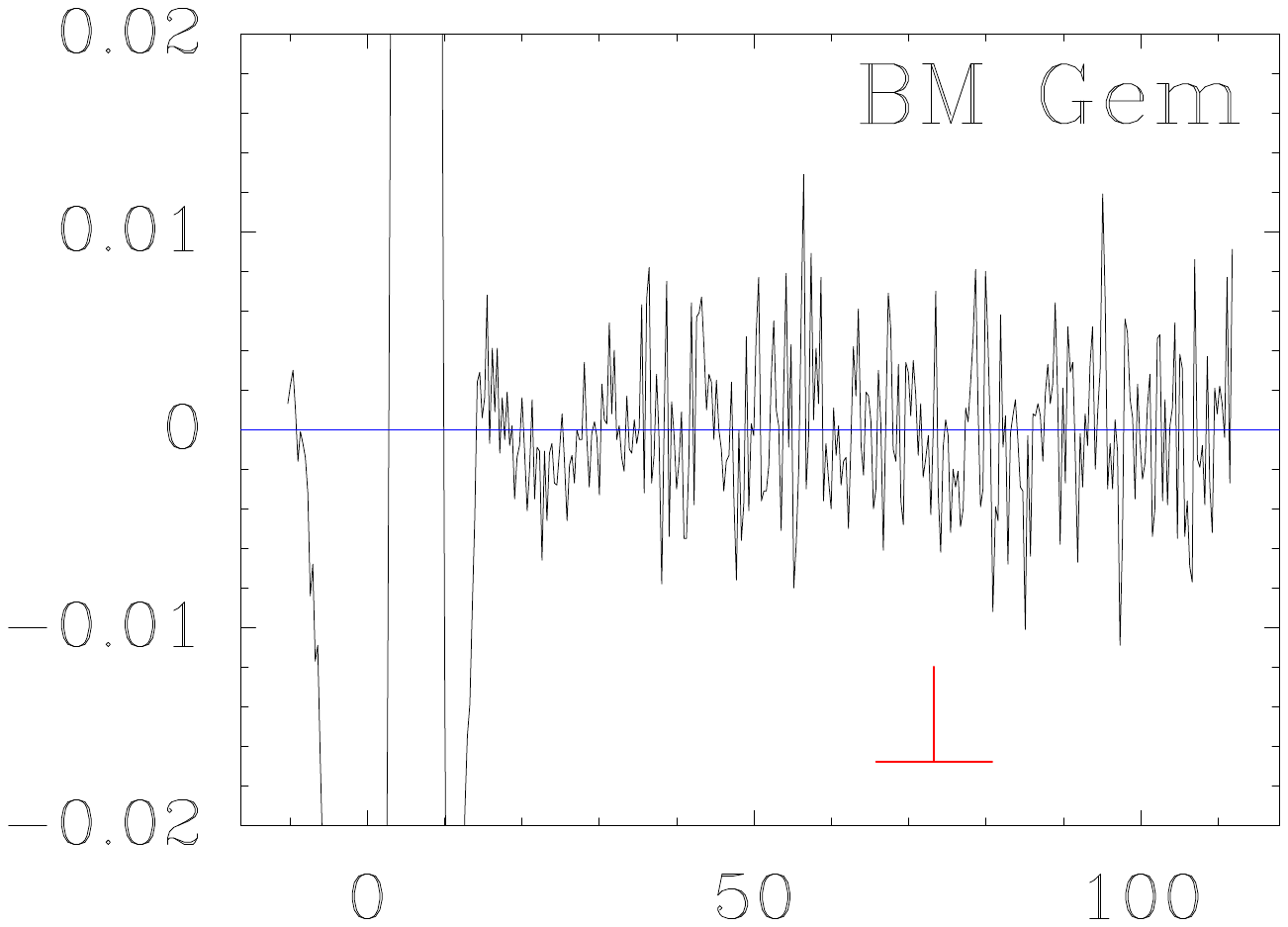}
\\ \vspace{-0.35cm}
\includegraphics[width=4.75cm]{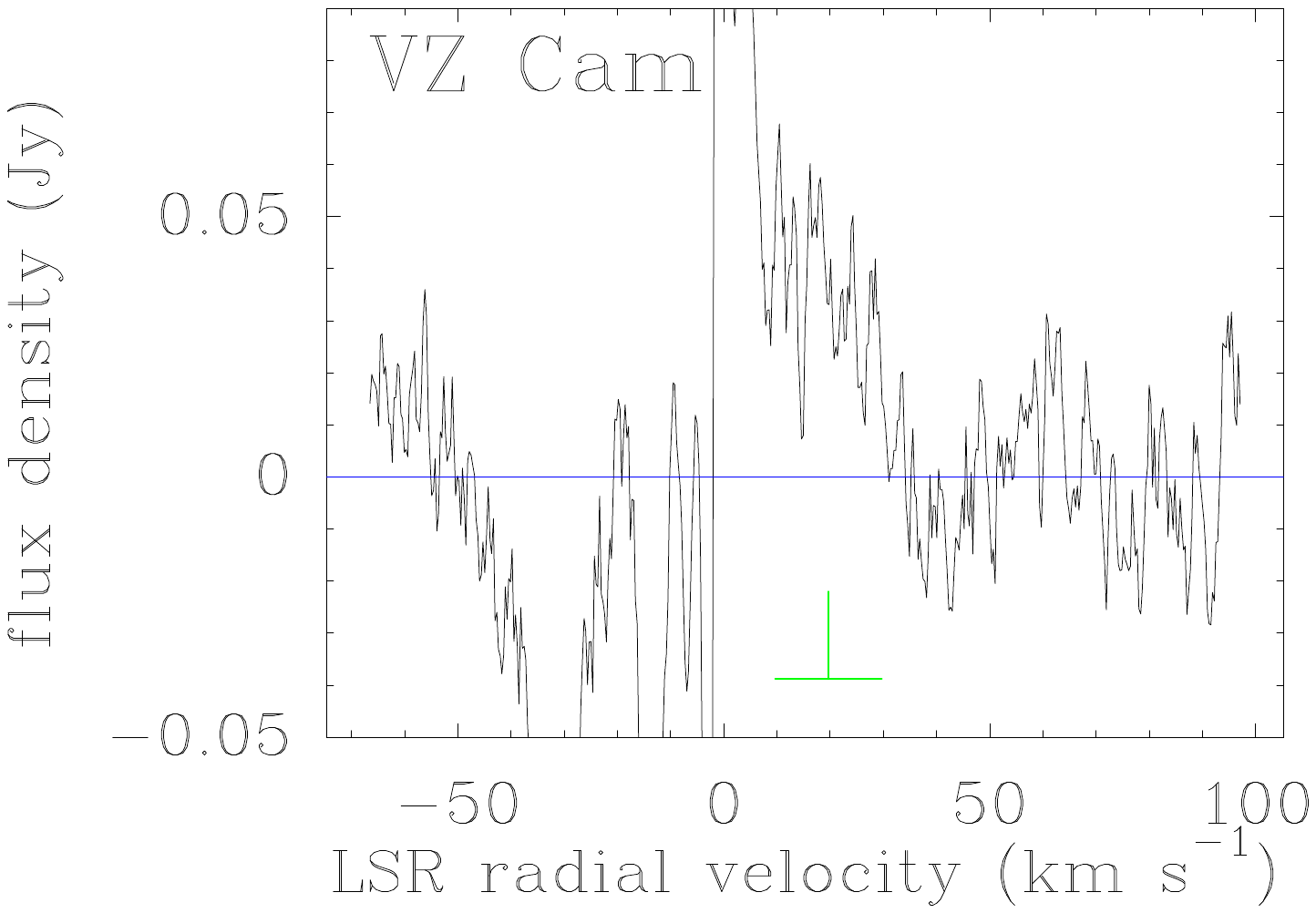}\hspace{-0.7cm}
\includegraphics[width=4.75cm]{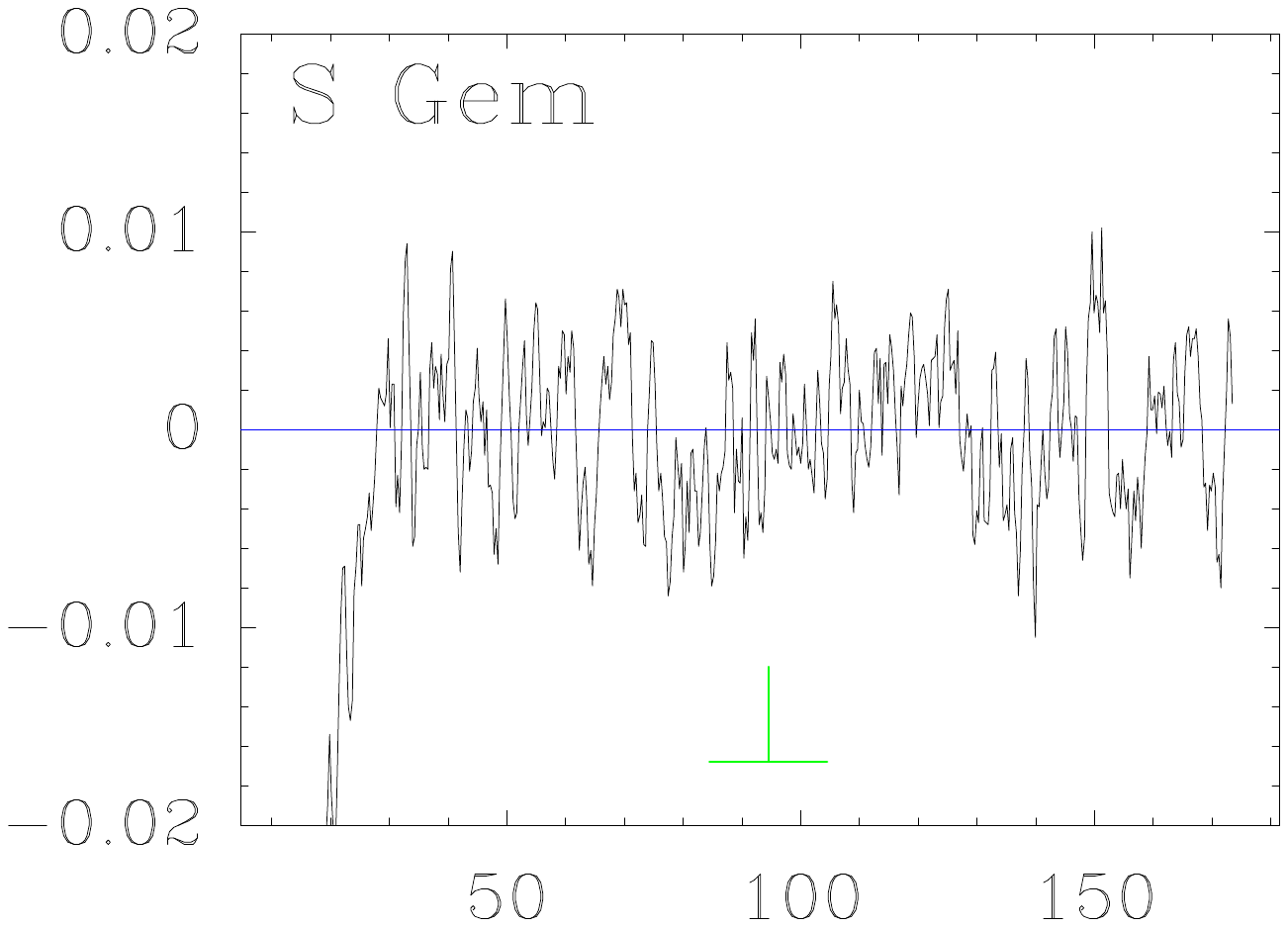}\hspace{-0.7cm}
\includegraphics[width=4.75cm]{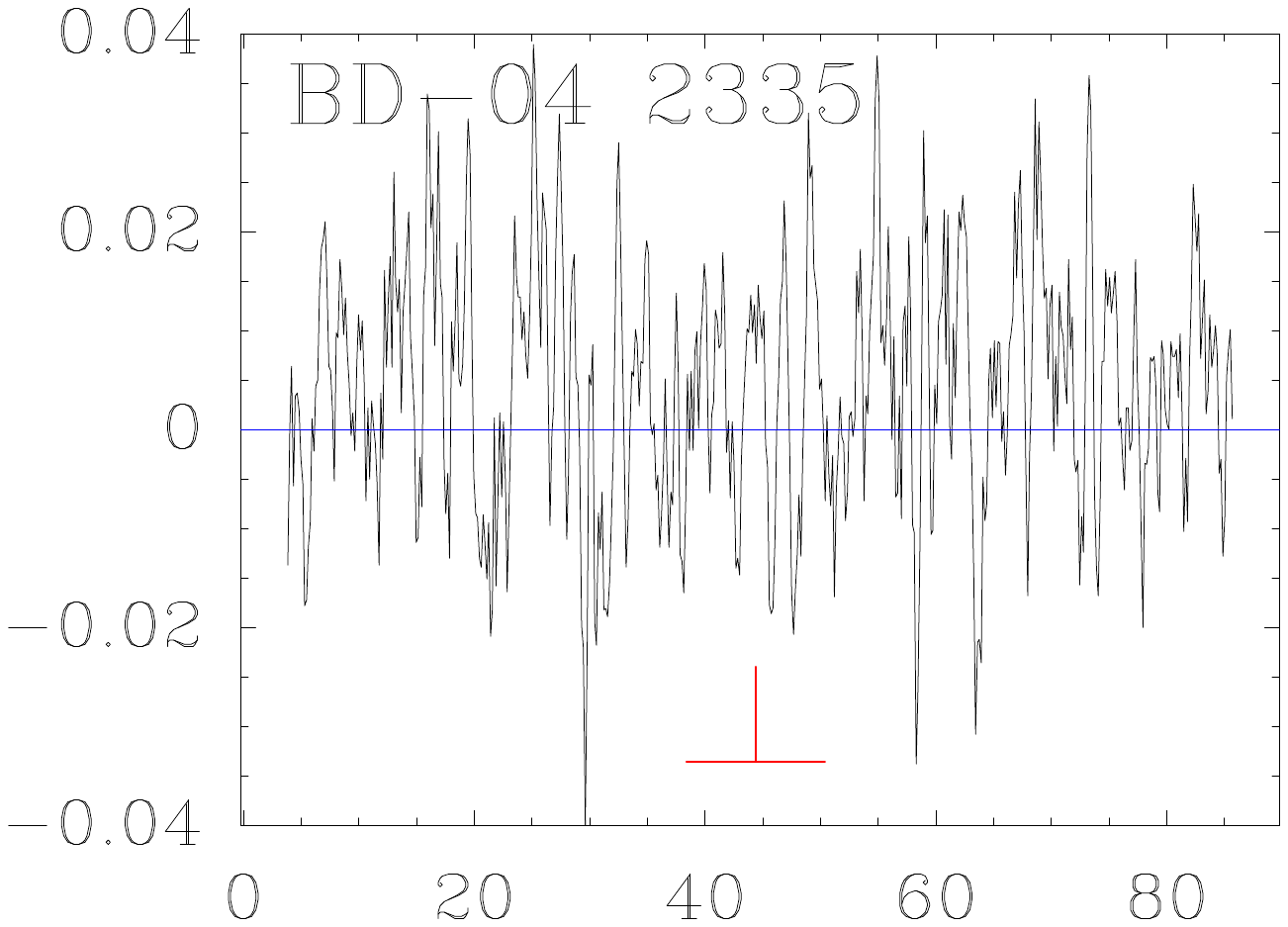}\hspace{-0.7cm}
\includegraphics[width=4.75cm]{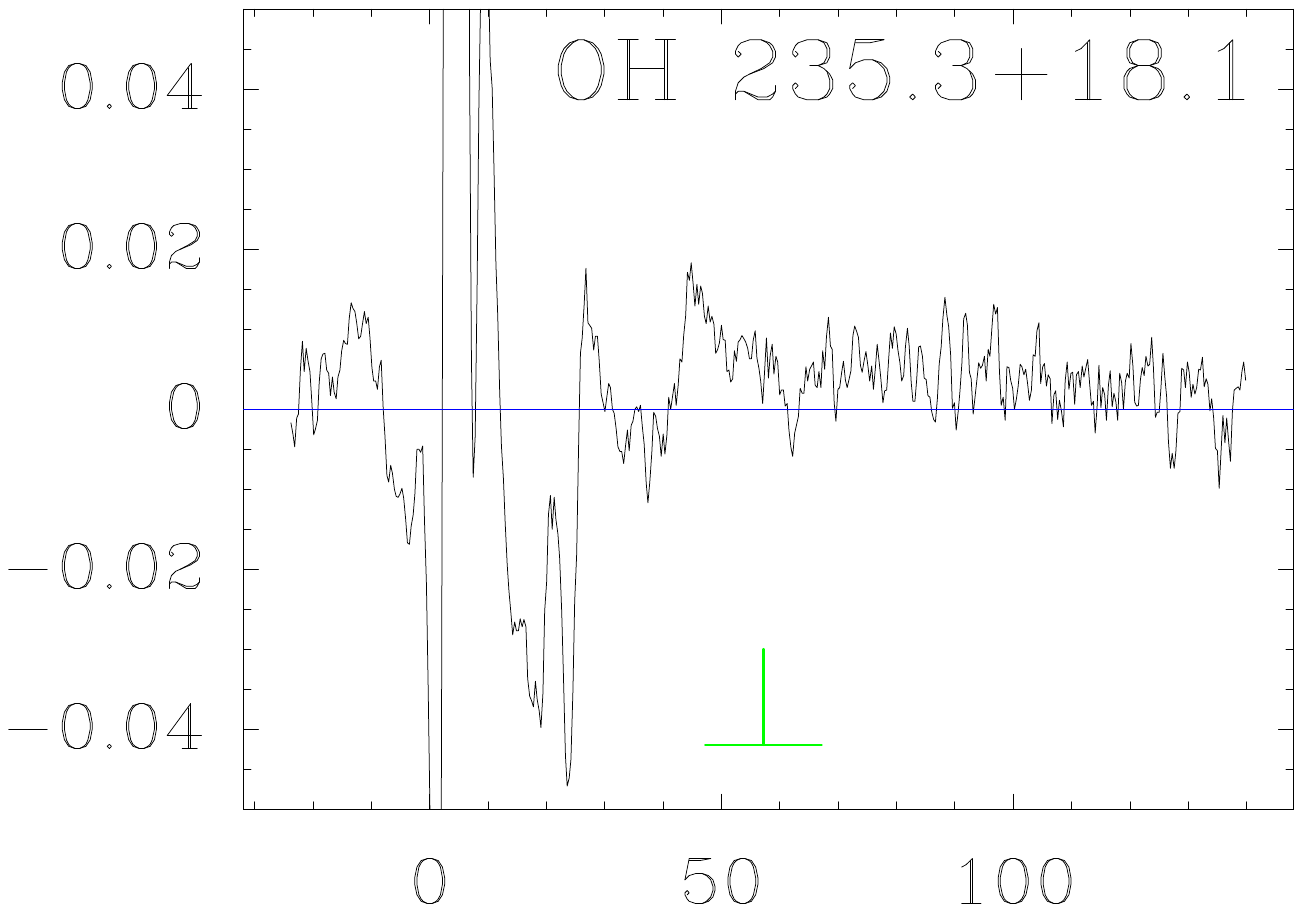}
% 
% \vspace{-3cm}
  \caption{{\bf a.} Upper limits.
Shown is flux density of the 21-cm \HI\ line emission, $S_{\rm HI}$, in Jy, 
as a function of radial velocity in the LSR reference frame, \VLSR, in \kms.
All spectra are peak profiles (see Section \ref{sec:sizes}). 
Spectra are shown only for the 70 objects with upper limits to their \HI\ line emission 
for which digital versions of the observations are available, i.e., 
those observed from 2001 onwards with the renovated NRT (thus excluding the 25 objects with the 
"{\footnotesize \it old data}" note in Table~\ref{table:upperlimits}).
The red vertical and horizontal lines indicate respectively the central CO or OH line velocity 
from the literature and its corresponding expansion velocity,
whereas green lines indicate other types of literature velocities (e.g., optical or SiO lines)
and an indicative expansion velocity of 10 \kms, i.e., the average measured value. 
The flat blue horizontal lines show the 0 Jy flux density level.
The plotted total velocity range is the full observed range for each object.}
\label{fig:spectralimits}
\end{figure*}

\begin{figure*}[ht]  % \label{fig:spectra_limits}  % Fig. 5b
\addtocounter{figure}{-1}
  \centering
\includegraphics[width=4.75cm]{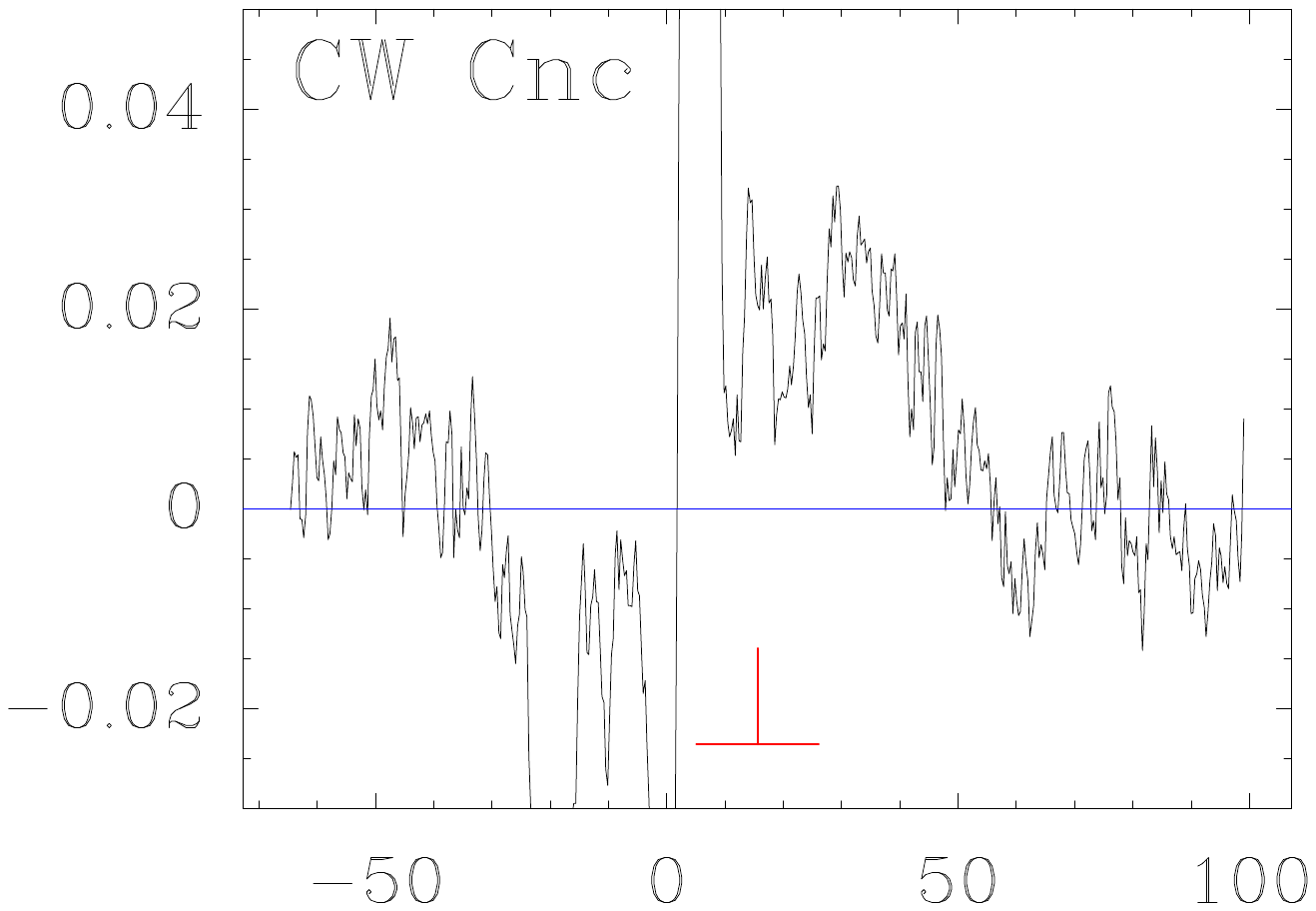}\hspace{-0.7cm}
\includegraphics[width=4.75cm]{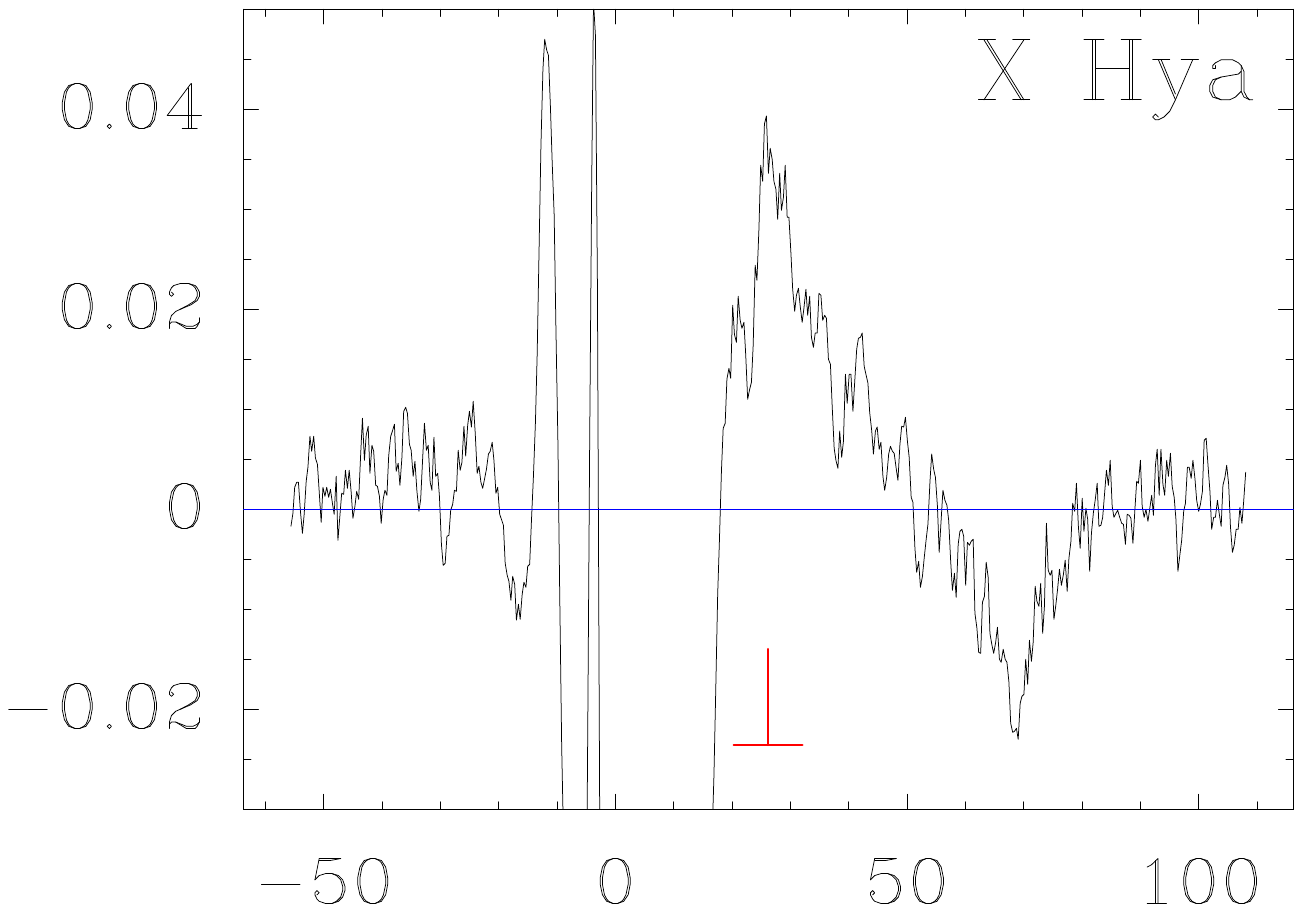}\hspace{-0.7cm}
\includegraphics[width=4.75cm]{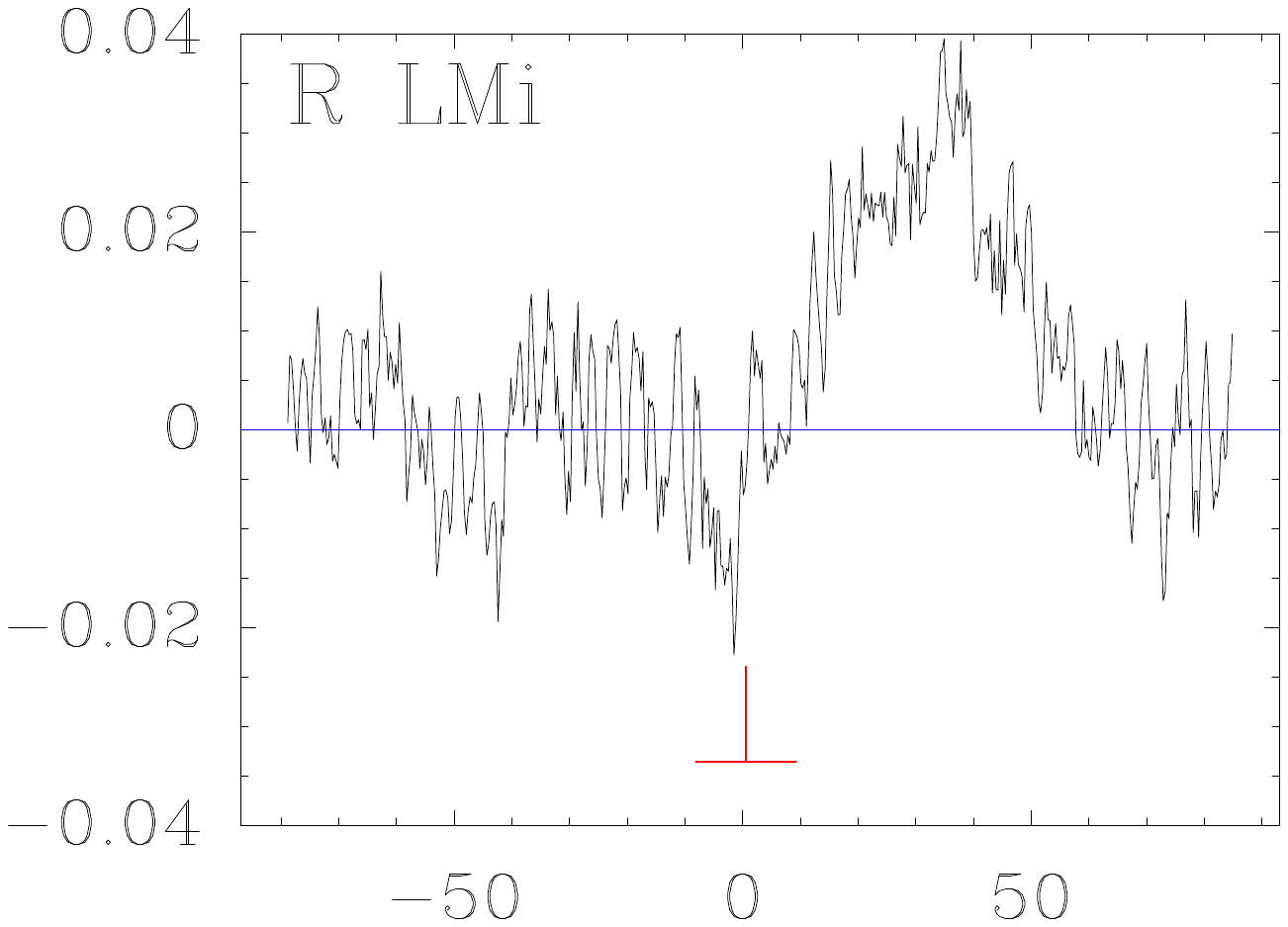}\hspace{-0.7cm}
\includegraphics[width=4.75cm]{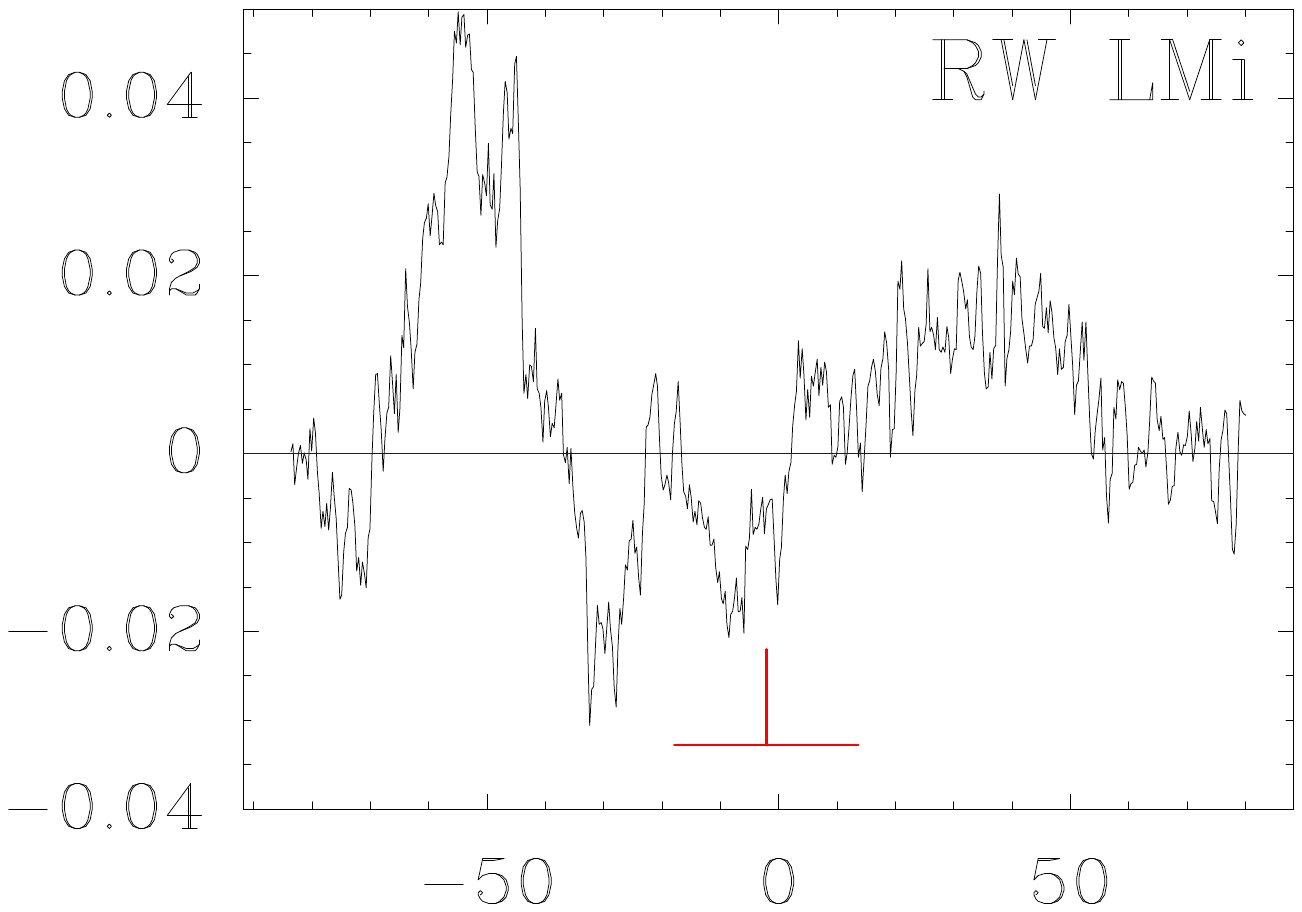}
\\ \vspace{-0.35cm}
\includegraphics[width=4.75cm]{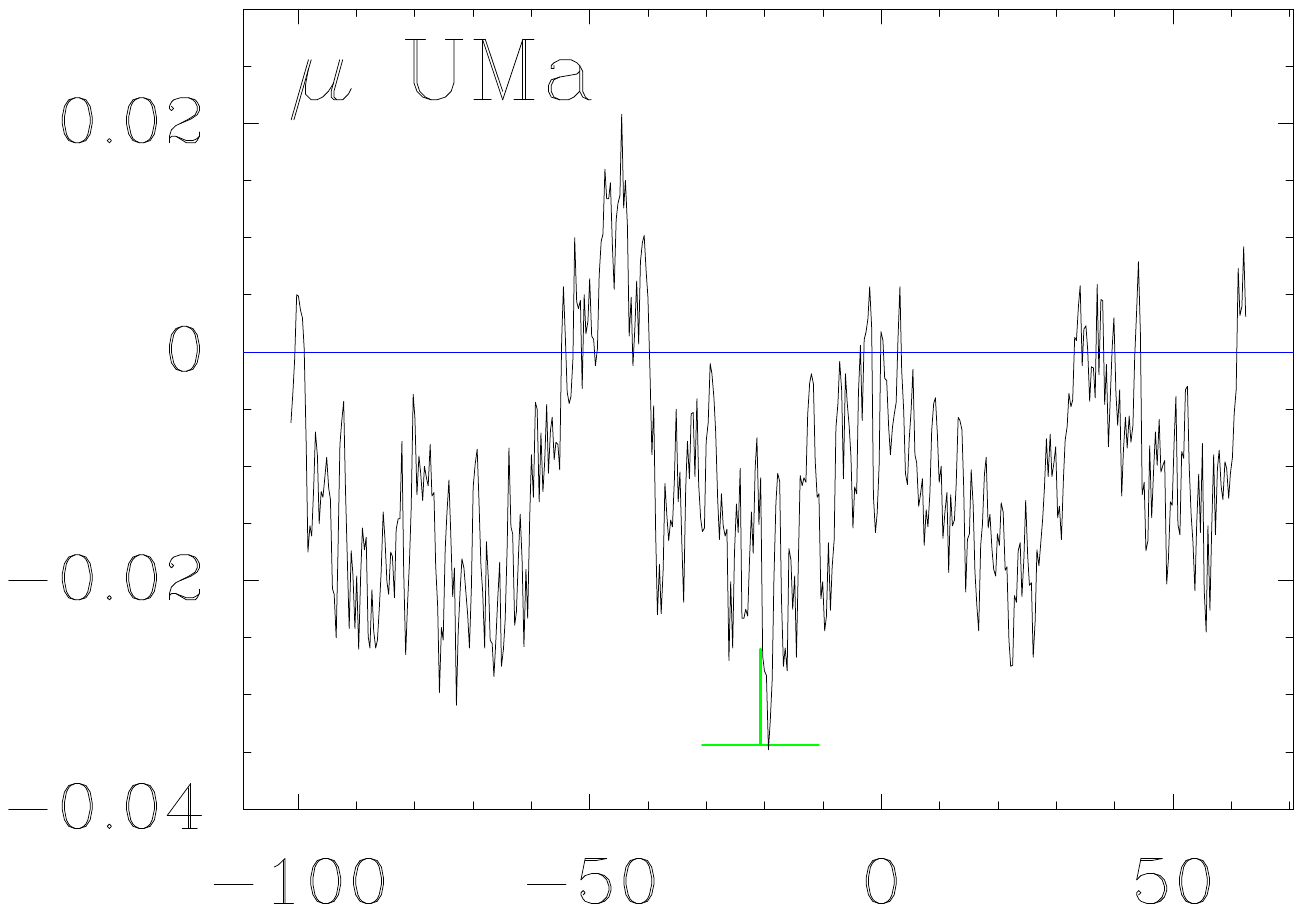}\hspace{-0.7cm}
\includegraphics[width=4.75cm]{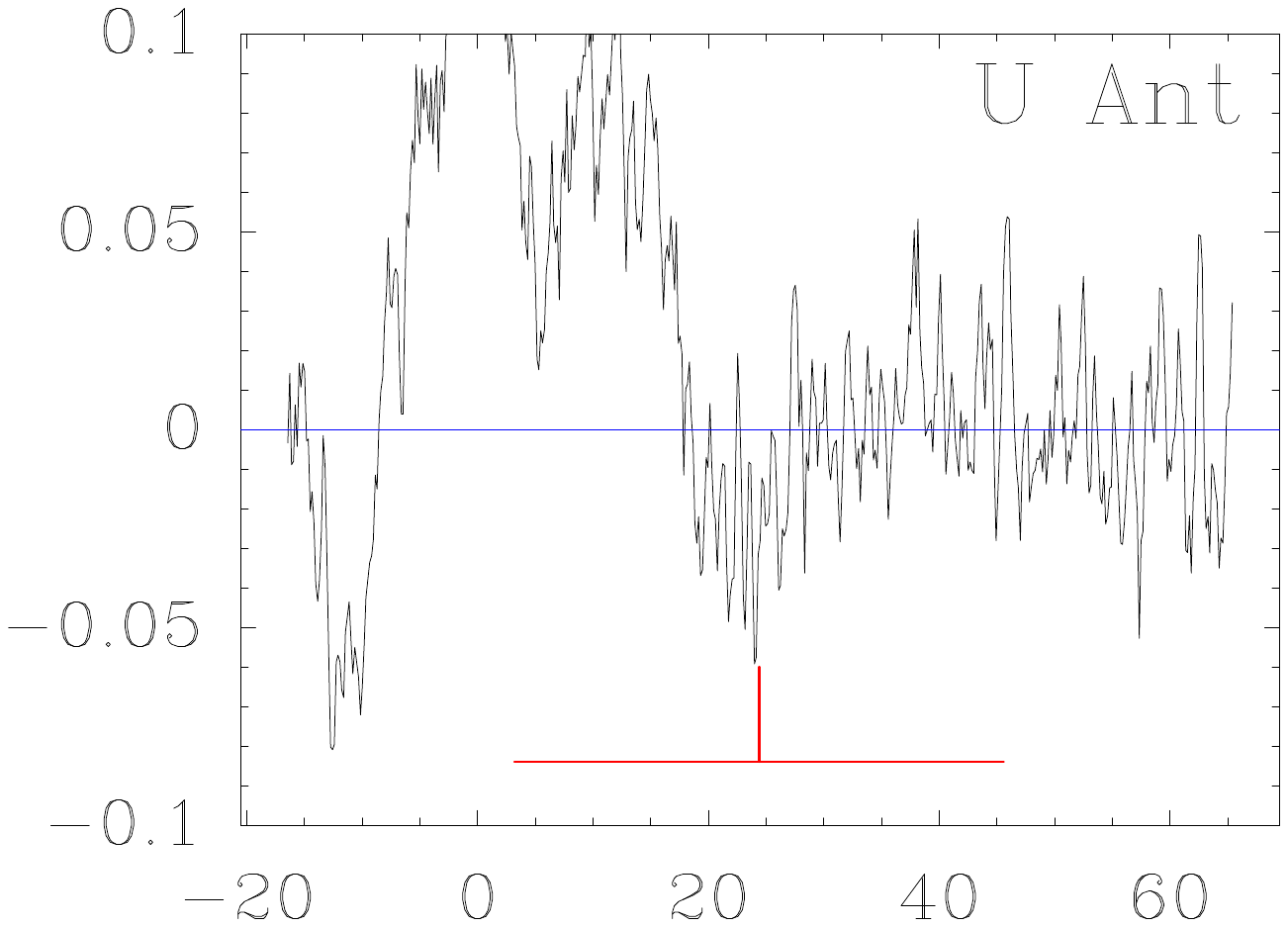}\hspace{-0.7cm}
\includegraphics[width=4.75cm]{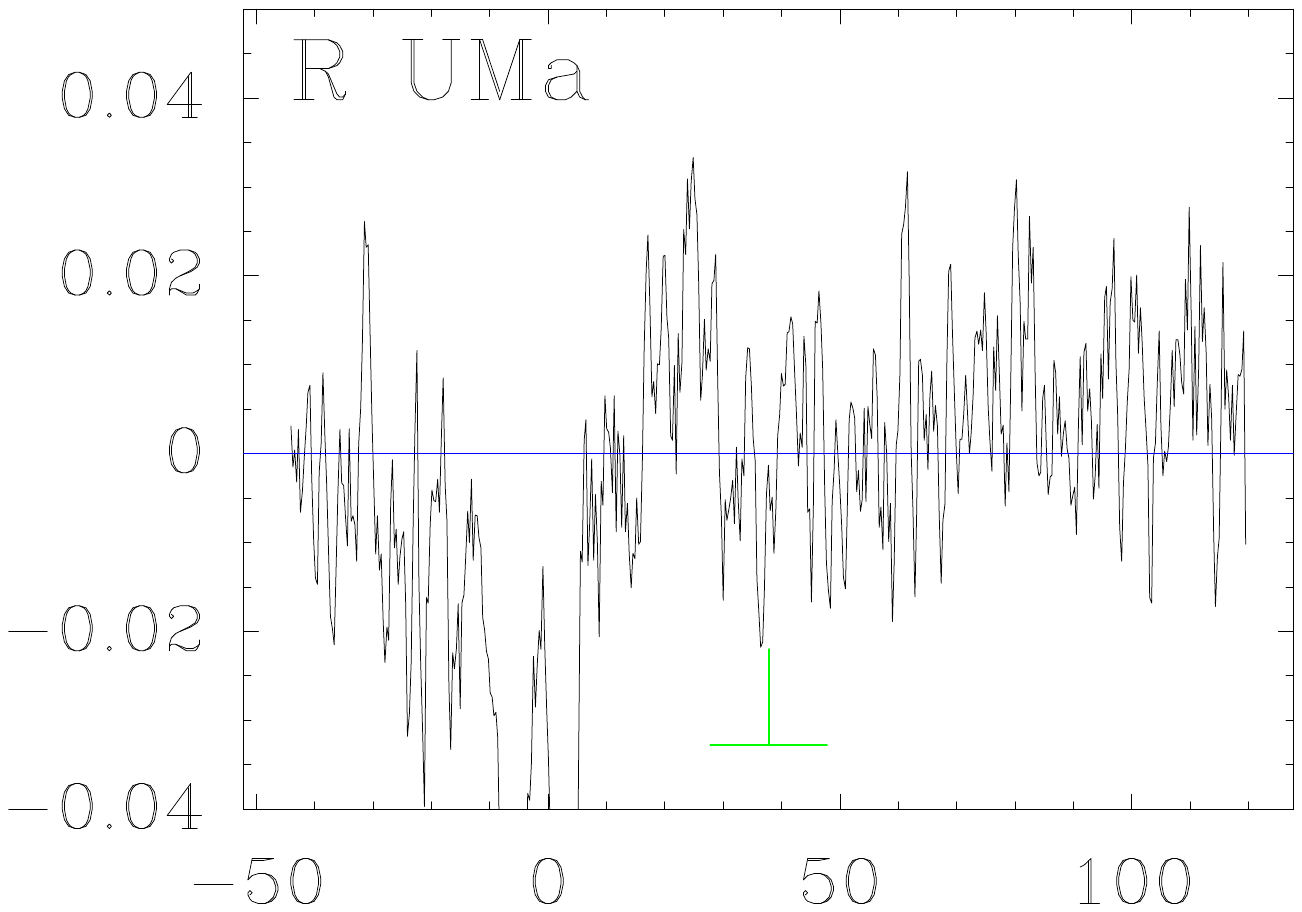}\hspace{-0.7cm}
\includegraphics[width=4.75cm]{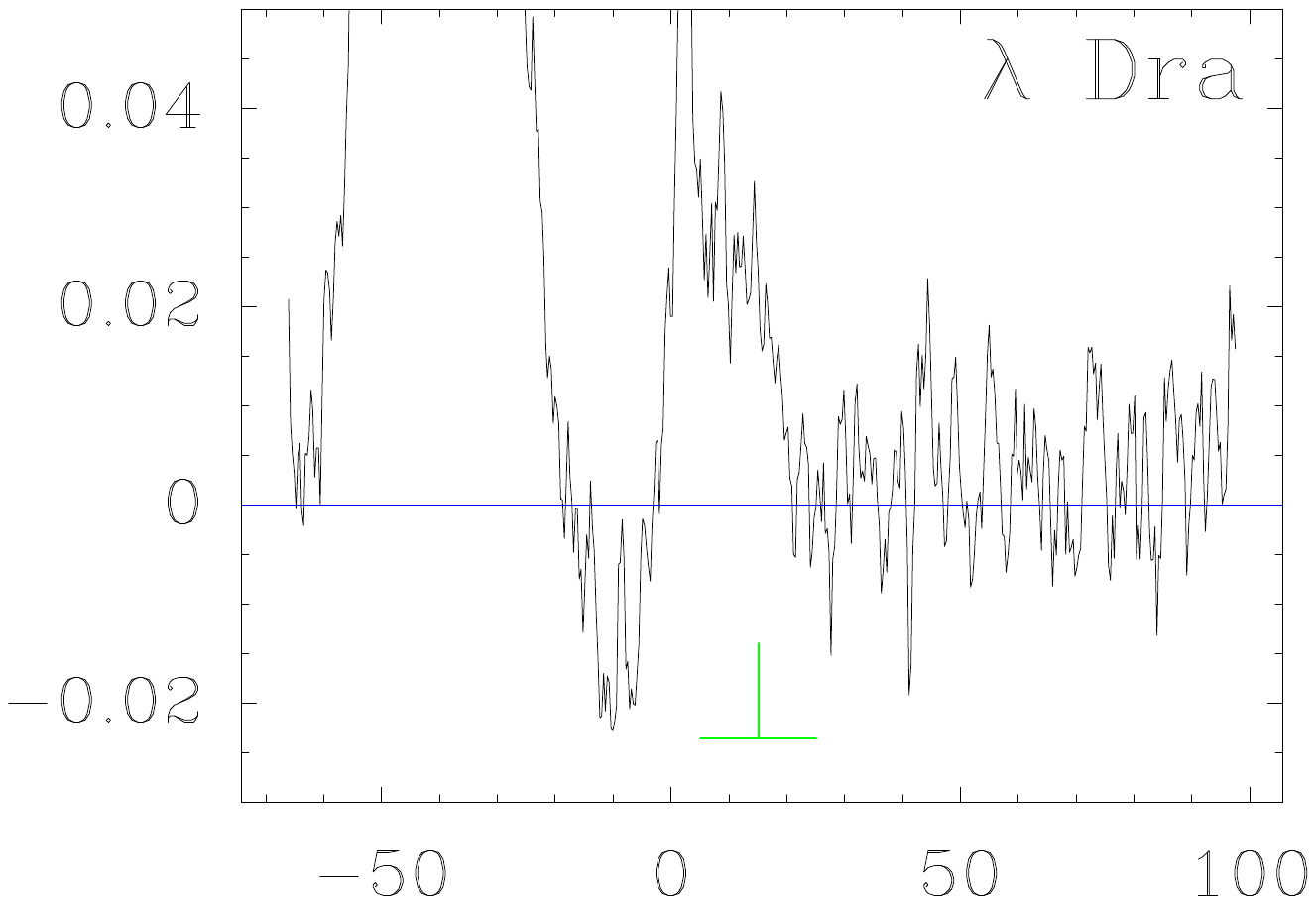}
\\ \vspace{-0.35cm}
\includegraphics[width=4.75cm]{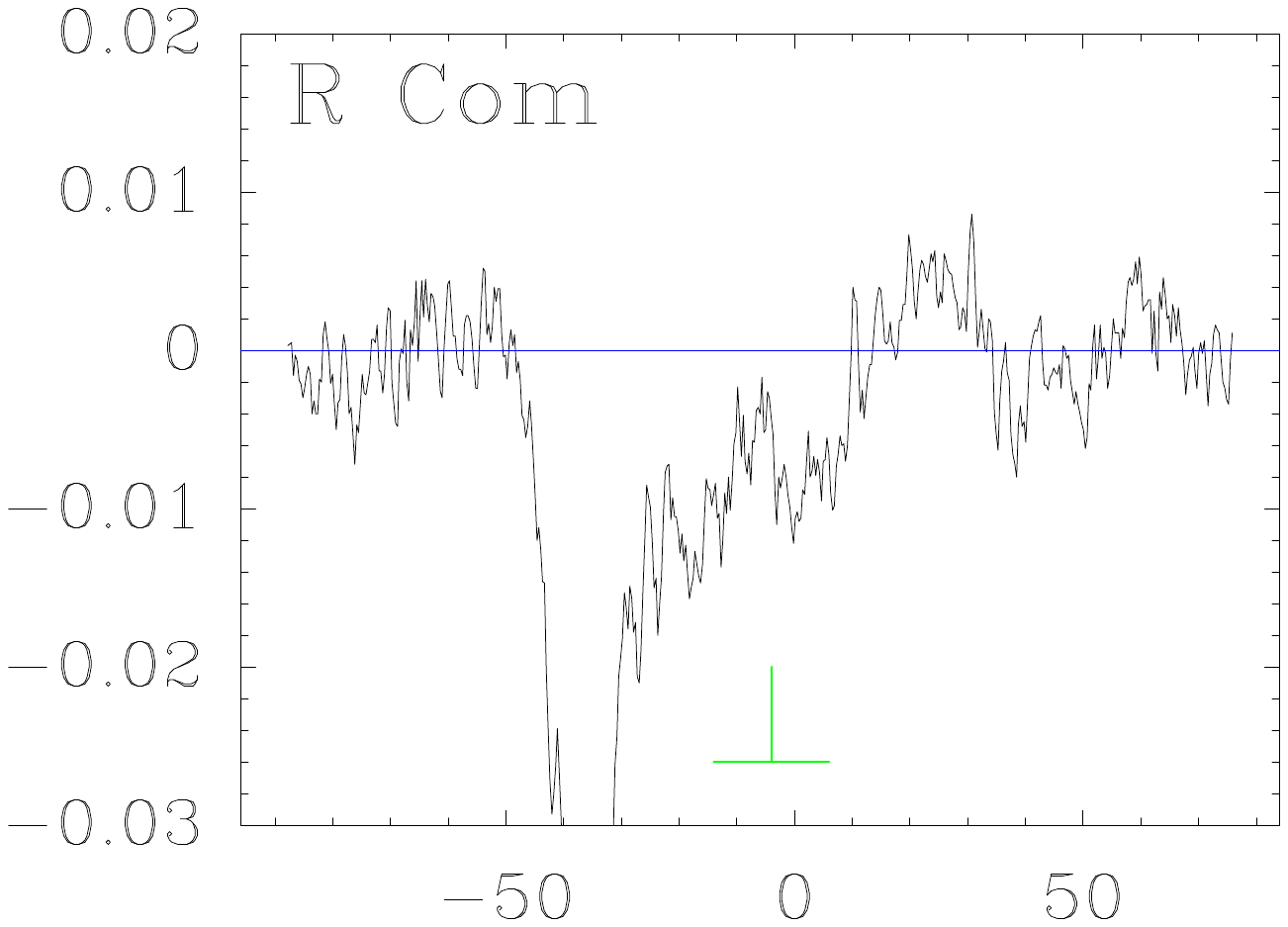}\hspace{-0.7cm}
\includegraphics[width=4.75cm]{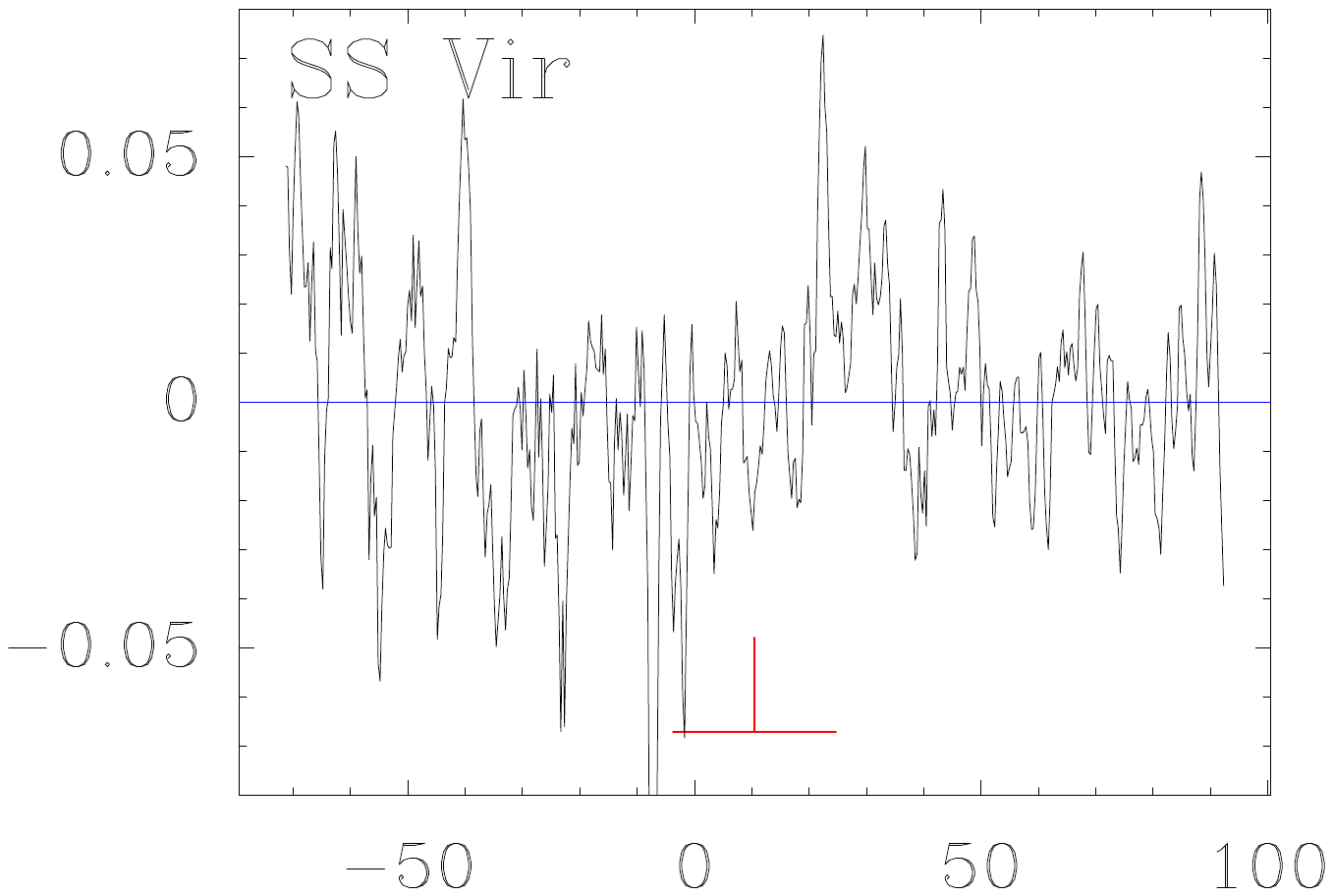}\hspace{-0.7cm}
\includegraphics[width=4.75cm]{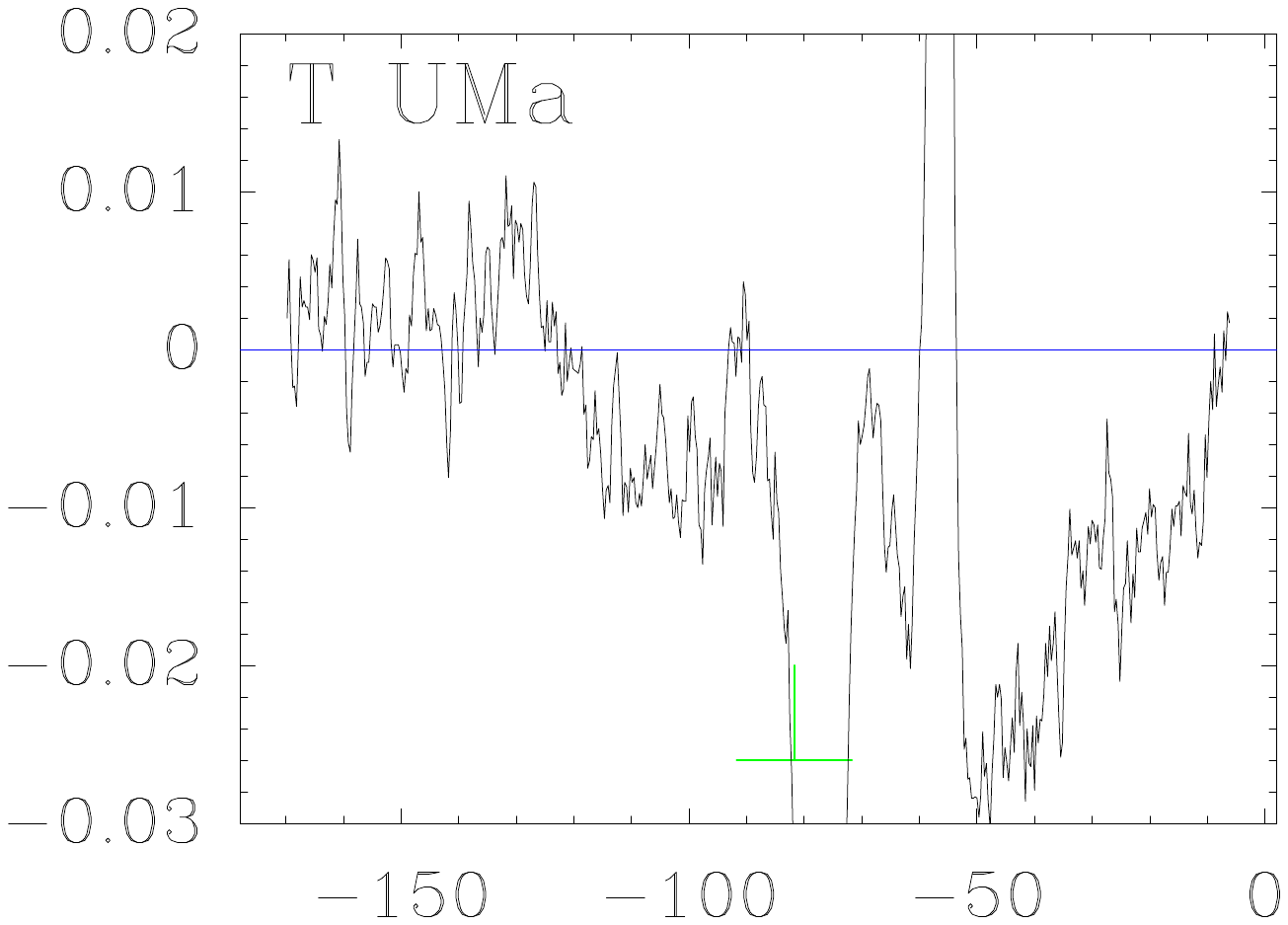}\hspace{-0.7cm}
\includegraphics[width=4.75cm]{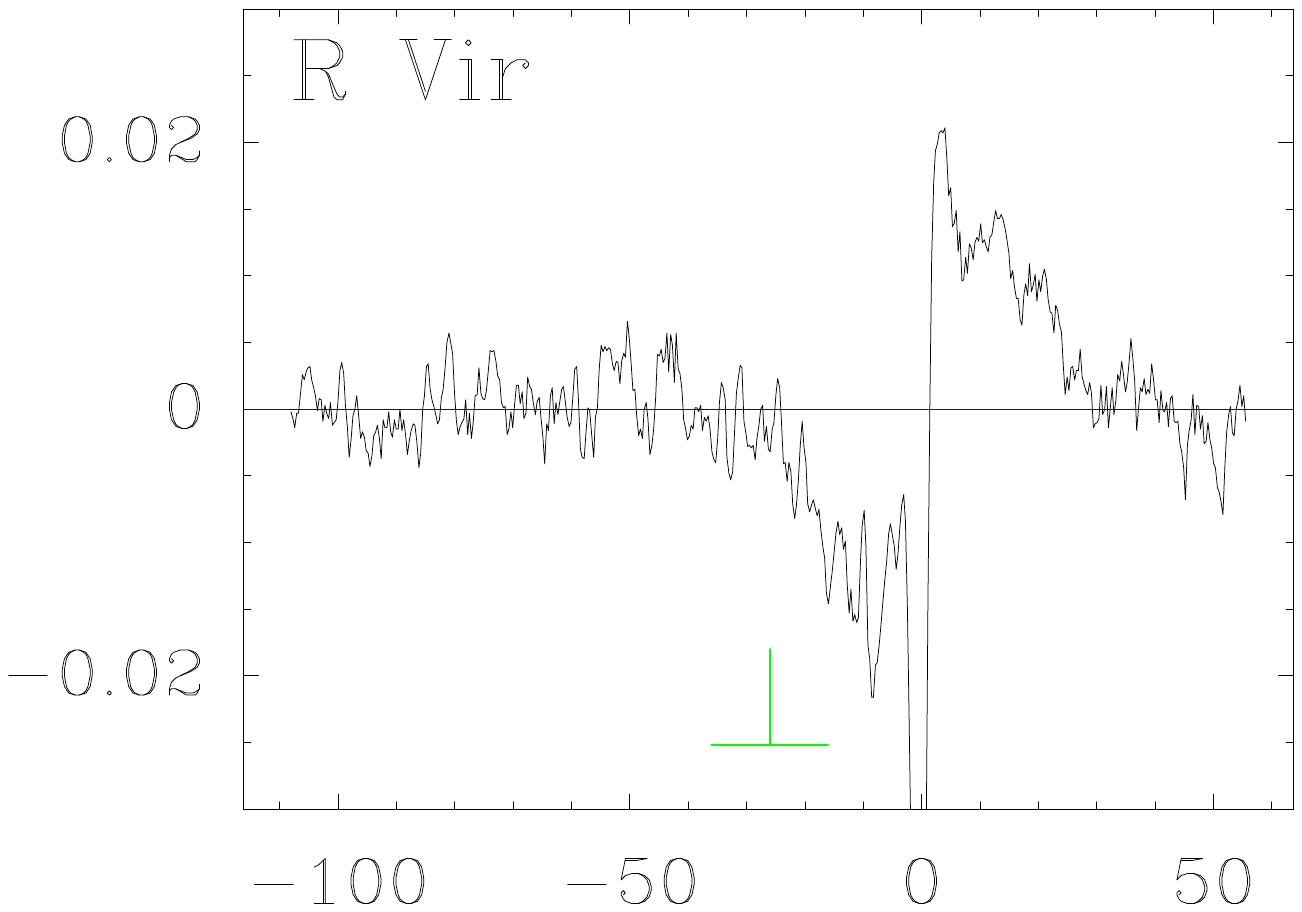}
\\ \vspace{-0.35cm}
\includegraphics[width=4.75cm]{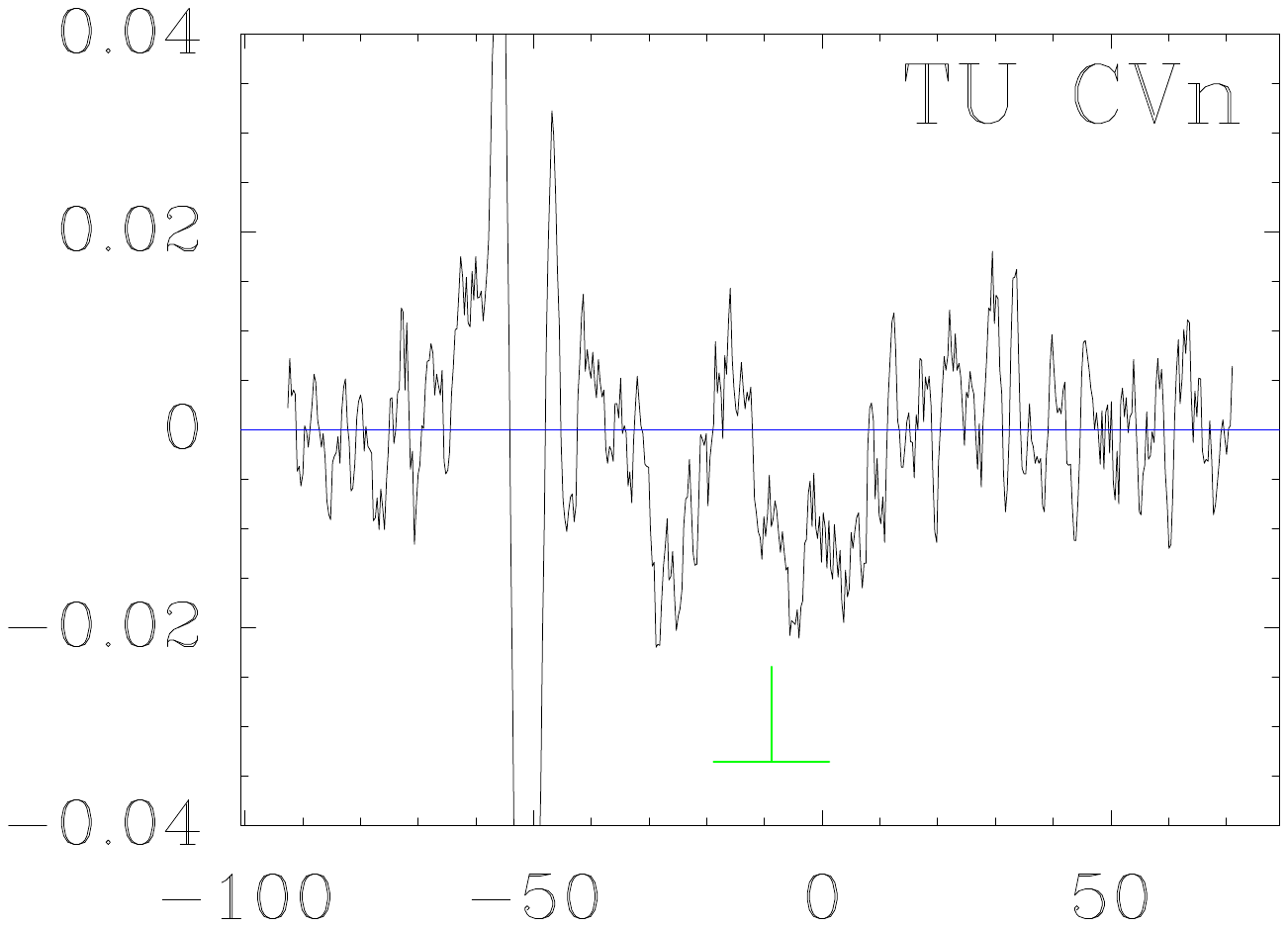}\hspace{-0.7cm}
\includegraphics[width=4.75cm]{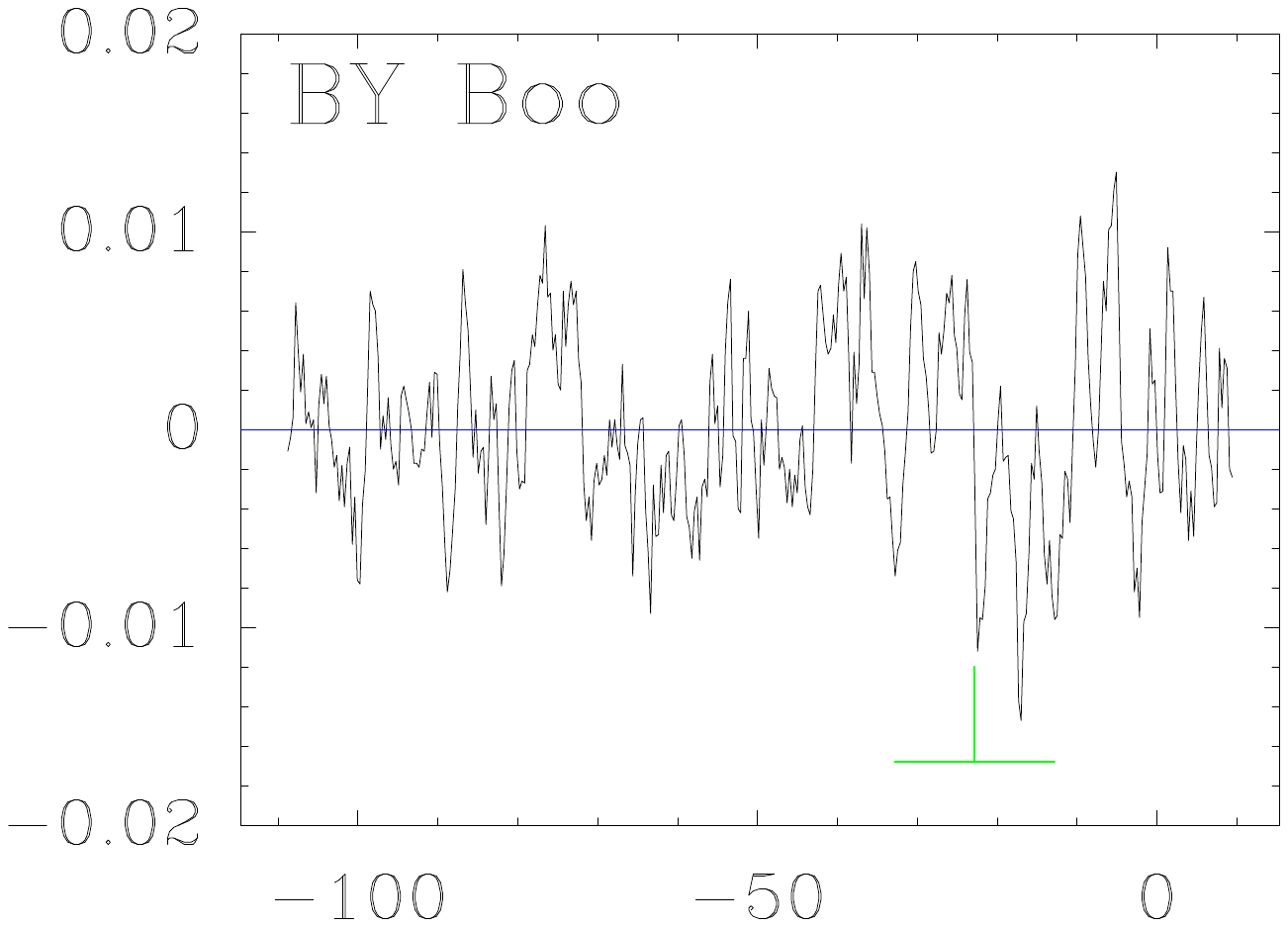}\hspace{-0.7cm}
\includegraphics[width=4.75cm]{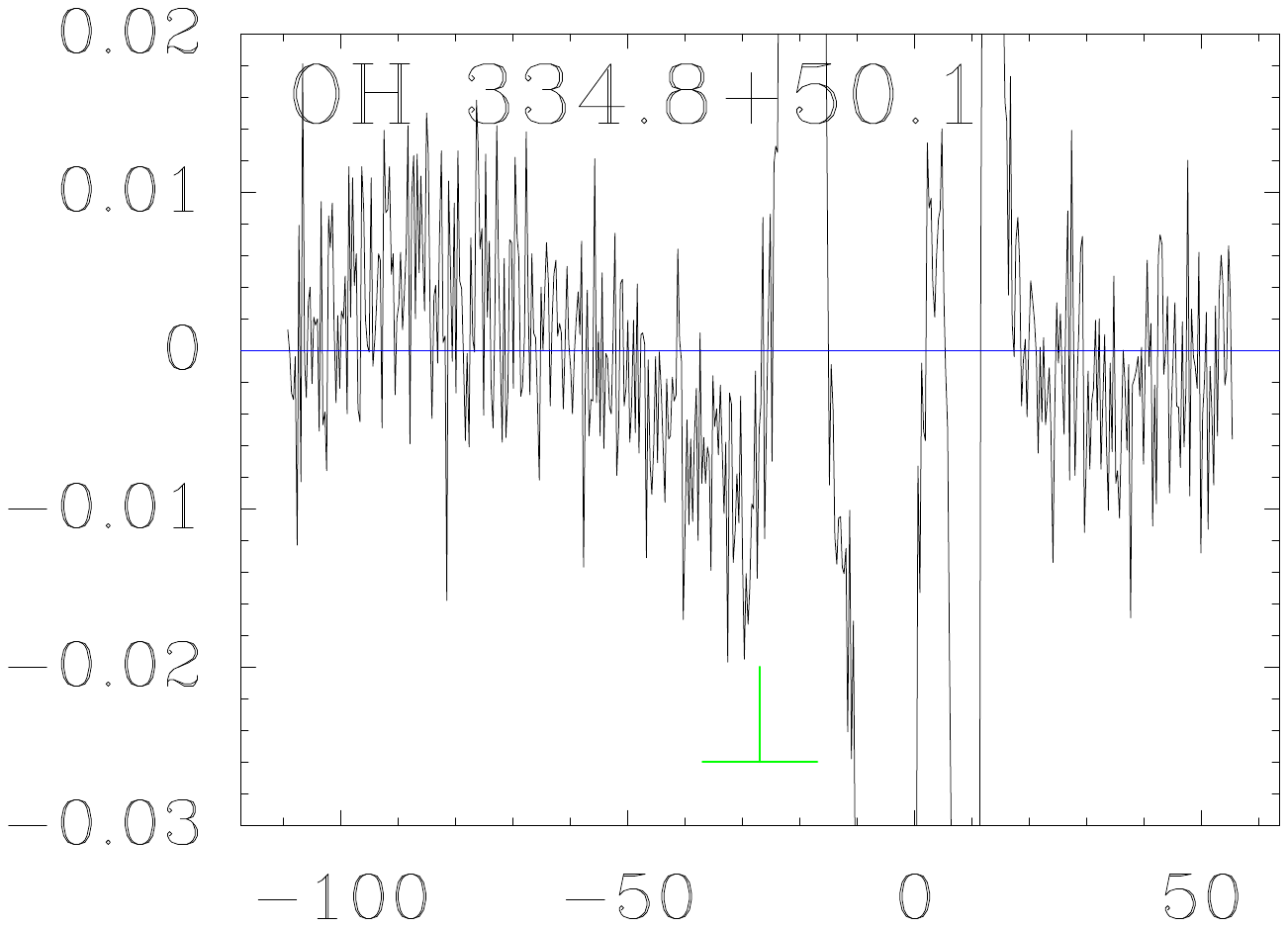}\hspace{-0.7cm}
\includegraphics[width=4.75cm]{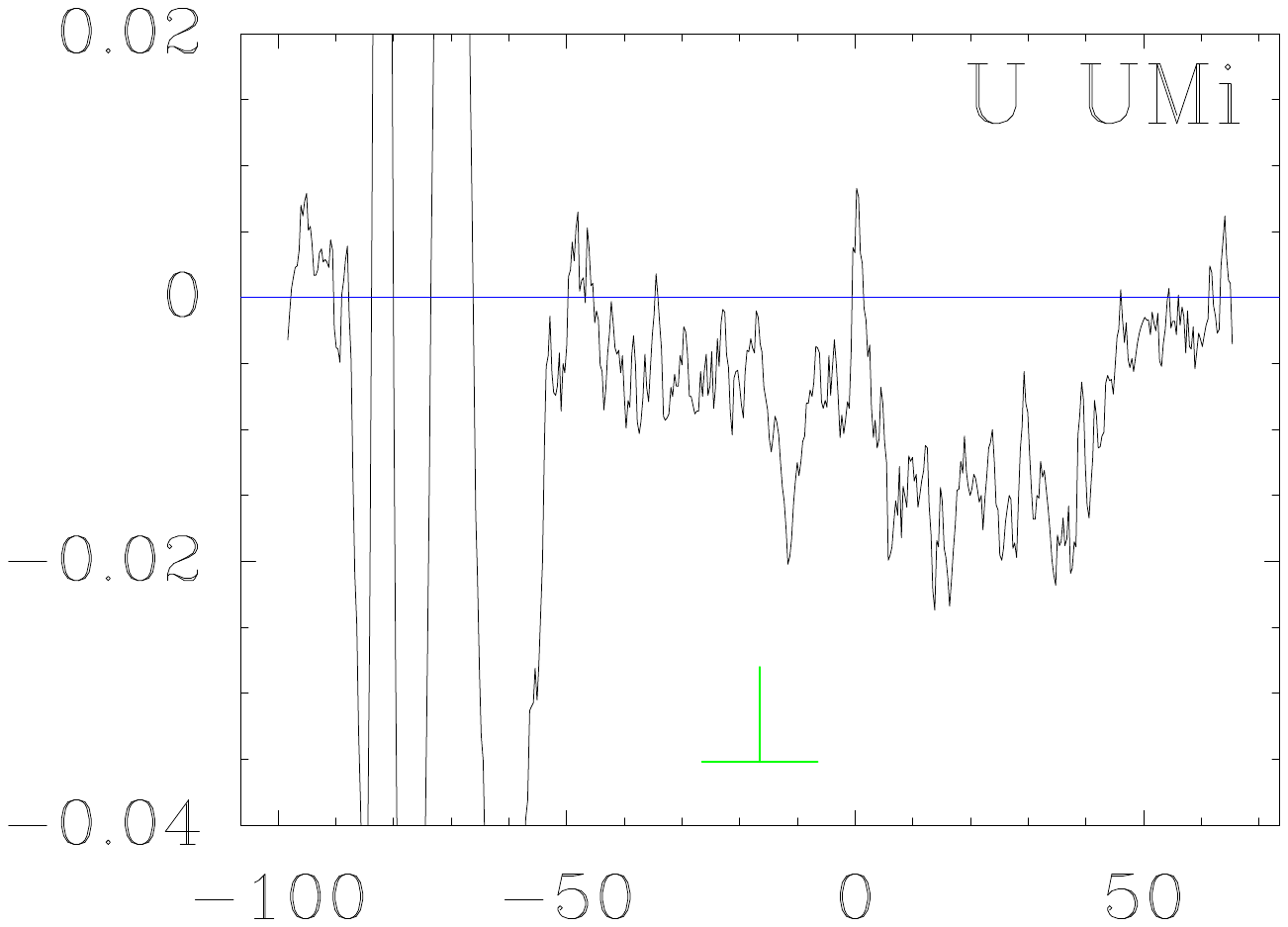}
\\ \vspace{-0.35cm}
\includegraphics[width=4.75cm]{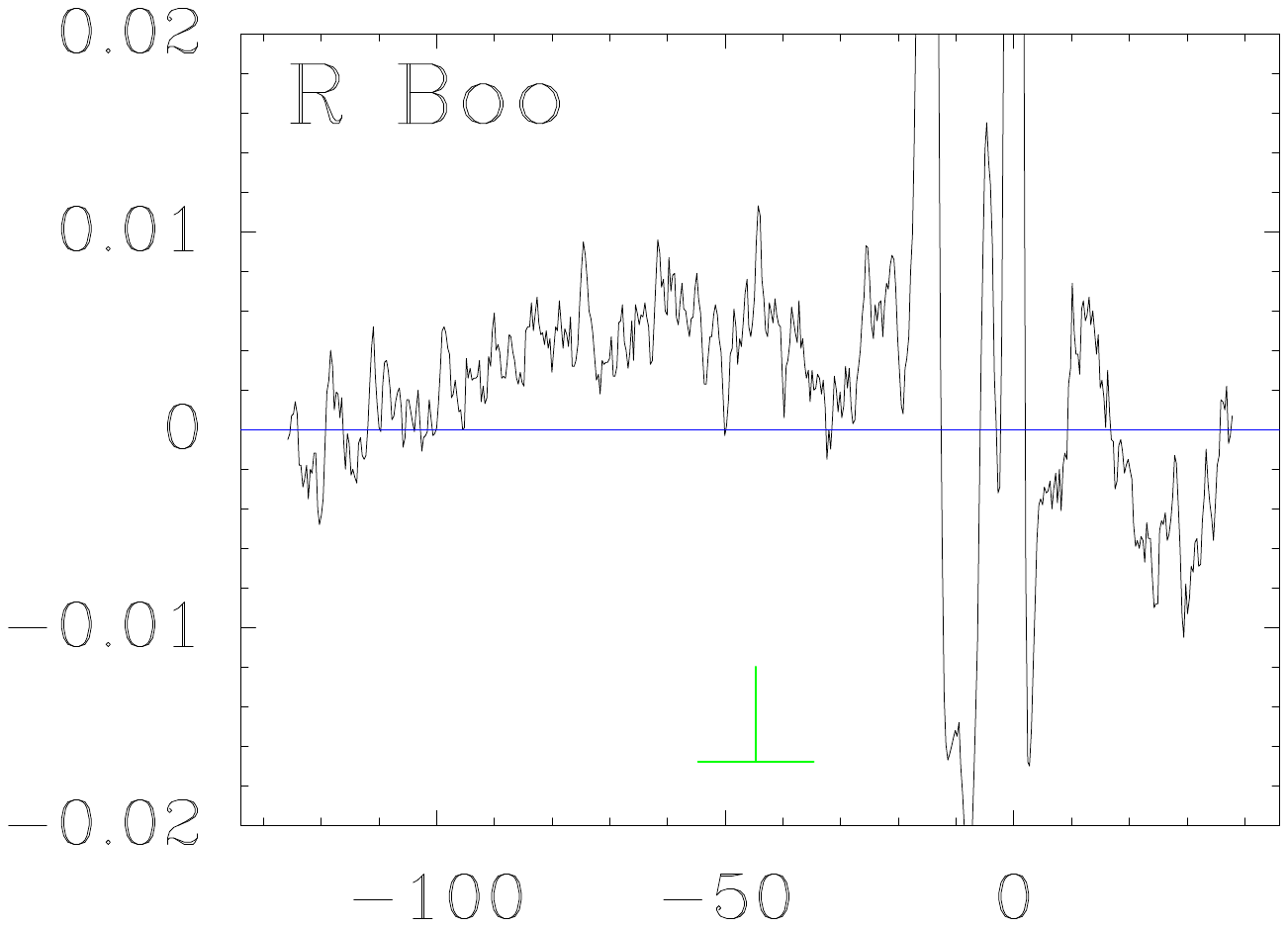}\hspace{-0.7cm}
\includegraphics[width=4.75cm]{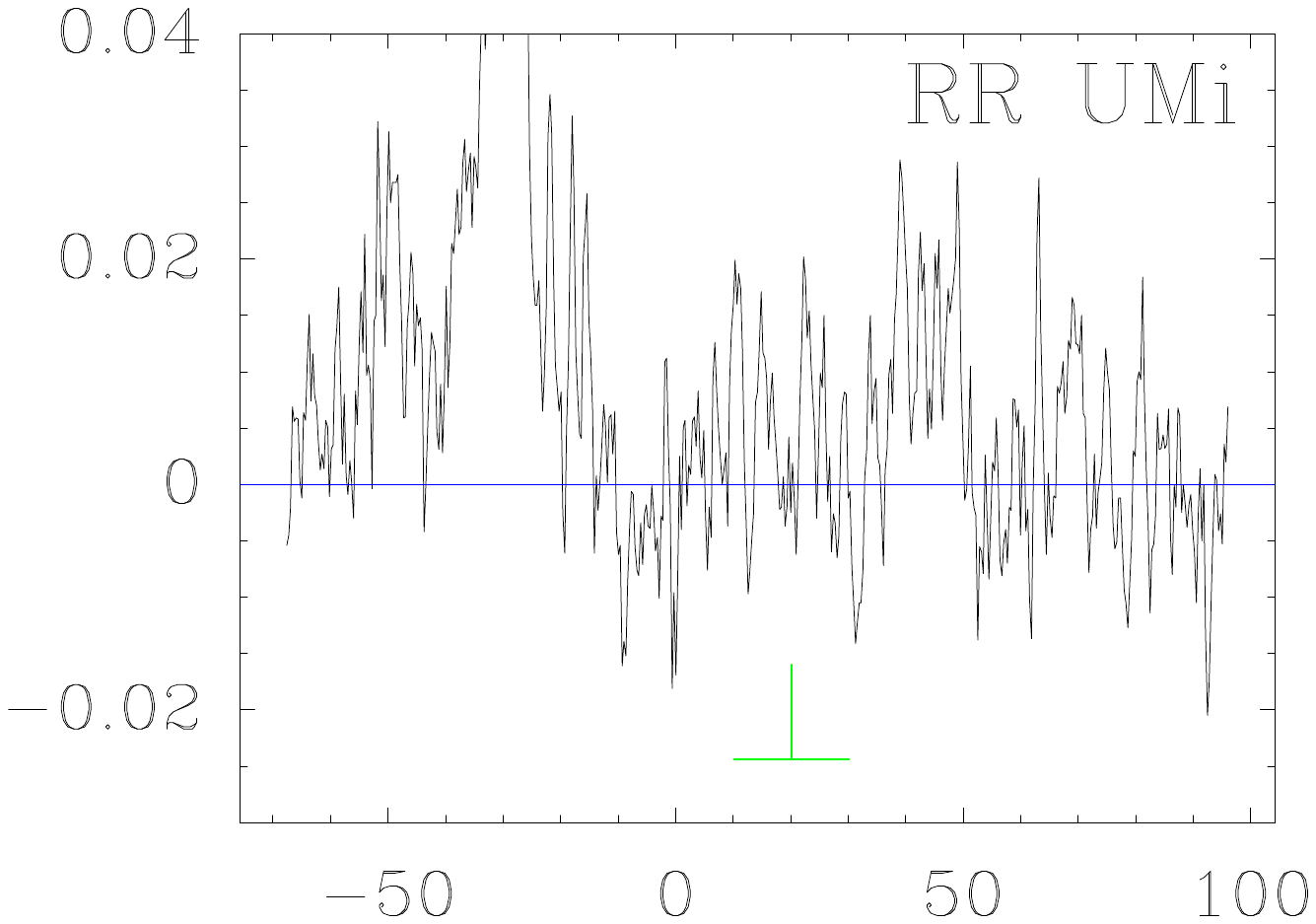}\hspace{-0.7cm}
\includegraphics[width=4.75cm]{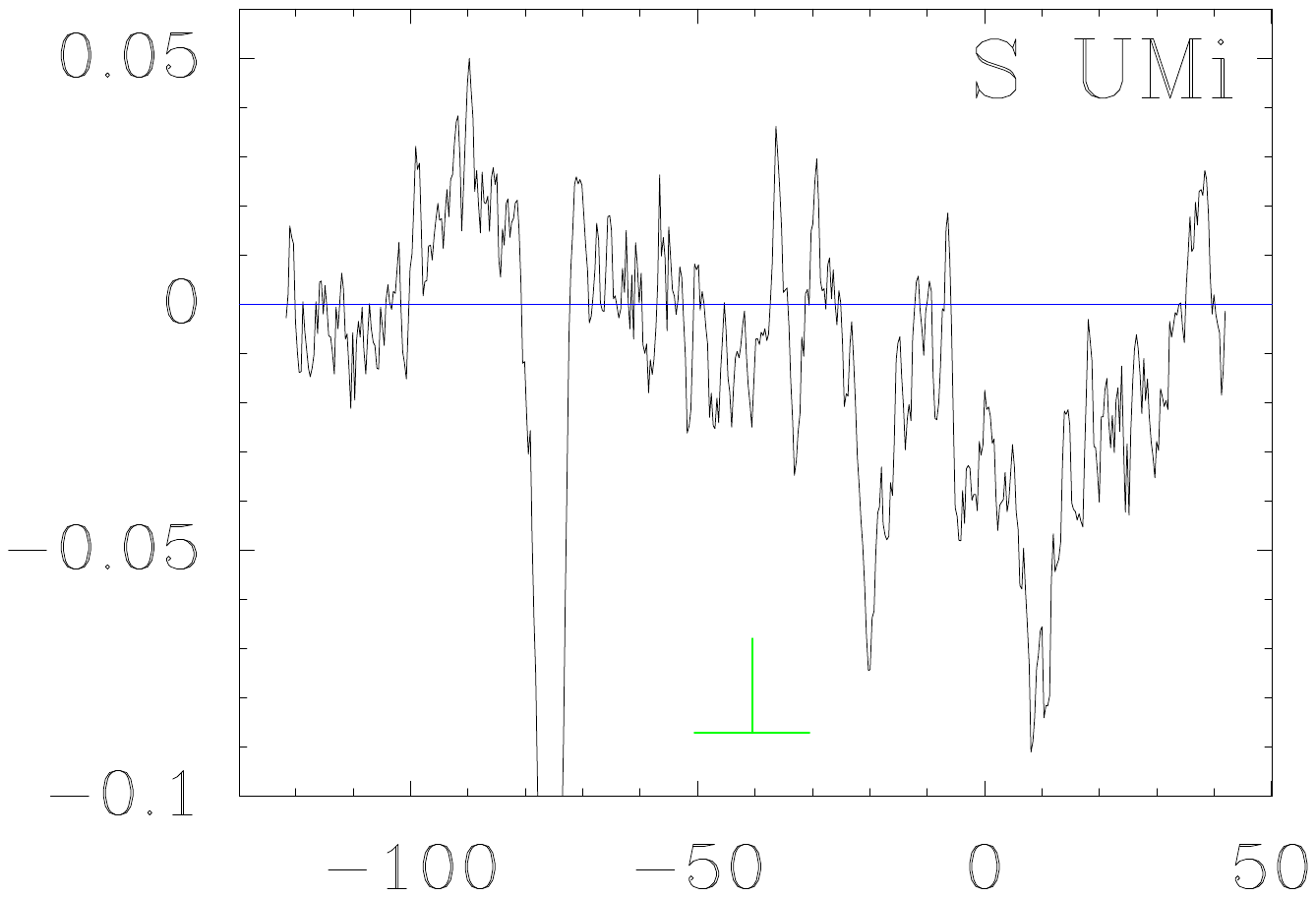}\hspace{-0.7cm}
\includegraphics[width=4.75cm]{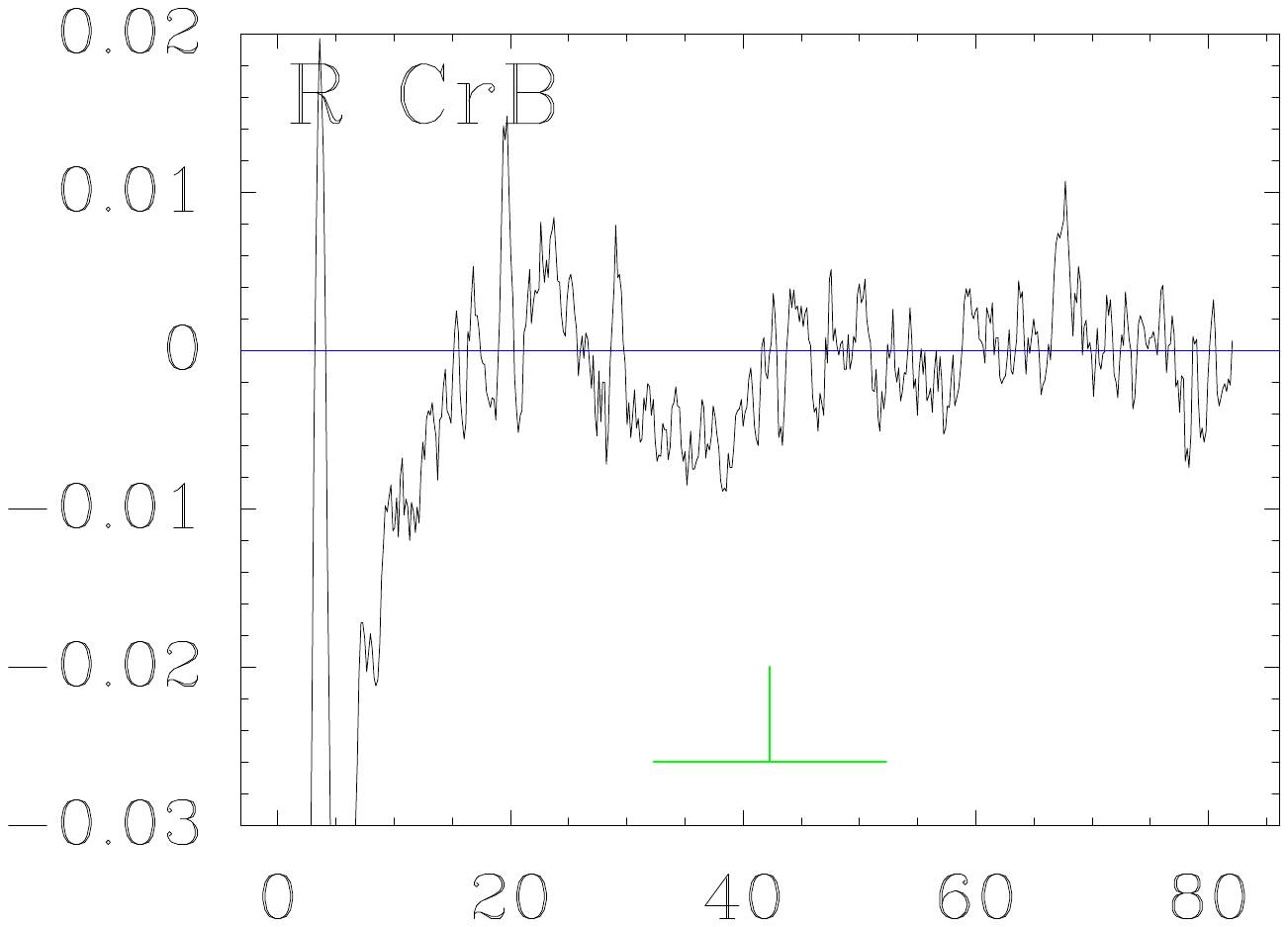}
\\ \vspace{-0.35cm}
\includegraphics[width=4.75cm]{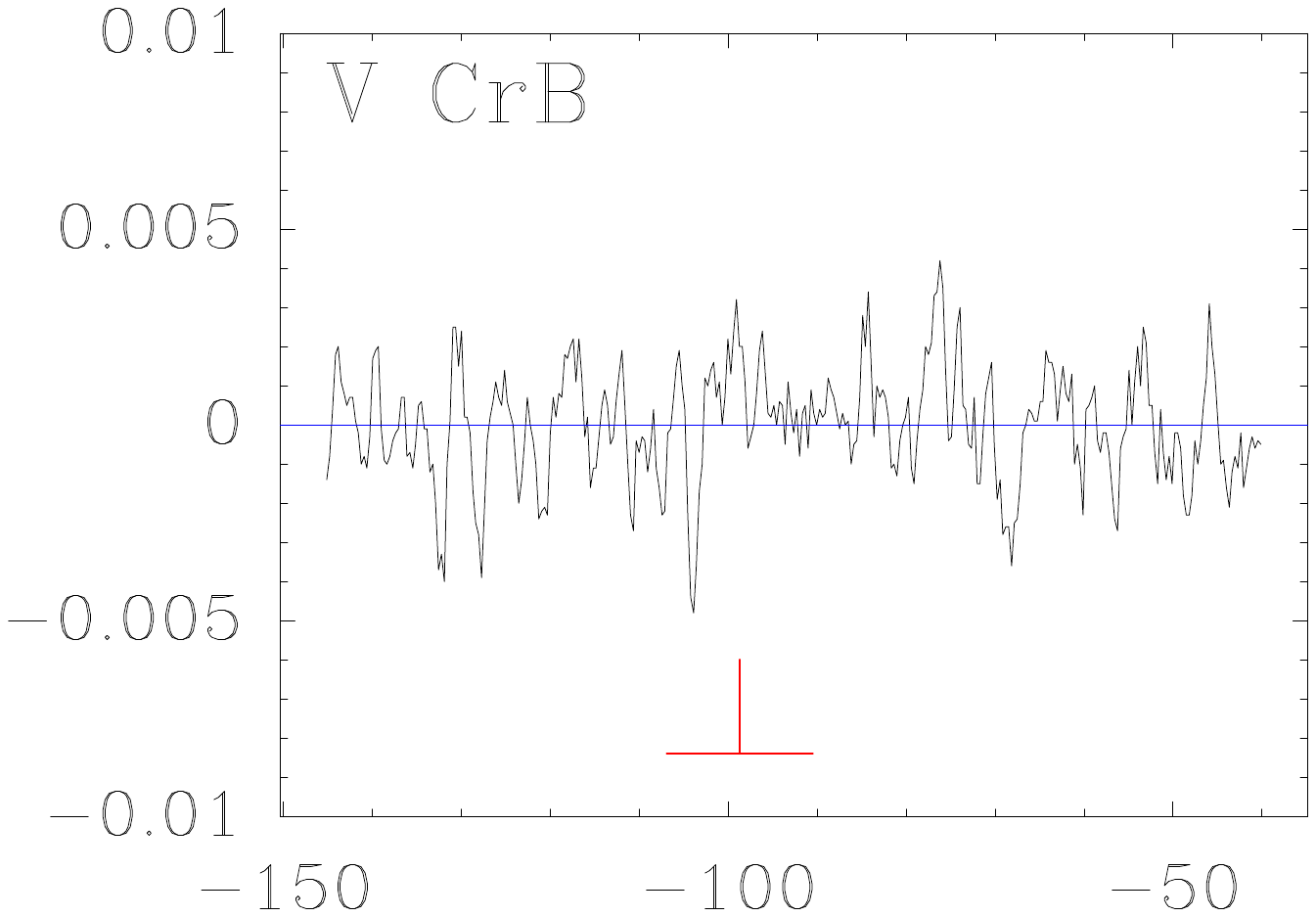}\hspace{-0.7cm}
\includegraphics[width=4.75cm]{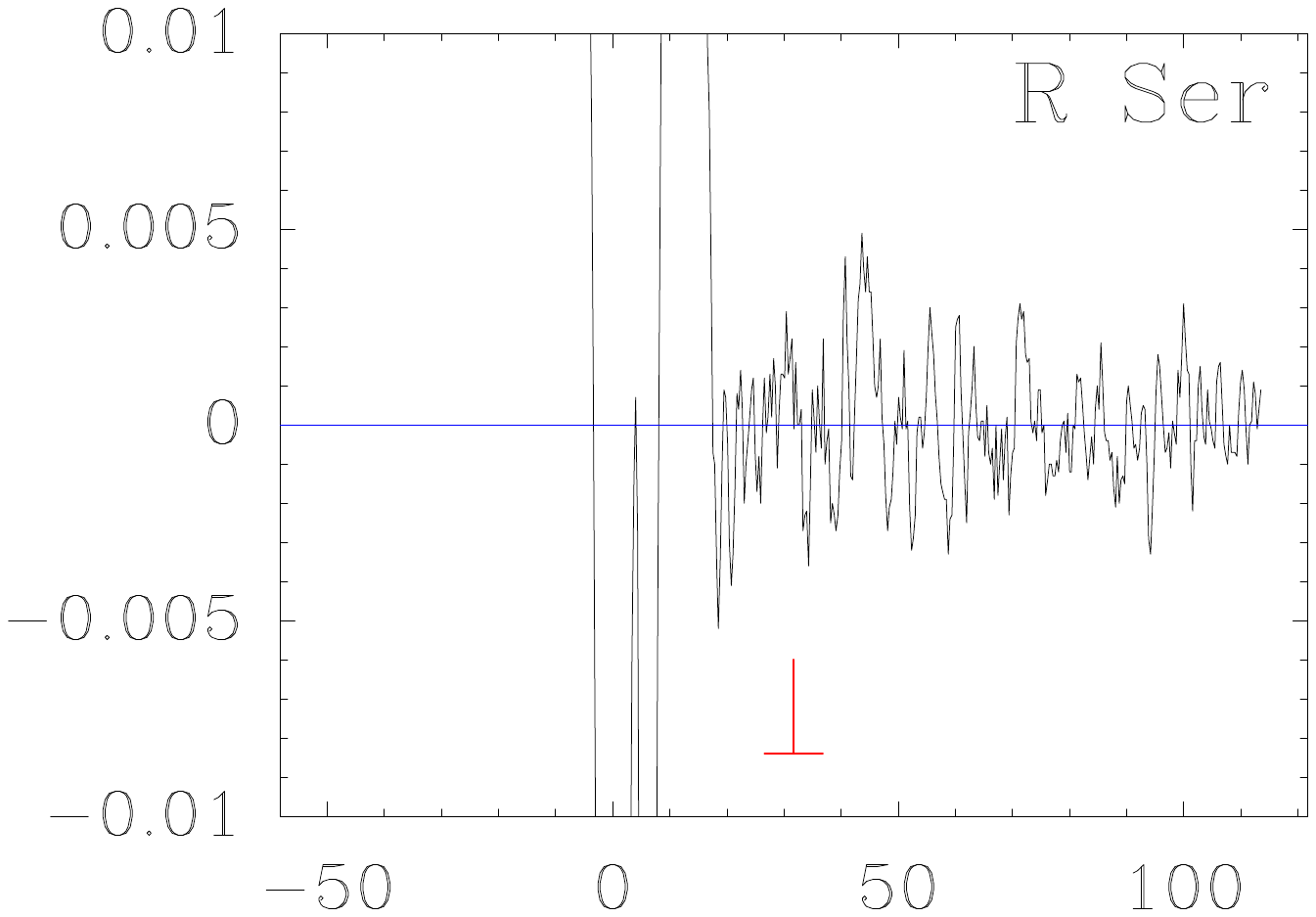}\hspace{-0.7cm}
\includegraphics[width=4.75cm]{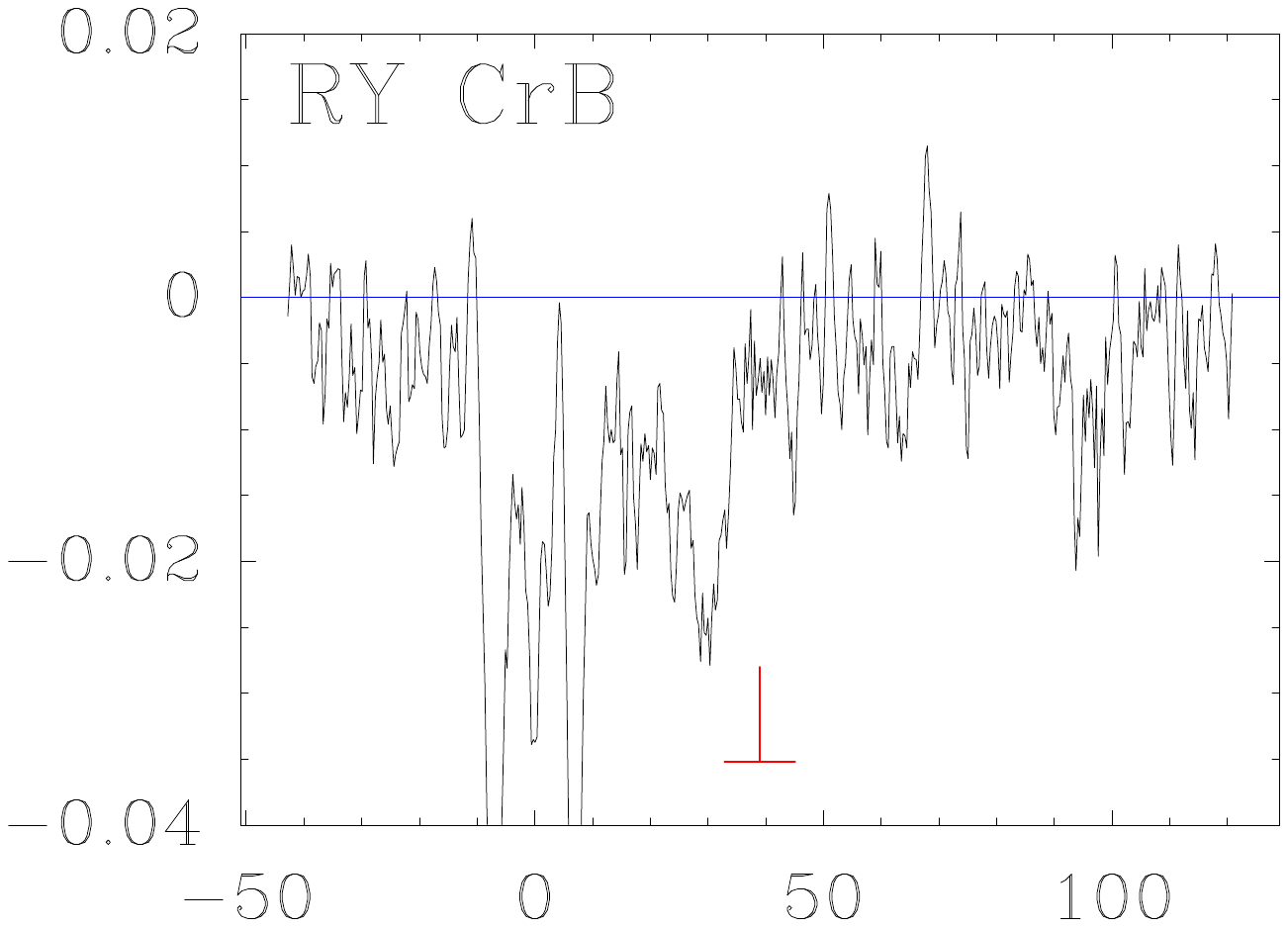}\hspace{-0.7cm}
\includegraphics[width=4.75cm]{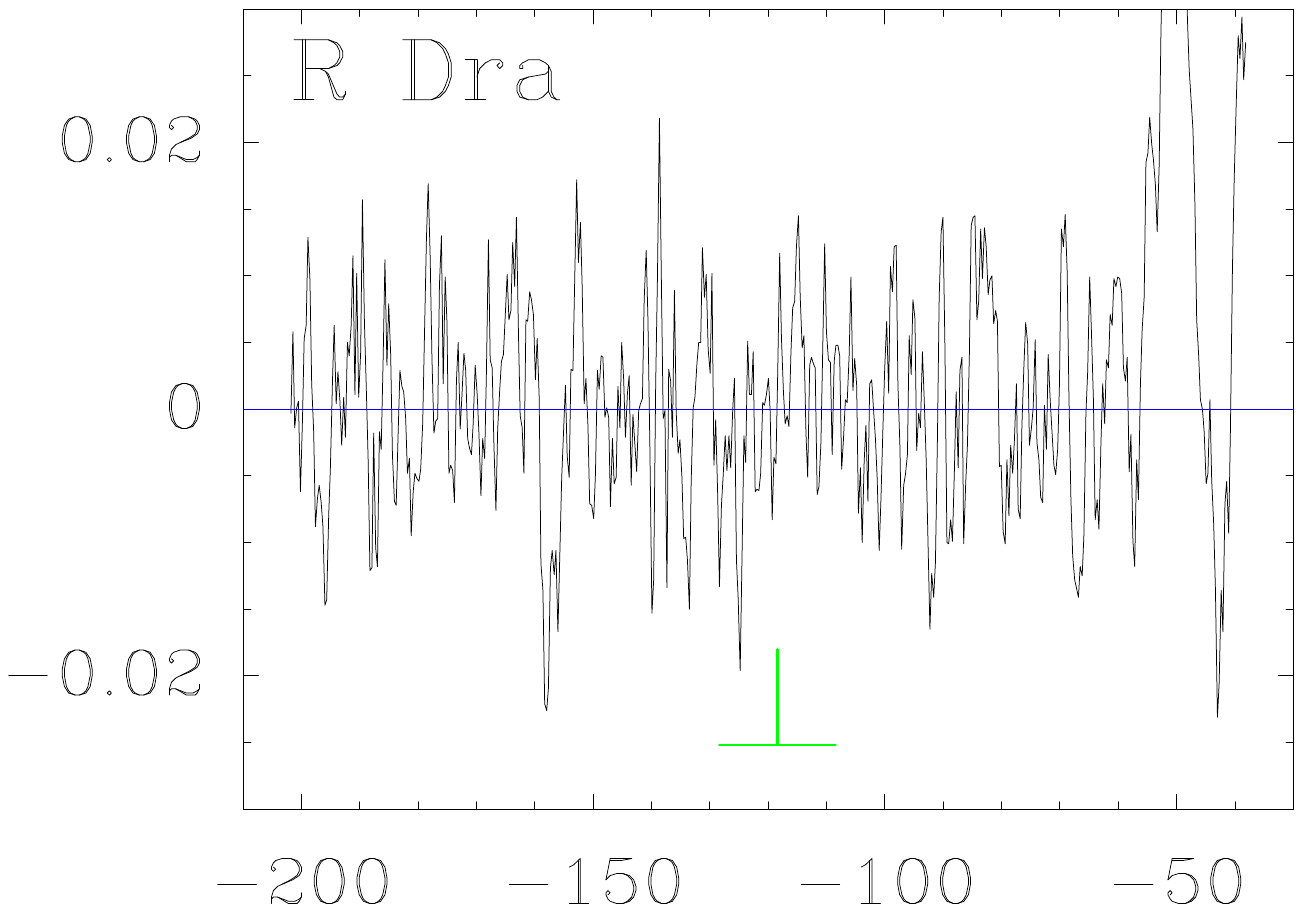}
\\ \vspace{-0.35cm}
\includegraphics[width=4.75cm]{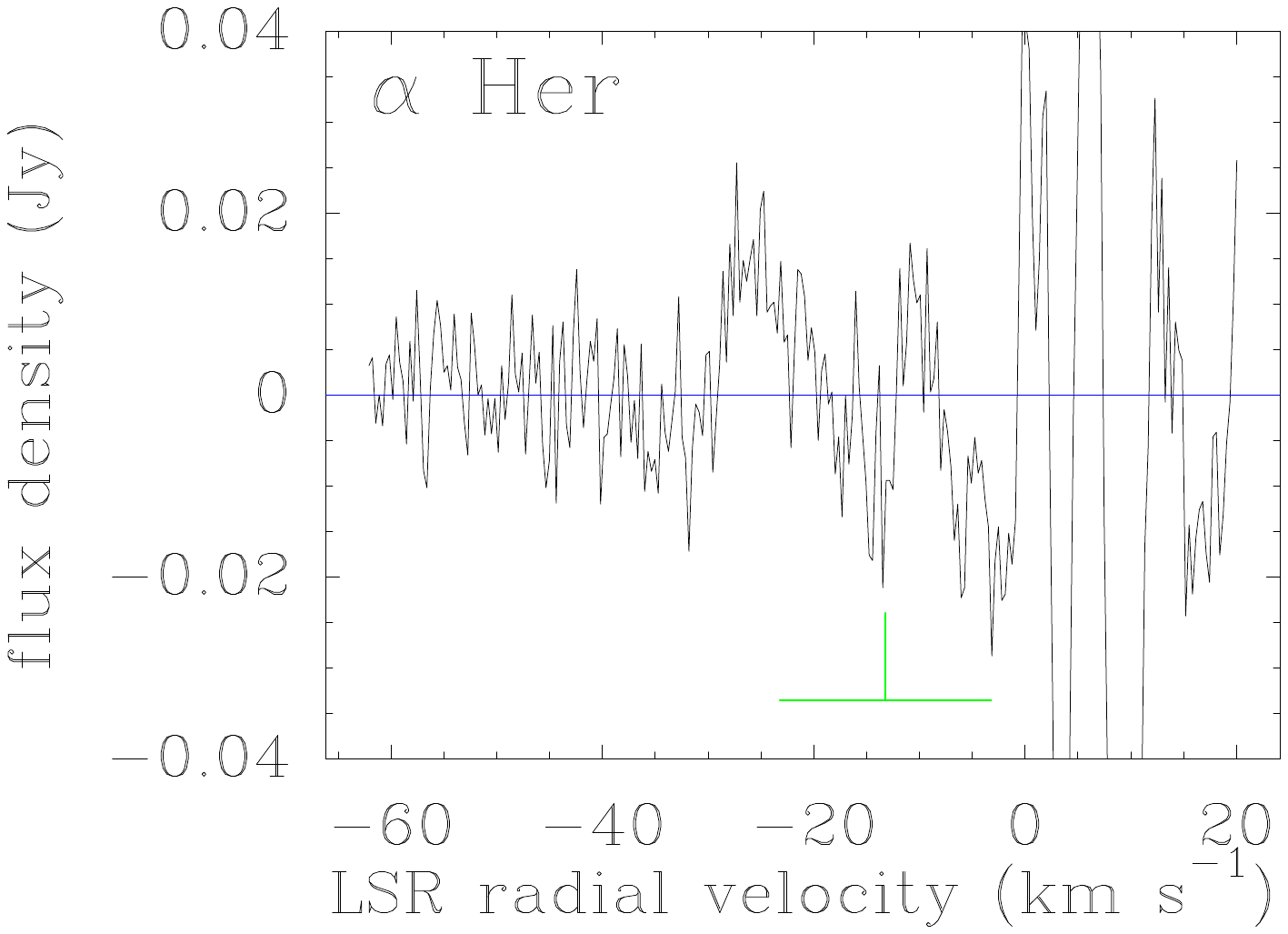}\hspace{-0.7cm}
\includegraphics[width=4.75cm]{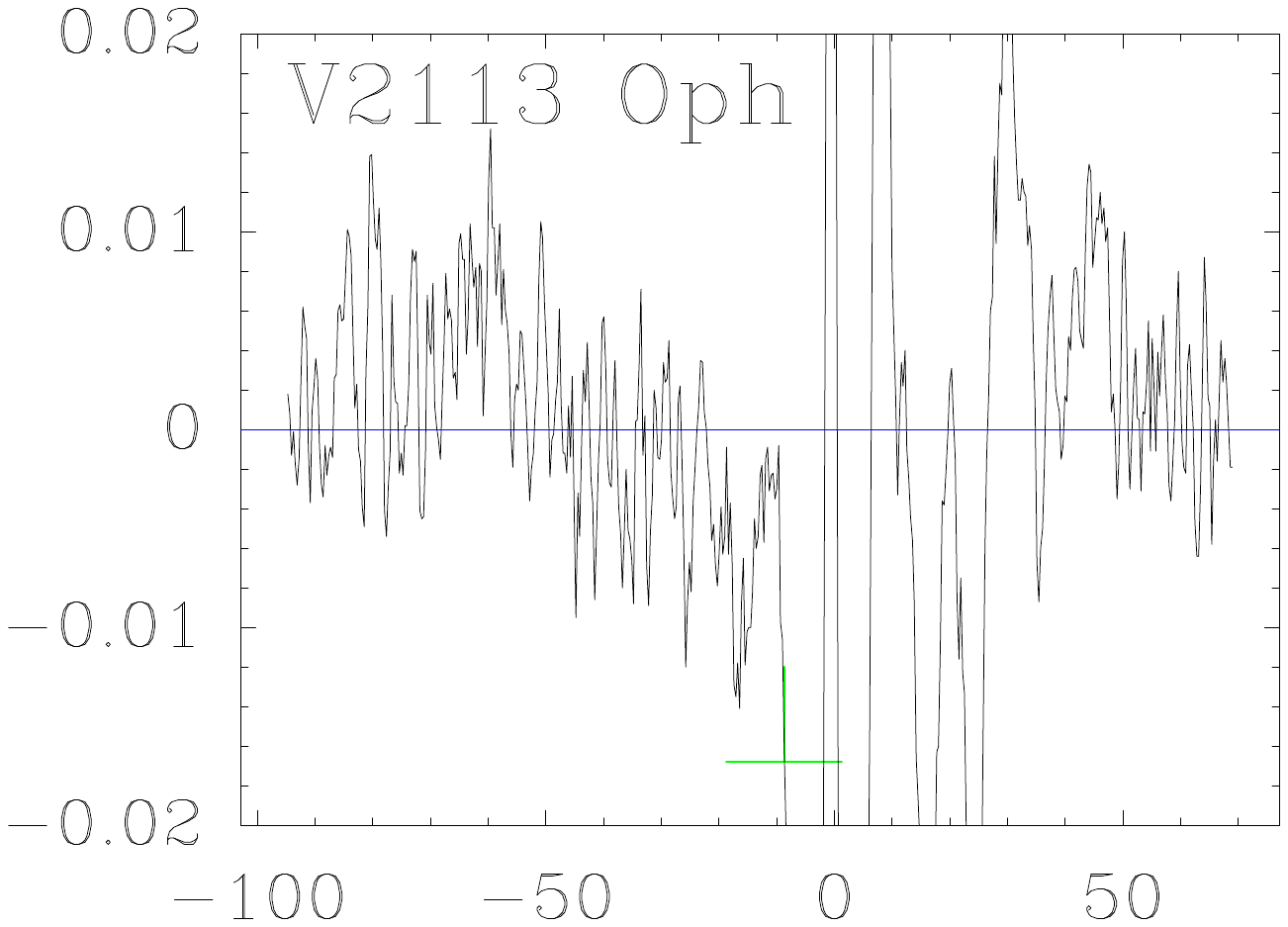}\hspace{-0.7cm}
\includegraphics[width=4.75cm]{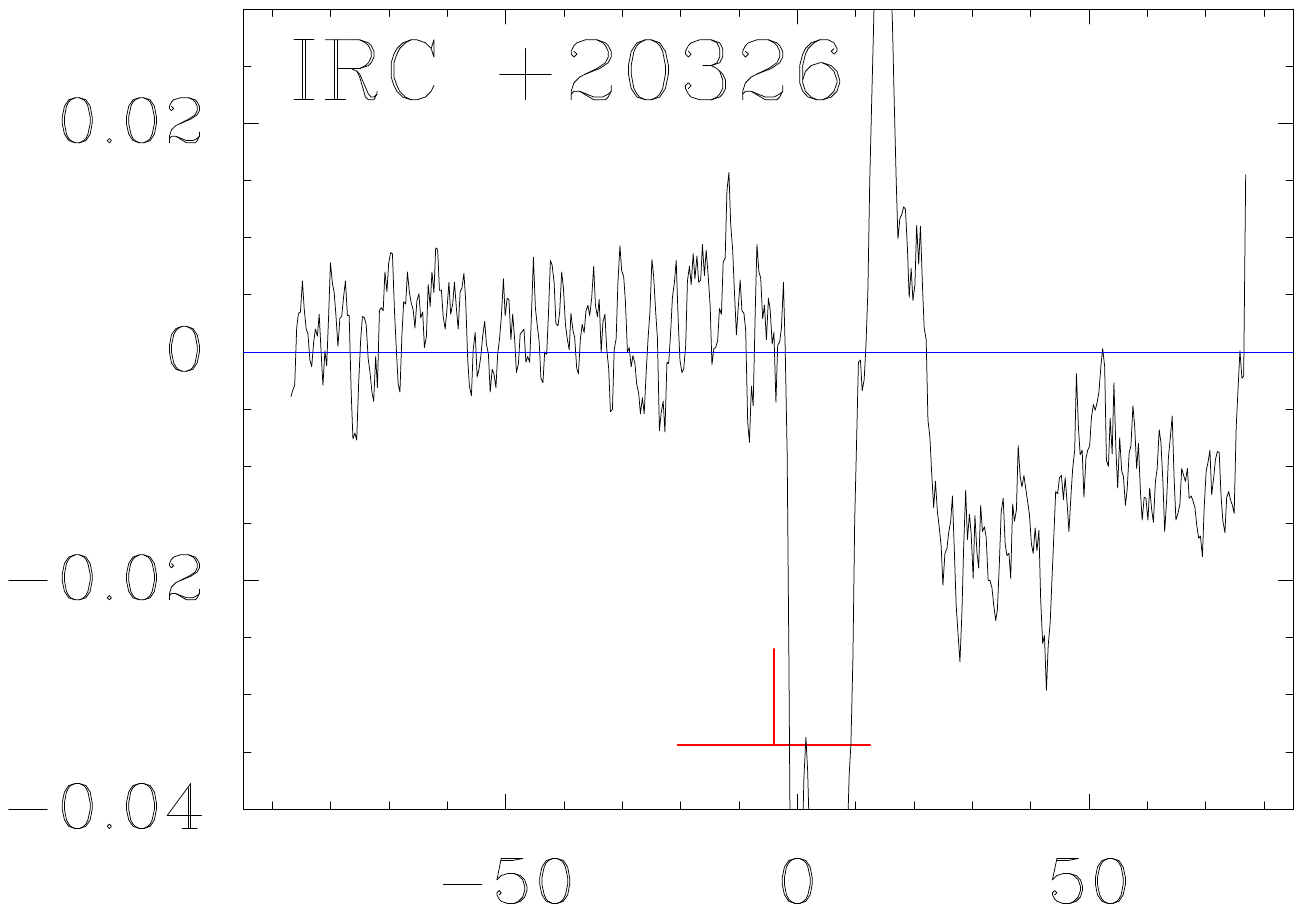}\hspace{-0.7cm}
\includegraphics[width=4.75cm]{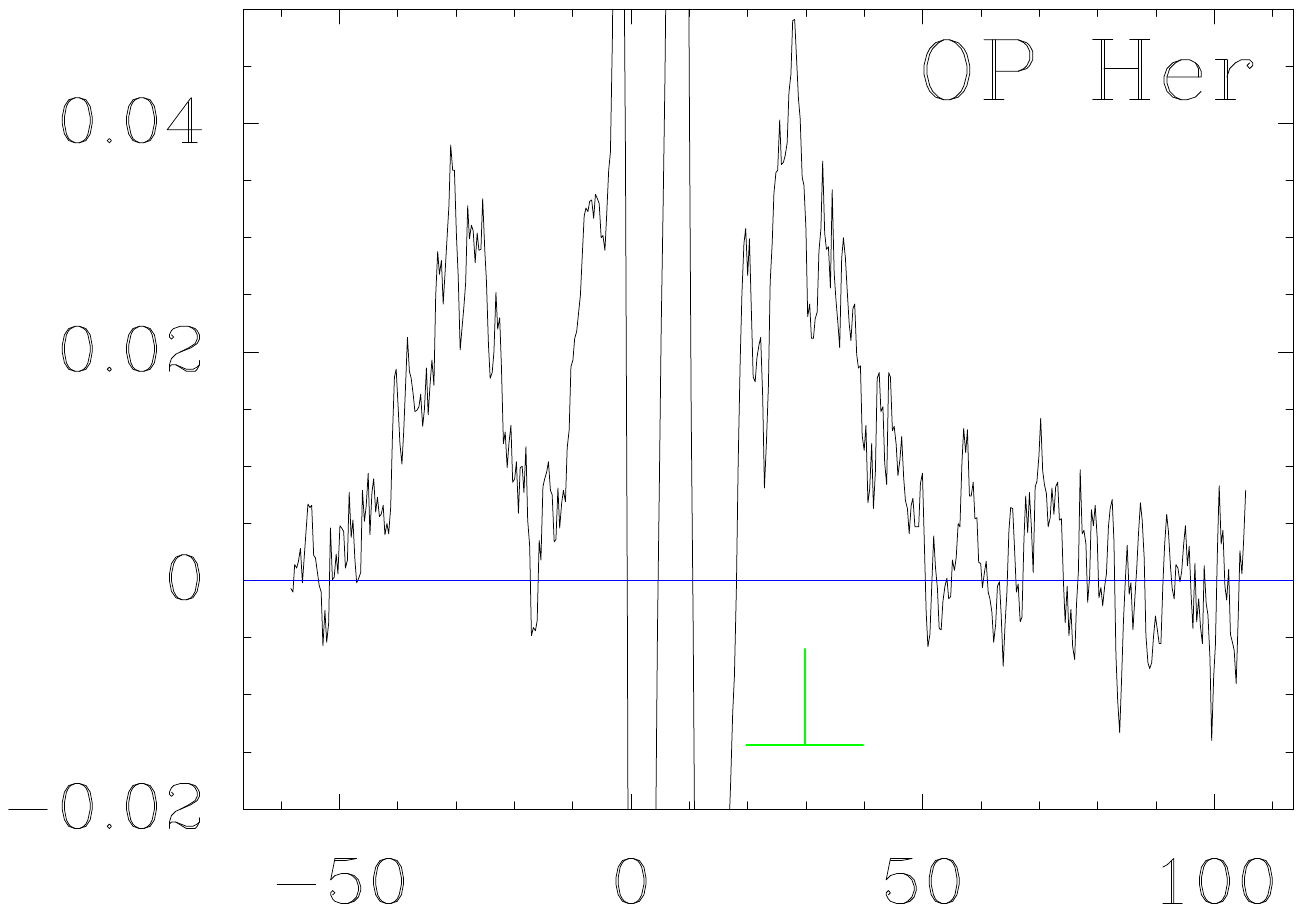}
%  \vspace{-2cm}
  \caption{{\bf b.} Upper limits -- continued. }
\end{figure*}

\begin{figure*}[ht] % \label{fig:spectra_limits}  % Fig. 5c
\addtocounter{figure}{-1}
%  \centering
%  \includegraphics[width=18cm]{upperLimits20240227_p3_col.pdf}
%  \includegraphics[width=18.35cm]{AGB_Legacy_specplots_limits_12_three_wrk1.png}
%
% 
\includegraphics[width=4.75cm]{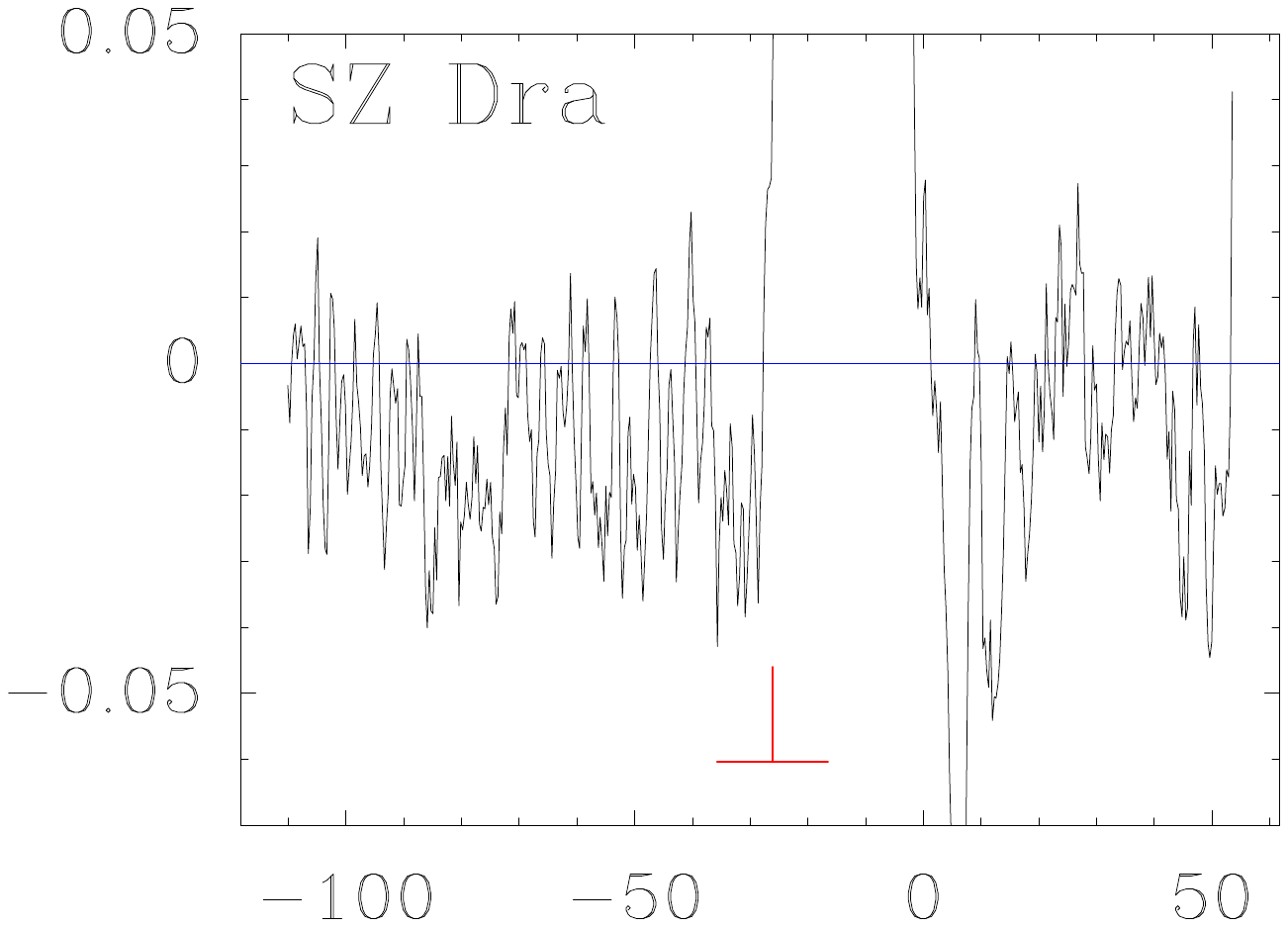}\hspace{-0.7cm}
\includegraphics[width=4.75cm]{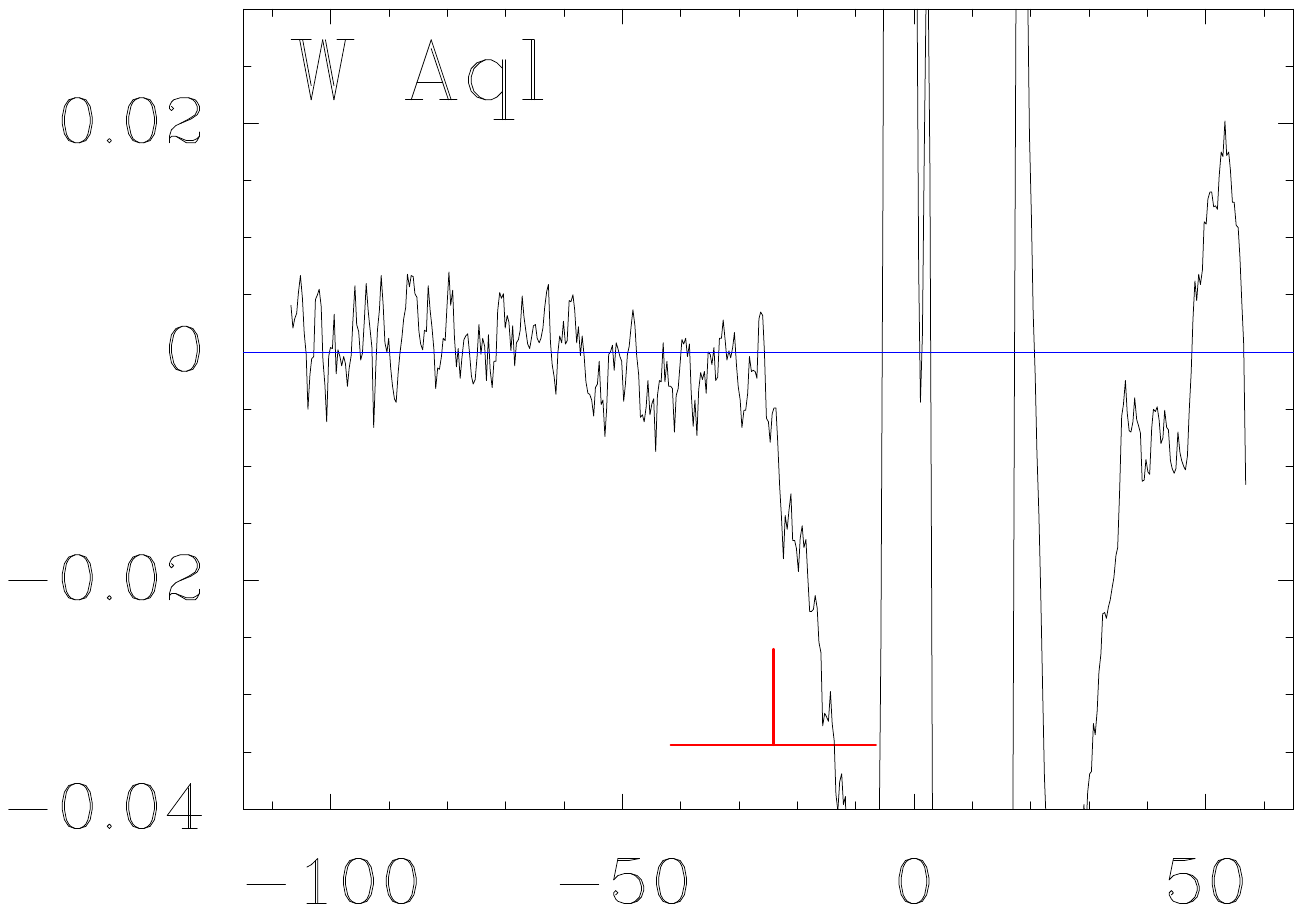}\hspace{-0.7cm}
\includegraphics[width=4.75cm]{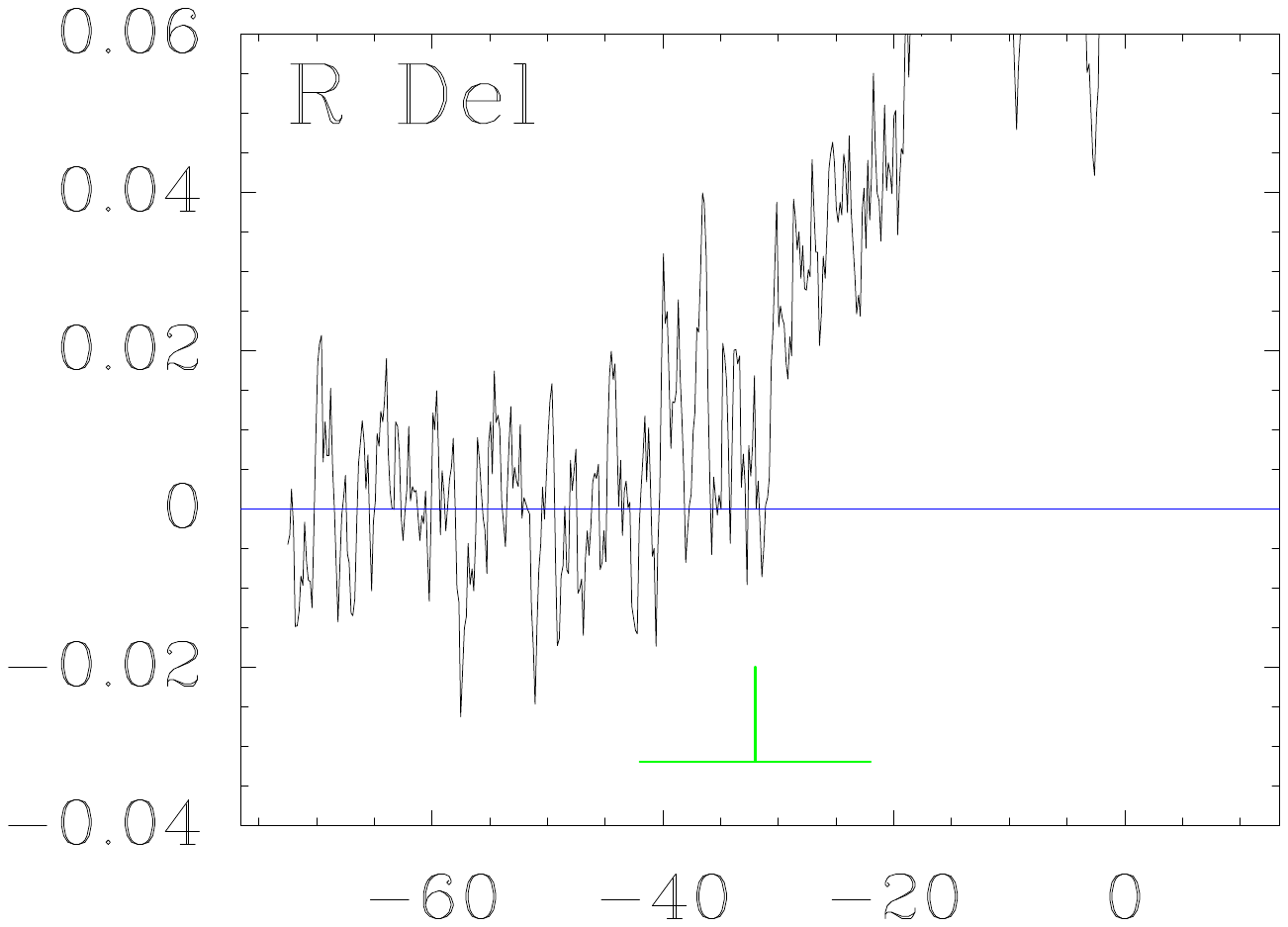}\hspace{-0.7cm}
\includegraphics[width=4.75cm]{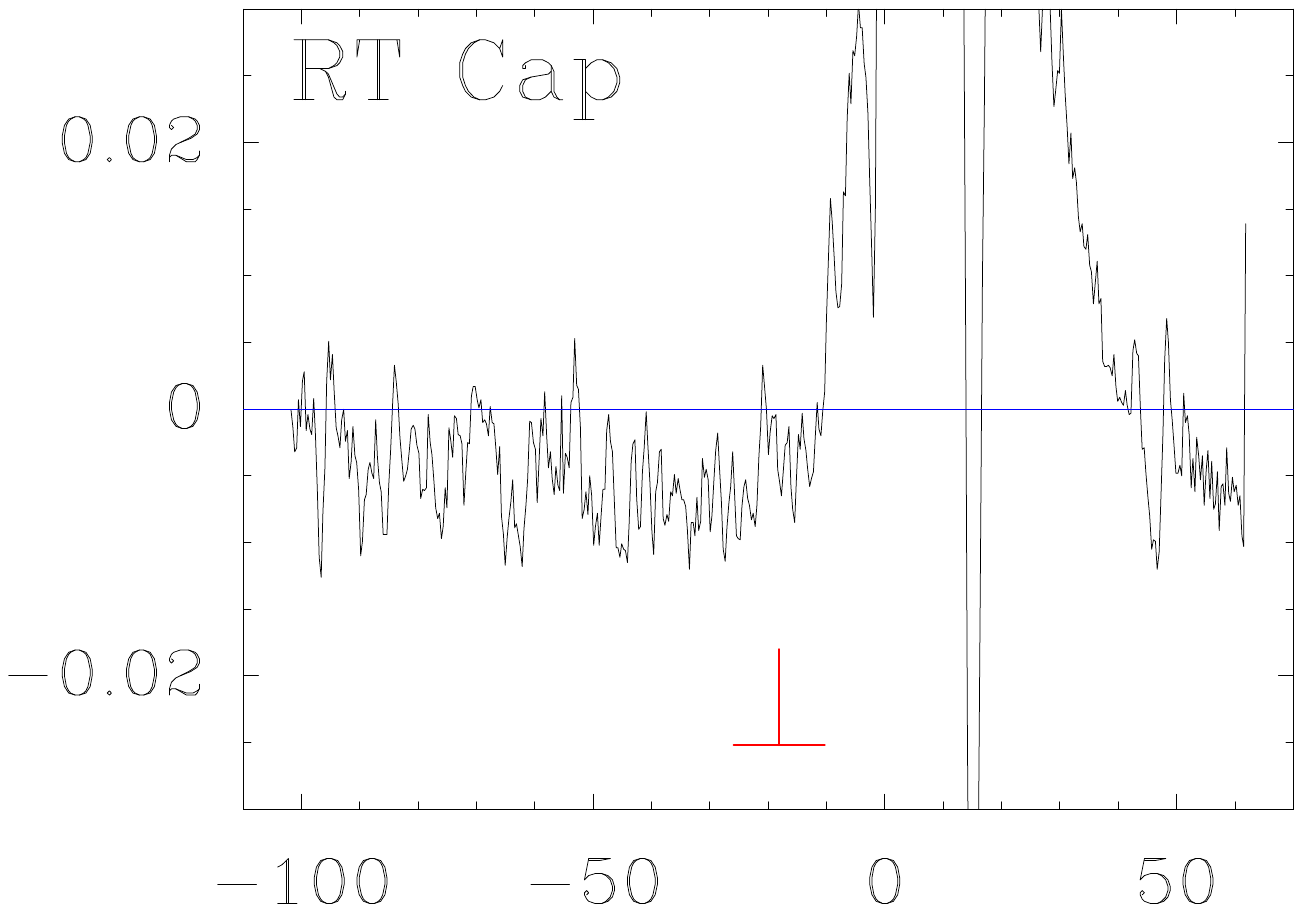}
\\ \vspace{-0.35cm}
\includegraphics[width=4.75cm]{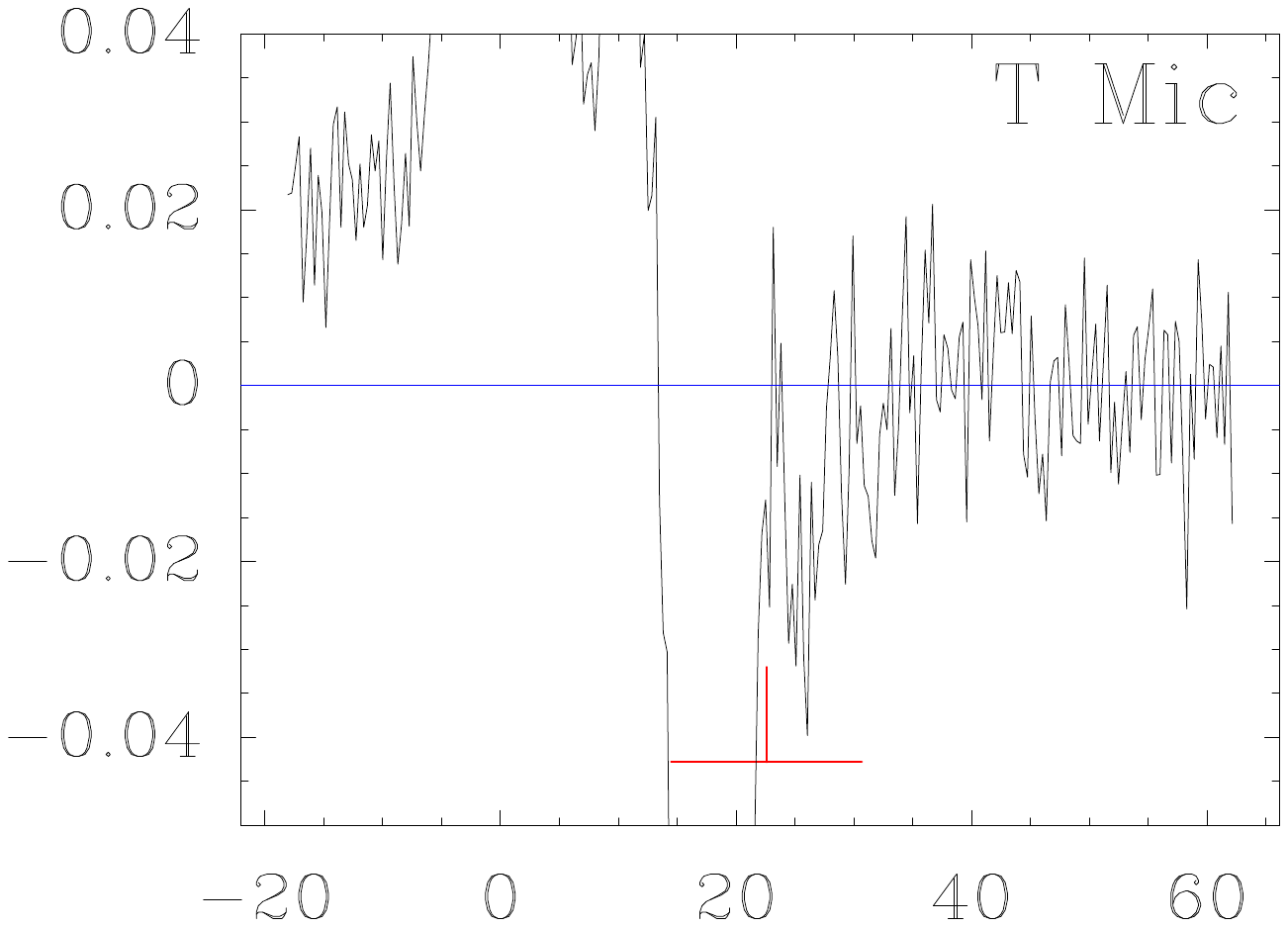}\hspace{-0.7cm}
\includegraphics[width=4.75cm]{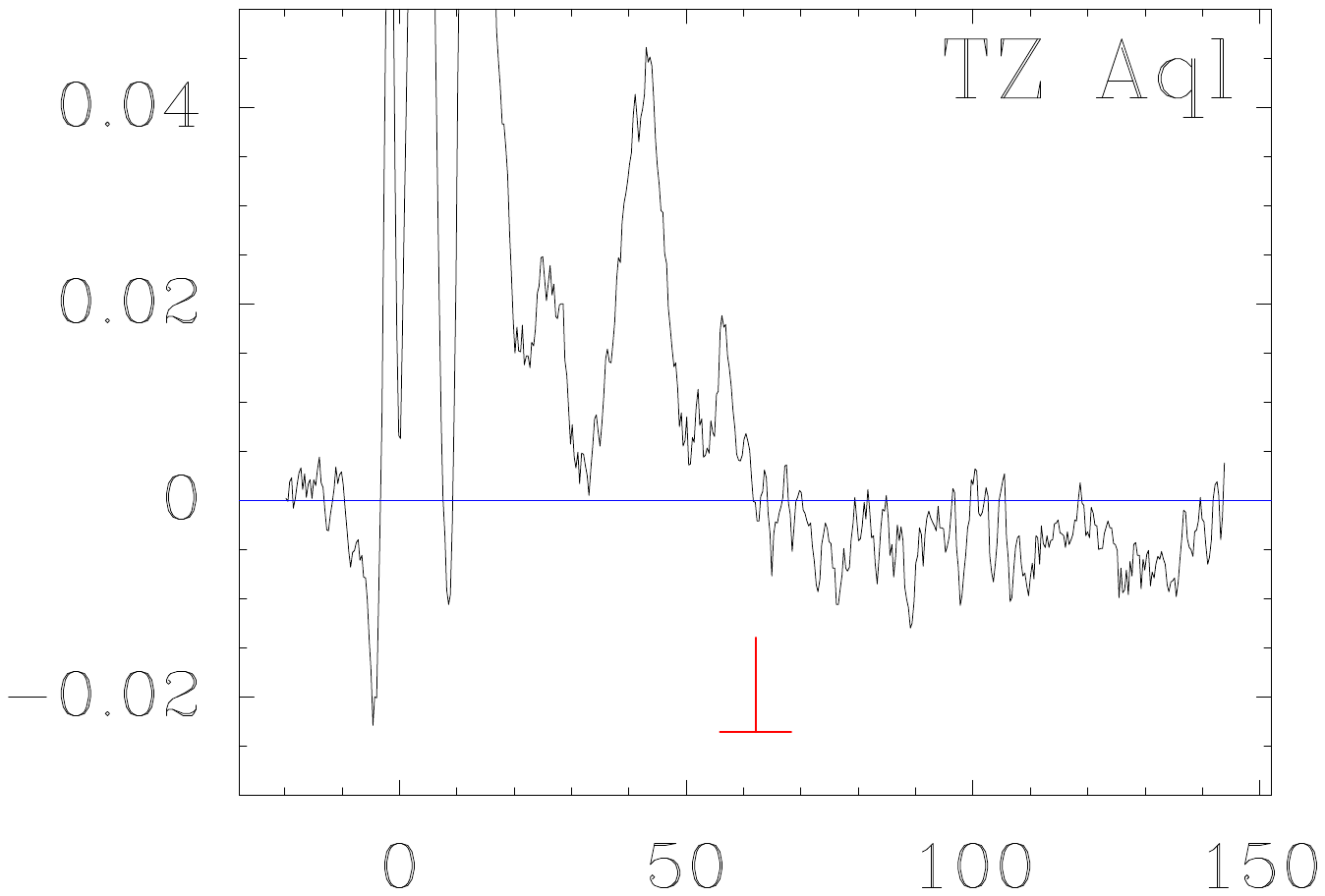}\hspace{-0.7cm}
\includegraphics[width=4.75cm]{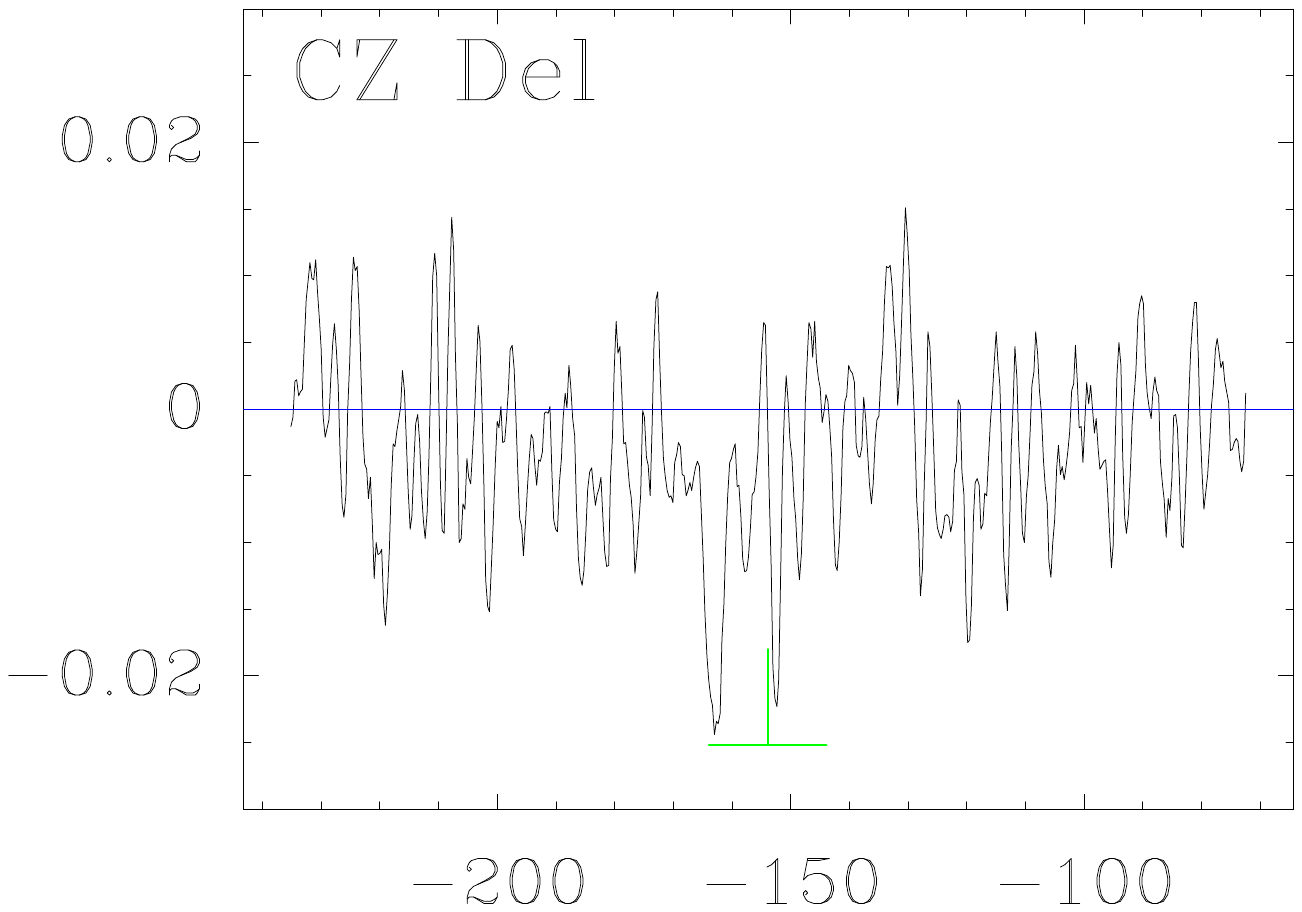}\hspace{-0.7cm}
\includegraphics[width=4.75cm]{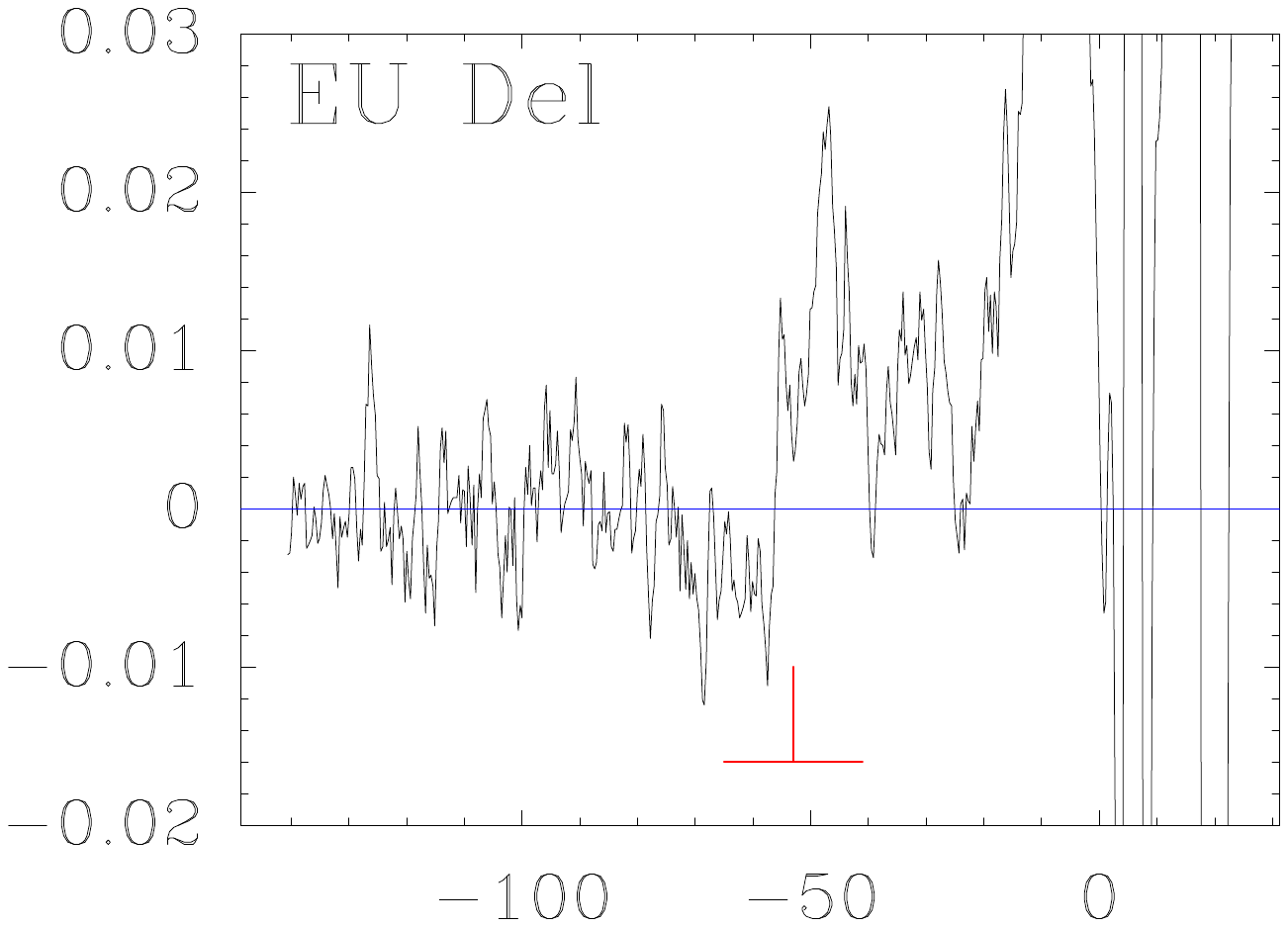}
\\ \vspace{-0.35cm}
\includegraphics[width=4.75cm]{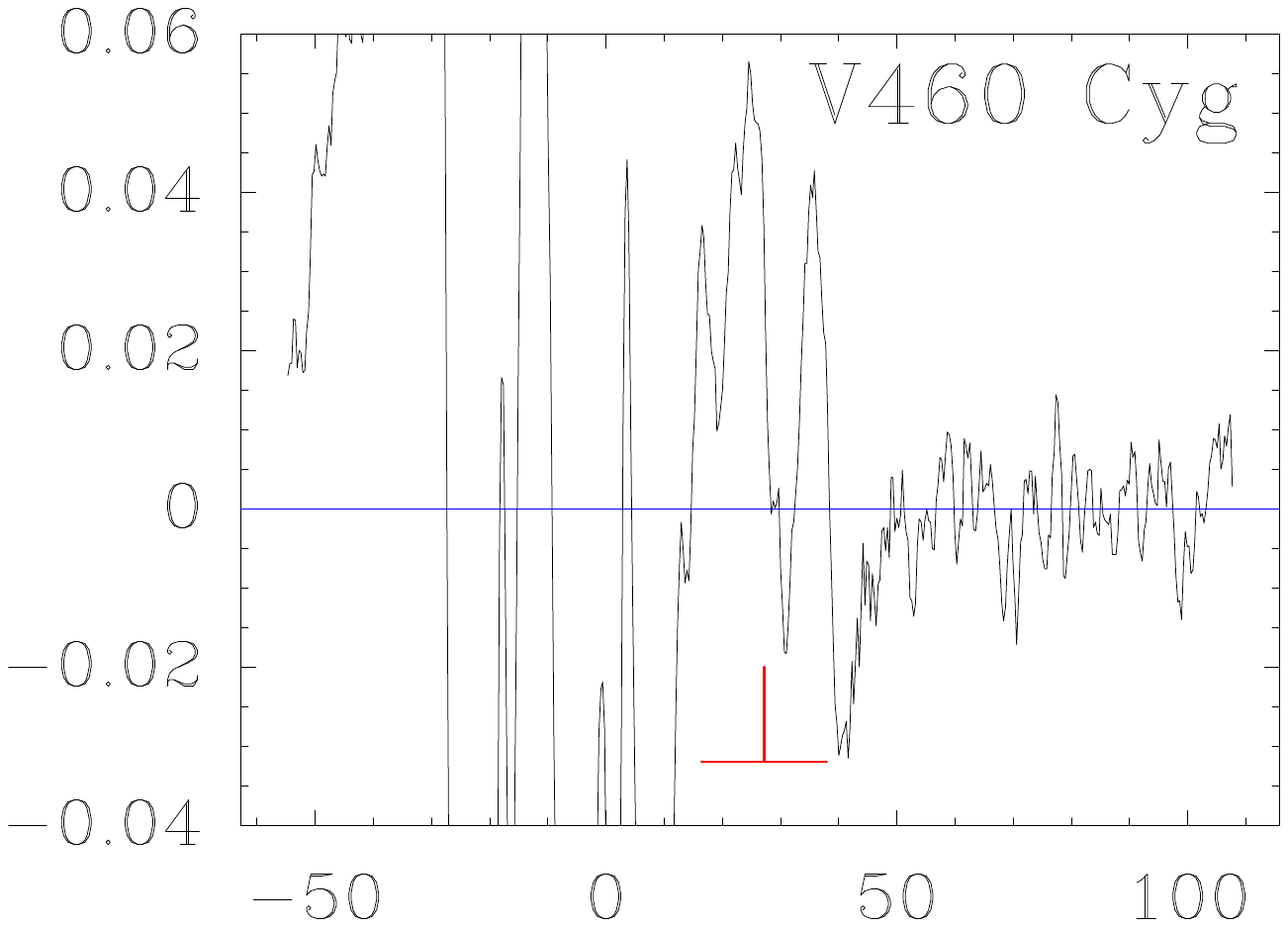}\hspace{-0.7cm}
\includegraphics[width=4.75cm]{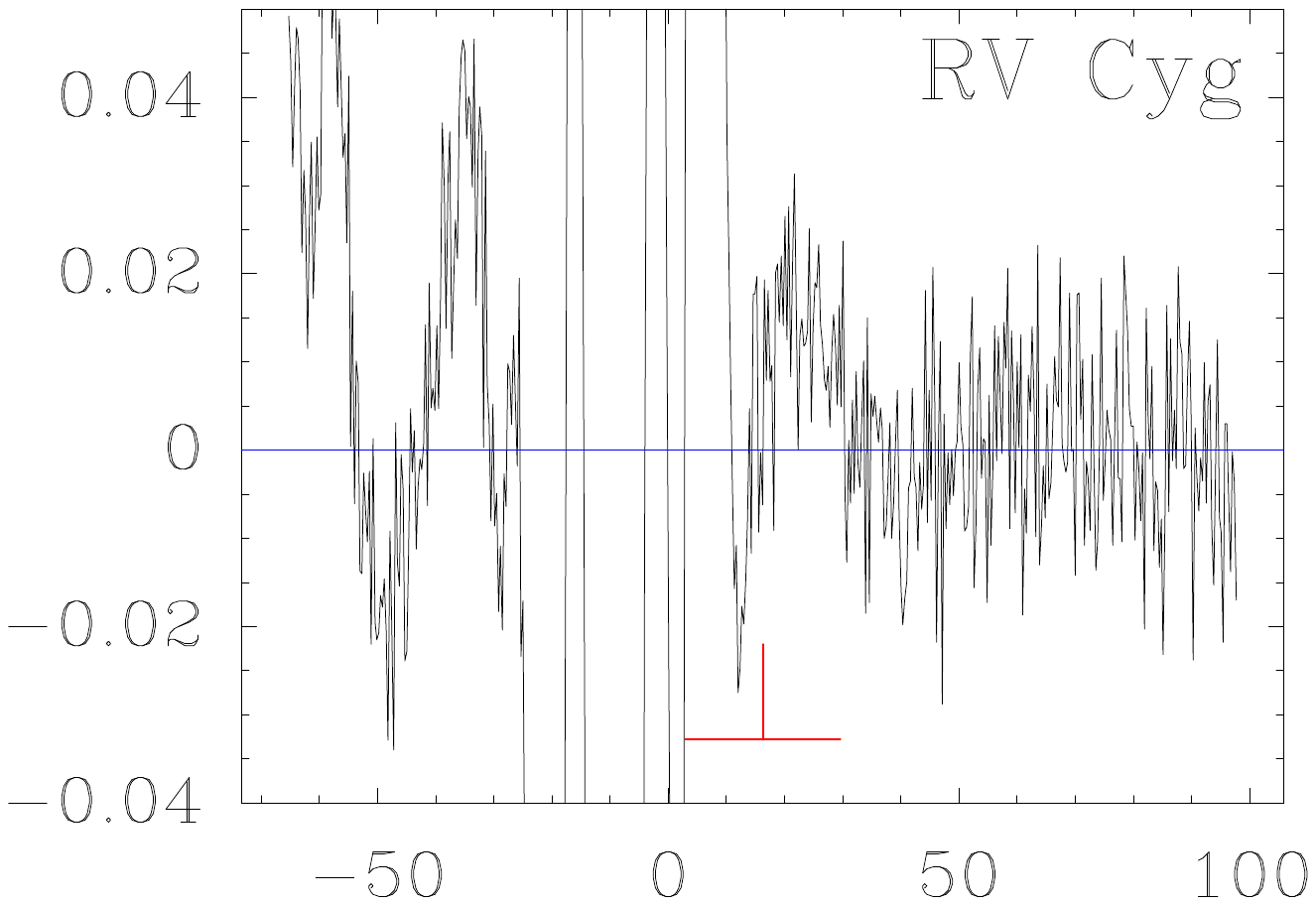}\hspace{-0.7cm}
\includegraphics[width=4.75cm]{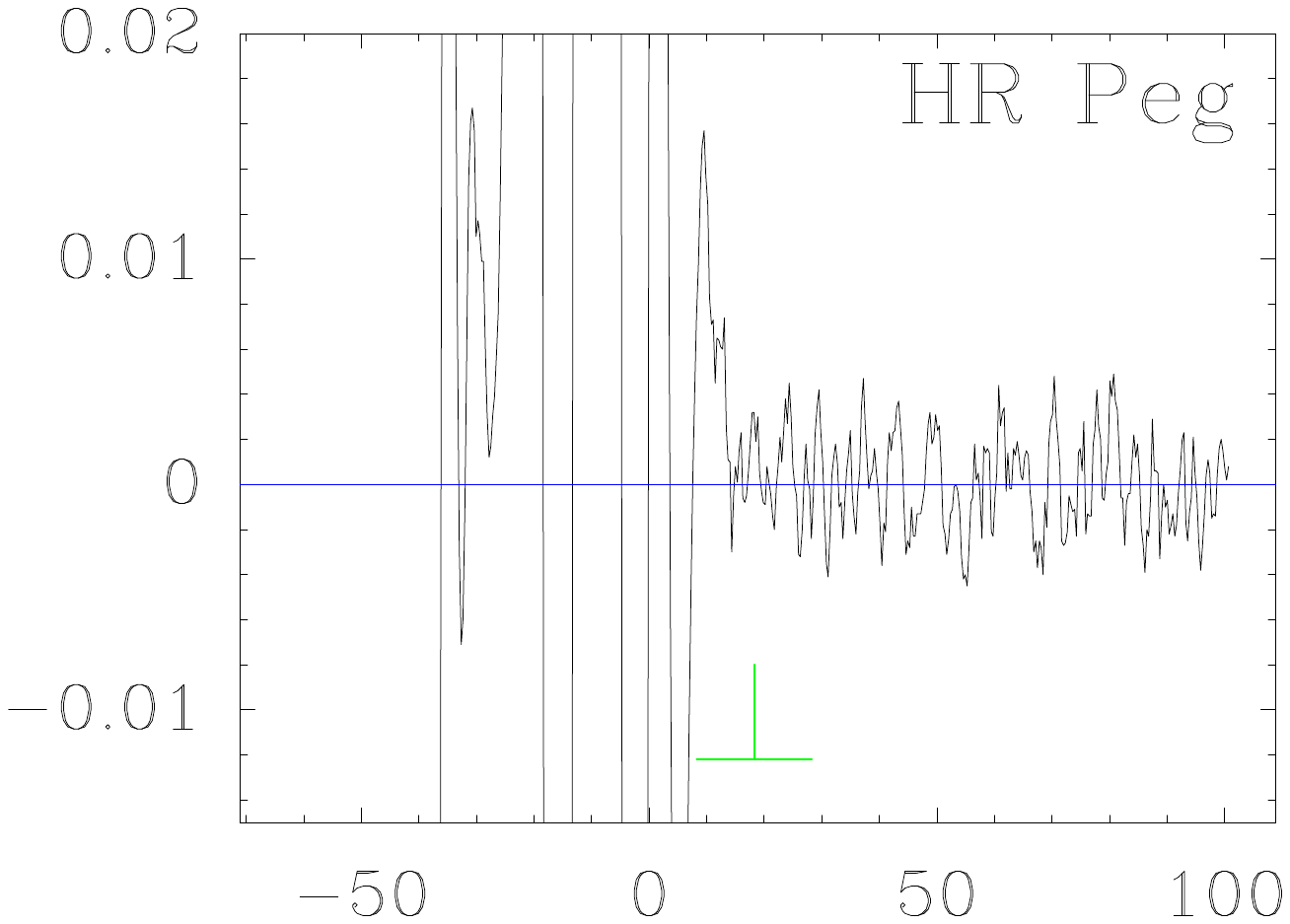}\hspace{-0.7cm}
\includegraphics[width=4.75cm]{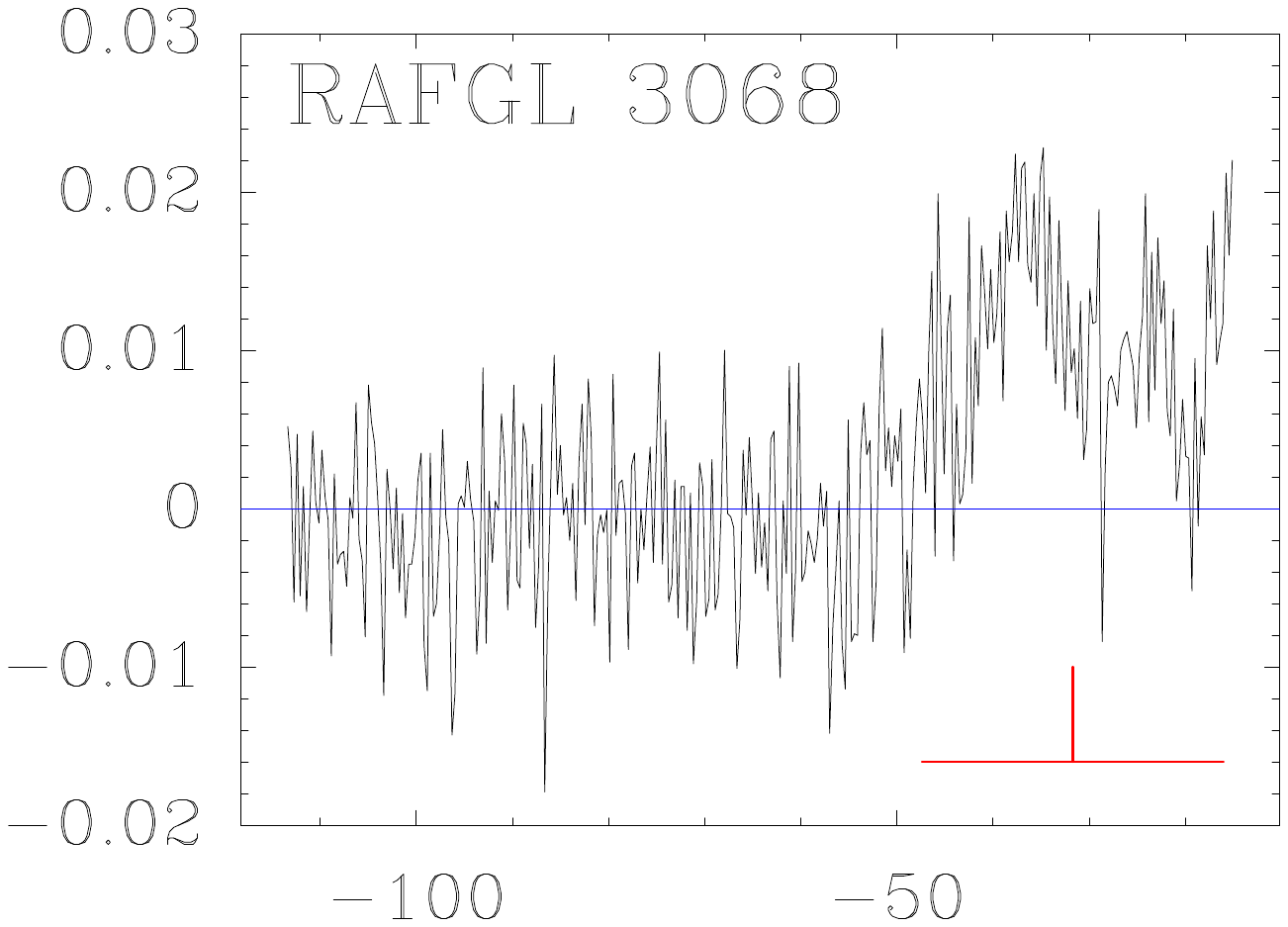}
\\ \vspace{-0.35cm}
\includegraphics[width=4.75cm]{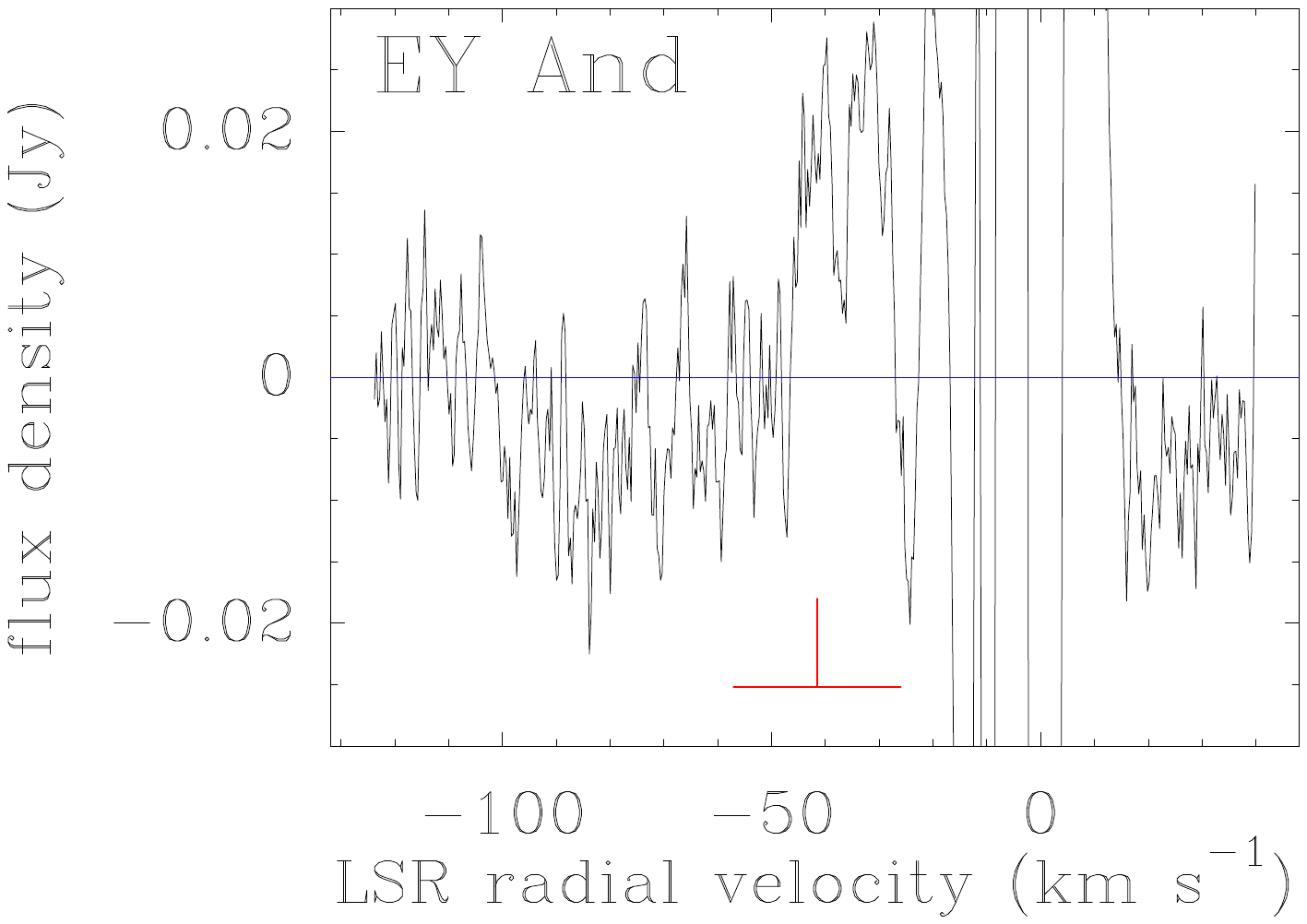}\hspace{-0.7cm}
\includegraphics[width=4.75cm]{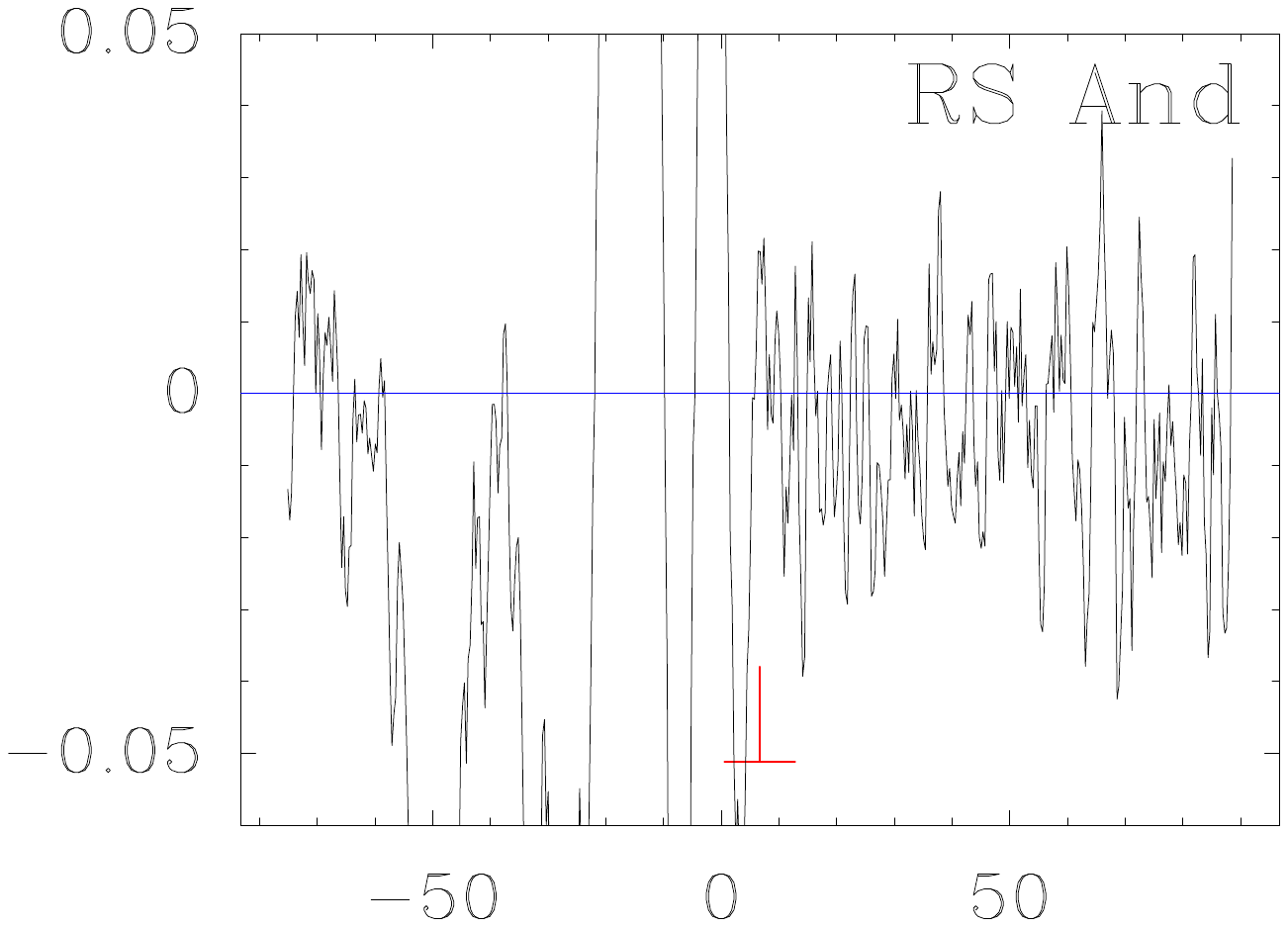}\hspace{-0.7cm}
%
%  \vspace{-2cm}
  \caption{{\bf c.} Upper limits -- continued. }
\end{figure*}

\begin{appendix}  % Appendix A

\section{Notes on individual objects}  \label{sec:Notes}

% \vspace{-5.8cm}

\noindent
Objects are listed first in order of their catalogue name (IRC, NGC and RAFGL), 
and then in alphabetical order of their constellation's name. 
% }
Stars which are clearly not AGBs are denoted by an $^n$ after their name and
those for which we consider their classification as AGB dubious are denoted by a $^d$.
\\

% \vspace{-6cm}

\noindent
% {\bf HD 56126:} As the $C1EW$ and $C2EW$ profiles are similar we used the averaged spectrum 
% for further analysis, and conclude that the source is unresolved. \\
% 
\noindent
{\bf IRC -10529: }
We adopted the distance of 270 pc used by \citet{Ra20} 
as it has no parallax listed in the Gaia EDR3. \\
\noindent
{\bf IRC +10216:} We now consider the NRT \HI\ observation to be confused
by a source located at an off-source position. It was previously classified 
as an NRT detection in {\it absorption} by \citet{Lebertre2001}, which the VLA imaging 
by \citet{Matthews2007} showed to be a likely artifact due to \HI\ emission detected with 
the VLA within an NRT off-source reference beam. 
However, subsequent imaging with the 100m Green Bank Telescope \citep{Matthews2015} revealed a 
low surface brightness \HI\ shell of about 20$'$ (0.8 pc) size centred on the star. \\
\noindent
{\bf NGC 6369$^n$}: Listed previously as a detection in \citet{Gerard2006}, we classified 
this planetary nebula as a possible detection, due to the low signal-to-noise ratio
of its broad (FWHM = 36 \kms) \HI\ profile.  \\
% 
% \noindent
% {\bf NGC 7293:} % was observed with E-W throw amplitudes up to 24$'$.
% As the maximum is reached between $C4EW$ and $C6EW$ we estimate its diameter as 44$'$. \\
%  
\noindent
{\bf RAFGL 865:}  We adopted the distance of 1580 pc used by \citet{Me06}
as it has no parallax listed in the Gaia EDR3. \\
\noindent
{\bf RAFGL 3068:} Classified as an upper limit, whereas it was previously 
listed as a detection in \citet{Gerard2006}. We adopted the distance of 980 pc used 
by \citet{An21} as it has no parallax listed in the Gaia EDR3. It is an AGB star transitioning 
towards the proto-Planetary Nebula phase \citep{Kim2017}. \\
\noindent % possible
{\bf RAFGL 3099}: Listed previously as an NRT upper limit by \citet{Gerard2006} and as a 
VLA non-detection by \citet{Matthews2013}, both with a $\sim$2 mJy limit. 
After 100 hours of NRT observations we classified this very low signal-to-noise 
\HI\ signal as a possible detection, with an \HI\ peak flux density of $\sim$7 mJy at 
the upper edge of the CO line FWHM. The average flux density we measured within the
Gaussian profile fit is 2$\pm$2 mJy, i.e., the previous NRT and VLA upper limit levels.
We adopted the distance of 2100 pc from \citet{Kn85}
as it has no parallax listed in the Gaia EDR3. \\
\noindent
{\bf W And:} Classified as confused, whereas it was previously classified as 
a detection in \citet{Gerard2011}. \\
\noindent
{\bf 56 Aql$^n$}: Classified as a possible detection due to the low signal-to-noise ratio
of the \HI\ profile, and the 6.4 \kms\ difference between the \HI\ and optical velocities; 
the latter has a published uncertainty of 2 \kms\ in the GCRV \citep{Wilson53}.  \\ %  -29.2 / -35.6$\pm$2.0
{\bf GY Aql:} Classified as a possible detection, as it lies on the edge of 
strong Galactic \HI\ lines, which obscure the lower half of the CO profile width. The NRT \HI\ profile
parameters are bound to be underestimates, and no diameter could be estimated. 
Although we could determine a total \HI\ profile, it is quite different from the peak profile. \\
\noindent
{\bf EP Aqr:} Classified as a possible detection, whereas it was previously 
listed as an NRT detection by \citet{Lebertre2004} and as a VLA detection by 
\citet{Matthews2007}, where arcminute-scale \HI\ structures were observed which 
were offset from the star's position by up to 10$'$. 
There could be two components in the NRT profile, but we have fitted only
a single Gaussian. \\
\noindent
{\bf R Aqr$^d$:} Classified as a clear detection.
%, which we fitted with two Gaussians. 
\citet{Matthews2007} did not detect an \HI\ source at the position of the continuum source, 
but do not exclude possible spatially extended \HI\ emission. 
%
% \noindent
% {\bf V Aqr:} As the spectra converge for $C3EW$ we estimate its diameter as 20$'$. \\
%  
% \noindent
% {\bf RV Boo:} As the spectra converge for $C3EW$ we estimate its diameter as 20$'$. \\
%  \noindent
{\bf RW Boo:} This is the only clear NRT detection with a significant difference
of 15 \kms\ between our \HI\ velocity of 10.1 \kms\ and the CO velocity
of -5.0 \kms\ published by \citet{Di19}. 
However, the CO line profile shows that their published central velocity of -5.0 \kms\
lies on the lower edge of the profile, whereas we estimate the central velocity 
as 6 \kms. The published CO expansion velocity of 34.6 \kms\ is in fact the FWHM 
of the line profile, so twice as large as our definition of \Vexp, whereas
the FWHM we estimate from their spectrum is only 18 \kms.
We therefore listed a literature velocity ($V$ lit.) of 6 \kms\ and a \Vexp\ of 9 km/s in Table 
\ref{table:cleardetsbasics}.
\\
%  
% \noindent
% {\bf R Cas:} As the spectra converge for $C2EW$ we estimate its diameter as 12$'$. \\
% The optical velocity of -15.0$\pm$0.7 \kms\ from \citet{Fa05} is significantly 
% different from our \HI\ velocity of 24.8 \kms\ and the CO line velocities of 25 \kms\
% from \citet{De10,Kn98,Ne98,Ny92}.
%
% \noindent
% {\bf AX Cep:} As the spectra converge at $C1EW$ we conclude that the source is unresolved. \\
%  
\noindent
{\bf $\mu$ Cep$^n$:} We adopted the distance of 641 pc determined by \citet{Mo19} 
as it has no parallax listed in the Gaia EDR3. 
Our \HI\ detection of this red supergiant, which is centered on 38 \kms, 
only shows the high-velocity half of the CO line profile; the lower part is confused by strong Galactic 
\HI\ signals. The single-dish CO(4-3) line profile by \citet{De10} is centered on 
$\sim$33 \kms, with a large $FWHM$ of $\sim$65 \kms. The NOEMA interferometric CO(2-1) line 
imaging by \citet{Mo19} shows a clumpy CO distribution, with emission in the range 
-8 to 58 \kms. A high-resolution optical spectrum obtained by \citet{Mo19} shows 
that the radial LSR velocity of the central star is 32.7 \kms. 
% TBD: add results of new VLA observations?
\\
\noindent
{\bf T Cep:} \citet{Ra09} noted that all six of their CO line profiles were
asymmetric, with stronger emission on the red side. They saw this as an indication
of an asymmetric outflow, as the lines are all optically thin. \\
\noindent
% {\bf $o$ Cet:} % Was observed with E-W throw amplitudes up to 24$'$.
% As the spectra converge at $C3EW$ we estimate its diameter as 20$'$.
% As peak profile we used the average of the $C3EW$, $C4EW$ and $C6EW$ spectra. \\
%
\noindent
{\bf T Cet:} % Was observed with E-W throw amplitudes up to 16$'$.
% As the spectra converge at $C3EW$ we estimate its diameter as 12$'$ ??.
% TBD:  should be 20' ??
The total profile shows a strong peak at 19 \kms\ which is not present in the 
peak profile; it remains to be seen if it is associated with the CSE. \\
\noindent  % possible
{\bf S CMi}: Listed previously as a detection in \citet{Gerard2006}, whereas we have classified 
it as a possible detection, unresolved by the NRT, as the new observations made pointing $11'$
north and south of the star indicate there may be confusion by ISM \HI\ emission along the line-of-sight. 
\\
\noindent  % detected
{\bf U CMi}: Listed previously as detection in \citet{Gerard2006} and as possible detection here; 
the case for possible confusion needs to be investigated further. 
We could however measure a total \HI\ profile for the detection.  \\
% Eric: POSSIBLE DETECTION 0.04 jy Vlsr 40.3 km.s HPW 11.7 km/s
% PUBLISHED IN GERARD&LE BERTRE, 2006 AJ SUSPICIOUS  C= 1/2N + 1/2S  COULD BE CONFUSION  
% ALSO THE DIFFERENT OFFSET DO NOT CONVERGE MORE ANALYSIS IS REQUIRED
% 
\noindent
% TBD: check throw amplitudes used
{\bf RS Cnc:} As the $C1EW$ and $C2EW$ profiles are rather similar we used the averaged spectrum 
for further analysis, and conclude that the source diameter is 4$'$. 
VLA \HI\ observations \citet{Matthews2007} show a compact (80$''$ diameter) source centred on
the star plus a 6$'$ long filament. \\
\noindent  
{\bf W Cnc}: Classified as a possible detection due to the low signal-to-noise ratio
of the \HI\ profile. \\
\noindent  
{\bf T Com}: Classified as a possible detection due to confusion by Galactic
\HI\ at the lower-velocity end of the profile.  \\
% We adopted a literature \VLSR\ value of 18 \kms\ for this source,
% being the mean value for the two strong 1612 OH maser lines observed by
% \citet{Si89}. Their 1667 MHz OH line data, and the data from \citet{Li89} show
% only the high-velocity line at $\sim$23 \kms.
% Its reported optical velocities are 20.4$\pm$2.6 \kms\ \citep{Fa05} and 22.8 \kms\ \citep{Sm65},
% and its SiO and H$_2$0 maser velocities are higher still, 
% 27.0 \kms\ \citet{Be90} and 28.0-30.8 \kms.
% from Tables: 20.4$\pm$2.6  Fa05 opt ; 22.8 Sm65 opt ; 28.3 30.8 SiO 28.0 H20 Ki13 ; 
%  23.1 and 12.9  Si89 OH ; 29.5 Ki13 SiO ; 27.0 Be90 SiO 
% 
\noindent
{\bf Y CVn:} It has the highest measured signal-to-noise ratio among our \HI\ CSE detections.
We have listed its diameter as 16$'$ as the E-W profiles first converge at a throw amplitude 
of 8$'$, but there is an indication of a weak extension out to 12$'$ radius, suggesting 
a faint envelope between 6$'$ and 10$'$ radius.
The profile is clearly asymmetric with an excess on its low-velocity side that is present 
down to 6.5 \kms\ in both the peak and total profiles. \\
\noindent
{\bf Z Cyg}:  Previously listed as an upper limit in \citet{Gerard2006}, we classified it 
as a possible detection as the low signal-to-noise \HI\ signal is centred exactly on the 
CO line detection by \citet{Jo98}. The fitted \HI\ line width (FWHM = 17: \kms) is
considerably broader than that of the CO line (9 \kms), but the Gaussian fit is not very
precise due to the weakness of the line signal. \\
\noindent
{\bf AH Dra}: Classified as a possible detection as it has a quite low signal-to-noise ratio. 
\\
\noindent
{\bf S Dra}: Classified as a possible detection as it has a low signal-to-noise ratio,
and as its lower-velocity edge may be contaminated by Galactic \HI. 
\\
\noindent
{\bf UX Dra}: Classified as a possible detection as its lower-velocity half may be contaminated 
by Galactic \HI. 
\\
\noindent
{\bf RY Dra:} The source is very extended, with an estimated diameter of 44$'$,
in agreement with the far-infrared IRAS source size measured at 60 and 100 $\mu$m
wavelength by \citet{Yo93}. \\
%  
% \noindent
% {\bf R Gem:} As the $C1EW$ and $C2EW$ profiles are similar we used the averaged spectrum 
% for further analysis, and conclude that the source is unresolved. \\
% 
\noindent
{\bf $\alpha$1 Her$^d$:} Classified as an upper limit, whereas it was previously 
classified as a detection in \citet{Gerard2006}. 
We adopted the distance of 110 pc based on its HIPPARCOS parallax \citep{Va07} 
as it has no parallax listed in the Gaia EDR3. \\
\noindent
{\bf OP Her:} Classified as an upper limit, whereas it was previously 
classified as a detection in \citet{Gerard2011}. \\
\noindent
{\bf X Her:} There is a hint of two Gaussian components in the spectrum,
but we made a fit using only one Gaussian. \\
% As the spectra converge at $C3EW$ we estimate its diameter as 20$'$. 
% 
% {\bf AK Hya:} % Was observed with E-W throw amplitudes up to 24$'$.
% As the spectra converge at $C3EW$ we estimate its diameter as 20$'$.
% As peak profile we used the average of the $C3EW$ and $C4EW$ spectra.
%
% \noindent
% {\bf R Hya:} As the spectra converge at $C3EW$ we estimate its diameter as 20$'$. \\
%   
% \noindent
% {\bf RT Hya:} % Was observed with E-W throw amplitudes up to 16$'$.
% As the spectra converge at $C3EW$ we estimate its diameter as 20$'$.
% As peak profile we used the average of the $C3EW$ and $C4EW$ spectra. \\
% 
% \noindent
% {\bf RV Hya:} As the $C1EW$ and $C2EW$ profiles are similar we used the averaged spectrum 
% for further analysis, and conclude that the source is unresolved. \\
% 
\noindent
% TBD: determine maximum throw amplitude used
{\bf U Hya:} % Was observed with E-W throw amplitudes up to XX$'$. 
% As the spectra converge at $C1.5EW$ ?? we estimate its diameter as 8$'$.
Although it is an SRb carbon star, like Y CVn, their mass loss histories appear 
to be very different, in the sense that U Hya has a 50 times smaller \HI\ mass 
and that the 2$'$ radius detached ring seen in the far-infrared maps of \citet{Cox12a, Cox12b} 
is most likely due to a short episode of enhanced mass-loss rate \citet{Waters94}. \\
\noindent
{\bf W Hya:} Classified as confused, whereas it was previously 
classified as a detection in \citet{Gerard2006}. 
No \HI\ was detected by \citet{Ha93} in VLA imaging observations with an rms noise level 
of 0.0055 Jy/beam. \\
%
% {\bf R Lep:} TBD: Eric's comment - VLA sees no central \HI\ ; from unpublished VLA 2014 observations \\
% Eric: CONFUSED PROFILES DO NOT CONVERGE   NO SAFE NRT UPPER LIMIT
% THE VLA SEES NO CENTRAL HI SOURCE ....WHERE IS THE HI???
% Eric's comment was based on a mail from Lynn on the 2014 VLA observations:
% Lynn : no HI at centre, perhaps S; confusion high
% 
% \noindent
% {\bf RX Lep:} % % was observed with E-W throw amplitudes up to 16$'$.
% As the spectra converge at $C3EW$ we estimate its diameter as 20$'$.
% As peak profile we used the average of the $C3EW$ and $C4EW$ spectra.
% The \VLSR\ from \citet{Ke99}, 10 \kms, as measured for 2 different CO lines,
% is significantly different from both our \HI\ velocity, 29.2 \kms, and the published optical 
% velocity, 27.7 \kms\ \citep{Gontcharov06}. It is intriguing to note that the central velocity 
% of the published CO spectra is 29 \kms\ on a heliocentric velocity scale, corresponding 
% exactly to the \HI\ and optical LSR velocities. This might indicate a mix-up between LSR and 
% heliocentric velocities in the CO spectra, but in the two other cases (AK Hya, SV Peg) no 
% difference was noted between our \HI\ spectra and the CO detections from \citet{Ke99}. 
%
\noindent
{\bf AF Leo :} As the maximum is almost reached at $C2EW$ but the $C2EW$ spectrum is a bit 
higher, we estimate its diameter as 16$'$.
As peak profile we used the average of the $C3EW$ and $C4EW$ spectra. \\
\noindent
{\bf $\delta$02 Lyr$^d$:} Classified as confused, whereas it was previously 
classified as detected in \citet{Gerard2006}. \\
\noindent
{\bf $\alpha$ Ori$^n$:} We now consider the NRT \HI\ observation of this red supergiant to be confused, 
whereas it was previously classified as a detection based on a VLA detection by \citet{Bo87} 
and on detections with both NRT and VLA by \citet{Lebertre2012} which shows a detached \HI\ shell 
of about 4$'$ diameter. \\
% Eric: DETECTED AND PUBLISHED   (SEE TH. LE BERTRE ET AL.2012)  BUT STRONG CONFUSION  
% Thibaut: alpha Ori was detected at the VLA (Bowers & Knapp, 1987; Le Bertre et al. 2012).
%
% \noindent
% {\bf R Peg:} % was observed with E-W throw amplitudes up to XX$'$.
% As the spectra converge at $C3EW$ we estimate its diameter as 20$'$.  \\
%  
\noindent  % possible
{\bf SV Peg}:  Classified as a possible detection as the profiles do not converge well. 
Furthermore, the lower-velocity edge of the CO profile width is contaminated by strong Galactic \HI.  \\
% 11.6 / 5.7, 21.4
%  
\noindent  % possible
{\bf V Peg:} We adopted the distance of 1380 pc used by \citet{Be96} 
as the Gaia EDR3 parallax is very uncertain, 0.022$\pm$0.092 mas \\
% IRAS 21585
\noindent
{\bf $\rho$ Per$^d$:} Classified as an upper limit, whereas it was previously 
classified as a detection in \citet{Gerard2006}. \\
\noindent
{\bf SV Psc:} As the $C1EW$ to $C4EW$ spectra are similar we used the averaged spectrum 
for further analysis, and conclude that the source is unresolved.
\citet{Wi03} noted that both their CO line profiles show two components
with different outflow velocities, of 2 and 11 \kms, respectively. The mass loss rate of
1.0 10$^{-6}$ \Msunpyear\ we listed in Table \ref{table:cleardetsbasics} is for the 
broader component. \\
\noindent
{\bf TX Psc:} Imaging observations with the VLA by \citet{Matthews2013} 
identified Galactic \HI\ confusion in the 7-8 \kms\ \VLSR\ range, a line signal 
which had contaminated the NRT spectra of \citet{Gerard2006}. 
Separating genuine CSE emission from the confusing emission feature is difficult 
in the NRT data. When using throw amplitudes up to 4$'$ ($C1EW$), we observe 
an \HI\ line centred on 12 \kms, in agreement with the VLA detection of the CSE. 
% However we miss part of the flux detected at the VLA probably because the 
% CSE extends beyond a 4$'$ radius.
As the NRT spectra converge between $C1.5EW$ and $C2.5EW$ we estimate its 
diameter as 8$'$, which is larger than the \am{6}{2} measured at the VLA, 
but this may be due to the confusing line signal which increases for 
larger throw amplitudes.  \\
% The optical velocity of -9.2$\pm$0.9 \kms\ from \citet{Gontcharov06} is significantly 
% different from our \HI\ velocity of 11.5 \kms, which is consistent with the mean CO line velocity of 12.6 \kms\
% from \citet{He89b}, \citet{He90}, \citet{Ny92} and \citet{Ol93}. 
%
\noindent  % possible
{\bf WX Psc:} We adopted the distance of 740 pc used by \citet{va90} 
as it has no parallax listed in the Gaia EDR3. \\
\noindent
{\bf Y Scl}: We have classified this clear \HI\ line as a possible detection due to 
the 4 \kms\ difference between the \HI\ and CO line velocities, which is significant 
when compared to the CO line FWHM of 14.6 \kms\ \citep{Ke99}. 
% HI 24.7 10.3  ; CO 28.6  7.3 Ke99 
\\
\noindent
{\bf $\kappa$ Ser$^n$}: Classified as a possible detection as the upper edge of
the \HI\ profile is contaminated by strong Galactic \HI. 
\\
\noindent
{\bf S Ser}: Classified as a possible detection as the lower-velocity half of
the \HI\ profile is contaminated by Galactic \HI, and as it is much narrower 
(FWHM = 2.6 \kms) than the average 20 \kms\ FWHM line width of the CO line detections 
of our sample (no CO line detected has been published of the source). 
\\
\noindent  % possible
{\bf WX Ser:} Classified as a possible detection on the edge of strong Galactic \HI\ lines,
which obscure the lower-velocity 60\% of the CO profile width. The NRT \HI\ profile
parameters are bound to be underestimates, and no diameter could be estimated. \\
\noindent  % possible
{\bf AQ Sgr:} Classified as a possible detection, and estimate its diameter 
as 4$'$. NRT observations show the presence of a confusing source to the north which appears around 18 \kms, 
i.e., about 2.5 \kms\ lower than the profile peak velocity, and which may cause the 
asymmetry in the profile shown here. However, the confusion appears negligible 
out to E-W throw amplitudes of 6$'$ and is only clearly present at 10$'$. \\
\noindent
{\bf V1942 Sgr:} High signal-to-noise source that was studied previously with the
NRT and VLA \citet{Matthews2013}. It was observed with throw amplitudes from 
1.3$'$ to 48$'$. The spectra start to converge at $C1.5EW$ (source diameter of 8$'$), 
but there appears to be \HI\ out to a diameter of 28$'$. 
We do not find evidence in our NRT spectra for the previously reported \citep{Libert2010a}
\HI\ emission pedestal between $-$29 and $-$37 \kms\ at the $\sim$6 mJy level,
and neither did \citet{Matthews2013} in their VLA observations.
The non-detection of the broad component is not surprising given the rms
noise levels in the VLA data and our NRT spectra (3.6 mJy for our peak spectrum 
and 27 mJy for our total spectrum).
\\
%  
% \noindent
% {\bf V1943 Sgr:} As the spectra converge at $C3EW$ we estimate its diameter as 20$'$. \\
%
\noindent
{\bf IK Tau:} Not detected by either the VLA \citet{Matthews2013} with an
rms of 1.4 mJy/beam nor at the NRT, with an rms of 3.7 mJy.  \\
% Eric: UPPER LIMIT 0.02 Jy saved in gt038777  CLEAN ON RED SIDE  Vlsr=33 km/s
% VLA: UPPER LIMIT O.005 Jy near 33 km/s (width 18 km/s see Matthews et al. 2013,
%
\noindent
{\bf Y UMa:}
\citet{Matthews2013} fitted two components to their VLA \HI\ profile.
The first has a considerably smaller width and higher peak flux density. 
The fitted values for the components are, respectively: 
\VHI\ = 16.4, 17.2 \kms; $FWHM$ = 3.2, 9.2 \kms; \Speak\ = 0.10, 0.03 mJy. 
We cannot discern the broader component in our NRT profile, which is not surprising
as its VLA peak value is only 2.3 times the 0.013 Jy rms noise level of the NRT spectrum.
We fitted only a single Gaussian to our NRT profile, with
\VHI\ = 16.8 \kms; $FWHM$ = 5.4 \kms, and \Speak\ = 0.10 Jy.
These values agree well with the overall parameters which we determined from the VLA profile:
\VHI\ = 16.6 \kms; $FWHM$ = 4.9 \kms, and \Speak\ = 0.13 Jy. \\
\noindent
{\bf V UMi}: Classified as a possible detection as the \HI\ profile has a low
signal-to-noise ratio, and as it is considerable narrower (FWHM = 4.9 \kms) than the 
average 20 \kms\ FWHM line width of the CO line detections of our sample
(no CO line detected has been published of the source). 
\\
%
% \noindent
% {\bf AY Vir:} As the spectra converge at $C1EW$ we conclude that the source is unresolved. \\
%   
% \noindent
% {\bf BK Vir:} As the spectra converge at $C2EW$ we estimate its diameter as 10$'$. \\
%
% \noindent
% {\bf RT Vir:} As the spectra converge at $C3EW$ we estimate its diameter as 20$'$. \\
%  
\noindent  % possible
{\bf RW Vir:} Classified as a possible detection as the $C1EW$ to $C4EW$ spectra 
do not converge well. \\
%  
% \noindent
% {\bf VY UMa:} As the spectra converge at $C1.5EW$ ?? we estimate its diameter as 8$'$. \\
%  
% \noindent
% {\bf Y UMa:} % was observed with E-W throw amplitudes up to 24$'$. 
% As the spectra converge at $C2EW$, we estimate its diameter as 12$'$. \\
% 

% \noindent
% {\bf :}  \\

\end{appendix}

% \newpage 

\onecolumn

\begin{landscape}
\begin{longtable}{llrlllrrrrrllrl} % Tab 1 
\caption{\label{table:cleardetsbasics} Clear NRT \HI\ detections: basic data} \\ 
\hline \hline
Name & RA & Dec & Type & Spec. & ref. & \Teff\ & ref. &$d$ & $V$ lit. & \Vexp\ & ref. & line & $\dot{\it M}$ & ref. \\
    & \multicolumn{2}{c}{(J2000.0)} &  &  & & (K) & & (pc) & {\footnotesize (km/s)} & {\footnotesize (km/s)} &   &  & {\footnotesize (\Msun/yr)} &  \\
\hline
\endfirsthead
% \caption{Clear NRT \HI\ detections: basic data -- continued.}\\
% \hline\hline
% Name & RA & Dec & Type & Spec. & ref. & $d$ & $V$ lit. & \Vexp\ & ref. & line & $\dot{\it M}$ & ref. \\
%     & \multicolumn{2}{c}{(J2000.0)} &  &  &  & (pc) & {\footnotesize (km/s)} & {\footnotesize (km/s)} &   &  & {\footnotesize (\Msun/yr)} &  \\
% \hline
% \endhead
\hline
\endfoot
% name       &   RA     &     Dec   & type & spec   & ref        & Teff & ref &     d  &   Vlit  &  Vexp  & ref &  line&   Mdot        &   ref     \\ 
SV And       & 00 04 20 &  40 06 36 & Mira & M5-7e       & Ke74           & ...  &       & 1106 & -88.9 &  ... & Ch96 & SiO & ...           &      \\ % -82.4$\pm$2.1 PCRV opt ; -88.6 Eric ; -88.4/-89.5 Ch89 SiO
T Cet$^{\star}$ & 00 21 46 & -20 03 29 & SRc & M5-6Se    & Ke54           & 2788 & De10  & 270 &  22.0 &  5.5 & Ra09 & CO  & 5.1 10$^{-8}$ & Ra09 \\  % 24.1$\pm$0.9 PCRV opt ; 20.9 Eric ; 22.0: Gr98 ; NO CO He90 ; Teff reference?
SV Psc$^{\star}$ & 01 46 35 &  19 05 05 & SRb  & M5   & {\tiny SIMBAD}        & ...  &       &  393 &   6.5 & 12.0 & Wi03 & CO  & 1.7 10$^{-7}$ & Wi03  \\ %  12.1 & Fe72 opt ; 7.T Eric ; IRAS 01438+1850 complex CO profile Wi03 
$o$ Cet      & 02 19 21 & -02 58 39 & Mira & {\footnotesize M5-9IIIe+DA} & Sk14 & 2550 & Th13 & 92 &  46.4 &  4.8 & Yo95 & CO  & 2.8 10$^{-8}$ & Wi03 \\ % 53.6$\pm$0.6 ; 46.5 0.0 (3-2) 46.4 0.0 (4-3) Yo95 CO ; 46.0 0.2 (3-2) Kn98 CO ; Ny92 ; 46.4 (3-2) Ke03 ; 46.4 De10 ; 46.3 Kn82 ; 45.4 Gr99 ; 47 Ma90 ; 47.2 ALMA CO imaging Khouri 2016 ; Teff Thirumalai A., Heyl J. S., 2012, MNRAS, Letters (arXiv: 1206.0044)
RX Lep       & 05 11 23 & -11 50 57 & SRb  & M6III       & Ho99           & 3339 & Du98 &  149 &  28.9 &  3.9 & Li08 & CO  & 1.6 10$^{-7}$ & Li08  \\ %  27.7$\pm$2.0 GCRV opt ; NO CO He90 ; -5.0 He89 ?? ; 28.9 Eric ; Teff (2) 3300 K Fluks, M.A., et al. 1994, A&AS 105, 311 ; 3339 K Dumm T., Schild H., 1998, New Astronomy 3, 137
R Gem        & 07 07 21 &  22 42 13 & Mira & S3.5-6.5/6e & Ke80           & ... &  &  848 & -59.0 & 11.0 & Kn98 & CO  & 6.3 10$^{-7}$ & Ra09 \\ %  -56.7$\pm$0.8 PCRV opt ; -59.0 0.6 (2-1) Kn98 ; -59.2 He90 ; -59.1 0.1 Sa95
HD 56126$^n$ & 07 16 10 &  09 59 48 & {\footnotesize post-AGB} & F0/5Ia  & Hr89 & ... &  & 2204 &  73.0 & 10.7 & Kn98 & CO  &  ...          &        \\
\end{longtable} 
\end{landscape} 
% \twocolumn
% ADD:  = \citet{},
\newpage

\begin{longtable}{lrrrrrrl} % Tab 2 
\caption{\label{table:cleardetsHI} Clear NRT \HI\ detections: \HI\ data} \\ 
\hline \hline
Name & \VHI\ & {\footnotesize $FWHM$} & \Speak\  & {\footnotesize diam.} & \FHI\ & \MHI\ & \HI\ ref.  \\
     & {\footnotesize (km/s)} & {\footnotesize (km/s)} & {\footnotesize (Jy)} & ($'$) & {\footnotesize (Jy km/s)} & (\Msun) &    \\
\hline
\endfirsthead
\caption{continued.}\\
\hline\hline
Name & \VHI\ & {\footnotesize $FWHM$} & \Speak\  & {\footnotesize diam.} & \FHI\ & \MHI\ & \HI\ ref.  \\
     & {\footnotesize (km/s)} & {\footnotesize (km/s)} & {\footnotesize (Jy)} & ($'$) & {\footnotesize (Jy km/s)} & (\Msun) &    \\
\hline
\endhead
\hline
\endfoot
% & & & & & & & \\
% HI diameter statistics: <4 7 ; 4  ; 10  ; 12  ; 44 
% name                 VHI G  FWHM G  Speak    diam   FHI     MHI      HI ref  
SV And$^P$           & -87.5 &  5.5 & 0.007 &   ... & 0.036 & 0.0113 &   \\  % OK P
T Cet$^{T,{\star}}$  &  19.2 &  7.7 & 0.167 &    12 & 0.889 & 0.0166 &   \\  % OK Tot
SV Psc$^{P,{\star}}$ &   7.2 &  5.3 & 0.010 &  $<$4 & 0.052 & 0.0021 &   \\  % OK P
$o$ Cet$^T$          &  44.9 &  6.1 & 0.090 &    20 & 0.453 & 0.0009 & {\footnotesize Bo88 Ge06 Ma08}\\  % OK Tot
RX Lep$^T$           &  29.2 &  2.9 & 0.420 &    20 & 1.205 & 0.0069 & {\footnotesize Li10a Ma13} \\  % OK Tot
R Gem$^P$            & -58.1 &  9.7 & 0.015 &  $<$4 & 0.122 & 0.0224 &   \\  % OK P
HD 56126$^{P,n}$     &  75.3 & 11.0 & 0.006 &  $<$4 & 0.035 & 0.0435 &   \\  % OK P 
RT Hya$^T$           &  30.4 &  7.1 & 0.229 &    20 & 1.970 & 0.0333 &   \\
\end{longtable} 
\newpage

\begin{landscape}
\begin{longtable}{llrlllrrrrrlll} % Tab 3
\caption{\label{table:possdetsbasics} Possible NRT \HI\ detections: basic data} \\ 
\hline \hline
Name & RA & Dec & Type & Spec. & ref. & \Teff\ & ref. &$d$ & $V$ lit. & \Vexp\ & $\dot{\it M}$ & line & ref. \\
    & \multicolumn{2}{c}{(J2000.0)} &  &  & & (K) & & (pc) & {\footnotesize (km/s)} & {\footnotesize (km/s)} & {\footnotesize (\Msun/yr)}  &  &  \\
\hline
\endfirsthead
\hline
\endfoot
% name            RA          Dec     type       spec         ref      Teff    ref    d     Vlit    Vexp   ref   line              ref  
S CMi$^{\star}$ & 07 32 43 &  08 19 05 & Mira  & M7-8e      & Ke74   &  ... &      &  418 &   51.7 &  4.7 & 6.2 10$^{-8}$ & CO  & Yo95  \\ %   53.6$\pm$4.8 PCRV opt ; 51.7 0.2 (3-2) Yo95 ; 51.5 1.4 (3-2) Kn98
U CMi$^{\star}$ & 07 41 20 &  08 22 49 & Mira  & M4e &{\tiny SIMBAD} &  ... &      & 1489 &   41.9 &  ... & ...           & opt & PCRV \\ % confused? ; 41.9$\pm$4.7 PCRV opt ; NO SiO Patel 1992 ; NO OH Lewis 1990 ; no masers Be90 ; 40.3 Eric
W Cnc        & 09 09 53 &  25 14 54 & Mira  & M6.5-9e       & Ke74   &  ... &      &  527 &   36.8 &  4.2 & 1.7 10$^{-8}$ & CO & Yo95  \\ % probable ; 42.6$\pm$4.7 PCRV opt ; 36.8 0.2 (3-2) Yo95
RW Vir$^{\star}$ & 12 07 15 & -06 45 56 & Lb    & M5II      & Ho99   &  ... &      &  455 &   21.0 &  8.0 & 4.2 10$^{-8}$ & CO & Ke98  \\ %
T Com        & 12 58 39 &  23 08 21 & Mira  & M8/9III:e     & Pe72   &  ... &      & 1810 &   18.0 &  5.5 &  ...          & OH & Si89  \\ % Maser : (OH/IR) ;20.4$\pm$2.6  Fa05 opt ; 22.8 Sm65 opt ; 28.3 30.8 SiO 28.0 H20 Ki13 ; 23.1 and 12.9  Si89 OH ; 29.5 Ki13 SiO ; 27.0 Be90 SiO 
V UMi        & 13 38 41 &  74 18 36 & SRb   & M5IIIab:      & Ya67   &  ... &      &  541 & -153.5 &  ... & ... & opt & {\footnotesize GCRV} \\ % -153.5$\pm$2.0  GCRV opt ; no masers Be90 ; -151.Eric
S Ser        & 15 21 39 &  14 18 53 & Mira  & M5e           & Pe15   & 2257 & va02 & 1303 &   21.1 &  ...        & ...         & SiO & Mc13 \\ % 27.3$\pm$4.7 PCRV opt ; 21.1/21.3 SiO Mc13 ; 19.7/19.0 Ch96 SiO ; 21.7 Eric
WX Ser$^{\star}$ & 15 27 47 &  19 33 52 & Mira  & M8.5      & So78   &  ... &      &  709 &   1.0  & 11.5 & 5.1 10$^{-7}$ & CO & Kn86  \\
\end{longtable} 
\end{landscape}

\begin{longtable}{lrrrrrrl} % Tab 4 
\caption{\label{table:possdetsHI} Possible NRT \HI\ detections: \HI\ data} \\ 
\hline \hline
Name & \VHI\ & {\footnotesize $FWHM$} & \Speak\  & {\footnotesize diam.} & \FHI\ & \MHI\ & \HI\ ref.  \\
     & {\footnotesize (km/s)} & {\footnotesize (km/s)} & {\footnotesize (Jy)} & ($'$) & {\footnotesize (Jy km/s)} & (\Msun) &    \\
\hline
\endfirsthead
\caption{continued.}\\
\hline\hline
Name & \VHI\ & {\footnotesize $FWHM$} & \Speak\  & {\footnotesize diam.} & \FHI\ & \MHI\ & \HI\ ref.  \\
     & {\footnotesize (km/s)} & {\footnotesize (km/s)} & {\footnotesize (Jy)} & ($'$) & {\footnotesize (Jy km/s)} & (\Msun) &    \\
\hline
\endhead
\hline
\endfoot
% name                     VHI G  FWHM G  Speak    diam   FHI     MHI      HI ref 
S CMi$^{P,{\star}}$     &   51.2 &  5.8  & 0.035 & $<$4 & 0.150 & 0.0067 & {\footnotesize Ge06} \\  % OK P
U CMi$^{T,{\star}}$     &   41.1 &  7.4  & 0.111 &   12 & 0.816 & 0.4631 & {\footnotesize Ge06} \\  % OK P
W Cnc$^{P,{\star}}$     &   33.3 & 17.1  & 0.011 &  ... & 0.136 & 0.0097 &  \\    % OK P           
RW Vir$^{P,{\star}}$    &   23.3 & 17.4  & 0.016 &  ... & 0.190 & 0.0101 &  \\   % OK P             
T Com$^{P,{\star}}$     &   16.0 & 15.5  & 0.023 & $<$4 & 0.342 & 0.2869 &  \\   % OK P    
% RT Vir: now classified as clear detection
V UMi$^{P,{\star}}$     & -151.2 &  4.9 & 0.013 &  ... & 0.038 & 0.0028 &  \\   % OK P
% RW Boo: now classified as clear detection
S Ser$^{P,{\star}}$     &   21.7 &  2.6  & 0.011 &  ... & 0.026 & 0.0112 &  \\   % OK P
WX Ser$^{P,{\star}}$    &   5.6: &$>$4.8 & 0.026 &  ... & $>$0.0821 & $>$0.0106 &  \\
\end{longtable} 

\newpage

\begin{landscape}
\begin{longtable}{lrrlllllrrrllllll} % Tab 5
\caption{\label{table:upperlimits} Upper limits to NRT \HI\ lines} \\ 
\hline \hline
Name & RA \,\,\,\,\,\,\, & Dec\,\,\,\,\,\,\, & Type & Spec. & ref. & \Teff\ & ref. & $d$ & $V$ lit. & \Vexp\ & $\dot{\it M}$ & ref. & line & \Speak\ & notes & \HI\ ref.\\
     & \multicolumn{2}{c}{(J2000.0)} &  &  &  & (K) &  &  (pc) & {\footnotesize (km/s)}  & {\footnotesize (km/s)} & {\footnotesize (\Msun/yr)} & & & {\footnotesize (Jy)} &  &  \\
\hline
\endfirsthead
\caption{Upper limits to NRT \HI\ lines -- continued.}\\
\hline\hline
% name     &           RA        &        DEC    &    var type      &    spec  &    ref   & Teff & ref & d &     Vlit   &   Vexp    & Mdot &   ref   &    line      &  Speak      &  HI refs
Name & RA \,\,\,\,\,\,\, & Dec\,\,\,\,\,\,\, & Type & Spec. & ref. & \Teff\ & ref. & $d$ & $V$ lit. & \Vexp\ & $\dot{\it M}$ & ref. & line & \Speak\ & notes & \HI\ ref.\\
% name     &           RA        &        DEC    &      type      &    spec  &    ref   & Teff & ref & d &     Vlit   &   Vexp    & Mdot &   ref   &    line      &  Speak      &  HI refs
     & \multicolumn{2}{c}{(J2000.0)} &  &  &  & (K) &  &  (pc) & {\footnotesize (km/s)}  & {\footnotesize (km/s)} & {\footnotesize (\Msun/yr)} & & & {\footnotesize (Jy)} &  &  \\
\hline
\endhead
\hline
\endfoot
%
% name                 RA                DEC        var type       spec      ref      Teff  ref  d      Vlit         Vexp      Mdot      ref       line         Speak        HI refs                         notes    name     par mas     
Z Peg           &   00 00 06.6   &    25 53 11   &   Mira       &  M7-8e  &  Ke74  &   ...  &   & 658 &   -29.1   &    ...   &   ...  &   Ch96   &   SiO      &   0.05    &   {\footnotesize \it old data}    & \\ %old data  -26.5$\pm$4.6 PCRV opt ; Be90 no masers ; -27.1 SiO Mc13 ; -29.2/-29.1 Ch96 SiO Z Peg           1.5204     
KU And          &   00 06 52.8   &    43 05 02   &   Mira       &  M10  &  Bi80  &   ...  &        & 536 &   -20.5   &   19.6   &  3.4 10$^{-6}$  &   Kn98   &   CO       &   0.05    &   {\footnotesize \it old data}     & \\ % Be90 no masers ; -21.4   Fe00  opt ; -20.5 1.8 (1-0) Kn98 ; -20.0 Ny92 CO  ; -20.0 Ne98 ; -21 Ma90KU And               
S Scl           &   00 15 22.3   &   -32 02 43   &   Mira       &  M7/8IIIe  &  Ho82  &   ...  &  & 496 &    35.0   &    ...   &   ...  &   Be90   &   H$_2$O   &   0.05    &   {\footnotesize \it old data}  & \\ % old data  28.9$\pm$4.5 PCRV opt ; 3.3/6.9 SiO Mc13 ; 8.6/8.3 Ch96 SiO ; NO CO Gr99 S Scl           2.0156     
VX And          &   00 19 54.0   &    44 42 34   &   SRa        &  C-J4.5  &  Ba96  & 2455 &  Be01  & 634 &    14.4   &   14.0   &  6.0 10$^{-8}$  &   Ol93   &   CO       &   0.01    &   {\footnotesize red side}  & \\ % Teff 2455 Be01 ; 15.1$\pm$2.0 GCRV opt ; 14.4 Eric OK ; NO CO Ol87 VX And               
NGC 246$^n$     &   00 47 03.3   &   -11 52 19   &  (PN)        &  PG1159  &  Me88  &   ...  &  & 556 &   -51.1   &    ...   &   ...  &   Sc83   &   opt      &   0.005   &                   & \\
\end{longtable}

% \vspace{1cm}
\newpage

\begin{longtable}{llrlllrllll} % Tab 6 
\caption{\label{table:confused} Confused NRT \HI\ spectra} \\ 
\hline \hline
Name & RA \,\,\,\,\,\,\, & Dec\,\,\,\,\,\,\, & Type & Spec. & ref. & $V$ lit. & ref. & line & notes & \HI\ ref.\\
     & \multicolumn{2}{c}{(J2000.0)} &  &  &  & {\footnotesize (km/s)} & & & & \\
\hline
\endfirsthead
\caption{Confused NRT \HI\ spectra -- continued.}\\
\hline\hline
Name & RA \,\,\,\,\,\,\, & Dec\,\,\,\,\,\,\, & Type & Spec. & ref. & $V$ lit. & ref. & line & notes & \HI\ ref.\\
     & \multicolumn{2}{c}{(J2000.0)} &  &  &  & {\footnotesize (km/s)} & & & & \\
\hline
\endhead
\hline
\endfoot
% name                 RA             DEC      type              spec      ref       VLSR lit  V ref  line   notes HI refs
%
WZ Cas          &  00 01 15.9  &   60 21 19  &  SRb            & C-N7III: & Ke93 &  -25.8  &  Ol93  &  CO      &  &     \\ % -25.3$\pm$2.0 GCRV opt ; NO CO Ol87 ; NO CO Wo91  
YY Psc          &  00 01 57.6  &  -06 00 51  &  Lb:            & M3III & Ke89 &  -12.7  &  PCRV  &  opt     &  &     \\ % -12.7$\pm$0.5 PCRV opt ; Be90 no masers  
T Cas           &  00 23 14.3  &   55 47 33  &  Mira           & M7-9e & Ke74 &   -5.7  &  Yo95  &  CO      &  &     \\ % -4.7$\pm$2.0 GCRV opt ; -9.3 0.2 (3-2) -5.7 0.6 (4-3) Yo95 ; -1.0 Be90 SiO ; -6 Ma90   
R And           &  00 24 01.9  &   38 34 37  &  Mira           & S5-7/4-5e & Ke80 &  -15.3  &  Kn98  &  CO      &  &     \\ % -1.5 PCRV opt ; -15.0 Be90 SiO ; -15.3 5.4 (3-2) Kn98 ; -16 Ma90  
AQ And          &  00 27 31.7  &   35 35 14  &  SRb            & S5-7/4-5e & Ba96 &   -9.7  &  {\footnotesize GCRV} &  opt     &      \\ % Be90 no masers   
TV Psc          &  00 28 02.9  &   17 53 35  &  SR             & M3III &{\tiny SIMBAD} &    6.5  &  Fa09  &  opt     &  &     \\ % 6.5$\pm$0.4 Fa09 opt ; Be90 no masers : NO CO He90  
W Cas           &  00 54 53.8  &   58 33 49  &  Mira           & C9,1e & De76 &  -36.7  &  Yo95  &  CO      &  &     \\ % V also in CS draft paper ; -34.2$\pm$2.7 PCRV ; Be90 no masers ; -36.7 0.3 (3-2) Yo95 
HV Cas          &  01 11 03.5  &   53 43 40  &  Mira           & C4,3e & Ha75 &  -18.9  &  Ol93  &  CO      &      \\
\end{longtable} 
\end{landscape}

\end{document}